\documentclass[12pt]{article}

\usepackage[margin=1in]{geometry}

\usepackage{setspace}

\usepackage{graphicx}

\usepackage{enumitem}

\usepackage{microtype}
\usepackage{subfigure}
\usepackage{booktabs}
\usepackage{comment}
\usepackage{wrapfig}
\usepackage{arydshln}
\usepackage{enumitem}
\usepackage{multirow}
\usepackage{url}
\usepackage{natbib}
\usepackage{authblk}

\RequirePackage[colorlinks,citecolor=blue,linkcolor=blue,urlcolor=blue,pagebackref]{hyperref}

\usepackage{amsmath}
\usepackage{amssymb}
\usepackage{mathtools}
\usepackage{amsfonts}
\usepackage{bm}
\usepackage{bbm}

\usepackage{algorithm}
\usepackage{algorithmic}

\def\eqref#1{equation~\ref{#1}}

\def\1{\bm{1}}

\def\rmd{{\mathrm{d}}}

\def\bmone{{\bm{1}}}
\def\bmmu{{\bm{\mu}}}
\def\bmnu{{\bm{\nu}}}

\def\bmP{{\bm{P}}}

\def\bmlambda{{\bm{\lambda}}}
\def\bmmu{{\bm{\mu}}}
\def\bmnu{{\bm{\nu}}}

\DeclareMathAlphabet{\mathsfit}{\encodingdefault}{\sfdefault}{m}{sl}
\SetMathAlphabet{\mathsfit}{bold}{\encodingdefault}{\sfdefault}{bx}{n}

\def\calA{{\mathcal{A}}}
\def\calB{{\mathcal{B}}}

\def\calE{{\mathcal{E}}}
\def\calF{{\mathcal{F}}}
\def\calG{{\mathcal{G}}}

\def\calL{{\mathcal{L}}}
\def\calM{{\mathcal{M}}}
\def\calN{{\mathcal{N}}}

\def\calP{{\mathcal{P}}}

\def\calR{{\mathcal{R}}}
\def\calS{{\mathcal{S}}}

\def\calU{{\mathcal{U}}}
\def\calV{{\mathcal{V}}}

\def\calY{{\mathcal{Y}}}

\def\bbE{{\mathbb{E}}}

\def\bbN{{\mathbb{N}}}

\def\bbP{{\mathbb{P}}}
\def\bbQ{{\mathbb{Q}}}
\def\bbR{{\mathbb{R}}}

\def\bbZ{{\mathbb{Z}}}

\newcommand{\Var}{\mathrm{Var}}

\DeclareMathOperator*{\argmax}{arg\,max}
\DeclareMathOperator*{\argmin}{arg\,min}

\newcommand{\p}[1]{\left(#1\right)}
\newcommand{\sqb}[1]{\left[#1\right]}
\newcommand{\cb}[1]{\left\{#1\right\}}

\newcommand{\bigp}[1]{\big(#1\big)}
\newcommand{\bigsqb}[1]{\big[#1\big]}
\newcommand{\bigcb}[1]{\big\{#1\big\}}

\newcommand{\Bigp}[1]{\Big(#1\Big)}
\newcommand{\Bigsqb}[1]{\Big[#1\Big]}
\newcommand{\Bigcb}[1]{\Big\{#1\Big\}}

\newcommand{\abs}[1]{\left|#1\right|}

\newcommand{\norm}[1]{\left\|#1\right\|}
\newcommand{\bignorm}[1]{\big\|#1\big\|}

\def\Regret{{\mathrm{Regret}}}

\usepackage{amsthm}

\theoremstyle{plain}

\newtheorem{theorem}{Theorem}[section]
\newtheorem{lemma}[theorem]{Lemma}

\newtheorem{corollary}[theorem]{Corollary}
\newtheorem{proposition}[theorem]{Proposition}

\newtheorem{definition}{Definition}[section]
\newtheorem{assumption}{Assumption}[section]
\newtheorem*{example}{Example}
\newtheorem*{remark}{Remark}

\usepackage[textsize=tiny]{todonotes}
\usepackage{multirow}
\usepackage{wrapfig}
\usepackage{subfigure}

\renewcommand{\eqref}[1]{(\ref{#1})}

\usepackage{xcolor}
\newcount\Comments  
\newcommand{\kibitz}[2]{\ifnum\Comments=1\textcolor{#1}{#2}\fi}

\usepackage[capitalize,noabbrev]{cleveref}

\usepackage{mathrsfs}  

\allowdisplaybreaks

\title{Minimax and Bayes Optimal Best-Arm Identification}

\author{Masahiro Kato\thanks{Email: \texttt{mkato-csecon@g.ecc.u-tokyo.ac.jp}}$\,$}

\affil{The University of Tokyo}

\date{\today}

\begin{document}

\maketitle 

\begin{abstract}
This study investigates minimax and Bayes optimal strategies for fixed-budget best-arm identification. We consider an adaptive procedure consisting of a sampling phase followed by a recommendation phase. Within this framework, we design an adaptive experiment to efficiently identify the best arm, defined as the one with the highest expected outcome. In our proposed strategy, the sampling phase consists of two stages. The first stage is a pilot phase, in which we allocate samples uniformly across arms to eliminate clearly suboptimal arms and to estimate outcome variances. Before entering the second stage, we solve a Gaussian minimax game, which yields a sampling policy and a decision rule. In the second stage, samples are allocated according to this policy. After the sampling phase, the procedure enters the recommendation phase, where we select an arm using the decision rule. We prove that this single strategy is simultaneously asymptotically minimax and Bayes optimal for the simple regret, and we establish upper bounds that coincide exactly with our lower bounds, including the constant terms. The lower bounds hold against every adaptive experiment and for every fixed number of arms, and the strategy attains them without knowing the outcome distributions or the prior.
\end{abstract}

{\flushleft{{\bf Keywords:} Best-arm identification; minimax optimality; Bayes optimality; Neyman allocation; multi-armed bandits; adaptive experimental design; machine learning; econometrics; statistics; treatment choice}}

\section{Introduction}
We investigate the problem of fixed-budget \emph{best-arm identification} \citep[BAI,][]{Audibert2010bestarm}, an instance of adaptive experimental design for identifying the arm with the highest expected outcome. This problem is also known by various names across disciplines, including ordinal optimization \citep{Chen2000simulationbudget} and adaptive experimental design for policy choice \citep{Kasy2021adaptivetreatment}.

An adaptive experimental algorithm in BAI usually consists of two phases: the sampling phase and the recommendation phase \citep{Kaufmann2016complexity}. Given a total of $T$ rounds, the strategy selects one arm at each round using the observations collected so far. After the final round, it recommends an arm using all collected observations.

In this study, we propose an asymptotically minimax and Bayes optimal strategy for BAI. First, we determine the exact asymptotic minimax and Bayes lower bounds for the simple regret for every fixed finite number of arms $K\ge2$ over the class of all adaptive experiments. The outcomes follow mean-parameterized canonical exponential families on a compact mean interval under the structural and moment conditions in Section~\ref{sec:parametricdist}. The constant in the minimax lower bound is $\Gamma\p{\bmP}$, which is characterized by sequential Gaussian games. For distinct arms $a$ and $b$, the pairwise tie set is $\cb{\bmmu\in\calM^K\colon \mu_a=\mu_b=\max_{c\in[K]}\mu_c}$. It consists of the mean vectors for which arms $a$ and $b$ have the same largest mean. For every prior satisfying Assumption~\ref{asm:uniformcontinuity}, the constant in the Bayes lower bound is determined by the prior density on these pairwise tie sets. In the Bernoulli case, the matching lower bound establishes the optimality of the leading constant conjectured in \citet[Eq.~(6)]{Komiyama2023rateoptimal}.

We refer to our proposed strategy as Sequential Minimax Allocation with Screening (SMAS). This single strategy attains both minimax and Bayes lower bounds with matching constants. SMAS begins with a pilot phase whose fraction of the total budget converges to zero as $T\to\infty$. This phase removes clearly suboptimal arms and estimates the outcome variances. When at least three arms remain, SMAS uses a sequential allocation rule that is updated after every observation. When two arms remain, it uses plug-in Neyman allocation. The sampling proportions are proportional to the estimated standard deviations. The strategy uses the seven constants in Assumption~\ref{asm:structural_envelope} that give common variance bounds, a common derivative bound, and common constants in the two exponential-moment conditions. It is not supplied with the families of outcome distributions for the individual arms or their variance functions. It estimates the variances needed for allocation from the pilot observations.

\subsection{Setup}
\label{sec:prob}
We formulate the problem as follows.
There are $K$ arms, and each arm $a \in [K] \coloneqq \{1,2,\dots, K\}$ has a potential outcome $Y_a \in \calY$, where $\calY \subseteq \bbR$ denotes the outcome space. Each potential outcome $Y_a$ follows a (marginal) distribution $P_{a, \mu_a}$ parameterized by $\mu_a \in \calM$, where $\calM \subset \bbR$ is a parameter space. For the parameter vector $\bmmu \coloneqq (\mu_1, \mu_2, \dots, \mu_K) \in \calM^K$, let $\bmP_{\bmmu} \coloneqq (P_{1, \mu_1}, P_{2, \mu_2}, \dots, P_{K, \mu_K})$ be the bandit instance at $\bmmu$, the $K$-tuple of arm distributions. The parameter $\mu_a \coloneqq \bbE_{P_{a,\mu_a}}\sqb{Y_a}$ is the mean of arm $a$.

Whenever an arm-valued $\argmax$ or $\argmin$ is not unique, we choose the arm with the smallest index.

Under a bandit instance $\bmP_{\bmmu}$, our objective is to identify the best arm
\[
    a^*_{\bmmu} = \argmax_{a \in [K]} \mu_a,
\]
through an adaptive experiment in which observations are collected from $\bmP_{\bmmu}$ according to our strategy.

\paragraph{Adaptive experiment}
Let $T$ denote the total sample size, also referred to as the budget. Assume there exist i.i.d. random vectors $(Y_{1,t},\ldots,Y_{K,t})_{t\in[T]}$ such that
for each $a\in[K]$ and each $t\in[T]$, $Y_{a,t}\sim P_{a,\mu_a}$. We consider an adaptive experimental procedure consisting of two phases:
\begin{enumerate}
    \item \textbf{Sampling phase}: For each $t \in [T] \coloneqq \{1, 2, \dots, T\}$:
    \begin{itemize}
        \item An arm $A_t \in [K]$ is selected based on the past observations $\{(A_s, Y_s)\}_{s=1}^{t-1}$.
        \item The corresponding outcome $Y_t$ is observed, where $ Y_t \coloneqq \sum_{a \in [K]} \mathbbm{1}[A_t = a]Y_{a, t}$.
    \end{itemize}
    \item \textbf{Recommendation phase}: At the end of the experiment ($t = T$), based on the observed outcomes $\{(A_t, Y_t)\}_{t=1}^{T}$, we recommend an arm $\widehat{a}_T \in [K]$ as an estimate of the best arm $a^*_{\bmmu}$.
\end{enumerate}

Our task is to design a strategy $\delta$ that determines how arms are selected during the sampling phase and how the final arm is recommended. We define a strategy through the conditional laws governing the sampling decisions and the final recommendation. This definition includes randomized rules.

Throughout this study, we equip $[K]$ with the power-set sigma-algebra $2^{[K]}$ and $\calY\subseteq\bbR$ with its Borel sigma-algebra. For each $t\ge0$, let $\mathsf H_t\coloneqq\p{[K]\times\calY}^{t}$ be equipped with the corresponding product sigma-algebra, where $\mathsf H_0$ consists of the empty history. A probability kernel from a measurable space $\mathsf H$ to $[K]$ assigns a probability distribution on $[K]$ to each $h\in\mathsf H$. This distribution varies measurably with $h$.

\begin{definition}[Strategy and the class of all adaptive experiments]
\label{def:strategy}
Fix a budget $T\in\bbN$. A strategy $\delta_T$ consists of sampling kernels $Q_{t,T}$ from $\mathsf H_{t-1}$ to $[K]$ for $t\in[T]$ and a recommendation kernel $Q_T^{\mathrm{rec}}$ from $\mathsf H_T$ to $[K]$. Given the observed history $h_{t-1}\in\mathsf H_{t-1}$, the strategy selects the arm $A_t^\delta$ according to $Q_{t,T}\p{h_{t-1},\cdot}$. Given the full history $h_T\in\mathsf H_T$, it recommends the arm $\widehat a_T^\delta$ according to $Q_T^{\mathrm{rec}}\p{h_T,\cdot}$. Let $\calA_T$ denote the class of all such strategies, with no further restrictions. Let $\calA\coloneqq\prod_{T\in\bbN}\calA_T$ denote the class of all adaptive experiments, whose elements are sequences $\delta=\p{\delta_T}_{T\in\bbN}$.
\end{definition}

Every strategy in $\calA_T$ can be represented without changing its sampling or recommendation kernels: its sampling actions and final recommendation are measurable functions of the observed history and a random element $U$ generated before the experiment and independent of the outcomes. Lemma~\ref{lem:randomization_representation} gives the construction.

Throughout this study, a \emph{strategy} is a complete adaptive experiment in the original bandit model. It includes all sampling kernels and the final recommendation kernel. A \emph{policy} is a rule used in an auxiliary sequential game or in the main phase of SMAS. At each round, a policy specifies a distribution over the available arms as a function of the observed history or state. After the policy's last round, its \emph{terminal kernel} assigns a probability distribution to the recommended arm.

The class $\calA_T$ contains every nonadaptive, plug-in, and fully sequential design at budget $T$. A deterministic rule is the special case in which every kernel assigns probability one to a single arm at each history. For simplicity, we omit the dependence on $\delta$ when it is clear from the context.

\paragraph{Regret}
We measure the performance of a strategy $\delta$ by the expected simple regret, defined by
\begin{align*}
    \Regret^\delta_T\p{\bmP_{\bmmu}}
    \coloneqq
    \bbE_{\bmP_{\bmmu}}\sqb{\mu_{a^*_{\bmmu}} - \mu_{\widehat{a}_T^{\delta}}}.
\end{align*}
This quantity is the mean gap between the best arm and the recommended arm, averaged over the experiment. It also equals the expected welfare loss from deploying the recommended arm in an independent future population. Let $Y^{\mathrm{new}}=\p{Y_a^{\mathrm{new}}}_{a\in[K]}$ be independent of the experiment and satisfy $\bbE_{\bmP_{\bmmu}}\sqb{Y_a^{\mathrm{new}}}=\mu_a$. Then, it holds that $\Regret^\delta_T\p{\bmP_{\bmmu}}=\bbE_{\bmP_{\bmmu}}\sqb{Y_{a^*_{\bmmu}}^{\mathrm{new}}-Y_{\widehat a_T^\delta}^{\mathrm{new}}}$. We refer to the expected simple regret as the simple regret or the regret.

\paragraph{Notation}
Let $\bbP_{\bmP_{\bmmu}}$ and $\bbE_{\bmP_{\bmmu}}$ denote probability and expectation under $\bmP_{\bmmu}$. We abbreviate them as $\bbP_{\bmmu}$ and $\bbE_{\bmmu}$ when no ambiguity arises. We write $\Regret_T^\delta\p{\bmmu}$ for $\Regret_T^\delta\p{\bmP_{\bmmu}}$. For each arm $a$, let $\sigma_a^2\p{\mu_a}$ denote the variance under $P_{a,\mu_a}$ and let $\sigma_a\p{\mu_a}\coloneqq\sqrt{\sigma_a^2\p{\mu_a}}$. For random elements $Z_1,\ldots,Z_m$, write $\widetilde{\sigma}\p{Z_1,\ldots,Z_m}$ for the sigma-algebra they generate. Let $\calF_t=\widetilde{\sigma}\p{A_1,Y_1,\ldots,A_t,Y_t}$. For stochastic convergence, $X_T\xrightarrow{p}X$ and $X_T\xrightarrow{d}X$ denote convergence in probability and convergence in distribution as $T\to\infty$. The same notation is used with the displayed sequence index in place of $T$. In limit statements, $\to$ denotes deterministic convergence. Deterministic limits always display the index that tends to infinity or zero.

Using this representation, define $\calF_t^{U}\coloneqq\widetilde{\sigma}\p{U,A_1,Y_1,\ldots,A_t,Y_t}$. In Appendix~\ref{app:transported_pair}, $A_t$ is $\calF_{t-1}^{U}$-measurable and $\widehat a_T$ is $\calF_T^{U}$-measurable.

For each arm, we index outcomes by the number of times that arm has been selected. Let $N_{a,t}=\sum_{u\le t}\mathbbm 1[A_u=a]$. In this representation, we generate independent sequences $\p{Y_{a,i}}_{i\in[T]}$ with $Y_{a,i}\sim P_{a,\mu_a}$ and set $Y_t=Y_{a,N_{a,t}}$ on $\cb{A_t=a}$. This construction induces the same law for the observed history as the round-indexed formulation.

\subsection{Main Results}
We evaluate $\delta^{\mathrm{SMAS}}$ under the minimax and Bayes regret criteria. The minimax criterion takes the worst case over $\bmmu\in\calM^K$ for a fixed collection $\bmP$, consisting of one family of outcome distributions for each arm. The Bayes criterion averages the regret over the same mean space under a prior. The collection $\bmP$ is held fixed in both criteria.

For every fixed finite number of arms $K\ge2$, we develop a strategy that attains the exact asymptotic simple-regret constants over all adaptive experiments under both the minimax and Bayes criteria. We first derive asymptotic minimax and Bayes lower bounds for each fixed collection of mean-parameterized canonical exponential families satisfying Definition~\ref{def:mean_param}. The upper bounds additionally use Assumption~\ref{asm:structural_envelope} to construct one strategy for the model class. The lower bounds hold over all adaptive experiments, including fully sequential designs, and $\delta^{\mathrm{SMAS}}$ attains both constants.

Sequential Minimax Allocation with Screening (SMAS), denoted by $\delta^{\mathrm{SMAS}}$, begins with a pilot phase whose fraction of the total budget converges to zero as $T\to\infty$. This phase removes clearly suboptimal arms and estimates the outcome variances. When at least three arms remain, it uses a sequential allocation rule that is updated after every observation. When two arms remain, it uses plug-in Neyman allocation. The strategy also reserves observations for an independent final comparison. It uses only the seven constants $\underline v$, $\overline v$, $L_v$, $\lambda_0$, $C_0$, $\eta_0$, and $D_0$ specified in Assumption~\ref{asm:structural_envelope}. It does not use the families of outcome distributions for the individual arms, their variance functions, or the prior. Matching lower and upper bounds establish exact asymptotic minimax and Bayes optimality for every fixed finite number of arms.

The exact minimax constant $\Gamma\p{\bmP}$ is defined through the sequential Gaussian games in Section~\ref{sec:seq_game}. The exact Bayes constant is defined in Section~\ref{sec:exactbayes} through the density of the prior on the pairwise tie sets.

\paragraph{Summary of main theoretical results}
To illustrate the main results, suppose that $Y_a$ follows a Gaussian distribution with mean $\mu_a$ and variance $\sigma_a^2$. Let $\bmP$ denote the collection of the $K$ families of outcome distributions, and let $\calB\p{\bmP}$ be the bandit model obtained by varying $\bmmu\in\calM^K$ while holding those families fixed. In this Gaussian illustration, $\sigma_a^2\p{\mu}=\sigma_a^2$. The general model in Section~\ref{sec:parametricdist} also allows mean-dependent variances and includes Bernoulli outcomes on a compact subinterval of $\p{0,1}$.

We define the minimax and Bayes regret as follows:
\begin{itemize}
    \item \textbf{Minimax regret:} $\sup_{\bmmu \in \calM^K} \Regret^\delta_T(\bmmu)$.
    \item \textbf{Bayes regret:} $\int_{\bmmu \in \calM^K} \Regret^\delta_T(\bmmu)\rmd H(\bmmu)$, where $H$ is a prior distribution on $\calM^K$ with a bounded and uniformly continuous density.
\end{itemize}
We evaluate both criteria over the class of all adaptive experiments. For every fixed finite number of arms, the same strategy attains the corresponding leading constants under the minimax and Bayes criteria. Here, Bayes optimality refers to the integrated frequentist risk under each admissible prior. SMAS itself does not use the prior.

\paragraph{Asymptotic minimax optimality}
We show that $\delta^{\mathrm{SMAS}}$ is asymptotically minimax optimal:
\begin{align*}
    \limsup_{T\to\infty}\sqrt{T}\sup_{\bmmu\in\calM^K} \Regret_T^{\delta^{\mathrm{SMAS}}}(\bmmu)
    \leq \Gamma(\bmP)
    \leq \inf_{\delta\in\calA} \liminf_{T\to\infty}\sqrt{T}\sup_{\bmmu\in\calM^K} \Regret^\delta_T(\bmmu),
\end{align*}
where the infimum is over the class $\calA$ of all adaptive experiments and $\Gamma(\bmP)$ is the exact minimax constant defined through the finite sequential Gaussian games of Section~\ref{sec:seq_game}.
The lower bound holds over every adaptive experiment, and $\delta^{\mathrm{SMAS}}\in\calA$ attains it.

For $K=2$, the constant has the closed form $\Gamma(\bmP)=\sup_{\mu\in\calM}\p{\sigma_1(\mu)+\sigma_2(\mu)}c_{\mathrm{mm}}$. The corresponding binary Gaussian game is attained asymptotically by the deterministic Neyman allocation. SMAS implements this allocation using the pilot estimates of the standard deviations. For $K\ge3$, $\Gamma\p{\bmP}$ is defined by the sequential Gaussian game in Section~\ref{sec:seq_game}. This is an exact characterization of the constant rather than a closed-form formula.

\paragraph{Asymptotic Bayes optimality}
For every prior $H$ satisfying Assumption~\ref{asm:uniformcontinuity}, we also show that $\delta^{\mathrm{SMAS}}$ attains asymptotic Bayes optimality as follows:
\begin{align*}
    &\limsup_{T\to\infty} T\int_{\bmmu \in \calM^K} \Regret^{\delta^{\mathrm{SMAS}}}_T(\bmmu)\rmd H(\bmmu)\\
    &\leq C^{\mathrm{Bayes}}\p{\bmP,H} \leq \inf_{\delta\in\calA} \liminf_{T\to\infty} T \int_{\bmmu \in \calM^K}\Regret_T^\delta(\bmmu)\rmd H(\bmmu),
\end{align*}
where $C^{\mathrm{Bayes}}\p{\bmP,H}$ is the exact Bayes constant defined in Section~\ref{sec:exactbayes}, and the infimum is over the class $\calA$ of all adaptive experiments.
Thus, as $T\to\infty$, the scaled Bayes regret of SMAS converges to the exact Bayes constant $C^{\mathrm{Bayes}}\p{\bmP,H}$ (Corollary~\ref{cor:bayes_opt}), and the same strategy attains the exact constants in both criteria.

The competing experiments may depend on $\bmP$. Under the Bayes criterion, they may also depend on $H$. They may not depend on the unknown mean vector $\bmmu$. SMAS depends only on $K$, $\calM$, and the seven constants collected in $\mathfrak E$ in Assumption~\ref{asm:structural_envelope}. SMAS uses the same tuning throughout the model class and for all admissible priors. The first convergence is pointwise in $\bmP$, and the second is pointwise in $\p{\bmP,H}$. Corollary~\ref{cor:single_strategy} states the resulting quantifier order.

\paragraph{Main message}
Minimax regret is governed by near ties with gaps of order $T^{-1/2}$, whereas Bayes regret under regular priors is governed by pairwise near ties between the best and second-best arms. Although the two criteria are determined by different local mean vectors, one strategy attains both exact constants.

\paragraph{Organization}
The remainder of this study is organized as follows. In Section~\ref{sec:review}, we review related work. Section~\ref{sec:parametricdist} introduces a class of distributions that we consider. Section~\ref{sec:seq_game} describes the sequential Gaussian game, and Section~\ref{sec:smas} defines our strategy. Section~\ref{sec:minimax} treats the minimax criterion and Section~\ref{sec:bayes} the Bayes criterion, each presenting a lower bound and a matching upper bound. Section~\ref{sec:proof_structure} states the main steps of the four bounds and the order in which the limits are taken. In Section~\ref{sec:discussion}, we discuss related problems.

\section{Literature Review}
\label{sec:review}
This section reviews the criteria and results most closely related to this study. For a more detailed review, see Appendix~\ref{appdx:review}. The earliest BAI formulations appeared under the name \emph{ordinal optimization} \citep{Chen2000simulationbudget,Glynn2004largedeviations}, with an emphasis on non-adaptive sampling proportions and large-deviation criteria. Those analyses often characterize an optimal vector of sampling proportions using the outcome distributions of the arms. Later work incorporated the estimation of the target sampling proportions into the experimental design \citep{Audibert2010bestarm,Bubeck2011pureexploration}.

BAI is mainly studied in the fixed-confidence and fixed-budget settings \citep{Bubeck2011pureexploration,Audibert2010bestarm,Kaufmann2016complexity}. In the fixed-confidence setting, the error probability is controlled at a prescribed level and the sample size is minimized. In the fixed-budget setting, the budget is fixed in advance and the objective is to minimize the probability of misidentification or the simple regret. Exact instance-dependent results are well developed in the fixed-confidence setting. In the fixed-budget setting, the budget is fixed before the observations are collected, and its optimal error rate remains less completely characterized \citep{Kaufmann2020contributions}.

A central question in fixed-budget BAI is whether one strategy can attain the best distribution-dependent error exponent at every instance. \citet{Garivier2016optimalbest} conjectures an information-theoretic benchmark for the probability of misidentification $\bbP_{\bmmu}\p{\widehat{a}^\delta_T\neq a^*_{\bmmu}}$, but no strategy is known to match it at every instance. This difficulty is closely related to the reverse Kullback--Leibler divergence problem discussed by \citet{Kaufmann2020contributions}. \citet{Barrier2023onbest} develops upper and lower bounds on the average log-probability of misidentification for general distribution classes. Its distribution class includes all distributions on a bounded interval. The criterion in that study is the exponential scale of misidentification, whereas we study the leading constant of expected simple regret under minimax and Bayes criteria.

\citet{Ariu2021policychoice} gives an earlier impossibility result for distribution-dependent expected policy regret. Using the construction of \citet{Carpentier2016tightlower}, the result shows that, for any algorithm, there is a problem instance and an infinite subsequence of budgets along which the expected policy regret is slower than the oracle rate claimed by \citet{Kasy2021adaptivetreatment}. This lower bound provides the counterexample that refutes the claimed global optimality of exploration sampling. \citet{Degenne2023existence} later formulates uniform attainability through the notion of a complexity and proves nonexistence for several fixed-budget identification tasks, including two-armed Bernoulli BAI. The BAI impossibility result in \citet{Ariu2021policychoice} predates this general formulation. The contribution of \citet{Degenne2023existence} is the definition of complexity and its extension to a broader class of identification tasks. \citet{Wang2024uniformlyoptimal} further shows that, in two-armed Bernoulli A/B testing, no algorithm weakly improves on the fixed-instance exponent of uniform sampling at every instance and strictly improves on it at some instance. \citet{Goldberger2026fundamentallimitations} gives a related impossibility result for fixed-instance error exponents; that criterion differs from the minimax and Bayes expected simple regret studied here.

Fixed-instance analyses also contain positive dominance results. \citet{Imbens2025admissibilitycompletely} shows that batched elimination can strictly dominate completely randomized trials when there are at least three homoscedastic Gaussian arms. \citet{Wang2023bestarm} studies fixed-budget BAI from a large-deviation perspective, and \citet{Balagopalan2026fixedbudget} relates fixed-budget and fixed-confidence guarantees up to logarithmic factors. These studies concern fixed-instance performance on an exponential scale.

At a fixed $\bmmu$ with a unique best arm and positive gaps, the probability of misidentification and the simple regret have the same logarithmic decay rate. Writing $\Delta_{\min}(\bmmu)=\min_{a\neq a^*_{\bmmu}}\Delta_{\bmmu}(a)>0$ and $\Delta_{\max}(\bmmu)=\max_{a\in[K]}\Delta_{\bmmu}(a)<\infty$, neither of which depends on $T$, the identity $\Regret^\delta_T(\bmmu)=\sum_{a\in[K]}\Delta_{\bmmu}(a)\bbP_{\bmmu}\p{\widehat{a}^\delta_T=a}$ gives
\[
\Delta_{\min}(\bmmu)\bbP_{\bmmu}\p{\widehat{a}^\delta_T\neq a^*_{\bmmu}}
\le
\Regret^\delta_T(\bmmu)
\le
\Delta_{\max}(\bmmu)\bbP_{\bmmu}\p{\widehat{a}^\delta_T\neq a^*_{\bmmu}}.
\]
Taking $T^{-1}\log$ gives the same limit inferior and the same limit superior for the two quantities. Under minimax and Bayes criteria, the relevant gaps may shrink with $T$. The gap factor then contributes to the first-order risk, and the probability of misidentification and the simple regret have different leading behavior.

Subsequent work studies minimax and Bayes analyses, which avoid the pointwise attainability problem by changing the quantifier order. In the minimax criterion, the worst-case mean vector may depend on the budget. Under a prior, neighborhoods of the pairwise tie sets retain enough prior mass to determine the leading Bayes risk. The resulting expected simple regret is polynomial rather than exponential in $T$. Within the exponent criterion, \citet{Komiyama2022minimaxoptimal} studies a normalized minimax error exponent over fixed instances. That result remains on the exponential scale and addresses a different asymptotic quantity from the $T^{-1/2}$ minimax simple regret studied here.

For expected simple regret, \citet{Bubeck2011pureexploration} establishes the minimax rate. \citet{Komiyama2023rateoptimal} establishes the Bayes rate for Bernoulli bandits under regular priors and gives an explicit $1/T$ upper bound with a leading constant conjectured to be optimal. \citet{Atsidakou2023bayesianfixed} and \citet{Nguyen2024priordependent} also study Bayesian fixed-budget BAI, including prior-dependent sampling proportions and structured bandit models. In the Bernoulli case, our exact Bayes constant coincides with the constant in \citet[Eq.~(6)]{Komiyama2023rateoptimal}, and the matching lower bound establishes its optimality. Our model also allows heteroscedastic outcomes in mean-parameterized canonical exponential families.

Several studies obtain sharper minimax results under local or restricted formulations. \citet{Adusumilli2022neymanallocation} and \citet{Adusumilli2023risk} derive exact two-arm minimax results in local diffusion-limit formulations, where the Neyman allocation is optimal. \citet{Kato2024generalizedneyman} obtains matching lower and upper bounds in a small-gap formulation under a consistency condition on the competing strategies. \citet{Kato2023asymptoticallyoptimal} derives variance-dependent minimax simple-regret bounds under a local location-shift model and proposes a design attaining the $T^{-1/2}$ rate, although its lower and upper leading constants do not coincide. These studies connect local BAI with Gaussian experiments and variance-dependent arm selection.

For two arms, our minimax constant also reduces to the Neyman value. The present analysis takes the worst case over the global compact mean space and compares SMAS with all adaptive experiments. The analysis also allows the variance functions in the exponential-family model to depend on the mean. For three or more arms, the exact minimax constant is characterized by the sequential Gaussian game, whose sampling rule may respond to the observations. This characterization allows the lower and upper bounds to use the same value for every fixed finite number of arms.

We establish matching minimax and Bayes lower and upper bounds, including the leading constants, for every fixed finite number of arms. The loss is the expected simple regret. The outcomes follow mean-parameterized canonical exponential families satisfying the structural and moment conditions in Section~\ref{sec:parametricdist}, and the Bayes result applies to the priors in Assumption~\ref{asm:uniformcontinuity}. Both lower bounds range over all adaptive experiments. The competing strategies may use the families of outcome distributions for the arms and, under the Bayes criterion, the prior. SMAS is constructed using only $K$, $\calM$, and the seven constants in $\mathfrak E$ defined by Assumption~\ref{asm:structural_envelope}. It uses neither those families nor the prior. A single strategy attains the minimax $T^{-1/2}$ constant and the Bayes $T^{-1}$ constant.

Our proofs use change-of-measure arguments and LAN \citep{LeCam1972theoryofstatisics,LeCam1986asymptoticmethods}. Change-of-measure identities also underlie distribution-dependent analyses of cumulative reward maximization \citep{Lai1985asymptoticallyefficient} and fixed-confidence BAI \citep{Garivier2016optimalbest,Kaufmann2016complexity}. Related work uses LAN to analyze adaptive experiments \citep{Hirano2025asymptoticrepresentations,Adusumilli2022neymanallocation,Adusumilli2023risk,Adusumilli2025samplestopsampling,Armstrong2022asymptoticefficiency}. However, those studies first restrict the parameters to local alternatives. Such analyses do not by themselves control mean vectors with nonlocal gaps or eliminate suboptimal arms. In fixed-budget BAI, their exact optimality results are confined to two-arm problems or restricted classes of strategies. They therefore do not determine the value over all adaptive experiments for every fixed finite number of arms. Our upper bounds use screening to control arms with nonvanishing gaps before the retained arms are compared with the sequential Gaussian game.

Table~\ref{tab:regret} places our results beside representative results for fixed-budget BAI and cumulative reward maximization (CRM). The columns refer to the criterion used within each problem. The BAI distribution-dependent entry concerns fixed-instance performance, whereas the minimax and Bayes entries concern expected simple regret. The CRM entries concern cumulative regret.

\begin{table}[t]
\caption{Representative asymptotic optimality results in cumulative reward maximization (CRM) and fixed-budget BAI.}
\label{tab:regret}
\centering
\resizebox{0.97\textwidth}{!}{
\begin{tabular}{llccc}
\toprule
Problem & Result & Distribution-dependent & Minimax & Bayes \\
\midrule
\multirow{2}{*}{BAI}
& Exact asymptotic optimality
& Impossible in general \citep{Ariu2021policychoice,Degenne2023existence}
& This study
& This study \\
& Selected rate results
& \citep{Carpentier2016tightlower}
& \citep{Bubeck2011pureexploration,Kato2023asymptoticallyoptimal}
& \citep{Komiyama2023rateoptimal} \\
\midrule
\multirow{2}{*}{CRM}
& Exact asymptotic optimality
& \citep{Lai1985asymptoticallyefficient}
& --
& \citep{Lai1987adaptivetreatment} \\
& Selected rate results
& --
& \citep{Audibert2009minimaxpolicies}
& -- \\
\bottomrule
\end{tabular}
}
\end{table}

For BAI, the table separates impossibility results for uniform distribution-dependent optimality from exact minimax and Bayes results for expected simple regret. The CRM results provide earlier examples of exact asymptotic analysis for adaptive arm selection under distribution-dependent and Bayes criteria, although the loss is cumulative regret rather than terminal simple regret.

\section{Bandit Models}
\label{sec:parametricdist}
From this section onward, we analyze the minimax and Bayes optimality of SMAS against the competing experiments specified in Sections~\ref{sec:minimax} and \ref{sec:bayes}. We first define a class $\calP$ of outcome distributions and a collection $\bmP$ of arm families belonging to this class. We then define the bandit model $\calB\p{\bmP}$ induced by $\bmP$ and evaluate worst-case and average-case performance over its mean vectors.

We consider canonical exponential families for this class, which are typically defined as follows \citep{Garivier2016optimalbest}:
\[
\cb{ (P_\theta)_{\theta \in \Theta}
\colon
\frac{dP_\theta}{d\xi}(y) = \exp\bigp{ y \theta - b\bigp{\theta} }}.
\]
Here, $P_\theta$ is indexed by the natural parameter $\theta$, whereas $P_\mu$ elsewhere is indexed by the mean. The set $\Theta\subset\bbR$ is the natural-parameter space. The measure $\xi$ is a reference measure on $\calY$, and $b\colon\Theta\to\bbR$ is a convex and twice differentiable function.

In this study, however, we evaluate worst-case and average-case performance over the mean parameter and express the lower and upper bounds in terms of variances. The mean equals $\dot b(\theta)$, and the variance equals $\ddot b(\theta)$. We therefore define the distribution class in terms of the mean and variance. This section provides such a definition, introduces a collection of $K$ arm families, and defines the bandit model that the collection induces by varying the mean vector.

\subsection{Mean-Parameterized Canonical Exponential Families}
We define a class of mean-parameterized canonical exponential families. This class is essentially the standard canonical exponential family, but the present definition serves three purposes: (i) it parameterizes the distribution class by the mean, (ii) it makes the inverse Fisher information equal to the variance, and (iii) it guarantees the finite third moments required in our analysis. Common distributions such as Gaussian and Bernoulli are included in this class.

\begin{definition}[Mean-parameterized canonical exponential family]
\label{def:mean_param}
Fix a measurable outcome space $(\calY,\mathscr{A})$ and a compact interval $\calM\subset\mathbb R$.
For each arm $a\in[K]$, let $\xi_a$ be a $\sigma$-finite measure on $(\calY,\mathscr A)$ and let
$\sigma_a^2\colon\calM\to(0,\infty)$ be twice continuously differentiable.
We say that $\{P_{a,\mu}\colon\mu\in\calM\}$ belongs to $\calP(\sigma_a^2,\calM,\calY)$ if there exist
an open interval $\Theta_a\subset\mathbb R$, a strictly convex $C^3$ log-partition function $b_a\colon\Theta_a\to\mathbb R$,
and a $C^1$ map $\theta_a\colon\calM\to\Theta_a$ such that for all $\mu\in\calM$:
\begin{enumerate}
    \item $P_{a,\mu}\ll \xi_a$ with density
$\frac{dP_{a,\mu}}{d\xi_a}(y)=\exp\{y\theta_a(\mu)-b_a(\theta_a(\mu))\}$;
\item $\dot b_a(\theta_a(\mu))=\mu$;
\item $\ddot b_a(\theta_a(\mu))=\sigma_a^2(\mu)$;
\item (Uniform local exponential moment) There exist constants $\lambda_{0,a}>0$ and $C_a<\infty$ such that
\[
\sup_{\mu\in\calM}\bbE_{a,\mu}\bigsqb{\exp\p{\lambda (Y-\mu)}}\le \exp(C_a\lambda^2)
\qquad\forall |\lambda|\le \lambda_{0,a}.
\]
\item (Uniform square-exponential moment) There exist constants $\eta_{0,a}>0$ and $D_a<\infty$ such that
\[
\sup_{\mu\in\calM}\bbE_{a,\mu}\bigsqb{\exp\p{\eta (Y-\mu)^2}}\le D_a
\qquad\forall 0\le \eta\le \eta_{0,a}.
\]
\end{enumerate}
\end{definition}

\begin{remark}
The outcome space $\calY$ and the parameter space $\calM$ should be chosen to satisfy the conditions in Definition~\ref{def:mean_param}. For example, if the outcome $Y_a$ follows a Bernoulli distribution with support $\calY = \{0, 1\}$, we can choose $\calM=[\underline\mu,\overline\mu]\subset(0,1)$ with
$0<\underline\mu<\overline\mu<1$. If we instead choose $\calM = [0, 1]$, the variance equals $0$ at $\mu = 0$ and $\mu = 1$, which violates our definition of the class. In this case, the Fisher information does not exist at $\mu = 0$ and $\mu = 1$, since the Fisher information is given by $I(\mu) = \frac{1}{\mu(1 - \mu)}$.
\end{remark}

\begin{remark}[Scope of the moment conditions]
\label{rem:moment_scope}
The uniform square-exponential moment in condition~5 yields uniform concentration of the pilot variance estimator in Lemma~\ref{lem:pilot_variance_concentration}. It excludes some unbounded canonical exponential families. For example, if $Y$ follows a Poisson distribution with mean $\mu$, then $\bbE_{a,\mu}\sqb{\exp\p{\eta\p{Y-\mu}^2}}=\infty$ for every $\eta>0$. The definition includes Gaussian families with fixed variance, Bernoulli families with $\calM\subset\p{0,1}$, and families whose outcomes are uniformly bounded over $\calM$.

Condition~4 supplies the moment bounds used for the pilot means and the means computed from the reserved comparison observations, as well as for the local likelihood expansion and the two-arm comparison. Condition~5 is used for the empirical variance concentration and is not used in the lower-bound proofs. The differentiability assumptions on the variance functions support the Taylor expansion and the continuity arguments used in the minimax and Bayes proofs. The lower-bound theorems are stated for each fixed collection satisfying Definition~\ref{def:mean_param}. The upper-bound construction additionally uses the common bounds in Assumption~\ref{asm:structural_envelope}.
\end{remark}

\subsection{Bandit Model}
We distinguish a collection of arm families, the bandit model induced by that collection, and the class of collections satisfying Assumption~\ref{asm:structural_envelope} with the same seven constants. The minimax and Bayes criteria vary the mean vector within a fixed bandit model.

The first is a collection of arm families. For each arm $a\in[K]$, let $\sigma_a^2\colon\calM\to(0,\infty)$ be twice continuously differentiable and bounded away from $0$ and $\infty$, and let $\calP_a=\cb{P_{a,\mu}\colon\mu\in\calM}$ be a family in the class $\calP(\sigma_a^2,\calM,\calY)$ of Definition~\ref{def:mean_param}. We call $\calP_a$ the family of outcome distributions for arm $a$. We write
\[
\bmP=\p{\calP_a}_{a\in[K]},
\qquad
\bm\sigma^2=\p{\sigma_a^2}_{a\in[K]},
\]
and call $\bmP$ a collection of arm families with variance functions $\bm\sigma^2$. The collection $\bmP$ describes the true environment, is fixed throughout an experiment, and is not supplied to the strategy.

The second is the bandit model that $\bmP$ induces. Recall that for a mean vector $\bmmu=\p{\mu_a}_{a\in[K]}\in\calM^K$ the bandit instance is $\bmP_{\bmmu}=\p{P_{a,\mu_a}}_{a\in[K]}$, in which arm $a$ has mean $\mu_a$. Define
\[
\calB\p{\bmP}
\coloneqq
\cb{\bmP_{\bmmu}\colon\bmmu\in\calM^K}.
\]
Only the mean vector varies within $\calB\p{\bmP}$, while the arm families, and with them the variance functions $\bm\sigma^2$, are held fixed. The worst case of Section~\ref{sec:minimax} and the average of Section~\ref{sec:bayes} are both taken over $\bmmu\in\calM^K$, that is, over $\calB\p{\bmP}$ at a fixed $\bmP$. Neither a supremum nor an average is taken over the choice of $\bmP$.

\begin{example}[Bandit instances]
The bandit model $\calB\p{\bmP}$ allows heterogeneity across arms. For example, with $K=2$, arm $1$ may be Bernoulli$(\mu_1)$, so $\sigma_1^2(\mu)=\mu(1-\mu)$, while arm $2$ may be Gaussian with possibly constant variance $\sigma_2^2(\mu)\equiv\sigma_2^2$.
\end{example}

The third is the class of collections satisfying the same uniform bounds on the variance functions and the same uniform moment conditions.

\begin{assumption}[Uniform bounds on variance functions and exponential moments]
\label{asm:structural_envelope}
There are finite constants $0<\underline v<\overline v<\infty$, $L_v<\infty$, $\lambda_0>0$, $C_0<\infty$, $\eta_0>0$, and $D_0<\infty$ such that, for every arm $a\in[K]$ and every $\mu\in\calM$,
\[
\underline v\le\sigma_a^2(\mu)\le\overline v,
\qquad
\abs{\partial_\mu\sigma_a^2(\mu)}\le L_v,
\]
and the moment inequalities (4) and (5) of Definition~\ref{def:mean_param} hold with the common constants $\p{\lambda_0,C_0}$ in place of $\p{\lambda_{0,a},C_a}$ and $\p{\eta_0,D_0}$ in place of $\p{\eta_{0,a},D_a}$. We write
\begin{align}
\label{eq:envelope_tuple}
\mathfrak E\coloneqq\p{\underline v,\overline v,L_v,\lambda_0,C_0,\eta_0,D_0}.
\end{align}
Here, $\underline v$ and $\overline v$ are uniform lower and upper bounds on the variance functions $\sigma_a^2(\mu)$. The constant $L_v$ is a uniform upper bound on $\abs{\partial_\mu\sigma_a^2(\mu)}$. The pair $\p{\lambda_0,C_0}$ specifies the range $\abs{\lambda}\le\lambda_0$ and the upper bound $\exp\p{C_0\lambda^2}$ in the local exponential-moment condition. The pair $\p{\eta_0,D_0}$ specifies the range $0\le\eta\le\eta_0$ and the upper bound $D_0$ in the square-exponential-moment condition.
\end{assumption}

Because $\calM$ is compact and each $\sigma_a^2$ is twice continuously differentiable and strictly positive by Definition~\ref{def:mean_param}, each variance function is bounded above and away from zero on $\calM$, and its derivative is bounded. Since $K$ is finite, the arm-specific moment constants in Definition~\ref{def:mean_param} can also be replaced by common constants. Thus, for every collection $\bmP$ satisfying Definition~\ref{def:mean_param}, there is at least one choice of the seven constants $\mathfrak E$ for which Assumption~\ref{asm:structural_envelope} holds.

Algorithm~\ref{alg:smas} is defined using one choice of the seven constants $\mathfrak E$ satisfying Assumption~\ref{asm:structural_envelope}. Together with $K$ and $\calM$, these constants determine the tolerances, the finite grid used to round estimated standard deviations, and the budget thresholds. The seven constants need not be the smallest valid bounds. For the finite calculations below, the endpoints of $\calM$ may be supplied through rational lower and upper bounds of any prescribed positive width. We may also use smaller positive rational bounds for $\underline v$, $\lambda_0$, and $\eta_0$, and larger rational bounds for $\overline v$, $L_v$, $C_0$, and $D_0$. Different valid choices of $\mathfrak E$ may produce different tolerances, grids, and budget thresholds, but they do not change $\Gamma(\bmP)$ or $C^{\mathrm{Bayes}}\p{\bmP,H}$, which are defined directly from $\bmP$ and $H$. The strategy does not use the collection $\bmP$, the variance functions $\sigma_a^2(\cdot)$, or their pointwise values.

Fixing the seven constants $\mathfrak E$ determines the class of collections for which the strategy uses the same tolerances, grids, and budget thresholds. By condition 3 of Definition~\ref{def:mean_param}, each family $\calP_a$ determines its variance function, $\sigma_a^2(\mu)=\Var_{P_{a,\mu}}(Y)$ for $\mu\in\calM$. Thus, a collection $\bmP=\p{\calP_a}_{a\in[K]}$ determines $\bm\sigma^2=\p{\sigma_a^2}_{a\in[K]}$, and we define
\begin{align}
\label{eq:envelope_class}
\mathfrak P\p{\mathfrak E}
\coloneqq
\left\{
\begin{aligned}
\bmP=\p{\calP_a}_{a\in[K]}\colon\;&
\calP_a\in\calP\p{\sigma_a^2,\calM,\calY}
\ \text{for every }a\in[K],\\
&\bmP\ \text{satisfies Assumption~\ref{asm:structural_envelope} with }\mathfrak E
\end{aligned}
\right\}.
\end{align}
Here, the variance functions $\bm\sigma^2$ in \eqref{eq:envelope_class} are those determined by $\bmP$. The single strategy in Corollary~\ref{cor:single_strategy} is fixed from $K$, $\calM$, and $\mathfrak E$. It attains $\Gamma(\bmP)$ for every $\bmP\in\mathfrak P\p{\mathfrak E}$ and $C^{\mathrm{Bayes}}\p{\bmP,H}$ for every such $\bmP$ and every prior $H$ satisfying Assumption~\ref{asm:uniformcontinuity}.

The constants $\lambda_0$ and $C_0$ in $\mathfrak E$ also yield a uniform bound on the absolute third moments. Under Assumption~\ref{asm:structural_envelope}, condition 4 of Definition~\ref{def:mean_param} holds with common constants $\p{\lambda_0,C_0}$ for all arms. Applying Markov's inequality with $\lambda=\lambda_0$ and $\lambda=-\lambda_0$ gives
$\bbP_{a,\mu}\p{\abs{Y-\mu}>t}\le2\exp\p{C_0\lambda_0^2-\lambda_0t}$
for all $t\ge0$, $a\in[K]$, and $\mu\in\calM$. Tonelli's theorem then gives
\begin{align}
&\sup_{a\in[K],\mu\in\calM}\bbE_{a,\mu}\sqb{\abs{Y-\mu}^3}\nonumber\\
&=\sup_{a\in[K],\mu\in\calM}\int_0^{\infty}3t^2\bbP_{a,\mu}\p{\abs{Y-\mu}>t}\rmd t
\le6\exp\p{C_0\lambda_0^2}\int_0^{\infty}t^2\exp\p{-\lambda_0t}\rmd t
=\overline\Lambda_3,
\label{eq:third_moment_envelope}
\end{align}
where $\overline\Lambda_3\coloneqq12\lambda_0^{-3}\exp\p{C_0\lambda_0^2}$. Thus, $\overline\Lambda_3$ bounds the absolute third moments uniformly over the arms and over $\calM$. All later constants involving third moments, including those used in the comparison arguments, are defined using $\overline\Lambda_3$. Since $\overline\Lambda_3$ is determined by $\mathfrak E$, these constants do not depend on the arm families.

\section{Finite Sequential Gaussian Game}
\label{sec:seq_game}
The exact minimax constant is built from finite sequential Gaussian games. In these discrete-time experiments, the sampling rule may respond to every observation. The constant is not the value of a single game. We first let the horizon tend to infinity. We then let the radius $C$ of the parameter set $[-C,C]^{\calS}$ tend to infinity and finally maximize over the common mean $m$ of the tied arms. We describe the game here and record its value. Appendices~\ref{app:finite_grid_duality} and \ref{app:seq_value_bellman} give the formal construction and all proofs.

Let $\calS\subseteq[K]$ with $\abs{\calS}\ge2$, let $n\ge\abs{\calS}$, and let $s=(s_a)_{a\in\calS}\in(0,\infty)^{\calS}$ be a vector of standard deviations. Nature chooses a local parameter $u=(u_a)_{a\in\calS}\in\bbR^{\calS}$. At round $t$, a randomized history-dependent policy selects an arm $A_t\in\calS$ from a distribution determined measurably by the preceding observations and auxiliary randomness independent of the Gaussian noises. We refer to such a policy as a behavioral policy. Conditional on $A_t=a$, we observe
\[
X_t=\frac{u_a}{\sqrt n}+s_a Z_t,
\qquad Z_t\sim\calN(0,1),
\]
with the $Z_t$ independent across rounds. After $n$ rounds the policy recommends $\widehat A_n\in\calS$ and incurs the simple regret $\Delta_u^{\calS}(\widehat A_n)=\max_{a\in\calS}u_a-u_{\widehat A_n}$. The policy may update the distribution of $A_t$ after every observation. This makes the sampling rule fully sequential.

For the parameter set $[-C,C]^{\calS}$ with $C<\infty$, the value $\mathfrak G_{\calS,n}^{C}(s)$ is the minimax risk over all behavioral policies (Definition~\ref{def:seq_gaussian_game}). The limit as the number of rounds tends to infinity exists,
\[
\mathfrak g_{\calS}^{C}(s)=\lim_{n\to\infty}\mathfrak G_{\calS,n}^{C}(s)=\inf_{n\ge\abs{\calS}}\mathfrak G_{\calS,n}^{C}(s),
\]
and the global sequential value $\mathfrak G_{\calS}^{\uparrow}(s)=\sup_{C<\infty}\mathfrak g_{\calS}^{C}(s)$ is finite (Theorem~\ref{thm:seq_value_limit}).

We define the exact minimax constant of the bandit model $\calB\p{\bmP}$ induced by $\bmP$ by taking the supremum of the game on all $K$ arms over the common mean $m$ of the tied arms:
\[
\Gamma(\bmP)
\coloneqq
\sup_{m\in\calM}\mathfrak G_{[K]}^{\uparrow}\bigp{\p{\sigma_a(m)}_{a\in[K]}}.
\]
The finiteness of $\Gamma(\bmP)$ follows from Definition~\ref{def:global_seq_constant}. The constant depends on $\bmP$ only through the functions $m\mapsto\sigma_a(m)$ for $a\in[K]$. Thus, two collections in $\mathfrak P\p{\mathfrak E}$ with the same standard-deviation functions have the same exact minimax constant.

Least favorable instances are near ties, where the top means differ by $O(T^{-1/2})$. The competing arms therefore share a common limiting mean $m$, and their standard deviations are evaluated at that mean. Screening may retain any subset of locally competitive arms. By Lemma~\ref{lem:arm_set_monotonicity}, the game value does not decrease when arms are added. Hence, $\Gamma(\bmP)$ dominates the game value on every subset that screening may retain, as stated in Corollary~\ref{cor:gamma_subsets}, and no additional maximization over subsets is required.

For two arms, the value is available in closed form,
\[
\mathfrak G_{\cb{a,b}}^{\uparrow}(s_a,s_b)=\p{s_a+s_b}c_{\mathrm{mm}},
\qquad
c_{\mathrm{mm}}=\sup_{x\ge0}x\Phi(-x)\approx0.1699712,
\]
attained asymptotically by the deterministic Neyman allocation $N_a/n\to s_a/(s_a+s_b)$ as $n\to\infty$, together with the empirical-best recommendation, and no sampling rule that adapts to the observed outcomes improves the constant
(Lemma~\ref{lem:binary_sequential_value}). For $K=2$, this gives $\Gamma(\bmP)=\sup_{\mu\in\calM}\p{\sigma_1(\mu)+\sigma_2(\mu)}c_{\mathrm{mm}}$.

For $K\ge3$, the exact constant is the value of the sequential game over all adaptive experiments. The static value is at least as large because static policies form a subclass of adaptive policies. Appendix~\ref{sec:numerics} gives a three-arm example in which the inequality is strict.

\section{Sequential Minimax Allocation with Screening}
\label{sec:smas}
We now define Sequential Minimax Allocation with Screening (SMAS), denoted by $\delta^{\mathrm{SMAS}}$. The strategy attains $\Gamma\p{\bmP}$ and $C^{\mathrm{Bayes}}\p{\bmP,H}$. It estimates the outcome variances from the pilot sample and otherwise uses only $K$, $\calM$, and the constants in $\mathfrak E$ defined in \eqref{eq:envelope_tuple}. Algorithm~\ref{alg:smas} summarizes the strategy, and Appendix~\ref{app:upper_comparison} gives the full definition.

The strategy uses indices $j\ge8$. We set $\alpha_j=\gamma_j=j^{-2}$, $B_j=j^2$, $H_j=j^3$, $R_j=8j^2$, and $\omega_j=j^{-4}$. Definition~\ref{def:indexed_near_minimax_policies} fixes the remaining mesh and tolerance parameters at values in $(0,j^{-4}]$. For each $j$, the deterministic number $N_j$ is chosen so that every sample-size and error requirement attached to $j$ holds for all $T\ge N_j$; we call $N_j$ the budget threshold for $j$. The thresholds increase to infinity and determine $j_T=\max\cb{j\ge8\colon N_j\le T}$ for each budget $T$. The calculation performed before data collection first fixes the finite-horizon approximation and computes the policies and their derivative bounds. It then chooses the tolerances and the finite grid used to round the estimated standard deviations before determining the budget thresholds. Theorem~\ref{thm:finite_computation_smas} shows that every step terminates using only $K$, $\calM$, and $\mathfrak E$.

\begin{definition}[Indexed family of near-minimax policies]
\label{def:indexed_near_minimax_policies}
For each $j$ and each arm subset $\calS\subseteq[K]$ with $\abs{\calS}\ge2$, apply Lemma~\ref{lem:uniform_guarantee_box}. Use $C=R_j$, $\eta=\omega_j$, and standard-deviation vectors in $[\sqrt{\underline v}/2,2\sqrt{\overline v}]^{\calS}$. Let
\[
D_{2,j},\quad D_{3,j},\quad B_{0,j},\quad d_{0,j},\quad C_j^{\mathrm{cmp}},\quad N_j^0
\]
be the maxima of the corresponding constants over the finitely many subsets, so that one set of constants applies to every subset. They depend only on $R_j$, $\omega_j$, and the constants in $\mathfrak E$. This dependence enters through $\Lambda_{3,\mathrm{box}}=\max\cb{\overline\Lambda_3,\p{2\sqrt{\overline v}}^3\bbE\abs Z^3}$, whose second argument is the largest centered third absolute moment of a Gaussian observation with standard deviation in $[\sqrt{\underline v}/2,2\sqrt{\overline v}]$. None of these constants depends on the finite set $\calV_j$ of rounded standard-deviation values introduced below. For every $\calS$, every $s\in[\sqrt{\underline v}/2,2\sqrt{\overline v}]^{\calS}$, and every $n\ge N_j^0$, the finite calculation in Lemma~\ref{lem:uniform_guarantee_box} returns a policy $\pi_{j,n,\calS,s}$ satisfying the $\p{D_{2,j},D_{3,j},B_{0,j}}$ bounds and having a continuous state in $\bbR^{d_{0,j}}$. The constant $C_j^{\mathrm{cmp}}$ can be used in Lemma~\ref{lem:guaranteed_recovery} for every standard-deviation vector in the box, and
\[
\sup_{u\in[-R_j,R_j]^{\calS}}R_n^{G}\p{\pi_{j,n,\calS,s},u}\le\mathfrak g_{\calS}^{R_j}(s)+\omega_j.
\]

We first fix $R_j$ and $\omega_j$ and then apply the finite calculation in Lemma~\ref{lem:uniform_guarantee_box} to obtain $D_{2,j}$, $D_{3,j}$, $B_{0,j}$, $d_{0,j}$, $C_j^{\mathrm{cmp}}$, and $N_j^0$. The tolerances $r_j$ and $\zeta_j$ are chosen only after these constants have been computed. This ordering makes the definition noncircular. Let $m_j$ be the smallest positive integer such that
\[
2^{-m_j}
\le
\min\cb{j^{-4},\ \frac{1}{j\p{1+C_j^{\mathrm{cmp}}D_{2,j}}},\ r_{\mathrm{conc}}},
\]
and set $r_j=\zeta_j=2^{-m_j}$. Here, $r_{\mathrm{conc}}=1/\underline v$ is the largest relative tolerance used in the proof of the pilot-variance concentration bound \eqref{eq:pilot_variance_bad_probability}. Then, $r_j,\zeta_j\in(0,j^{-4}]$ and $r_j\le r_{\mathrm{conc}}$. Moreover,
\[
\p{1+C_j^{\mathrm{cmp}}D_{2,j}}\p{r_j+\zeta_j}\le2j^{-1}.
\]
We also write
\[
\rho_j\coloneqq4\p{r_j+\zeta_j}.
\]

Let $\calV_j$ be the multiplicative grid on $[\sqrt{\underline v}/2,2\sqrt{\overline v}]$ with ratio $1+\zeta_j$, together with its top endpoint. We call $\calV_j$ the standard-deviation grid; the rounded standard deviations used by the strategy take values in this finite set. The definition gives $r_j\le j^{-4}$ and $r_j\to0$ as $j\to\infty$. The additional restriction $r_j\le r_{\mathrm{conc}}$ affects only finitely many small values of $j$. It therefore does not change the convergence of the error caused by approximating the standard deviations. For every horizon $n\ge N_j^0$, the policies $\pi_{j,n,\calS,s}$ indexed by pairs $\p{\calS,s}$ with $\abs{\calS}\ge2$ and $s\in\calV_j^{\calS}$ form a finite collection. The indexed family for $j$ consists of these policies over all $n\ge N_j^0$. It depends only on $j$, $K$, $\calM$, and the constants in $\mathfrak E$.

The set of indices $\p{j,n,\calS,s}$ with $j\ge8$, $n\ge N_j^0$, $\abs{\calS}\ge2$, and $s\in\calV_j^{\calS}$ is countable because every grid $\calV_j$ is finite. For each index, the quantitative bound in Theorem~\ref{thm:finite_horizon_rate} determines a finite base horizon, Theorem~\ref{thm:finite_computation_smooth_policy} computes the corresponding smooth policy, and Lemma~\ref{lem:replicated_smooth_policy} extends it to horizon $n$. Lemma~\ref{lem:uniform_guarantee_box} combines these steps and computes the policy assigned to the index.

The pilot standard deviations are rounded to the finite grid $\calV_{j_T}$, and the strategy uses the policy assigned to the resulting index. Thus, no policy is computed while the experiment runs.
\end{definition}

For the selected index $j_T$, write $T_0=K\lfloor\alpha_{j_T}T/K\rfloor$ and $\tau_T=B_{j_T}/\sqrt T$. Stage~1 selects every arm exactly $T_0/K$ times, computes the empirical means $\widehat\mu_{a,T_0}$ and unbiased empirical variances $\widehat\sigma_{a,T_0}^2$, and forms the retained set
\[
\widehat{\calS}_T
=
\cb{a\in[K]\colon\widehat\mu_{a,T_0}+\tau_T\ge\max_{b\in[K]}\widehat\mu_{b,T_0}-\tau_T}.
\]
It also sets $\widehat m_T=\max_a\widehat\mu_{a,T_0}$ and constructs $\overline s_T$ by truncating the empirical variances, slightly increasing the resulting standard-deviation estimates, and rounding them upward on a finite multiplicative grid. The strategy reserves $T_g=K\lfloor\gamma_{j_T}T/K\rfloor$ observations for an independent final comparison and selects each arm exactly $T_g/K$ times during these rounds. These observations are interleaved with the main rounds and are not used to update the state variables of the main-phase policy. The remaining $n=T-T_0-T_g$ rounds form the main phase.

The main-phase rule depends on $\widehat{\calS}_T$. If $\abs{\widehat{\calS}_T}=1$, the strategy selects the retained arm in every main round and recommends it as the preliminary recommendation. If $\abs{\widehat{\calS}_T}=2$, write $\widehat{\calS}_T=\{a,b\}$ with $a<b$. The strategy uses plug-in Neyman allocation. It selects arm $a$ in $\lceil n\widehat w_{T,a}\rceil$ main rounds and selects arm $b$ in the remaining main rounds. It then recommends the arm with the larger main-sample mean as the preliminary recommendation. If $\abs{\widehat{\calS}_T}\ge3$, the strategy uses the near-minimax policy $\pi_{j_T,n,\widehat{\calS}_T,\overline s_T}$ fixed in Definition~\ref{def:indexed_near_minimax_policies}. After each main observation, it updates the continuous and finite memory states through \eqref{eq:quasi_log_odds_update}. The policy kernel selects the arm for the next round, and the terminal kernel recommends the preliminary recommendation.

Let $\widetilde a_T$ be the preliminary recommendation and let $\widehat a_T^{g}$ be the arm with the largest empirical mean among the reserved comparison observations. The final comparison rule recommends $\widehat a_T^{g}$ when its empirical mean exceeds that of $\widetilde a_T$ by at least $H_{j_T}/\sqrt T$. Otherwise, it recommends $\widetilde a_T$.
\begin{algorithm}[t]
\caption{Sequential Minimax Allocation with Screening (SMAS)}
\label{alg:smas}
\begin{algorithmic}[1]
\STATE Compute $j_T$ and the grids and indexed policies required at budget $T$ by Theorem~\ref{thm:finite_computation_smas}. Set $j=j_T$, $T_0=K\lfloor\alpha_jT/K\rfloor$, $T_g=K\lfloor\gamma_jT/K\rfloor$, and $\tau_T=B_j/\sqrt T$.
\STATE Select each arm exactly $T_0/K$ times and compute its empirical mean and empirical variance.
\STATE Form $\widehat{\calS}_T$ and $\widehat m_T$, and construct the rounded vector $\overline s_T$ as described above.
\STATE During the rounds reserved for the final comparison, select each arm exactly $T_g/K$ times and do not use these observations to update the state variables of the main-phase policy.
\IF{$\abs{\widehat{\calS}_T}=1$}
\STATE Select the retained arm in every main round and recommend it as $\widetilde a_T$.
\ELSIF{$\abs{\widehat{\calS}_T}=2$}
\STATE Use plug-in Neyman allocation and select each of the two arms according to the resulting proportions. After the main rounds, recommend the arm with the larger main-sample mean as $\widetilde a_T$.
\ELSE
\STATE \label{line:indexed_policy} Use the policy $\pi_{j,n,\widehat{\calS}_T,\overline s_T}$ fixed in Definition~\ref{def:indexed_near_minimax_policies}, and initialize its states.
\FOR{each main round}
\STATE Select the arm according to the distribution returned by the current policy kernel, observe $Y$, update the state by \eqref{eq:quasi_log_odds_update}, and evaluate the policy kernel at the updated state.
\ENDFOR
\STATE Recommend $\widetilde a_T$ according to the terminal kernel.
\ENDIF
\STATE Apply the final comparison rule to $\widetilde a_T$ and the empirical leader based on the comparison observations, and recommend $\widehat a_T$.
\end{algorithmic}
\end{algorithm}

Every sampling decision may depend on the full history, so $\delta^{\mathrm{SMAS}}$ is an adaptive experiment. The pilot screens clearly suboptimal arms, and the main-phase policy updates its arm-selection probabilities after every observation. Neither phase uses the variance functions.

\section{Asymptotic Minimax Optimality}
\label{sec:minimax}
We evaluate the worst-case regret over the class $\calA$ of all adaptive experiments in Definition~\ref{def:strategy}. Competing strategies may use the arm families but not the unknown mean vector $\bmmu$.

\begin{remark}[The criterion is the limit of the finite-budget value]
\label{rem:finite_budget_value}
Let
\[
V_T^{\mathrm{mm}}\coloneqq\inf_{\delta_T\in\calA_T}\sup_{\bmmu\in\calM^K}\Regret_T^{\delta_T}\p{\bmmu},
\qquad
V_T^{H}\coloneqq\inf_{\delta_T\in\calA_T}\int_{\bmmu\in\calM^K}\Regret_T^{\delta_T}\p{\bmmu}\rmd H\p{\bmmu}.
\]
Because $\calA=\prod_{T\in\bbN}\calA_T$, it holds that
\[
\inf_{\delta\in\calA}\liminf_{T\to\infty}\sqrt T\sup_{\bmmu\in\calM^K}\Regret_T^\delta\p{\bmmu}
=
\liminf_{T\to\infty}\sqrt TV_T^{\mathrm{mm}},
\]
and the analogous identity holds for $TV_T^H$.

For every $\delta\in\calA$ and every $T$, the definition of $V_T^{\mathrm{mm}}$ gives
\[
\sqrt T\sup_{\bmmu\in\calM^K}\Regret_T^\delta\p{\bmmu}
\ge
\sqrt T V_T^{\mathrm{mm}}.
\]
Taking the limit inferior and then the infimum over $\delta$ gives one inequality. For the converse, choose for each $T$ a strategy $\delta_T^{\mathrm{mm}}\in\calA_T$ such that
\[
\sup_{\bmmu\in\calM^K}\Regret_T^{\delta_T^{\mathrm{mm}}}\p{\bmmu}
\le
V_T^{\mathrm{mm}}+T^{-2}.
\]
The sequence $\delta^{\mathrm{mm}}=\p{\delta_T^{\mathrm{mm}}}_{T\in\bbN}$ belongs to $\calA$. Therefore,
\[
\inf_{\delta\in\calA}\liminf_{T\to\infty}
\sqrt T\sup_{\bmmu\in\calM^K}\Regret_T^\delta\p{\bmmu}
\le
\liminf_{T\to\infty}\sqrt T V_T^{\mathrm{mm}},
\]
because $\sqrt T\,T^{-2}\to0$ as $T\to\infty$. For the analogous equality involving $TV_T^H$, choose separately, for each $T$, a strategy whose Bayes regret is at most $V_T^H+T^{-2}$. After multiplication by $T$, the additional term is $T^{-1}$ and converges to zero as $T\to\infty$.
\end{remark}

The exact minimax value equals the game constant $\Gamma(\bmP)$ of Section~\ref{sec:seq_game} for every fixed finite number of arms.

\begin{theorem}[Asymptotic minimax lower bound]
\label{thm:minimax_lower}
Fix $K\ge2$, an outcome space $\calY$, and a parameter space $\calM$ with $\min\calM<\max\calM$, and let $\bmP=(\calP_a)_{a\in[K]}$ be a true collection of arm families satisfying Definition~\ref{def:mean_param}, with variance functions $\bm\sigma^2=(\sigma_a^2)_{a\in[K]}$ and induced bandit model $\calB\p{\bmP}$. The collection $\bmP$ is fixed, and the supremum below is over the mean vectors of $\calB\p{\bmP}$. Then
\[
\inf_{\delta\in\calA}\liminf_{T\to\infty}\sqrt T\sup_{\bmmu\in\calM^K}\Regret_T^{\delta}(\bmmu)
\ge
\Gamma(\bmP),
\]
where $\Gamma(\bmP)$ is the global sequential minimax constant of Definition~\ref{def:global_seq_constant}.
\end{theorem}

The lower bound holds over all adaptive experiments. Appendix~\ref{app:lower_comparison} proves it by comparing the original local experiment with the sequential Gaussian game through posterior Bellman recursions.

\begin{theorem}[Asymptotic worst-case upper bound]
\label{thm:minimax_upper}
Under the assumptions of Theorem~\ref{thm:minimax_lower} and Assumption~\ref{asm:structural_envelope}, the deterministic budget thresholds and the policies indexed by $j$ in Algorithm~\ref{alg:smas} can be fixed using only $K$, $\calM$, and the constants in $\mathfrak E$ so that
\[
\limsup_{T\to\infty}\sqrt T\sup_{\bmmu\in\calM^K}\Regret_T^{\delta^{\mathrm{SMAS}}}(\bmmu)
\le
\Gamma(\bmP).
\]
\end{theorem}

Appendix~\ref{app:upper_comparison} proves the upper bound by combining the smooth policies constructed in Appendix~\ref{app:smooth_guaranteed} with concentration bounds for the pilot mean and variance estimates and a comparison of the original and Gaussian transition operators for one observation. Although the strategy does not know the variance functions, it attains the leading constant determined by them.

\begin{corollary}[Asymptotic minimax optimality]
\label{cor:minimax_opt}
Under the assumptions of Theorem~\ref{thm:minimax_upper}, for every fixed finite $K\ge2$,
\[
\lim_{T\to\infty}\sqrt T\sup_{\bmmu\in\calM^K}\Regret_T^{\delta^{\mathrm{SMAS}}}(\bmmu)
=
\Gamma(\bmP)
=
\inf_{\delta\in\calA}\liminf_{T\to\infty}\sqrt T\sup_{\bmmu\in\calM^K}\Regret_T^{\delta}(\bmmu),
\]
Thus, the strategy $\delta^{\mathrm{SMAS}}$, whose construction does not use the arm families or their variance functions, attains the exact minimax value over all adaptive experiments.
\end{corollary}

\begin{proof}
Since $\delta^{\mathrm{SMAS}}\in\calA$, Theorem~\ref{thm:minimax_lower} gives the lower bound $\Gamma(\bmP)$ for its limit inferior. Theorem~\ref{thm:minimax_upper} gives the same quantity as an upper bound for its limit superior. Hence, the limit exists and equals $\Gamma(\bmP)$.

Theorem~\ref{thm:minimax_lower} gives the same lower bound for every strategy in $\calA$, while $\delta^{\mathrm{SMAS}}$ attains it. Therefore, the infimum over $\calA$ also equals $\Gamma(\bmP)$.
\end{proof}

For $K=2$, the value is $\Gamma(\bmP)=\sup_{\mu\in\calM}\p{\sigma_1(\mu)+\sigma_2(\mu)}c_{\mathrm{mm}}$ by Lemma~\ref{lem:binary_sequential_value}. For each fixed common mean $\mu$ of the two tied arms, the binary Gaussian game is attained asymptotically by the deterministic Neyman allocation. SMAS realizes this allocation by using the plug-in Neyman proportions.

Remark~\ref{rem:finite_budget_value} and Corollary~\ref{cor:minimax_opt} also give $\lim_{T\to\infty}\sqrt TV_T^{\mathrm{mm}}=\Gamma(\bmP)$.

\section{Asymptotic Bayes Optimality}
\label{sec:bayes}
This section establishes matching lower and upper bounds for the Bayes regret of SMAS.

\subsection{Exact Bayes Constant}
\label{sec:exactbayes}
Let $H$ be a prior distribution on $\calM^K$ satisfying the following condition.

\begin{assumption}[Prior with a bounded uniformly continuous density]
\label{asm:uniformcontinuity}
The prior $H$ admits a density $h$ with respect to Lebesgue measure on $\calM^K$, and $h$ is bounded and uniformly continuous on the compact set $\calM^K$.
\end{assumption}
Assumption~\ref{asm:uniformcontinuity} fixes two constants of the prior, the density bound and the modulus of uniform continuity,
\begin{align}
\label{eq:prior_constants}
\overline h\coloneqq\sup_{\bmmu\in\calM^K}h(\bmmu)<\infty,
\qquad
\omega_H(s)\coloneqq
\sup\Bigcb{
\abs{h(\bmmu)-h(\bmlambda)}\colon
\substack{\bmmu,\bmlambda\in\calM^K,\\
\max_{a\in[K]}\abs{\mu_a-\lambda_a}\le s}
},
\end{align}
so that $\omega_H$ is nondecreasing, $\omega_H\le2\overline h$, and $\omega_H(s)\downarrow0$ as $s\downarrow0$. Both depend on the prior alone and never on the horizon.

All prior integrals below use the joint density $h$. No conditional density is required.

For each $a\in[K]$, let $\bmmu_{\setminus\{a\}} = (\mu_b)_{b\in[K]\setminus\{a\}} \in \calM^{K-1}$.
For each $a\in[K]$, define the index and mean of the best remaining arm, with ties broken by the fixed rule:
\[
b^*_{\setminus \{a\}}\coloneqq \argmax_{b\in[K]\setminus\{a\}} \mu_b,
\qquad
m_a(\bmmu_{\setminus \{a\}})\coloneqq \mu_{b^*_{\setminus \{a\}}}.
\]
For $\bmmu_{\setminus\{a\}}\in\calM^{K-1}$ and $\mu\in\calM$, write $h(\bmmu_{\setminus\{a\}},\mu_a=\mu)$ for the density $h$ evaluated at the point of $\calM^K$ whose $a$th coordinate is $\mu$ and whose remaining coordinates are $\bmmu_{\setminus\{a\}}$, and, for distinct $a\neq b$, write $h(\bmmu_{-ab},\mu_a=x,\mu_b=y)$ for the analogous evaluation at a pair of coordinates.

We define the exact Bayes constant as
\begin{align}
C^{\mathrm{Bayes}}\p{\bmP,H}
&\coloneqq
\frac{1}{4}\sum_{a\in[K]}
\int_{\calM^{K-1}}
\Bigp{
\sigma_a\p{m_a(\bmmu_{\setminus \{a\}})}
+\sigma_{b^*_{\setminus \{a\}}}\p{m_a(\bmmu_{\setminus \{a\}})}
}^2\nonumber\\
&\qquad\times
h\Bigp{\bmmu_{\setminus \{a\}},\mu_a=m_a(\bmmu_{\setminus \{a\}})}
\rmd\bmmu_{\setminus \{a\}}.
\label{eq:Cbayes_def}
\end{align}
The constant depends on $\bmP$ through the standard deviations at common mean values and on $H$ through the density on the pairwise tie sets.

\subsection{Asymptotic Bayes Lower Bound}
We now show a Bayes lower bound. Let $H$ be a prior distribution on $\calM^K$. Every prior integral below is taken against the density $h$ of Assumption~\ref{asm:uniformcontinuity}. In particular, the proof of Theorem~\ref{thm:bayes_lower} evaluates the prior density at points satisfying $\mu_a=\mu_b=m$, written as $h(\bmmu_{-ab},\mu_a=m,\mu_b=m)$. These points lie in $\calM^K$, where $h$ is defined and continuous.
\begin{theorem}[Asymptotic Bayes lower bound]
\label{thm:bayes_lower}
Assume the same outcome model as in Theorem~\ref{thm:minimax_lower}, and let the prior $H$ satisfy Assumption~\ref{asm:uniformcontinuity}. Then, for the class $\calA$ of all adaptive experiments,
\[
\inf_{\delta\in\calA}\liminf_{T\to\infty}T\int_{\bmmu\in\calM^K}\Regret^\delta_T\p{\bmmu}\rmd H\p{\bmmu}
\ge
C^{\mathrm{Bayes}}\p{\bmP,H},
\]
where $C^{\mathrm{Bayes}}\p{\bmP,H}$ is given in \eqref{eq:Cbayes_def}.
\end{theorem}

The constant $C^{\mathrm{Bayes}}\p{\bmP,H}$ is governed by pairwise near ties between the best and second-best arms. The factor $\sigma_a(m)+\sigma_b(m)$ arises because the common mean $m$ of the tied arms is unknown and is averaged under the prior. The proof compares a local instance in which arm $a$ has the larger mean with a second local instance in which arm $b$ has the larger mean. The common mean in the second instance is shifted so that each selection of either arm contributes the same standardized information. A change of variables in the integral over $m$ then maps the second comparison back to the original pair. A pointwise comparison at a known $m$ would have a different constant, as stated in Remark~\ref{rem:loc_invariance_needed}.

\subsection{Asymptotic Bayes Upper Bound}
The lower bound allows competing experiments to depend on the arm families and the prior. SMAS attains the same constant using only $K$, $\calM$, and the constants in $\mathfrak E$. Appendix~\ref{app:bayes_upper_alladapt} gives the proof.

\begin{theorem}[Asymptotic Bayes upper bound]
\label{thm:bayes_upper}
Under the assumptions of Theorem~\ref{thm:bayes_lower} and Assumption~\ref{asm:structural_envelope}, the deterministic budget thresholds and the policies indexed by $j$ in Algorithm~\ref{alg:smas} can be fixed without using the variance functions or the prior, using the same choice as in Theorem~\ref{thm:minimax_upper}, so that
\[
\limsup_{T\to\infty}T\int_{\bmmu\in\calM^K}\Regret^{\delta^{\mathrm{SMAS}}}_T\p{\bmmu}\rmd H\p{\bmmu}
\le
C^{\mathrm{Bayes}}\p{\bmP,H}.
\]
\end{theorem}

\paragraph{Proof sketch for Theorem~\ref{thm:bayes_upper}}
After multiplying the Bayes regret by $T$, only mean vectors with a near tie between two arms contribute to the limit. On the event where the pilot mean and variance estimates satisfy their concentration bounds, the set of mean vectors for which at least three arms are retained has prior mass of order $O\p{\tau_T^2}$, and their contribution vanishes as $T\to\infty$. The remaining leading term comes from the two-arm plug-in Neyman rule and yields the factor $\frac14\p{\sigma_a\p{m}+\sigma_{b^*_{\setminus \{a\}}}\p{m}}^2$ after integrating over the mean vectors on which the two relevant arms have the same largest mean.

The policies used when at least three arms remain after screening are needed for the minimax result. Under a prior satisfying Assumption~\ref{asm:uniformcontinuity}, mean vectors with three or more near-tied arms make a vanishing contribution after the Bayes regret is multiplied by $T$. The Bayes constant is therefore determined by the two-arm sampling rule. A prior-averaged bound controls the additional regret caused by the final comparison rule.

The Bayes lower bound in Theorem~\ref{thm:bayes_lower} and the matching upper bound in Theorem~\ref{thm:bayes_upper} imply asymptotic Bayes optimality.

\begin{corollary}[Asymptotic Bayes optimality]
\label{cor:bayes_opt}
Under the conditions of Theorems~\ref{thm:bayes_lower} and \ref{thm:bayes_upper}, we have
\begin{align*}
&\lim_{T\to\infty}T\int_{\bmmu \in \calM^K} \Regret^{\delta^{\mathrm{SMAS}}}_T(\bmmu)\rmd H(\bmmu)\\
&=
C^{\mathrm{Bayes}}\p{\bmP,H}
=
\inf_{\delta\in\calA}\liminf_{T\to\infty}T\int_{\bmmu\in\calM^K}\Regret_T^{\delta}(\bmmu)\rmd H(\bmmu).
\end{align*}
Therefore, the strategy $\delta^{\mathrm{SMAS}}$, whose construction uses only $K$, $\calM$, and $\mathfrak E$, attains the exact Bayes value over the class of all adaptive experiments.
\end{corollary}

\begin{proof}
Since $\delta^{\mathrm{SMAS}}\in\calA$, the lower bound of Theorem~\ref{thm:bayes_lower} gives
\[
\liminf_{T\to\infty}T\int_{\bmmu\in\calM^K}\Regret_T^{\delta^{\mathrm{SMAS}}}(\bmmu)\rmd H(\bmmu)
\ge
C^{\mathrm{Bayes}}\p{\bmP,H},
\]
Theorem~\ref{thm:bayes_upper} gives the reverse inequality for the limit superior. Combining the two bounds gives the first equality.

Theorem~\ref{thm:bayes_lower} also bounds the asymptotic Bayes risk of every strategy in $\calA$ below by $C^{\mathrm{Bayes}}\p{\bmP,H}$. Since $\delta^{\mathrm{SMAS}}$ attains this constant, the infimum over $\calA$ equals $C^{\mathrm{Bayes}}\p{\bmP,H}$.
\end{proof}

As in the minimax case, the exact Bayes constant is the limit of the finite-budget Bayes value. At every budget $T$, it holds that $V_T^{H}\le\int_{\bmmu\in\calM^K}\Regret_T^{\delta^{\mathrm{SMAS}}}(\bmmu)\rmd H(\bmmu)$. Theorem~\ref{thm:bayes_upper} therefore bounds $\limsup_{T\to\infty}TV_T^{H}$ above by $C^{\mathrm{Bayes}}\p{\bmP,H}$. Remark~\ref{rem:finite_budget_value} and Theorem~\ref{thm:bayes_lower} bound $\liminf_{T\to\infty}TV_T^{H}$ below by the same constant. Hence, $\lim_{T\to\infty}TV_T^{H}=C^{\mathrm{Bayes}}\p{\bmP,H}$.

The following corollary states the common tuning and the order of the quantifiers.

\begin{corollary}[One strategy for every model and every prior]
\label{cor:single_strategy}
Fix a finite $K\ge2$, a measurable outcome space $\calY$, a compact interval $\calM$ with $\min\calM<\max\calM$, and constants $\mathfrak E=\p{\underline v,\overline v,L_v,\lambda_0,C_0,\eta_0,D_0}$ satisfying Assumption~\ref{asm:structural_envelope}. There is a strategy $\delta^{\mathrm{SMAS}}\in\calA$, namely Algorithm~\ref{alg:smas} with the sequences indexed by $j$ in Appendix~\ref{app:upper_comparison}, the policies fixed in Definition~\ref{def:indexed_near_minimax_policies}, and the budget thresholds \eqref{eq:budget_thresholds}, all fixed from $K$, $\calM$, and $\mathfrak E$. For every collection $\bmP\in\mathfrak P\p{\mathfrak E}$, the first equality below holds. For every such $\bmP$ and every prior $H$ satisfying Assumption~\ref{asm:uniformcontinuity}, the second equality holds:
\begin{align*}
&\lim_{T\to\infty}\sqrt T\sup_{\bmmu\in\calM^K}\Regret_T^{\delta^{\mathrm{SMAS}}}(\bmmu)=\Gamma(\bmP),\\
&\lim_{T\to\infty}T\int_{\bmmu\in\calM^K}\Regret_T^{\delta^{\mathrm{SMAS}}}(\bmmu)\rmd H(\bmmu)=C^{\mathrm{Bayes}}\p{\bmP,H}.
\end{align*}
The first convergence is pointwise in $\bmP$, and the second is pointwise in $\p{\bmP,H}$.
\end{corollary}

\begin{proof}
The sequences indexed by $j$, the tolerances $r_j$ and $\zeta_j$, and the standard-deviation grid $\calV_j$ are fixed from $K$, $\calM$, and $\mathfrak E$. For every admissible index $\p{j,n,\calS,s}$, Definition~\ref{def:indexed_near_minimax_policies} fixes one policy before $\bmP$ and $H$ are specified. All constants involving third moments depend on $\mathfrak E$ through $\overline\Lambda_3$ in \eqref{eq:third_moment_envelope}. The budget thresholds \eqref{eq:budget_thresholds} are fixed from the same data and satisfy the requirements of both upper bounds. Thus, Algorithm~\ref{alg:smas} defines one strategy independently of $\bmP$ and $H$. Corollary~\ref{cor:minimax_opt} gives the first limit for every $\bmP\in\mathfrak P\p{\mathfrak E}$, and Corollary~\ref{cor:bayes_opt} gives the second limit for every such $\bmP$ and every prior $H$ satisfying Assumption~\ref{asm:uniformcontinuity}.
\end{proof}

The main results and the analysis of the Gaussian game involve several closely related values and constants; Table~\ref{tab:notation} collects them together with their defining references.

\begin{table}[t]
\caption{Symbols, values, and constants appearing in the main results.}
\label{tab:notation}
    \centering
    \scalebox{0.85}{
    \begin{tabular}{c|l|c}
    \toprule
        Symbol & Meaning & Defined in \\
        \hline
        $\bmP$ & Collection of arm families, fixed and not supplied to the strategy & Section~\ref{sec:parametricdist} \\
        \hline
        $\calB\p{\bmP}$ & Bandit model induced by $\bmP$, the mean vector varying over $\calM^K$ & Section~\ref{sec:parametricdist} \\
        \hline
        $\mathfrak E$ & Constants in Assumption~\ref{asm:structural_envelope} & \eqref{eq:envelope_tuple} \\
        \hline
        $\mathfrak P\p{\mathfrak E}$ & Collections of arm families satisfying Assumption~\ref{asm:structural_envelope} with the constants $\mathfrak E$ & \eqref{eq:envelope_class} \\
        \hline
        $\mathfrak G_{\calS,n}^{C}(s)$ & Finite sequential Gaussian value on subset $\calS$, horizon $n$, and parameter set $[-C,C]^{\calS}$ & Definition~\ref{def:seq_gaussian_game} \\
        \hline
        $\mathfrak g_{\calS}^{C}(s)$ & Limit $\lim_{n\to\infty}\mathfrak G_{\calS,n}^{C}(s)$ at fixed radius $C$ & Theorem~\ref{thm:seq_value_limit} \\
        \hline
        $\mathfrak G_{\calS}^{\uparrow}(s)$ & Global sequential value $\sup_{C<\infty}\mathfrak g_{\calS}^{C}(s)$ & Theorem~\ref{thm:seq_value_limit} \\
        \hline
        $\Gamma(\bmP)$ & Exact minimax constant $\sup_{m\in\calM}\mathfrak G_{[K]}^{\uparrow}\p{\p{\sigma_a(m)}_{a\in[K]}}$ & Definition~\ref{def:global_seq_constant} \\
        \hline
        $c_{\mathrm{mm}}$ & Binary minimax constant $\sup_{x\ge0} x\Phi(-x)\approx 0.1699712$ & Lemma~\ref{lem:binary_sequential_value} \\
        \hline
        $C^{\mathrm{Bayes}}\p{\bmP,H}$ & Exact Bayes constant & \eqref{eq:Cbayes_def} \\
        \hline
        $V_T^{\mathrm{mm}}$ & Finite-budget minimax value at budget $T$ & Remark~\ref{rem:finite_budget_value} \\
        \hline
        $V_T^{H}$ & Finite-budget Bayes value at budget $T$ under the prior $H$ & Remark~\ref{rem:finite_budget_value} \\
    \bottomrule
    \end{tabular}
    }
\end{table}

\section{Structure of the Proofs}
\label{sec:proof_structure}
This section summarizes the main steps of the proofs and the order of limits.

For the minimax lower bound, Lemma~\ref{lem:s_arm_reduction} reduces the problem to a local $\calS$-arm experiment. Lemmas~\ref{lem:original_posterior_markovization} and \ref{lem:posterior_markovization} express the original and Gaussian experiments through posterior Bellman recursions. Lemmas~\ref{lem:smooth_gaussian_subsolution}, \ref{lem:local_logodds_expansion}, and \ref{lem:one_step_operator_comparison} then compare the two Bellman values. Lemma~\ref{lem:finite_grid_saddle} provides a maximizing finite-grid prior, and Lemma~\ref{lem:grid_extension} and Theorem~\ref{thm:seq_value_limit} recover the sequential Gaussian-game constant. We first fix $\calS$, the common mean $m$, the radius $C$, and the grid $\calU$, and then let $T\to\infty$. After that limit, we send the grid mesh to zero and the radius to infinity, and finally maximize over the common mean $m$ of the tied arms. The arm set remains fixed at $\calS=[K]$.

The minimax and Bayes upper bounds use the same pilot phase and screening rule, and both approximate the standard deviation of the scaled difference between the two sample means. In the minimax proof, the strategy uses a policy from the indexed family in Definition~\ref{def:indexed_near_minimax_policies} when at least three arms remain and the two-arm sampling rule when two arms remain. The Bayes proof instead shows that mean vectors with three or more near-tied arms have negligible prior mass and analyzes the retained pair consisting of the two largest-mean arms by averaging the two-arm risk bound under the prior. For each fixed $j$, the scaled regret is bounded by the corresponding minimax or Bayes constant plus terms that vanish as $T\to\infty$ and terms that vanish as $j\to\infty$. The budget thresholds \eqref{eq:budget_thresholds} define the index $j_T$ used at budget $T$, and all these error terms vanish along this sequence.

For the Bayes lower bound, we restrict the prior integral to neighborhoods of the pairwise tie sets. Lemmas~\ref{lem:info_completion} and \ref{lem:array_mclt} yield the testing affinity used in Lemma~\ref{lem:pairwise_testing_lb}. We first take $T\to\infty$ with the cutoff $U$ on the local displacement, the interval of possible common means, and the remaining means fixed. We then extend that interval to $\calM$ and let $U\to\infty$.

\section{Discussion}
\label{sec:discussion}
\subsection{Minimax Optimal Strategy}
The minimax constant is determined by mean vectors for which the relevant gaps are of order $O\p{T^{-1/2}}$. Screening does not determine this local constant. In SMAS, it controls mean vectors with nonvanishing gaps and connects the problem over the full compact mean space to the sequential Gaussian game used in the minimax analysis.

\subsection{Computing the Gaussian-Game Policy}
\label{sec:computing_gaussian_policy}
The policies and budget thresholds in Definition~\ref{def:indexed_near_minimax_policies} are computed before data collection. Theorem~\ref{thm:finite_horizon_rate} gives a quantitative bound for the finite-horizon Gaussian game. This bound determines a finite horizon for each index $j$. The parameter grids and the grids $\calV_j$ used to round the standard-deviation estimates are finite, and Theorem~\ref{thm:finite_computation_smooth_policy} computes the corresponding smooth policies.

The fixed-horizon calculation uses a finite net of priors and backward induction in the posterior log-odds state. It restricts the state to finite boxes and partitions the Gaussian increments into finitely many intervals. Gaussian tail bounds control the omitted tails. Finite weighted sums approximate the remaining expectations, and Bernstein polynomials approximate the action-value functions on the state boxes. The proof of Theorem~\ref{thm:finite_computation_smooth_policy} controls these errors and computes the second- and third-order derivative bounds of the smooth kernels.

The finite matrix game selects a mixture of the computed policies. The mixture chooses one policy before round $1$ and follows it throughout the experiment. Lemma~\ref{lem:replicated_smooth_policy} extends a fixed-horizon policy to the required main-phase horizon. Each budget threshold is the first integer satisfying a finite list of monotone inequalities. Theorem~\ref{thm:finite_computation_smas} combines these calculations. During data collection, SMAS rounds the estimated standard deviations to $\calV_{j_T}$ and evaluates the kernels assigned to the resulting index.

\subsection{Bayes Optimal Strategy}
\label{sec:bayes_opt_strategy}
For Bayes optimality alone, the sequential policy used when three or more arms remain after screening is unnecessary. If $\abs{\widehat{\calS}_T}=2$, the main phase uses the Neyman allocation for the retained pair and the empirical-best recommendation. If $\abs{\widehat{\calS}_T}\ge3$, it selects every retained arm equally often.

On the event where the pilot mean estimate satisfies its concentration bound, three or more retained arms require a triple near tie. Such mean vectors have prior mass of order $O\p{\tau_T^2}$, and their contribution vanishes after the Bayes regret is multiplied by $T$ and the screening bound is applied. The exact Bayes constant is therefore determined by mean vectors with a near tie between the two largest-mean arms, as shown in Appendix~\ref{app:bayes_upper_alladapt}.

\subsection{LAN Approaches in Existing Studies}
\label{sec:discussion_com}
Applying LAN to all arms clarifies the local decision problem, and in the two-armed case it yields exact Neyman-allocation results \citep{Adusumilli2022neymanallocation,Adusumilli2023risk}. If every mean is parameterized from the outset as $m+u_a/\sqrt n$, the analysis contains only local alternatives. An arm whose mean remains a fixed positive distance below the best mean is outside this parameterization, so LAN or diffusion arguments alone cannot show that the arm is removed. In our construction, large-deviation concentration bounds screen such suboptimal arms and connect the problem over the full compact mean space to the local Gaussian game that determines the exact constant.

For example, \citet{Lai1987adaptivetreatment} and \citet{Adusumilli2023risk} both study Bayes-optimal allocation for cumulative regret, but they obtain different strategies. \citet{Lai1987adaptivetreatment} combines local asymptotic analysis with large-deviation bounds on the number of selections of suboptimal arms. These bounds separate arms with fixed positive gaps from local alternatives and make the sampling proportions assigned to the former converge to zero. By contrast, \citet{Adusumilli2023risk} restricts the parameter space to local alternatives whose mean differences are of order $n^{-1/2}$. Under this restriction, no arm has a fixed positive gap, and the optimal strategy is different. Both studies derive strategies that attain the corresponding lower bounds, including the leading constants.\footnote{\citet{Adusumilli2023risk} formulates the local problem through a diffusion limit. \citet{Lai1987adaptivetreatment} does not use this formulation.}

Bandit lower bounds often begin with an exact change-of-measure identity \citep{Lai1985asymptoticallyefficient,Kaufmann2016complexity}. LAN expresses the corresponding likelihood-ratio comparison under local alternatives, and Le Cam's third lemma transfers limiting distributions between nearby experiments. Our lower bounds use these likelihood-ratio arguments in the local multi-arm and pairwise comparisons.

\subsection{Uniform Arm Selection for Bernoulli Outcomes}
For Bernoulli outcomes, the strategy $\delta^{\mathrm{unif}}$ in Appendix~\ref{app:bernoulli_uniform} omits variance estimation and selects every retained arm equally often during the main phase. Proposition~\ref{prop:bernoulli_uniform_upper} shows that it attains the exact Bayes constant.

The Bayes constant is governed by pairwise near ties, where the two relevant means differ by $O\p{T^{-1/2}}$. For Bernoulli outcomes, every arm has the variance function $\sigma^2\p{\mu}=\mu\p{1-\mu}$. Along a pairwise near tie, the two standard deviations therefore differ by $O\p{T^{-1/2}}$.

Under equal sampling, the standard deviation of the scaled difference between the two sample means is $\sqrt{2\p{\sigma^2\p{\mu_a}+\sigma^2\p{\mu_b}}}$, whereas under Neyman allocation the corresponding value is $\sigma\p{\mu_a}+\sigma\p{\mu_b}$. The difference between their squares is $\p{\sigma\p{\mu_a}-\sigma\p{\mu_b}}^2=O\p{T^{-1}}$ along a pairwise near tie. Uniform arm selection among the retained arms therefore attains the same leading Bayes constant without estimating the variances. This conclusion concerns the Bayes criterion.

\begin{proposition}[Exact Bayes constant under uniform arm selection for Bernoulli outcomes]
\label{prop:bernoulli_constant}
Suppose $P_{a,\mu}=\mathrm{Bernoulli}(\mu)$ for all $a\in[K]$ so that $\calM=[\underline\mu,\overline\mu]\subset(0,1)$ with
$0<\underline\mu<\overline\mu<1$,
and $\sigma_a^2(\mu)=\mu(1-\mu)$.
Let the prior $H$ satisfy Assumption~\ref{asm:uniformcontinuity} with density $h$.
Write $\mu_{*\setminus \{a\}}\coloneqq \max_{b\neq a}\mu_b$.
Then the exact Bayes constant $C^{\mathrm{Bayes}}\p{\bmP,H}$ in \eqref{eq:Cbayes_def} equals
\[
C_{\mathrm{opt}}
\coloneqq
\sum_{a\in[K]} \int_{\calM^{K-1}}
\mu_{*\setminus \{a\}}\bigp{1-\mu_{*\setminus \{a\}}} h\bigp{\bmmu_{\setminus \{a\}},\mu_a=\mu_{*\setminus \{a\}}}\rmd\bmmu_{\setminus \{a\}},
\]
which is the corresponding Bernoulli-prior constant on the restricted parameter space $\calM$.
\end{proposition}

\section{Conclusion}
In this study, we developed an asymptotically minimax and Bayes optimal strategy for the simple regret, with matching lower and upper bounds over the class of all adaptive experiments and for every fixed finite number of arms $K\ge2$. The exact minimax constant $\Gamma(\bmP)$ is obtained by first letting the number of rounds in the finite sequential Gaussian games tend to infinity, then letting the radius $C$ of the parameter set $[-C,C]^{[K]}$ tend to infinity, and finally maximizing over the common mean of the tied arms. For every prior $H$ satisfying Assumption~\ref{asm:uniformcontinuity}, the exact Bayes constant $C^{\mathrm{Bayes}}\p{\bmP,H}$ is governed by pairwise near ties between the best and second-best arms. SMAS removes clearly suboptimal arms in a pilot phase whose fraction of the budget converges to zero as $T\to\infty$, estimates the outcome variances from the pilot observations, and selects arms during the remaining rounds using a sequential rule when at least three arms remain and plug-in Neyman allocation when two arms remain.

\clearpage

\begin{appendix}

Throughout the appendices we work with a collection of arm families $\bmP\in\mathfrak P\p{\mathfrak E}$ that is fixed once and for all and, wherever a prior appears, with a prior $H$ satisfying Assumption~\ref{asm:uniformcontinuity} that is likewise fixed. We accordingly abbreviate the exact Bayes constant $C^{\mathrm{Bayes}}\p{\bmP,H}$ of \eqref{eq:Cbayes_def} by $C^{\mathrm{Bayes}}$, and its pairwise terms by $C^{\mathrm{Bayes}}_{ab}$.

\section{Strict Improvement from Sequential Sampling in the Three-Arm Game}
\label{sec:numerics}
For two arms, the deterministic Neyman allocation attains the sequential Gaussian value. For three arms, the inclusion of static policies in the class of behavioral policies can be strict. We give one example with $\calS=\cb{1,2,3}$ and $s=(1,1,1)$. Appendix~\ref{app:numerics} gives the calculation.

\paragraph{The two values}
Within Definition~\ref{def:seq_gaussian_game}, call a behavioral policy static when its sampling counts $\p{N_a}_{a\in\calS}$ are generated before round $1$ from a distribution measurable with respect to the external randomization alone. The policy may select the arms in any order that realizes those counts. Its terminal kernel remains unrestricted and may use all $n$ observations. Write $\mathfrak G_{\calS,n}^{C,\mathrm{stat}}(s)$ for the minimax risk over static policies against local parameters in $[-C,C]^{\calS}$, and define
\[
\mathfrak G_{\calS}^{\uparrow,\mathrm{stat}}(s)
\coloneqq
\sup_{C<\infty}\ \liminf_{n\to\infty}\ \mathfrak G_{\calS,n}^{C,\mathrm{stat}}(s).
\]
Static policies are behavioral policies, so
\[
\mathfrak G_{\calS}^{\uparrow}(s)
\le
\mathfrak G_{\calS}^{\uparrow,\mathrm{stat}}(s).
\]

For a prior $p$ on the local parameters, let $B(p,f)$ denote the Bayes risk under the static sampling-proportion vector $f$ and the best terminal rule. For a policy $\pi$ whose sampling decisions may depend on previous observations, let $R^{\infty}(u,\pi)$ denote its risk in the limit experiment of Appendix~\ref{app:num_limit}. With $\Delta^{\circ}$ denoting the open simplex, put
\[
B^{\star}(p)\coloneqq\inf_{f\in\Delta^{\circ}}B(p,f),
\qquad
R^{\star}(\pi)\coloneqq\sup_{u\in\bbR^{\calS}}R^{\infty}(u,\pi).
\]
Lemmas~\ref{lem:num_static_lower} and \ref{lem:num_seq_upper} give
\[
B^{\star}(p)
\le
\mathfrak G_{\calS}^{\uparrow,\mathrm{stat}}(s),
\qquad
\mathfrak G_{\calS}^{\uparrow}(s)
\le
R^{\star}(\pi).
\]
Thus, one prior and one policy whose later sampling decisions use earlier observations are enough to prove strict inequality.

\paragraph{The prior and the two-stage policy}
Let $p$ assign probability $1/3$ to each of the three gap vectors
\[
(0,-2,-2),
\qquad
(-2,0,-2),
\qquad
(-2,-2,0).
\]
Let $\pi^{\mathrm{two}}$ denote the following two-stage policy, whose second-stage allocation depends on the first-stage sample means. It first selects each arm with sampling proportion $1/5$. It then removes the arm with the smallest pooled sample mean. The remaining sampling proportion is divided equally between the two retained arms, so each retained arm has final sampling proportion $2/5$. After the sampling phase, the policy chooses the retained arm with the larger pooled sample mean.

\begin{proposition}[Strict comparison between sequential and static sampling]
\label{prop:strict_sequential_static}
For $\calS=\cb{1,2,3}$ and $s=(1,1,1)$,
\[
\mathfrak G_{\calS}^{\uparrow}(s)
\le
0.641
<
0.644
\le
\mathfrak G_{\calS}^{\uparrow,\mathrm{stat}}(s).
\]
\end{proposition}
\begin{proof}
For the prior above, Appendix~\ref{app:num_spec} recursively divides the parameter domain into finitely many rectangles. It evaluates each rectangle using directed rounding: lower bounds are rounded downward and upper bounds are rounded upward. The calculation gives
\[
0.644\le B^{\star}(p).
\]
For the stated two-stage policy, the calculation also gives
\[
R^{\star}\p{\pi^{\mathrm{two}}}\le0.641.
\]
Combining these bounds with Lemmas~\ref{lem:num_static_lower} and \ref{lem:num_seq_upper} yields
\[
\mathfrak G_{\calS}^{\uparrow}(s)
\le
R^{\star}\p{\pi^{\mathrm{two}}}
\le
0.641
<
0.644
\le
B^{\star}(p)
\le
\mathfrak G_{\calS}^{\uparrow,\mathrm{stat}}(s).
\]
\end{proof}

Table~\ref{tab:k3_static_vs_sequential} reports the bounds used in Proposition~\ref{prop:strict_sequential_static} and the numerical values found by a finer search over the continuous parameters.

\begin{table}[t]
\centering
\caption{Static and sequential values for the homoscedastic three-arm comparison.}
\label{tab:k3_static_vs_sequential}
\begin{tabular}{lcc}
\toprule
Quantity & Bound & Value from a finer numerical search \\
\midrule
Bayes risk under static sampling, $B^{\star}(p)$ & $\ge 0.644$ & $0.6451437085$ \\
Worst-case risk of $\pi^{\mathrm{two}}$, $R^{\star}\p{\pi^{\mathrm{two}}}$ & $\le 0.641$ & $0.6227561146$ \\
\bottomrule
\end{tabular}
\end{table}

Proposition~\ref{prop:strict_sequential_static} uses these two quantities to establish the strict comparison; the table does not report the exact value of $\mathfrak G_{\calS}^{\uparrow}(1,1,1)$. The Bayes risk $B(p,f)$ is minimized near equal sampling. The reported static allocation is $(1/3,1/3,1/3)$. The largest risk of the specified two-stage policy occurs near a mean vector for which the two inferior arms have the same gap, with both gaps approximately $1.94935$.

\section{Detailed Literature Review}
\label{appdx:review}
The earliest BAI formulation appeared under the name \emph{ordinal optimization} \citep{Chen2000simulationbudget,Glynn2004largedeviations}, focusing on non-adaptive optimal designs via large-deviation principles.
That literature often assumes that an experimenter knows which arms to select and how often to select them to attain optimality, which requires knowledge of the arms' outcome distributions.
Beginning in the 2010s, BAI was formulated by explicitly addressing the estimation of the optimal sampling rule \citep{Audibert2010bestarm,Bubeck2011pureexploration}.

BAI is typically studied in two settings: the fixed-confidence setting and the fixed-budget setting. In the fixed-confidence setting, we first fix a target error probability $\bbP_{\bmmu}\p{\widehat{a}^\delta_T \neq a^*_{\bmmu}}$, while the sample size $T$ is left unspecified. The strategy selects arms until the probability of misidentification is theoretically guaranteed to be below a pre-specified threshold. This setting is closely related to sequential hypothesis testing. By contrast, fixed-budget BAI aims to minimize the misidentification probability $\bbP_{\bmmu}\p{\widehat{a}^\delta_T \neq a^*_{\bmmu}}$ or the simple regret $\Regret^\delta_T(\bmmu)$ given a fixed sample size $T$. In this study, we focus solely on the fixed-budget setting and refer to it simply as BAI.

\paragraph{Performance measures and uncertainty evaluation}
In BAI, two main performance metrics have been used: the misidentification probability $\bbP_{\bmmu}\p{\widehat{a}^\delta_T \neq a^*_{\bmmu}}$ and the simple regret $\Regret^\delta_T(\bmmu)$. They satisfy the following identity:
\[
\Regret^\delta_T(\bmmu)
=
\sum_{a \in [K]} \Delta_{\bmmu}(a) \bbP_{\bmmu}\p{ \widehat{a}^\delta_T = a },\]
where $\Delta_{\bmmu}(a) \coloneqq \mu_{a^*_{\bmmu}} - \mu_a$ denotes the gap in expected outcomes between the best arm and arm $a$.

Optimality under the probability of misidentification and the simple regret depends on how uncertainty about the bandit instance $\bmP_{\bmmu}$ is evaluated. Throughout, the collection of arm families $\bmP$ is held fixed, so only the mean vector $\bmmu\in\calM^K$ varies. We use three evaluation criteria:
\begin{itemize}
    \item \textbf{Distribution-dependent analysis:} Performance is evaluated at a fixed mean vector $\bmmu\in\calM^K$, that is, at a fixed bandit instance $\bmP_{\bmmu}$.
    \item \textbf{Minimax analysis:} Performance is evaluated under the worst case over $\bmmu\in\calM^K$ by taking the supremum of the risk.
    \item \textbf{Bayesian analysis:} Performance is evaluated by averaging the risk over $\bmmu\in\calM^K$ against a prior $H$.
\end{itemize}
The last two criteria therefore range over the bandit model $\calB\p{\bmP}$ of Section~\ref{sec:parametricdist} for the fixed collection of arm families $\bmP$. Neither criterion optimizes or averages over the choice of $\bmP$.

\paragraph{Distribution-dependent analysis}
In a distribution-dependent analysis of BAI, the misidentification probability and the simple regret are evaluated on the exponential scale in $T$, through $\frac{1}{T}\log \bbP_{\bmmu}\p{\widehat{a}^\delta_T \neq a^*_{\bmmu}}$ and $\frac{1}{T}\log \Regret^\delta_T(\bmmu)$.
At a fixed $\bmmu$ with a unique best arm and positive gaps, the two logarithmic rates coincide. The identity above bounds the simple regret below by $\Delta_{\min}(\bmmu)\allowbreak\bbP_{\bmmu}\p{\widehat{a}^\delta_T \neq a^*_{\bmmu}}$. It bounds the regret above by $\Delta_{\max}(\bmmu)\allowbreak\bbP_{\bmmu}\p{\widehat{a}^\delta_T \neq a^*_{\bmmu}}$. Both gap constants are positive, finite, and independent of $T$. Applying $\frac1T\log$ to the two bounds removes the gap constants. Thus, at the level of fixed-instance logarithmic rates, it suffices to study the misidentification probability $\bbP_{\bmmu}\p{\widehat{a}^\delta_T \neq a^*_{\bmmu}}$. This equivalence of rates fails when the gaps shrink with $T$, as they do under the criteria of this study.

Lower bounds for this probability have been developed by \citet{Kaufmann2014complexity,Kaufmann2016complexity}, extending the classical bounds for regret minimization \citep{Lai1985asymptoticallyefficient,Burnetas1996optimaladaptive}. \citet{Degenne2023existence} asks whether a task admits a complexity, that is, an error exponent attained at every instance by one algorithm, and shows two things: if such a complexity exists it is the one determined by the static oracle proportions, of the kind considered by \citet{Glynn2004largedeviations}, and for several fixed-budget tasks, including two-armed Bernoulli BAI, no such complexity exists.

For two-armed Gaussian problems with known variances, \citet{Kaufmann2014complexity,Kaufmann2016complexity} show that Neyman allocation, under which the sampling proportions are proportional to the standard deviations, is optimal. They also show that when outcomes follow a one-parameter exponential family and the number of arms is two, uniform arm selection is nearly optimal. When variances are unknown, \citet{Kato2025neymanallocation} proves that for two-armed Gaussian problems with unknown variances, Neyman allocation with adaptive variance estimation remains optimal in a local regime where the mean gap is small, while \citet{Wang2024uniformlyoptimal} establishes, for two-armed Bernoulli problems, that uniform arm selection is undominated, in the sense that no algorithm weakly improves on its fixed-instance error exponent at every instance and strictly improves on it at some instance.

For two-armed bandits under more general settings, as well as for bandits with $K \ge 3$ arms, the existence of optimal designs long remained unclear \citep{Kaufmann2020contributions}. While strategies that match the lower bounds have been identified in the fixed-confidence setting \citep{Garivier2016optimalbest}, such strategies have not been found in the fixed-budget setting.
In this setting, there are various technical challenges, including the \emph{reverse Kullback–Leibler (KL) divergence} problem \citep{Kaufmann2020contributions}. \citet{Kasy2021adaptivetreatment} claims to resolve the question by adapting top-two Thompson sampling, originally proposed for fixed-confidence BAI by \citet{Russo2020simplebayesian}. However, \citet{Ariu2021policychoice} identifies a technical issue in the proof and provides a counterexample based on a different lower bound from \citet{Carpentier2016tightlower}.
Subsequent work provides both negative and positive results. On the negative side, \citet{Degenne2023existence} rules out a single algorithm attaining the best fixed-instance error exponent at every instance for several identification tasks. In two-armed Bernoulli A/B testing, \citet{Wang2024uniformlyoptimal} rules out an algorithm that matches uniform sampling everywhere and strictly improves on it somewhere. See also the open problem posed by \citet{Qin2022openproblem}. On the positive side, \citet{Imbens2025admissibilitycompletely} shows that batched elimination designs can strictly dominate completely randomized trials with three or more homoscedastic Gaussian arms, which makes a nonadaptive design inadmissible under an efficiency-exponent criterion. That result is a dominance result, not an impossibility result, and it points the same way as our $K\ge3$ analysis, in which the sampling rule is allowed to respond to every observation.

\paragraph{Minimax and Bayesian analysis}
This study focuses on minimax and Bayesian criteria. These criteria assess performance under uncertainty about $\bmmu$, and they use a different quantifier order from the distribution-dependent criterion: the worst-case mean vector may depend on the budget, and a prior may place mass near ties. The negative results recorded above are therefore neither evaded nor contradicted here, because they concern a fixed instance and an exponential scale, whereas the leading term of the criteria below is polynomial in $T$.

Under these criteria, misidentification probability and regret lead to different conclusions. We begin by explaining the reason for this divergence. For simplicity, consider a sequence $\p{\bmmu_T}_{T\in\bbN}$ of two-armed mean vectors for which arm $1$ is the best arm at every budget. We suppose for this heuristic that the error probability obeys an exponential upper bound with some constant $C>0$, so that
\[
\Regret^\delta_T(\bmmu_T) = \Delta_{\bmmu_T}(2) \cdot \bbP_{\bmmu_T} \p{ \widehat{a}^\delta_T = 2 } \leq \Delta_{\bmmu_T}(2) \cdot \exp \p{ - C T \Delta_{\bmmu_T}(2)^2 }.
\]
This bound yields three cases:
\begin{itemize}
    \item If $\Delta_{\bmmu_T}(2)$ converges to zero at a rate slower than $1/\sqrt{T}$ as $T\to\infty$, then $T\Delta_{\bmmu_T}(2)^2 \to \infty$ as $T\to\infty$. The displayed bound is therefore $o(1/\sqrt{T})$.
    \item If $\Delta_{\bmmu_T}(2) = C_1/\sqrt{T}$ for some constant $C_1 > 0$, then $\exp(- C T \Delta_{\bmmu_T}(2)^2) = \exp(- C C_1^2)$ is constant in $T$. The displayed bound is therefore $C_2/\sqrt{T}$ with $C_2 = C_1\exp(-CC_1^2)$. This is an upper bound of order $1/\sqrt{T}$, not an asymptotic equality.
    \item If $\Delta_{\bmmu_T}(2)$ converges to zero at a rate faster than $1/\sqrt{T}$ as $T\to\infty$, then $\Regret^\delta_T(\bmmu_T) \le \Delta_{\bmmu_T}(2) = o(1/\sqrt{T})$ holds, since the probability is at most one.
\end{itemize}
The first two cases use that schematic exponential upper bound, while the third uses only that a probability is at most one. This calculation therefore gives a heuristic explanation of why the scale $\Delta_{\bmmu_T}(2) = C_1 / \sqrt{T}$ governs the worst-case and Bayesian analyses, and it rests on a schematic exponential bound rather than on a bound that holds for every strategy. It is not by itself a reduction to local alternatives. The lower bounds of Sections~\ref{sec:minimax} and \ref{sec:bayes} establish that reduction rigorously, for the stated model and for the whole class of adaptive experiments.

Minimax rate-optimal designs for simple regret are given in \citet{Bubeck2011pureexploration}, whereas Bayes rate-optimal designs are proposed by \citet{Komiyama2023rateoptimal}. These results achieve optimal convergence rates, but exact constant matching between upper and lower bounds remains unresolved in general.

Our contribution addresses this gap. We derive tight minimax and Bayes lower bounds, including exact constants, and construct a single adaptive design whose simple regret asymptotically attains these bounds. Table~\ref{tab:regret} summarizes existing results and our contributions.

\section{Finite Sequential Gaussian Game and Finite-Grid Duality}
\label{app:finite_grid_duality}
This appendix records an auxiliary sequential experiment, the finite sequential Gaussian game, together with two decision-theoretic facts about its finite-grid restriction: a sufficiency reduction to the reference posterior and a minimax duality. In contrast with a static Gaussian game that fixes all arm selections before any observation, the sequential game below may change the selected arm after every observation, which is the feature used when passing from fixed to adaptive experiments. The duality uses the following classical minimax theorem in a finite-dimensional form. No additional compactness argument is needed.

\begin{theorem}[Kneser--Fan minimax theorem]
\label{thm:kneser_fan}
Let $X$ be a nonempty compact convex subset of a Hausdorff topological vector space and let $Y$ be a nonempty convex subset of a vector space, carrying no topology. Let $f:X\times Y\to\bbR$ be such that $x\mapsto f(x,y)$ is convex and lower semicontinuous for every $y\in Y$, and $y\mapsto f(x,y)$ is concave for every $x\in X$. Then
\[
\inf_{x\in X}\ \sup_{y\in Y}f(x,y)=\sup_{y\in Y}\ \inf_{x\in X}f(x,y),
\]
and the infimum on the left is attained.
\end{theorem}

The theorem is due to \citet{Kneser1952surun} and \citet{Fan1953minimaxtheorems}. See also \citet{Lecam2000asymptoticsin}.

\subsection{The Game and Its Finite-Grid Restriction}

\begin{definition}[Finite sequential Gaussian game]
\label{def:seq_gaussian_game}
Fix a nonempty arm set $\calS\subseteq[K]$ with $\abs{\calS}\ge2$, a horizon $n\ge\abs{\calS}$, and standard deviations $s=\p{s_a}_{a\in\calS}\in(0,\infty)^{\calS}$. Nature chooses a mean vector $u=\p{u_a}_{a\in\calS}\in\bbR^{\calS}$. Write $\calF_t=\widetilde{\sigma}\p{A_1,X_1,\ldots,A_t,X_t}$ for the observed history. At round $t$, a behavioral policy $\pi$ selects an action $A_t\in\calS$ from a distribution measurable with respect to $\calF_{t-1}$ and an external randomization. We write the regular conditional law of $A_t$ given $\calF_{t-1}$ as $\alpha_t\p{\cdot\mid\calF_{t-1}}$. Conditional on $A_t=a$, the policy observes
\[
X_t=\frac{u_a}{\sqrt n}+s_a Z_t,
\qquad
Z_t\sim\calN(0,1),
\]
with the $Z_t$ independent. After round $n$ the policy recommends $\widehat A_n\in\calS$ from a terminal kernel $q_n\p{\cdot\mid\calF_n}$ and incurs the simple regret
\[
\Delta_u^{\calS}\p{\widehat A_n}=\max_{a\in\calS}u_a-u_{\widehat A_n}.
\]
The per-parameter risk is
\[
R_n^{G}\p{\pi,u}
\coloneqq
\bbE_{u,s}^{\pi}\sqb{\Delta_u^{\calS}\p{\widehat A_n}}.
\]

Behavioral policies are closed under randomization performed before the experiment. Let $\p{\pi^{(i)}}_{i\in I}$ be a countable family of behavioral policies, and select an index according to a distribution on $I$ before round $1$. The policy that follows the selected $\pi^{(i)}$ is again behavioral. The selected index is part of the external randomization and remains available throughout the experiment. The game has perfect recall, so each round conditions on the entire past of the selected policy. Thus, the mixture selects $A_t$ from a distribution measurable with respect to $\calF_{t-1}$ and the external randomization. Its terminal choice also has the form required in this definition. This is the elementary direction of Kuhn's equivalence between mixed and behavioral strategies under perfect recall.

The selected index is independent of the outcome noise. Therefore, at every parameter point, the risk of the mixture is the corresponding convex combination of the risks of the $\pi^{(i)}$. This fact makes the risk set convex in Step 1 of the proof of Lemma~\ref{lem:finite_grid_saddle}. For a compact radius $C<\infty$, the compact sequential value is
\[
\mathfrak G_{\calS,n}^{C}(s)
\coloneqq\inf_{\pi}\sup_{u\in[-C,C]^{\calS}}R_n^{G}\p{\pi,u}.
\]
\end{definition}

Fix in addition a finite grid $\calU=\cb{u^{(1)},\ldots,u^{(L)}}\subset[-C,C]^{\calS}$ and write, for a behavioral policy $\pi$,
\[
r_\ell(\pi)\coloneqq R_n^{G}\p{\pi,u^{(\ell)}}
=\bbE_{u^{(\ell)},s}^{\pi}\sqb{\Delta_{u^{(\ell)}}^{\calS}\p{\widehat A_n}},
\qquad
\mathfrak G_{\calS,n}^{\calU}(s)\coloneqq\inf_{\pi}\max_{\ell\in[L]}r_\ell(\pi),
\]
We call $r(\pi)=\p{r_\ell(\pi)}_{\ell\in[L]}$ the risk vector of $\pi$ on $\calU$. The second quantity is the finite-grid value.

\subsection{The Reference Posterior}

Fix a reference prior $r=\p{r_1,\ldots,r_L}$ on $\calU$ with $r_\ell>0$ for every $\ell$. The reference posterior $\Pi_t^r=\p{\Pi_{t,\ell}^r}_{\ell\in[L]}$ is the posterior distribution of the hidden index given $\calF_t$ under $r$. Since the action law is common to all hidden points and cancels from every likelihood ratio,
\[
\Pi_{t,\ell}^r
\propto
r_\ell\prod_{i=1}^t\varphi_{s_{A_i}^2}\p{X_i-\frac{u_{A_i}^{(\ell)}}{\sqrt n}},
\]
where $\varphi_{\sigma^2}$ is the $\calN(0,\sigma^2)$ density and the proportionality is in $\ell$. In particular $\Pi_t^r$ is a measurable function of $\calF_t$, and the one-observation Bayes update is a fixed measurable map $F_{t+1}$,
\[
\Pi_{t+1}^r=F_{t+1}\p{\Pi_t^r,A_{t+1},X_{t+1}},
\qquad
\p{F_{t+1}\p{p,a,x}}_\ell
\propto
p_\ell\,\varphi_{s_a^2}\p{x-\frac{u_a^{(\ell)}}{\sqrt n}},
\]
which does not depend on the policy. We write $P_{\ell,t}^{\pi}$ for the law of $\calF_t$, and $P_\ell^{\pi}$ for the full path law of $\p{A_1,X_1,\ldots,A_n,X_n,\widehat A_n}$, under $u^{(\ell)}$ and policy $\pi$, and $M_t^r=\sum_{\ell=1}^L r_\ell P_{\ell,t}^{\pi}$ for the reference-mixture law.

\subsection{Sufficiency and Duality}

\begin{lemma}[Sufficiency of the posterior under a full-support reference prior]
\label{lem:full_support_posterior_sufficiency}
Fix the finite grid $\calU$, a standard-deviation vector $s$, and a reference prior $r$ with $r_\ell>0$ for every $\ell$. For any behavioral policy $\pi$ there is a policy $\pi^r$ that is Markov in $\p{t,\Pi_t^r}$ such that, for every hidden point $\ell\in[L]$ and every time $t$, the one-time marginal law of the reference posterior and the law of the recommendation are preserved:
\[
\calL_{u^{(\ell)},s}^{\pi^r}\p{\Pi_t^r}
=\calL_{u^{(\ell)},s}^{\pi}\p{\Pi_t^r}
\quad(0\le t\le n),
\qquad
\calL_{u^{(\ell)},s}^{\pi^r}\p{\widehat A_n}
=\calL_{u^{(\ell)},s}^{\pi}\p{\widehat A_n}.
\]
In particular every coordinate risk is preserved, $R_n^{G}\p{\pi^r,u^{(\ell)}}=R_n^{G}\p{\pi,u^{(\ell)}}$ for every $\ell\in[L]$.
\end{lemma}

\begin{remark}
The full joint law of the posterior process $\p{\Pi_0^r,\ldots,\Pi_n^r}$, and a fortiori that of the raw history and actions, is in general not preserved: under $\pi$ the process $\p{\Pi_t^r}$ need not be Markov, because $\pi$ may act on features of the history that are ancillary given $\Pi_t^r$. Only the one-time marginals of the posterior and the recommendation law are preserved. These quantities determine the risks $R_n^G\p{\pi,u^{(\ell)}}$ at the grid points $u^{(\ell)}$, which are the quantities used below.
\end{remark}

\begin{proof}
Throughout, abbreviate $\calL_\ell^{\pi}=\calL_{u^{(\ell)},s}^{\pi}$ and $\calL_\ell^{\pi^r}=\calL_{u^{(\ell)},s}^{\pi^r}$.
Dominance and Bayes' formula give $dP_{\ell,t}^{\pi}/dM_t^r=\Pi_{t,\ell}^r/r_\ell$. This Radon--Nikodym derivative is a measurable function of $\Pi_t^r$. The Fisher--Neyman factorization therefore shows that $\Pi_t^r$ is sufficient for the hidden index. Hence, for every bounded measurable functional $\varphi$ of the history through time $t$,
\begin{align}
\bbE_{P_{\ell,t}^{\pi}}\sqb{\varphi\mid\Pi_t^r}
=\bbE_{M_t^r}\sqb{\varphi\mid\Pi_t^r}
\qquad\text{for every }\ell.
\label{eq:seq_suff}
\end{align}
All conditional expectations are evaluated for fixed measurable versions, and \eqref{eq:seq_suff} holds $M_t^r$-almost surely, hence $P_{\ell,t}^{\pi}$-almost surely for every $\ell$, these laws being mutually absolutely continuous. Define the action kernel of $\pi^r$ at time $t+1$ and its terminal kernel by
\[
\alpha_{t+1}^r\p{\cdot\mid\Pi_t^r}
=\bbE_{M_t^r}\sqb{\alpha_{t+1}\p{\cdot\mid\calF_t}\mid\Pi_t^r},
\qquad
q_n^r\p{\cdot\mid\Pi_n^r}
=\bbE_{M_n^r}\sqb{q_n\p{\cdot\mid\calF_n}\mid\Pi_n^r},
\]
fixed as measurable maps of $\Pi_t^r$ respectively $\Pi_n^r$ and defined arbitrarily on the null set where the conditioning is undefined, which by \eqref{eq:seq_suff} equal the corresponding $P_\ell^{\pi}$-conditional expectations for every $\ell$ simultaneously.

Fix $\ell$ and $t$. Conditioning on $\calF_t$ and then on $\Pi_t^r$, for every bounded measurable $\psi$,
\begin{align*}
&\bbE_{P_\ell^{\pi}}\sqb{\psi\p{\Pi_{t+1}^r}\mid\Pi_t^r}\\
&=\sum_{a\in\calS}\alpha_{t+1}^r\p{a\mid\Pi_t^r}
\int\psi\p{F_{t+1}\p{\Pi_t^r,a,x}}\,
d\calN\p{u_a^{(\ell)}/\sqrt n,s_a^2}(x)
=\p{Q_{t,\ell}\psi}\p{\Pi_t^r}.
\end{align*}
Here, $F_{t+1}$ is the fixed Bayes update. Conditional on $\calF_t$, the action has law $\alpha_{t+1}\p{\cdot\mid\calF_t}$. Conditional on $A_{t+1}=a$ and the hidden point $\ell$, the observation is independent of $\calF_t$ and has distribution $\calN\p{u_a^{(\ell)}/\sqrt n,s_a^2}$. Equation~\eqref{eq:seq_suff} then reduces the conditional action law to $\alpha_{t+1}^r$. The right-hand side depends on the history only through $\Pi_t^r$. Taking expectations yields the marginal recursion
\[
\calL_\ell^{\pi}\p{\Pi_{t+1}^r}
=\int Q_{t,\ell}\p{\cdot\mid z}\,d\calL_\ell^{\pi}\p{\Pi_t^r=z}.
\]
The policy $\pi^r$ selects $A_{t+1}$ according to $\alpha_{t+1}^r\p{\cdot\mid\Pi_t^r}$, observes the same Gaussian, and applies the same $F_{t+1}$, so its posterior marginal obeys the identical recursion with the identical kernel $Q_{t,\ell}$. Both recursions start from the common deterministic value $\Pi_0^r$. Induction on $t$ gives $\calL_\ell^{\pi}\p{\Pi_t^r}=\calL_\ell^{\pi^r}\p{\Pi_t^r}$ for every $t$ and every $\ell$.

Finally, by \eqref{eq:seq_suff}, the conditional law of $\widehat A_n$ given $\Pi_n^r$ is $q_n^r\p{\cdot\mid\Pi_n^r}$ under $P_\ell^{\pi}$. The policy $\pi^r$ uses the same terminal kernel. Hence, we have
\[
\calL_\ell^{\pi}\p{\widehat A_n}
=\int q_n^r\p{\cdot\mid z}\,d\calL_\ell^{\pi}\p{\Pi_n^r}
=\int q_n^r\p{\cdot\mid z}\,d\calL_\ell^{\pi^r}\p{\Pi_n^r}
=\calL_\ell^{\pi^r}\p{\widehat A_n}.
\]
Only the single conditioning on $\Pi_t^r$ is used, so no Markov property of $\p{\Pi_t^r}$ under $\pi$ is claimed. Since $\Delta_{u^{(\ell)}}^{\calS}\p{\widehat A_n}$ depends only on $\widehat A_n$, the risks at the points $u^{(\ell)}$ coincide.
\end{proof}

\begin{lemma}[Finite-grid minimax duality]
\label{lem:finite_grid_saddle}
Fix $\calS$, the finite grid $\calU=\cb{u^{(1)},\ldots,u^{(L)}}\subset[-C,C]^{\calS}$, a standard-deviation vector $s$, and a horizon $n$. Let
\[
\mathfrak R\coloneqq\cb{\p{r_1(\pi),\ldots,r_L(\pi)}\colon\pi\text{ a behavioral policy}}\subseteq\bbR^L
\]
be the risk set. Then the finite-grid game has a value,
\[
\mathfrak G_{\calS,n}^{\calU}(s)
=\inf_{\pi}\max_{\ell\in[L]}r_\ell(\pi)
=\max_{p\in\calP([L])}\inf_{\pi}\sum_{\ell=1}^L p_\ell r_\ell(\pi),
\]
the maximum over priors is attained, and for every $\varepsilon>0$ there is a behavioral policy whose worst-case coordinate risk is at most the value plus $\varepsilon$.
\end{lemma}

\begin{remark}
The lemma asserts the value identity, the existence of a maximizing prior, and the existence of $\varepsilon$-minimax policies. It does not assert the existence of an exact minimax policy, which is unnecessary for the arguments below.
\end{remark}

\begin{proof}
\emph{Step 1 (the risk set is convex and bounded).}
Each loss $\Delta^{\calS}_{u^{(\ell)}}\p{a}=\max_{b\in\calS}u_b^{(\ell)}-u_a^{(\ell)}$ lies in $[0,2C]$ because $u^{(\ell)}\in[-C,C]^{\calS}$, so $\mathfrak R\subseteq[0,2C]^L$ is bounded. Given two behavioral policies $\pi^{(1)},\pi^{(2)}$ and $\lambda\in[0,1]$, choose one of the two policies before the first observation, with probabilities $\lambda$ and $1-\lambda$. The resulting policy is behavioral. Its coordinate risk at $\ell$ is $\lambda r_\ell\p{\pi^{(1)}}+(1-\lambda)r_\ell\p{\pi^{(2)}}$. Hence, $\mathfrak R$ is convex. Its closure $\overline{\mathfrak R}$ is therefore a nonempty compact convex subset of $\bbR^L$.

\emph{Step 2 (the Bayes value of a prior).}
For $p\in\calP([L])$ set $\phi(p)\coloneqq\inf_\pi\sum_{\ell}p_\ell r_\ell(\pi)=\inf_{x\in\mathfrak R}\langle p,x\rangle$, which is finite because $\mathfrak R$ is nonempty and bounded. Because $x\mapsto\langle p,x\rangle$ is continuous and $\mathfrak R$ is dense in $\overline{\mathfrak R}$,
\[
\phi(p)=\inf_{x\in\mathfrak R}\langle p,x\rangle=\min_{x\in\overline{\mathfrak R}}\langle p,x\rangle.
\]

\emph{Step 3 (finite-dimensional duality).}
Apply the Kneser--Fan minimax theorem (Theorem~\ref{thm:kneser_fan}) with $X=\overline{\mathfrak R}$, a nonempty compact convex subset of the Hausdorff topological vector space $\bbR^L$, with $Y=\calP([L])$, a nonempty convex set on which no topology is imposed, and with the payoff $f(x,p)=\langle p,x\rangle$, which is convex and continuous in $x$ for each fixed $p$ and concave in $p$ for each fixed $x$. The theorem gives
\[
\min_{x\in\overline{\mathfrak R}}\ \sup_{p\in\calP([L])}\langle p,x\rangle
=\sup_{p\in\calP([L])}\ \min_{x\in\overline{\mathfrak R}}\langle p,x\rangle,
\]
with the outer minimum on the left attained. The supremum on the right is a maximum, because $p\mapsto\min_{x\in\overline{\mathfrak R}}\langle p,x\rangle$ is concave and upper semicontinuous on the compact simplex $\calP([L])$.

\emph{Step 4 (translation to policies).}
For every $x\in\bbR^L$ one has $\sup_{p\in\calP([L])}\langle p,x\rangle=\max_{\ell\in[L]}x_\ell$. Since $x\mapsto\max_\ell x_\ell$ is continuous and $\mathfrak R$ is dense in $\overline{\mathfrak R}$,
\[
\min_{x\in\overline{\mathfrak R}}\max_{\ell}x_\ell
=\inf_{x\in\mathfrak R}\max_{\ell}x_\ell
=\inf_\pi\max_{\ell}r_\ell(\pi)
=\mathfrak G^{\calU}_{\calS,n}(s).
\]
Combining with Steps 2 and 3,
\[
\mathfrak G^{\calU}_{\calS,n}(s)
=\max_{p\in\calP([L])}\min_{x\in\overline{\mathfrak R}}\langle p,x\rangle
=\max_{p\in\calP([L])}\phi(p)
=\max_{p\in\calP([L])}\inf_\pi\sum_{\ell}p_\ell r_\ell(\pi).
\]
This proves the asserted value identity and shows that the maximizing prior is attained. Finally, the definition of the infimum in $\inf_\pi\max_\ell r_\ell(\pi)$ furnishes, for every $\varepsilon>0$, a behavioral policy whose worst-case coordinate risk is at most $\mathfrak G^{\calU}_{\calS,n}(s)+\varepsilon$.
\end{proof}

\section{Sequential Gaussian Game Value and Bellman Recursion}
\label{app:seq_value_bellman}
This appendix builds on the finite sequential Gaussian game of Appendix~\ref{app:finite_grid_duality}. It establishes the limit of $\mathfrak G_{\calS,n}^{C}(s)$ as the horizon goes to infinity for fixed $C$, then takes the supremum of that limit over $C$, and derives the posterior Bellman recursion for the Bayes value under a full-support prior. These results are used in the comparison arguments below. Throughout, $\calS\subseteq[K]$ is a nonempty arm set with $\abs{\calS}\ge2$, $s=\p{s_a}_{a\in\calS}\in(0,\infty)^{\calS}$, and the game, its risks $R_n^{G}\p{\pi,u}$, the simple regret $\Delta_u^{\calS}$, and the value $\mathfrak G_{\calS,n}^{C}(s)$ on $[-C,C]^{\calS}$ are those of Definition~\ref{def:seq_gaussian_game}.

\subsection{Global Sequential Value and the Minimax Constant}

\begin{lemma}[Uniform finite upper bound]
\label{lem:seq_uniform_finite}
For every $\calS$ and $s$ there is a finite constant $M_{\calS}(s)$ such that
\[
\sup_{C<\infty}\ \sup_{n\ge2\abs{\calS}}\mathfrak G_{\calS,n}^{C}(s)\le M_{\calS}(s).
\]
\end{lemma}
\begin{proof}
Fix $n\ge2\abs{\calS}$ and consider the nonadaptive policy that selects the arms in a fixed balanced order, so that every arm is selected $N_a\ge n/\p{2\abs{\calS}}$ times, and chooses the arm with the largest sample mean, ties broken by the fixed rule of Section~\ref{sec:prob}. Fix $u\in\bbR^{\calS}$, let $a^\star$ be a best arm, and write $d_b=u_{a^\star}-u_b\ge0$. Under $u$ the sample means of arms $a^\star$ and $b$ are independent Gaussians, and an incorrect choice of $b$ requires the sample mean of $b$ to exceed that of $a^\star$, so
\[
\bbP_{u,s}\p{\widehat A_n=b}
\le
\Phi\p{-\frac{d_b}{\sqrt{2\abs{\calS}\p{s_{a^\star}^2+s_b^2}}}}.
\]
Hence
\begin{align*}
&\bbE_{u,s}\sqb{\Delta_u^{\calS}\p{\widehat A_n}}
\le
\sum_{b\ne a^\star}d_b
\Phi\p{-\frac{d_b}{\sqrt{2\abs{\calS}\p{s_{a^\star}^2+s_b^2}}}}
\le
\abs{\calS}\sqrt{2\abs{\calS}\overline{s}_{\calS}^2}\sup_{x\ge0}x\Phi(-x),\\
&\overline{s}_{\calS}^2\coloneqq\max_{a,b\in\calS}\p{s_a^2+s_b^2},
\end{align*}
using $d_b\Phi\p{-d_b/\kappa}=\kappa\p{d_b/\kappa}\Phi\p{-d_b/\kappa}\le\kappa\sup_{x\ge0}x\Phi(-x)$ with $\kappa=\sqrt{2\abs{\calS}\p{s_{a^\star}^2+s_b^2}}$. The right-hand side is finite and independent of $u$, $C$, and $n$, and bounds $\mathfrak G_{\calS,n}^{C}(s)$ because the balanced policy is admissible on every box $[-C,C]^{\calS}$.
\end{proof}

\begin{theorem}[Limit over the number of rounds at a fixed radius]
\label{thm:seq_value_limit}
For every $\calS$, $s$, and $C<\infty$, the limit
\[
\mathfrak g_{\calS}^{C}(s)\coloneqq\lim_{n\to\infty}\mathfrak G_{\calS,n}^{C}(s)
=\inf_{n\ge\abs{\calS}}\mathfrak G_{\calS,n}^{C}(s)
\]
exists, and the global sequential value
\[
\mathfrak G_{\calS}^{\uparrow}(s)\coloneqq\sup_{C<\infty}\mathfrak g_{\calS}^{C}(s)
\]
is finite.
\end{theorem}
\begin{proof}
Fix $q\ge\abs{\calS}$ and a horizon $N\ge q$. Put $m_N=\lfloor N/q\rfloor\ge1$ and $c_N=\sqrt{m_Nq/N}\in(0,1]$. Use only the first $m_Nq$ observations. Given a behavioral $q$-round policy $\pi_q$, form an $N$-round policy as follows. Each time $\pi_q$ selects an arm, the $N$-round policy selects that arm for $m_N$ consecutive rounds and supplies the normalized aggregate to $\pi_q$. During the remaining $N-m_Nq<q$ rounds, it selects an arbitrary fixed arm in $\calS$ and discards the outcomes. Its terminal recommendation is the recommendation of $\pi_q$. Thus, it is a horizon-$N$ behavioral policy whose regret equals the regret of $\pi_q$ in the aggregated experiment. If the parameter in the $N$-round game is $u$ and the $q$-round policy selects arm $a$, then the aggregate has law
\[
\frac1{\sqrt{m_N}}\sum_{j=1}^{m_N}X_{a,j}
=\frac{c_Nu_a}{\sqrt q}+s_aZ,
\qquad Z\sim\calN(0,1).
\]
The $q$ aggregates are conditionally independent given the arms selected by the $q$-round policy. Applying $\pi_q$ to the parameter $c_Nu$ in the $q$-round game and using the positive homogeneity $\Delta_u^{\calS}=c_N^{-1}\Delta_{c_Nu}^{\calS}$ gives, for every $u$,
\[
R_N^{G}\p{\pi,u}=c_N^{-1}R_q^{G}\p{\pi_q,c_Nu}.
\]
Since $c_N\le1$, one has $c_Nu\in[-C,C]^{\calS}$ whenever $u\in[-C,C]^{\calS}$, so taking suprema over the box and then the infimum over $\pi_q$ yields
\[
\mathfrak G_{\calS,N}^{C}(s)\le c_N^{-1}\mathfrak G_{\calS,q}^{C}(s).
\]
As $N\to\infty$, $c_N\to1$, hence $\limsup_{N\to\infty}\mathfrak G_{\calS,N}^{C}(s)\le\mathfrak G_{\calS,q}^{C}(s)$. Taking the infimum over $q$ gives
\[
\limsup_{N\to\infty}\mathfrak G_{\calS,N}^{C}(s)
\le\inf_{q\ge\abs{\calS}}\mathfrak G_{\calS,q}^{C}(s)
\le\liminf_{N\to\infty}\mathfrak G_{\calS,N}^{C}(s),
\]
This proves that the limit exists and equals the infimum. Finiteness of $\mathfrak G_{\calS}^{\uparrow}(s)$ follows from Lemma~\ref{lem:seq_uniform_finite}, since $\mathfrak g_{\calS}^{C}(s)\le M_{\calS}(s)$ for every $C$.
\end{proof}

\begin{lemma}[Basic properties of the global sequential value]
\label{lem:seq_game_properties}
The compact finite-horizon values satisfy the scaling identity $\mathfrak G_{\calS,n}^{C}(cs)=c\mathfrak G_{\calS,n}^{C/c}(s)$ for every $c>0$, are coordinatewise nondecreasing in $s$, and are nondecreasing in $C$. Consequently, the global sequential value $\mathfrak G_{\calS}^{\uparrow}$ is positively homogeneous, $\mathfrak G_{\calS}^{\uparrow}(cs)=c\mathfrak G_{\calS}^{\uparrow}(s)$, and coordinatewise nondecreasing in $s$. It also satisfies the following continuity bound under relative perturbations: if $(1-r)s_a\le s_a'\le(1+r)s_a$ for every $a$ with $0<r<1$, then
\[
(1-r)\mathfrak G_{\calS}^{\uparrow}(s)
\le
\mathfrak G_{\calS}^{\uparrow}(s')
\le
(1+r)\mathfrak G_{\calS}^{\uparrow}(s).
\]
At every fixed radius $C<\infty$, the limiting value $\mathfrak g_{\calS}^{C}$ in Theorem~\ref{thm:seq_value_limit} has the same three properties: it satisfies $\mathfrak g_{\calS}^{C}(cs)=c\mathfrak g_{\calS}^{C/c}(s)$, is coordinatewise nondecreasing in $s$ and nondecreasing in $C$, and, under the same hypothesis on $s'$,
\begin{align}
(1-r)\mathfrak g_{\calS}^{C}(s)
\le
\mathfrak g_{\calS}^{C}(s')
\le
(1+r)\mathfrak g_{\calS}^{C}(s).
\label{eq:g_mult_continuity}
\end{align}
\end{lemma}
\begin{proof}
For the scaling identity, consider the game with standard deviations $cs$ and radius $C$. An observation from arm $a$ under parameter $u\in[-C,C]^{\calS}$ is $X_t=u_a/\sqrt n+cs_aZ_t$. Dividing every observation by $c$ gives $X_t/c=\p{u_a/c}/\sqrt n+s_aZ_t$. This is an invertible transformation of the data. As $u$ ranges over $[-C,C]^{\calS}$, the transformed parameter $u/c$ ranges over $[-C/c,C/c]^{\calS}$. Since $\Delta_u^{\calS}=c\Delta_{u/c}^{\calS}$, the risk at $u$ in the $cs$-game equals $c$ times the risk at $u/c$ in the transformed $s$-game. Taking the infimum over policies and the supremum over the box gives $\mathfrak G_{\calS,n}^{C}(cs)=c\mathfrak G_{\calS,n}^{C/c}(s)$.

For monotonicity in $s$, the game with standard deviations $s$ can reproduce the noisier game with $s_a'\ge s_a$. Whenever arm $a$ is selected, add an independent $\calN\p{0,(s_a')^2-s_a^2}$ variable to the observation. Any policy for the noisier game then runs on the less noisy game with the same regret. Hence, $\mathfrak G_{\calS,n}^{C}(s)\le\mathfrak G_{\calS,n}^{C}(s')$.

Monotonicity in $C$ holds because the box $[-C,C]^{\calS}$ grows with $C$. Passing first to the limit as $n\to\infty$ and then to the supremum over $C$ preserves the three properties and gives homogeneity and monotonicity of $\mathfrak G_{\calS}^{\uparrow}$. At a fixed $C$, the limit as $n\to\infty$ gives the scaling identity and the two monotonicities for $\mathfrak g_{\calS}^{C}$. For multiplicative continuity, monotonicity and the scaling identity give
\[
\mathfrak G_{\calS}^{\uparrow}(s')
\ge
\mathfrak G_{\calS}^{\uparrow}\p{(1-r)s}
=(1-r)\mathfrak G_{\calS}^{\uparrow}(s),
\qquad
\mathfrak G_{\calS}^{\uparrow}(s')
\le
\mathfrak G_{\calS}^{\uparrow}\p{(1+r)s}
=(1+r)\mathfrak G_{\calS}^{\uparrow}(s),
\]
which is the claim. The same two steps at a fixed radius give
\begin{align*}
&\mathfrak g_{\calS}^{C}(s')
\ge
\mathfrak g_{\calS}^{C}\p{(1-r)s}
=(1-r)\mathfrak g_{\calS}^{C/(1-r)}(s)
\ge(1-r)\mathfrak g_{\calS}^{C}(s),\\
&\mathfrak g_{\calS}^{C}(s')
\le
(1+r)\mathfrak g_{\calS}^{C/(1+r)}(s)
\le(1+r)\mathfrak g_{\calS}^{C}(s),
\end{align*}
using $C/(1-r)\ge C\ge C/(1+r)$ and the monotonicity of $\mathfrak g_{\calS}^{C}$ in $C$, which is \eqref{eq:g_mult_continuity}.
\end{proof}

\begin{lemma}[Uniform horizon convergence on compact sets of standard-deviation vectors]
\label{lem:uniform_horizon_variance_box}
Fix $\calS$, $C<\infty$, and $0<\underline s<\overline s<\infty$. Then
\[
\sup_{s\in[\underline s,\overline s]^{\calS}}
\abs{\mathfrak G_{\calS,n}^{C}(s)-\mathfrak g_{\calS}^{C}(s)}
\longrightarrow0
\qquad(n\to\infty).
\]
\end{lemma}
\begin{proof}
By the scaling identity and the two monotonicities of Lemma~\ref{lem:seq_game_properties}, if $(1-r)s\le s'\le(1+r)s$ coordinatewise then
\[
\mathfrak G_{\calS,n}^{C}(s')
\ge
\mathfrak G_{\calS,n}^{C}\p{(1-r)s}
=(1-r)\mathfrak G_{\calS,n}^{C/(1-r)}(s)
\ge(1-r)\mathfrak G_{\calS,n}^{C}(s).
\]
The symmetric inequality is $\mathfrak G_{\calS,n}^{C}(s')\le(1+r)\mathfrak G_{\calS,n}^{C/(1+r)}(s)\le(1+r)\mathfrak G_{\calS,n}^{C}(s)$, where we use $C/(1-r)\ge C\ge C/(1+r)$. The same two-sided bound holds for $\mathfrak g_{\calS}^{C}$ by \eqref{eq:g_mult_continuity}.

Lemma~\ref{lem:seq_uniform_finite} bounds $\mathfrak G_{\calS,n}^{C}$ and $\mathfrak g_{\calS}^{C}$ by a common constant on $[\underline s,\overline s]^{\calS}$. Hence, both are uniformly equicontinuous in the multiplicative metric on this box, with a modulus independent of $n$.

Fix $\varepsilon>0$. Choose $r>0$ so that the two-sided multiplicative slack contributes at most $\varepsilon$ uniformly, and take a finite multiplicative $r$-net $\cb{s^{(1)},\ldots,s^{(M_r)}}$ of $[\underline s,\overline s]^{\calS}$. Theorem~\ref{thm:seq_value_limit} gives $\mathfrak G_{\calS,n}^{C}\p{s^{(i)}}\to\mathfrak g_{\calS}^{C}\p{s^{(i)}}$ at each net point. Therefore, the maximum error over the finite net converges to $0$ as $n\to\infty$. Equicontinuity bounds the error at an arbitrary $s$ by the net error plus $\varepsilon$. Letting $n\to\infty$ and then $\varepsilon\downarrow0$ proves the claim.
\end{proof}

\begin{theorem}[Quantitative finite-horizon approximation on a finite parameter set]
\label{thm:finite_horizon_rate}
Fix $\calS$, $C<\infty$, a finite parameter set $\calU=\cb{u^{(1)},\ldots,u^{(L)}}\subset[-C,C]^{\calS}$, and $0<\underline w<\overline w<\infty$. There is a finite constant $A^{\mathrm{hor}}=A^{\mathrm{hor}}\p{\calS,C,\calU,\underline w,\overline w}$ such that, for every $s\in[\underline w,\overline w]^{\calS}$ and every $q\ge\abs{\calS}$,
\[
0
\le
\mathfrak G_{\calS,q}^{\calU}(s)
-
\inf_{r\ge\abs{\calS}}\mathfrak G_{\calS,r}^{\calU}(s)
\le
A^{\mathrm{hor}}q^{-1/4}.
\]
If $C$ is supplied with a rational upper bound and $\underline w$ with a positive rational lower bound, then a rational upper bound on $A^{\mathrm{hor}}$ can be computed from these bounds, $L$, and $\abs{\calS}$.
\end{theorem}
\begin{proof}
Fix a prior $p\in\calP([L])$ on $\calU$. For $a\in\calS$ and $\ell\in[L]$, put
\[
\overline u_a(\pi)=\sum_{r=1}^L\pi_ru_a^{(r)},
\qquad
\vartheta_{a,\ell}^{s}(\pi)
=
\pi_\ell\frac{u_a^{(\ell)}-\overline u_a(\pi)}{s_a}.
\]
The posterior probability vector satisfies the controlled diffusion below. Let $P_\Delta$ be the Euclidean projection of $\bbR^L$ onto the probability simplex $\calP([L])$, and extend $\vartheta_a^s$ to $\bbR^L$ by composition with $P_\Delta$. On a filtered probability space carrying a Brownian motion, consider the controlled diffusion
\[
\rmd\Pi_{t,\ell}=\vartheta_{A_t,\ell}^{s}(\Pi_t)\rmd W_t,
\qquad
\Pi_0=p.
\]
The extended coefficients are bounded and globally Lipschitz. Therefore, the equation has a unique strong solution. On the simplex, $\sum_\ell\vartheta_{a,\ell}^{s}=0$. Hence, $\sum_\ell\Pi_{t,\ell}=1$. Moreover, each coordinate has the form $\rmd\Pi_{t,\ell}=\Pi_{t,\ell}h_{t,\ell}\rmd W_t$ with bounded $h_{t,\ell}$. A coordinate that starts at zero remains zero, while a positive coordinate has the stochastic-exponential representation and cannot reach zero. Thus, $\Pi_t$ remains in $\calP([L])$.

The terminal Bayes loss is
\[
g(\pi)
=
\min_{b\in\calS}
\sum_{\ell=1}^L\pi_\ell\Delta_{u^{(\ell)}}^{\calS}(b).
\]
We next identify the constant-action transition of the diffusion. Let a hidden index $J$ have distribution $\pi$, and suppose that arm $a$ is selected for an interval of length $h$ in the Gaussian signal $\rmd Y_t=u_a^{(J)}\rmd t+s_a\rmd B_t$. If the observation increment over this interval is $D$, Bayes' formula gives
\[
\Pi_{t+h,\ell}
=
\frac{
\pi_\ell\exp\p{u_a^{(\ell)}D/s_a^2-\p{u_a^{(\ell)}}^2h/(2s_a^2)}
}{
\sum_{r=1}^L
\pi_r\exp\p{u_a^{(r)}D/s_a^2-\p{u_a^{(r)}}^2h/(2s_a^2)}
}.
\]
The Bayes posterior in this signal experiment solves the posterior diffusion for $\Pi_t$ defined above, with its innovation Brownian motion. The observation filtration and the filtration generated by the innovation process and the external randomization coincide: the innovation is obtained by centering the observations with the posterior mean, while the observations are recovered from the innovation equation. Thus, the admissible controls in the posterior diffusion are the observation-based controls. Uniqueness in law therefore identifies the posterior update given by Bayes' formula above as the constant-action transition of that diffusion. For $h=1/q$, the normalized increment $X=\sqrt qD$ has law $\calN\p{u_a^{(\ell)}/\sqrt q,s_a^2}$ under the hidden point $u^{(\ell)}$.

Thus, the diffusion sampled at the grid times has the same controlled transition kernel and terminal loss as the posterior-state formulation of the $q$-round Gaussian game. Any control that is constant on the grid intervals induces, through its regular conditional action laws at the grid times, a behavioral policy for this finite controlled Markov chain. Conditional on an endpoint increment, the Brownian bridge over the interval is independent of the hidden index. Any dependence of a later action on that bridge can therefore be reproduced by the external randomization allowed for a behavioral policy. Conversely, every behavioral policy of the $q$-round game defines such a piecewise-constant control.

The finite-horizon dynamic-programming argument of Lemma~\ref{lem:posterior_markovization} therefore gives equality of the two Bayes values. If some coordinates of $p$ are zero, they remain zero and the same argument applies on the corresponding face of the simplex.

The projection $P_\Delta$ is $1$-Lipschitz. Uniformly over $s\in[\underline w,\overline w]^{\calS}$, it follows that
\[
\norm{\vartheta_a^s(\pi)}
\le
\frac{2C}{\underline w},
\qquad
\operatorname{Lip}\p{\vartheta_a^s}
\le
\frac{C\p{2+\sqrt L}}{\underline w}.
\]
The extension of $g$ by composition with $P_\Delta$ satisfies
\[
\norm g_\infty\le2C,
\qquad
\operatorname{Lip}(g)\le2C\sqrt L.
\]
Put
\[
C_0^{\mathrm{hor}}
=
\frac{C\p{2+\sqrt L}}{\underline w},
\qquad
C_1^{\mathrm{hor}}
=
2C\sqrt L.
\]
These constants bound the coefficients and their spatial Lipschitz constants and the terminal loss and its spatial Lipschitz constant, respectively. The action set $\calS$ is finite, the coefficients are time-independent, and the running cost is zero. The controlled diffusion therefore satisfies the assumptions of \citet[Theorem~2.1]{Jakobsen2019improvedorder}. Apply that result to the maximization problem with terminal reward $-g$. If $V^p(s)$ denotes the unrestricted minimum Bayes loss and $V_q^p(s)$ denotes the minimum when the action is held constant on the $q$ equal time intervals, then
\[
0\le V_q^p(s)-V^p(s)\le A^{\mathrm{hor}}q^{-1/4}.
\]
The constant is uniform in the initial posterior $p$ and in $s\in[\underline w,\overline w]^{\calS}$ because $C_0^{\mathrm{hor}}$ and $C_1^{\mathrm{hor}}$ are uniform.

The quantitative proof of the cited theorem gives, before setting its smoothing parameter equal to $q^{-1/4}$, an upper bound of the form
\[
\overline C\p{\varepsilon+q^{-1/2}+q^{-1}\varepsilon^{-3}}.
\]
We now specify how to obtain an upper bound for $\overline C$. Repeat Propositions~2.1--2.3 and Steps~1--4 in the proof of the cited theorem with the bounds $C_0^{\mathrm{hor}}$ and $C_1^{\mathrm{hor}}$. The continuous-dependence estimates for the diffusions follow from the $L^2$ Doob inequality with constant $4$, It\^o isometry, and Gronwall inequality.

For the regularization, use the product of the time kernel proportional to $t^5(1-t)^5\mathbbm 1[0<t<1]$ and the spatial kernels proportional to $\p{1-x^2}^5\mathbbm 1[\abs x<1]$. These kernels are $C^4$, which is sufficient because only derivatives satisfying $2m+k\le4$ enter the proof. Their normalization constants are rational. Rational upper bounds for the $L^1$-norms of the required derivatives follow from the support volume and the sums of the absolute values of the polynomial coefficients.

Gaussian moments have rational upper bounds, and the exponential in the Gronwall bound has rational upper bounds from its Taylor series with a remainder bound. All suprema over actions are finite maxima. Following the cited proof in order and replacing each generic constant by a rational upper bound on the displayed right-hand side at that step yields a rational upper bound for $\overline C$ from $L$, $C_0^{\mathrm{hor}}$, and $C_1^{\mathrm{hor}}$. We use three times this upper bound as $A^{\mathrm{hor}}$, because $\varepsilon=q^{-1/4}$ makes every term in the preceding display at most $q^{-1/4}$.

Finite-grid minimax duality gives
\[
\mathfrak G_{\calS,q}^{\calU}(s)=\max_{p\in\calP([L])}V_q^p(s).
\]
Taking the maximum over $p$ in the preceding bound gives
\[
0
\le
\mathfrak G_{\calS,q}^{\calU}(s)-\max_{p\in\calP([L])}V^p(s)
\le
A^{\mathrm{hor}}q^{-1/4}.
\]
Every finite-horizon value is at least $\max_pV^p(s)$, and the upper bound converges to $\max_pV^p(s)$ as $q\to\infty$. Hence, $\max_pV^p(s)=\inf_{r\ge\abs{\calS}}\mathfrak G_{\calS,r}^{\calU}(s)$. This proves the claim.
\end{proof}

\begin{lemma}[Exact binary sequential value]
\label{lem:binary_sequential_value}
For $\calS=\cb{a,b}$,
\[
\mathfrak G_{\calS}^{\uparrow}\p{s_a,s_b}=\p{s_a+s_b}c_{\mathrm{mm}},
\qquad
c_{\mathrm{mm}}=\sup_{x\ge0}x\Phi(-x).
\]
\end{lemma}
\begin{proof}
For the upper bound, fix deterministic counts $N_a+N_b=n$. Select each arm the prescribed number of times and choose the arm with the larger sample mean. If $d=\abs{u_a-u_b}$, the difference between the two sample means is Gaussian with mean $\pm d/\sqrt n$ and variance $s_a^2/N_a+s_b^2/N_b=c_n^2/n$, where $c_n=\sqrt{n\p{s_a^2/N_a+s_b^2/N_b}}$. Thus, an incorrect recommendation has probability $\Phi\p{-d/c_n}$ and the risk is $d\Phi\p{-d/c_n}$. Its supremum over $d\ge0$ equals $c_nc_{\mathrm{mm}}$. Choosing $N_a/n\to s_a/\p{s_a+s_b}$ gives $c_n\to s_a+s_b$. Letting $C\to\infty$ then yields $\mathfrak G_{\calS}^{\uparrow}\p{s_a,s_b}\le\p{s_a+s_b}c_{\mathrm{mm}}$.

For the lower bound, fix $x>0$ and consider the two local parameters
\[
u^{+}=\p{xs_a,-xs_b},
\qquad
u^{-}=\p{-xs_a,xs_b},
\]
whose best arms differ and whose regret from a wrong recommendation is $x\p{s_a+s_b}$ under either point. Under any behavioral policy, the log-likelihood ratio of the complete Gaussian history under $u^{+}$ against $u^{-}$ is
\[
L_n=2x^2+\frac{2x}{\sqrt n}\sum_{t=1}^n\varepsilon_tZ_t,
\qquad
\varepsilon_t=\mathbbm{1}[A_t=a]-\mathbbm{1}[A_t=b],
\]
Selecting arm $a$ contributes
\[
\log\p{
\frac{\varphi_{s_a^2}\p{X_t-xs_a/\sqrt n}}
{\varphi_{s_a^2}\p{X_t+xs_a/\sqrt n}}
}
=
\frac{2xX_t}{s_a\sqrt n}.
\]
Substituting $X_t=xs_a/\sqrt n+s_aZ_t$ gives $2x^2/n+2xZ_t/\sqrt n$. Selecting arm $b$ gives $2x^2/n-2xZ_t/\sqrt n$.

Represent the policy's external randomization by a seed $U$. Write $\calG_0^U\coloneqq\widetilde{\sigma}(U)$ and $\calG_t^U\coloneqq\widetilde{\sigma}\p{U,A_1,X_1,\ldots,A_t,X_t}$ for $t\ge1$. Then, $\varepsilon_t$ is $\calG_{t-1}^U$-measurable, and $Z_t$ is a fresh standard normal independent of $\calG_{t-1}^U$. Hence,
$\bbE\sqb{\exp\p{i\lambda\varepsilon_tZ_t}\mid\calG_{t-1}^U}=\exp\p{-\lambda^2/2}$ for every $\lambda\in\bbR$. Iterating this identity over the $n$ rounds gives
$\bbE\sqb{\exp\p{i\lambda\sum_{t=1}^n\varepsilon_tZ_t}}=\exp\p{-n\lambda^2/2}$. Therefore, $\sum_{t=1}^n\varepsilon_tZ_t\sim\calN(0,n)$ exactly. The signs need not be independent of the noises.

It follows that $L_n\sim\calN\p{2x^2,4x^2}$ under $u^{+}$ and $L_n\sim\calN\p{-2x^2,4x^2}$ under $u^{-}$ for every behavioral policy. The equal-prior Bayes test between the two points has error probability $\Phi(-x)$. Thus, the average simple regret under the two-point prior is at least $x\p{s_a+s_b}\Phi(-x)$.

Both points lie in $[-C,C]^{\calS}$ once $C\ge x\max\cb{s_a,s_b}$. Therefore, for every $n$,
\[
\mathfrak G_{\calS,n}^{C}\p{s_a,s_b}
\ge
x\p{s_a+s_b}\Phi(-x).
\]
Let $n\to\infty$ and then let $C\uparrow\infty$. Taking the supremum over $x\ge0$ gives $\mathfrak G_{\calS}^{\uparrow}\p{s_a,s_b}\ge\p{s_a+s_b}c_{\mathrm{mm}}$.
\end{proof}

The value of the game does not decrease when arms are added. Therefore, screening does not increase the leading minimax constant: whenever at least two arms are retained, the game played on the retained arms is no harder than the game played on all $K$ arms, while a single retained arm incurs no main-phase regret.

\begin{lemma}[Arm-set monotonicity]
\label{lem:arm_set_monotonicity}
Let $\calS\subseteq\calS'\subseteq[K]$ with $\abs{\calS}\ge2$, let $s=\p{s_a}_{a\in\calS'}\in(0,\infty)^{\calS'}$, and write $s_{\calS}=\p{s_a}_{a\in\calS}$ for its restriction. Then
\[
\mathfrak G_{\calS,n}^{C}\p{s_{\calS}}\le\mathfrak G_{\calS',n}^{C}(s)
\]
for every compact radius $C<\infty$ and every horizon $n\ge\abs{\calS'}$. Consequently $\mathfrak g_{\calS}^{C}\p{s_{\calS}}\le\mathfrak g_{\calS'}^{C}(s)$ for every such $C$, and
\[
\mathfrak G_{\calS}^{\uparrow}\p{s_{\calS}}\le\mathfrak G_{\calS'}^{\uparrow}(s).
\]
\end{lemma}
\begin{proof}
Fix $C<\infty$, a horizon $n\ge\abs{\calS'}$, and an arm $b_0\in\calS$. For $u\in[-C,C]^{\calS}$ let $\overline u(u)\in[-C,C]^{\calS'}$ be the parameter that agrees with $u$ on $\calS$ and sets every additional coordinate equal to the lower endpoint $-C$,
\[
\overline u_a(u)=u_a\quad\p{a\in\calS},
\qquad
\overline u_c(u)=-C\quad\p{c\in\calS'\setminus\calS}.
\]
The value $-C$ assigned to the additional coordinates does not depend on $u$.

Let $\pi'$ be a behavioral policy of the $\calS'$-game with horizon $n$. Construct a horizon-$n$ behavioral policy $\pi$ of the $\calS$-game that runs an internal copy of $\pi'$. Whenever the copy selects an arm $a\in\calS$, $\pi$ selects $a$, consumes one actual arm selection, and feeds the observed outcome to the copy. Whenever the copy selects an arm $c\in\calS'\setminus\calS$, $\pi$ generates a surrogate outcome from the law $\calN\p{-C/\sqrt n,s_c^2}$ using the pre-experiment randomization seed $U$ of Section~\ref{sec:prob}, consumes no actual arm selection, and feeds the surrogate to the copy. Once the copy has completed its $n$ internal rounds, $\pi$ selects arbitrary arms in $\calS$ in the remaining rounds and discards their outcomes until exactly $n$ actual arm selections have been made, so that $\pi$ is a horizon-$n$ policy of the $\calS$-game. The surrogate law is determined by $C$, $n$, and $s_c$ alone, so every action of $\pi$ is measurable with respect to the real observed history and the seed, as Definition~\ref{def:seq_gaussian_game} requires, and $\pi$ never uses the unknown $u$.

An actual arm selection of $a\in\calS$ under the horizon-$n$ policy $\pi$ produces $X=u_a/\sqrt n+s_aZ$ with a fresh $Z\sim\calN(0,1)$. This is exactly the law of a selection of $a$ in the $\calS'$-game at $\overline u(u)$ with the same horizon $n$. Padding the schedule to $n$ actual arm selections makes $\pi$ an admissible horizon-$n$ policy of the $\calS$-game. The noise scalings agree because both games use the same horizon. A surrogate for $c\in\calS'\setminus\calS$ has law $\calN\p{-C/\sqrt n,s_c^2}$, which is exactly the observation law after selecting $c$ in the $\calS'$-game at $\overline u(u)$. In the $\calS'$-game the observation law depends on the arm and not on the round index, so it is immaterial that the copy's internal round index need not coincide with the index of the real round that serves it. Both the fresh noise variables and the surrogate outcomes are independent of the copy's past given the selected arm, so an induction on the copy's internal rounds shows that, for every $u\in[-C,C]^{\calS}$, the joint law of the copy's action-observation history and of its choice $\widehat A_n\in\calS'$ under $\pi$ at $u$ equals the corresponding law under $\pi'$ at $\overline u(u)$.

Let $\pi$ choose $\psi\p{\widehat A_n}$, where $\psi(a)=a$ for $a\in\calS$ and $\psi(c)=b_0$ for $c\in\calS'\setminus\calS$. Fix $u\in[-C,C]^{\calS}$. Since $\max_{a\in\calS}u_a\ge-C$, the additional coordinates set equal to $-C$ do not raise the maximum, and $\max_{a\in\calS'}\overline u_a(u)=\max_{a\in\calS}u_a$. For $a\in\calS$,
\[
\Delta_u^{\calS}\p{\psi(a)}=\max_{b\in\calS}u_b-u_a=\Delta_{\overline u(u)}^{\calS'}(a),
\]
while for $c\in\calS'\setminus\calS$, using $u_{b_0}\ge-C$ and $\overline u_c(u)=-C$,
\[
\Delta_u^{\calS}\p{\psi(c)}=\max_{b\in\calS}u_b-u_{b_0}\le\max_{b\in\calS}u_b+C=\Delta_{\overline u(u)}^{\calS'}(c).
\]
Hence $\Delta_u^{\calS}\p{\psi(a)}\le\Delta_{\overline u(u)}^{\calS'}(a)$ for every $a\in\calS'$, and taking expectations under the common law of $\widehat A_n$ gives $R_n^{G}\p{\pi,u}\le R_n^{G}\p{\pi',\overline u(u)}$. Setting the additional arms equal to $-C$ keeps the comparison within the same compact radius. Choosing smaller means would require comparing games with different radii before taking the supremum over $C$.

Taking the supremum over $u\in[-C,C]^{\calS}$ and using $\overline u(u)\in[-C,C]^{\calS'}$,
\[
\mathfrak G_{\calS,n}^{C}\p{s_{\calS}}
\le\sup_{u\in[-C,C]^{\calS}}R_n^{G}\p{\pi,u}
\le\sup_{v\in[-C,C]^{\calS'}}R_n^{G}\p{\pi',v}.
\]
Taking the infimum over $\pi'$ gives the first display. Theorem~\ref{thm:seq_value_limit} applies at radius $C$ to both games. Letting $n\to\infty$ therefore preserves the inequality and gives $\mathfrak g_{\calS}^{C}\p{s_{\calS}}\le\mathfrak g_{\calS'}^{C}(s)$. Taking the supremum over $C<\infty$ gives the last display.
\end{proof}

\begin{definition}[Global sequential minimax constant]
\label{def:global_seq_constant}
Fix a true collection of arm families $\bmP=\p{\calP_a}_{a\in[K]}$ satisfying Definition~\ref{def:mean_param}, and write $\sigma_a(m)$ for the standard deviation of arm $a$ at mean $m$. Define
\[
\Gamma(\bmP)
\coloneqq
\sup_{m\in\calM}\
\mathfrak G_{[K]}^{\uparrow}\p{\p{\sigma_a(m)}_{a\in[K]}}.
\]
\end{definition}

Since $\calM$ is a compact interval and each $\sigma_a$ is continuous on $\calM$ by Definition~\ref{def:mean_param}, the vectors $\p{\sigma_a(m)}_{a\in[K]}$ range over a compact subset of $(0,\infty)^{[K]}$. On this subset, $\mathfrak G_{[K]}^{\uparrow}$ is finite by Lemma~\ref{lem:seq_uniform_finite} and multiplicatively continuous by Lemma~\ref{lem:seq_game_properties}. Hence, the supremum is finite and $\Gamma(\bmP)<\infty$. Although we write $\Gamma(\bmP)$, the constant depends on the collection of arm families only through the standard deviations evaluated at common mean values $m\in\calM$. No other feature of the outcome distributions enters.
 The variance profiles that enter $\Gamma(\bmP)$ describe the true environment and are not supplied to a strategy.

\begin{corollary}[Uniform upper bound for the minimax constant]
\label{cor:uniform_gamma_bound}
Under Assumption~\ref{asm:structural_envelope}, every $\bmP\in\mathfrak P\p{\mathfrak E}$ satisfies
\[
\Gamma\p{\bmP}\le\overline\Gamma_K,
\qquad
\overline\Gamma_K\coloneqq2K\sqrt{K\overline v}\,c_{\mathrm{mm}}.
\]
\end{corollary}
\begin{proof}
For $s=\p{\sigma_a(m)}_{a\in[K]}$, Assumption~\ref{asm:structural_envelope} gives $\overline s_{[K]}^2\le2\overline v$ in the notation of Lemma~\ref{lem:seq_uniform_finite}. The balanced-policy bound in that lemma therefore gives
\[
\mathfrak G_{[K]}^{\uparrow}(s)
\le
K\sqrt{2K\overline s_{[K]}^2}\,c_{\mathrm{mm}}
\le
2K\sqrt{K\overline v}\,c_{\mathrm{mm}}.
\]
Taking the supremum over $m\in\calM$ proves the claim.
\end{proof}

Screening may leave any subset of locally competitive arms, and the game value on such a subset is what the upper bound must control. Arm-set monotonicity turns this into a statement about the full arm set alone.

\begin{corollary}[Subsets do not raise the constant]
\label{cor:gamma_subsets}
For every $\calS\subseteq[K]$ with $\abs{\calS}\ge2$ and every $m\in\calM$,
\[
\mathfrak G_{\calS}^{\uparrow}\p{\p{\sigma_a(m)}_{a\in\calS}}\le\Gamma(\bmP),
\]
and consequently
\[
\Gamma(\bmP)
=
\max_{\substack{\calS\subseteq[K]\\\abs{\calS}\ge2}}\ \sup_{m\in\calM}\
\mathfrak G_{\calS}^{\uparrow}\p{\p{\sigma_a(m)}_{a\in\calS}}.
\]
\end{corollary}
\begin{proof}
Fix $m\in\calM$ and apply Lemma~\ref{lem:arm_set_monotonicity} with $\calS'=[K]$ and $s=\p{\sigma_a(m)}_{a\in[K]}$, whose restriction to $\calS$ is $\p{\sigma_a(m)}_{a\in\calS}$ because both are evaluated at the same $m$. This gives $\mathfrak G_{\calS}^{\uparrow}\p{\p{\sigma_a(m)}_{a\in\calS}}\le\mathfrak G_{[K]}^{\uparrow}\p{\p{\sigma_a(m)}_{a\in[K]}}\le\Gamma(\bmP)$, which is the first display. Taking the supremum over $m$ and then the maximum over the finitely many $\calS$ shows that the right-hand side of the second display is at most $\Gamma(\bmP)$. It is at least $\Gamma(\bmP)$ because $\calS=[K]$ is admissible in the maximum and $K\ge2$.
\end{proof}

\begin{lemma}[Finite-grid extension]
\label{lem:grid_extension}
Let $\calU\subset[-C,C]^{\calS}$ be a finite $\rho$-net of $[-C,C]^{\calS}$ in the supremum norm, and write $\underline s=\min_{a\in\calS}s_a$. Then, for every behavioral policy $\pi$,
\[
\sup_{u\in[-C,C]^{\calS}}R_n^{G}\p{\pi,u}
\le
\max_{v\in\calU}R_n^{G}\p{\pi,v}
+2\rho+\frac{C\rho}{\underline s},
\]
and consequently
\[
\abs{\mathfrak G_{\calS,n}^{C}(s)-\mathfrak G_{\calS,n}^{\calU}(s)}
\le
2\rho+\frac{C\rho}{\underline s}.
\]
\end{lemma}
\begin{proof}
Fix $u\in[-C,C]^{\calS}$ and choose $v\in\calU$ with $\norm{u-v}_\infty\le\rho$. Let $\bbP_u^{\pi}$ and $\bbP_v^{\pi}$ be the laws of the complete history $\p{A_1,X_1,\ldots,A_n,X_n,\widehat A_n}$ under $\pi$ at $u$ and at $v$. Because $\pi$ is the same policy under both parameters, its sampling-action and recommendation factors cancel from the likelihood ratio. The log-likelihood ratio is therefore the sum of the per-round Gaussian log ratios. Taking its expectation under $\bbP_u^{\pi}$ gives
\[
\mathrm{KL}\p{\bbP_u^{\pi}\|\bbP_v^{\pi}}
=\frac1{2n}\bbE_{u,s}^{\pi}\sum_{t=1}^n\frac{\p{u_{A_t}-v_{A_t}}^2}{s_{A_t}^2}
\le\frac{\rho^2}{2\underline s^2}.
\]
Each Gaussian summand compares $\calN\p{u_a/\sqrt n,s_a^2}$ with $\calN\p{v_a/\sqrt n,s_a^2}$ and contributes $\p{u_a-v_a}^2/\p{2ns_a^2}$. There are $n$ rounds. We use the convention
\[
\norm{\bbP-\bbQ}_{\mathrm{TV}}
=
\frac12\int\abs{d\bbP-d\bbQ}.
\]
Pinsker's inequality gives
\[
\norm{\bbP_u^{\pi}-\bbP_v^{\pi}}_{\mathrm{TV}}
\le\sqrt{\tfrac12\mathrm{KL}\p{\bbP_u^{\pi}\|\bbP_v^{\pi}}}
\le\frac{\rho}{2\underline s}.
\]
On $[-C,C]^{\calS}$ the loss satisfies $0\le\Delta_v^{\calS}\p{\widehat A_n}\le2C$ and $\abs{\Delta_u^{\calS}(\cdot)-\Delta_v^{\calS}(\cdot)}\le2\norm{u-v}_\infty\le2\rho$ pointwise, so
\begin{align*}
R_n^{G}\p{\pi,u}
&=\bbE_{u,s}^{\pi}\sqb{\Delta_u^{\calS}\p{\widehat A_n}}\\
&\le\bbE_{u,s}^{\pi}\sqb{\Delta_v^{\calS}\p{\widehat A_n}}+2\rho\\
&\le\bbE_{v,s}^{\pi}\sqb{\Delta_v^{\calS}\p{\widehat A_n}}
+2C\norm{\bbP_u^{\pi}-\bbP_v^{\pi}}_{\mathrm{TV}}+2\rho.
\end{align*}
The right-hand side is at most $R_n^{G}\p{\pi,v}+2\rho+C\rho/\underline s$. Taking the maximum over $v\in\calU$ and then the supremum over $u$ gives the first display. The second follows by taking the infimum over $\pi$ and using $\calU\subset[-C,C]^{\calS}$, which gives $\mathfrak G_{\calS,n}^{\calU}(s)\le\mathfrak G_{\calS,n}^{C}(s)$ as well.
\end{proof}

\subsection{Posterior State, Gaussian Transition, and the Bellman Recursion}

Fix the finite grid $\calU=\cb{u^{(1)},\ldots,u^{(L)}}\subset[-C,C]^{\calS}$, a standard-deviation vector $s$, a horizon $n$, and a full-support prior $p\in\calP([L])$ with $p_\ell>0$ for every $\ell$. Using $u^{(L)}$ as reference, parametrize the posterior $\Pi_t^p$ of Appendix~\ref{app:finite_grid_duality} by its log odds
\begin{align*}
z_r&=\log\frac{\Pi_{t,r}^p}{\Pi_{t,L}^p},
&& r=1,\ldots,L-1,\\
p_r(z)&=\frac{e^{z_r}}{1+\sum_{q<L}e^{z_q}},
&& r=1,\ldots,L-1,\\
p_L(z)&=\frac1{1+\sum_{q<L}e^{z_q}}.
\end{align*}
Because $p$ has full support and each Gaussian likelihood is everywhere positive, $\Pi_t^p$ has full support at every $t$, so $z_t\in\bbR^{L-1}$ is well-defined along every trajectory. For an action $a$ and grid points $r,\ell$, set
\[
\beta_{a,r}=\frac{u_a^{(r)}-u_a^{(L)}}{s_a},
\qquad
b_{a,\ell,r}=\frac{\p{u_a^{(r)}-u_a^{(L)}}u_a^{(\ell)}-\tfrac12\p{\p{u_a^{(r)}}^2-\p{u_a^{(L)}}^2}}{s_a^2}.
\]
A direct computation of the Gaussian log-likelihood ratio shows that, under hidden point $u^{(\ell)}$, one observation from arm $a$ changes the log odds by
\begin{align}
\xi_{a,\ell}=\frac{\beta_a}{\sqrt n}Z+\frac{b_{a,\ell}}n,
\qquad Z\sim\calN(0,1),
\label{eq:gaussian_logodds_increment}
\end{align}
coordinatewise in $r$. The posterior-predictive (Bayes) transition operator and the terminal posterior loss are
\[
K_{a,n}f(z)=\sum_{\ell=1}^Lp_\ell(z)\bbE\sqb{f\p{z+\xi_{a,\ell}}},
\qquad
c_b(z)=\sum_{\ell=1}^Lp_\ell(z)\p{\max_{a\in\calS}u_a^{(\ell)}-u_b^{(\ell)}},
\]
and the unregularized Gaussian Bayes recursion is
\[
V_{n,n}^{G,p}(z)=\min_{b\in\calS}c_b(z),
\qquad
V_{t,n}^{G,p}(z)=\min_{a\in\calS}K_{a,n}V_{t+1,n}^{G,p}(z)
\qquad(0\le t<n).
\]

\begin{lemma}[Reduction to posterior-state policies]
\label{lem:posterior_markovization}
Fix the finite grid $\calU$, standard deviations $s$, horizon $n$, and a full-support prior $p\in\calP([L])$. Then the infimum of the Bayes risk over all behavioral policies equals the Bellman value:
\[
\inf_{\pi}\sum_{\ell=1}^Lp_\ell r_\ell(\pi)
=V_{0,n}^{G,p}\p{z(p)},
\qquad z(p)_r=\log\frac{p_r}{p_L},
\]
and the infimum is attained by a policy that is Markov in $\p{t,\Pi_t^p}$.
\end{lemma}
\begin{proof}
Take the reference prior in Lemma~\ref{lem:full_support_posterior_sufficiency} equal to $p$. That lemma replaces any behavioral policy $\pi$ by a policy $\pi^p$ that is Markov in $\p{t,\Pi_t^p}$ and preserves every coordinate risk, hence preserves the Bayes risk $\sum_\ell p_\ell r_\ell$. Therefore the infimum over all behavioral policies equals the infimum over policies Markov in $\p{t,\Pi_t^p}$, equivalently in $\p{t,z_t}$.

For such a Markov policy, write $\bbE_{M^p}$ for expectation under the reference-mixture law. Under this law, the hidden index $J$ has distribution $p$. Given $\p{z_t,A_{t+1}=a}$, the next observation has the mixture distribution $\sum_\ell p_\ell(z_t)\calN\p{u_a^{(\ell)}/\sqrt n,s_a^2}$ and updates $z_{t+1}=z_t+\xi_{a,J}$ as in \eqref{eq:gaussian_logodds_increment}. This is exactly the transition represented by $K_{a,n}$.

The Bayes risk equals $\bbE_{M^p}\sqb{\Delta_{u^{(J)}}^{\calS}\p{\widehat A_n}}$. Conditioning on $\calF_n$ replaces the terminal loss by its posterior expectation $c_{\widehat A_n}(z_n)$, giving $\bbE_{M^p}\sqb{c_{\widehat A_n}(z_n)}$.

The problem is a finite-horizon controlled Markov chain with state $z_t$, finite action set $\calS$, bounded costs, transition kernel $K_{a,n}$, and terminal cost $\min_{b\in\calS}c_b$.

By backward induction each $V_{t,n}^{G,p}$ is a bounded Borel function of $z$: the terminal cost $\min_{b\in\calS}c_b$ is continuous and bounded, taking values in $[0,2C]$, and $K_{a,n}$ maps bounded Borel functions to bounded Borel functions. The dynamic-programming principle for finite horizons then gives that the minimal Bayes risk over Markov policies is $V_{0,n}^{G,p}\p{z_0}$ with the displayed backward recursion, and, since $\calS$ is finite, the pointwise minimizer of $a\mapsto K_{a,n}V_{t+1,n}^{G,p}$ is attained and Borel measurable, being the argmin over finitely many Borel functions under the fixed tie-break of Section~\ref{sec:prob}, so a minimizing Markov policy exists. The initial state is the deterministic $z_0=z(p)$, which gives the claim.
\end{proof}

\begin{remark}
Lemma~\ref{lem:posterior_markovization} is stated for full-support priors, for which the log-odds chart and the recursion are everywhere defined. The comparison arguments use it only for such priors. When necessary, they approximate a boundary least-favorable prior from the interior. The transition $K_{a,n}$ is the Bayes posterior-predictive operator. The single-hidden-point operator used in the comparison arguments is introduced where it is first needed.
\end{remark}

\section{Minimax Lower Bound}
\label{app:lower_comparison}
This appendix proves the minimax lower bound at the level of $\Gamma\p{\bmP}$. The proof compares the original local experiment with the sequential Gaussian experiment through posterior Bellman recursions.

\subsection{Reduction to the \texorpdfstring{$\calS$}{S}-arm local experiment}

For $u\in\bbR^{\calS}$ and a budget $T$, consider the local mean vector $\bmmu_T(u)\in\bbR^K$ with
\[
\mu_{a,T}(u)=m+\frac{u_a}{\sqrt T}\quad(a\in\calS),
\qquad
\mu_{a,T}(u)=\underline\mu\quad(a\notin\calS),
\qquad
\underline\mu\coloneqq\min\calM<m.
\]
Since $m\in\operatorname{int}(\calM)$, there is $T_1\p{\calS,m,C}<\infty$ such that for $T\ge T_1$ and every $u\in[-C,C]^{\calS}$ one has $\bmmu_T(u)\in\calM^K$ and the largest coordinate of $\bmmu_T(u)$ is attained inside $\calS$, because $\max_{a\in\calS}\mu_{a,T}(u)\ge m-C/\sqrt T>\underline\mu$. For an adaptive experiment $\delta\in\calA$ with choice $\widehat A_T$, define the scaled original-experiment risk at local parameter $u$ by
\[
R_T^{P}\p{\delta,u}\coloneqq\sqrt T\,\Regret_T^{\delta}\p{\bmmu_T(u)}
=\bbE_{\bmmu_T(u)}^{\delta}\sqb{\max_{a\in[K]}\sqrt T\,\mu_{a,T}(u)-\sqrt T\,\mu_{\widehat A_T,T}(u)}.
\]
Under $\bmmu_T(u)$ arm $a\in\calS$ produces outcomes from $P_{a,m+u_a/\sqrt T}$, while every arm outside $\calS$ produces outcomes from the fixed law $P_{a,\underline\mu}$, which does not depend on $u$.

The $\calS$-arm local experiment is the sequential experiment in which only arms in $\calS$ are selected over a horizon $T$: at round $t$ a behavioral policy $\pi$ selects $A_t\in\calS$ from an $\calF_{t-1}$-measurable law and observes $Y_t\sim P_{A_t,m+u_{A_t}/\sqrt T}$, and after round $T$ it chooses $\widehat A_T\in\calS$ and incurs the scaled simple regret $\Delta_u^{\calS}\p{\widehat A_T}=\max_{a\in\calS}u_a-u_{\widehat A_T}$. Write $\widetilde R_T^{P}\p{\pi,u}\coloneqq\bbE_u^{\pi}\sqb{\Delta_u^{\calS}\p{\widehat A_T}}$ for its risk.

\begin{lemma}[Reduction to the $\calS$-arm experiment]
\label{lem:s_arm_reduction}
There is $T_2\p{\calS,m,C}\ge T_1$ such that, for every prior $p$ on $\calU$ and every $T\ge T_2$,
\[
\inf_{\delta\in\calA}\sum_{\ell=1}^L p_\ell R_T^{P}\p{\delta,u^{(\ell)}}
\ge
\inf_{\pi}\sum_{\ell=1}^L p_\ell\,\widetilde R_T^{P}\p{\pi,u^{(\ell)}},
\]
the right infimum running over all behavioral policies of the $\calS$-arm local experiment.
\end{lemma}
\begin{proof}
Fix a $K$-arm behavioral experiment $\delta$ and $T\ge T_1$. Construct an $\calS$-arm policy $\widetilde\pi$ that runs an internal copy of $\delta$. Whenever the copy selects an arm $a\in\calS$, the policy $\widetilde\pi$ selects $a$, feeds the observed outcome to the copy, and consumes one actual selection from $\calS$. Whenever the copy selects an arm $a\notin\calS$, the policy $\widetilde\pi$ generates an independent surrogate outcome from the fixed law $P_{a,\underline\mu}$ using its external randomization. It feeds this outcome to the copy without consuming an actual arm selection. The outside-arm laws do not depend on the hidden index $\ell$. Therefore, for every $\ell$, the law of the pair $\p{\text{copy history},\widehat A_T}$ under $\widetilde\pi$ at the local parameter $u^{(\ell)}$ equals the law of that pair under $\delta$ at $\bmmu_T\p{u^{(\ell)}}$.

The copy uses at most $T$ actual selections from $\calS$ and reproduces the choice law exactly at every hidden point. Let $\widetilde\pi_0$ be the resulting $\calS$-sampling policy whose terminal recommendation is the arm $\widehat A_T\in[K]$ produced by the copy. If the copy uses fewer than $T$ actual arm selections, $\widetilde\pi_0$ selects arbitrary arms in $\calS$ during the remaining rounds and discards their outcomes. The policy then makes exactly $T$ actual arm selections and is therefore a horizon-$T$ $\calS$-arm policy.

Then, $\sum_\ell p_\ell R_T^P\p{\delta,u^{(\ell)}}$ equals the Bayes risk of $\widetilde\pi_0$ under the $K$-arm scaled regret.

If $\widehat A_T\notin\calS$, replace it by a fixed arm $b_0\in\calS$. Recommending an arm outside $\calS$ has scaled regret at least $\sqrt T\p{\max_{a\in\calS}\mu_{a,T}(u)-\underline\mu}\ge\sqrt T\p{m-\underline\mu}-C$. The fixed replacement has scaled regret at most $\max_{a\in\calS}u_a-\min_{a\in\calS}u_a\le2C$. Choose $T_2\ge T_1$ so that $\sqrt{T_2}\p{m-\underline\mu}-C\ge2C$. For $T\ge T_2$, the replacement lowers the loss pointwise, uniformly over $u\in[-C,C]^{\calS}$. It produces an $\calS$-arm policy $\widetilde\pi$ that selects and recommends only arms in $\calS$. It satisfies $\sum_\ell p_\ell\widetilde R_T^{P}\p{\widetilde\pi,u^{(\ell)}}\le\sum_\ell p_\ell R_T^P\p{\delta,u^{(\ell)}}$. Taking the infimum over $\widetilde\pi$ on the right and over $\delta$ on the left gives the claim.
\end{proof}

\subsection{Original Posterior Transitions and Bellman Value}

Retain the finite grid $\calU$, the standard-deviation vector $s=s_{\calS}(m)$, the horizon $T$, and a full-support prior $p\in\calP([L])$, and use the log-odds chart $z\in\bbR^{L-1}$, the posterior weights $p_\ell(z)$, and the Gaussian increment $\xi_{a,\ell}$ of Appendix~\ref{app:seq_value_bellman}. For an arm $a$ and grid points $r,\ell$, the log-odds increment produced by one original observation $Y$ from arm $a$, between grid points $r$ and $L$, is
\[
\Xi_{a,r}^{P}(Y)=\ell_{a,m+u_a^{(r)}/\sqrt T}(Y)-\ell_{a,m+u_a^{(L)}/\sqrt T}(Y),
\qquad r=1,\ldots,L-1,
\]
where $\ell_{a,\mu}$ is the log-density of $P_{a,\mu}$ of Lemma~\ref{lem:score_fisher}. The original posterior-predictive operator and its single-hidden-point restrictions are
\begin{align*}
&K_{a,T}^{P}f(z)=\sum_{\ell=1}^L p_\ell(z)\,\bbE_{a,m+u_a^{(\ell)}/\sqrt T}\sqb{f\p{z+\Xi_a^{P}(Y)}},\\
&K_{a,T}^{P,\ell}f(z)=\bbE_{a,m+u_a^{(\ell)}/\sqrt T}\sqb{f\p{z+\Xi_a^{P}(Y)}},
\end{align*}
and the Gaussian operator $K_{a,T}\eqqcolon K_{a,T}^{G}$ of Appendix~\ref{app:seq_value_bellman} decomposes correspondingly as
\[
K_{a,T}^{G}f(z)=\sum_{\ell=1}^L p_\ell(z)\,K_{a,T}^{G,\ell}f(z),
\qquad
K_{a,T}^{G,\ell}f(z)=\bbE\sqb{f\p{z+\xi_{a,\ell}}}.
\]
The original-experiment and Gaussian Bayes values share the terminal posterior loss $c_b(z)$ of Appendix~\ref{app:seq_value_bellman}, which depends only on the state:
\[
V_{T,T}^{P}(z)=\min_{b\in\calS}c_b(z)=V_{T,T}^{G}(z),
\qquad
V_{t,T}^{P}(z)=\min_{a\in\calS}K_{a,T}^{P}V_{t+1,T}^{P}(z)
\qquad(0\le t<T),
\]
and $V_{t,T}^{G}$ is the recursion of Appendix~\ref{app:seq_value_bellman}, which as a function of $z$ does not depend on the initializing prior, so we drop the prior superscript on it. Both $K_{a,T}^{P}$ and $K_{a,T}^{G}$ are averaging operators: they are monotone and reproduce constants.

\begin{lemma}[Reduction of the original experiment to posterior-state policies]
\label{lem:original_posterior_markovization}
For every full-support prior $p\in\calP([L])$, the infimum of the $\calS$-arm original Bayes risk over all behavioral policies equals the Bellman value:
\[
\inf_{\pi}\sum_{\ell=1}^L p_\ell\,\widetilde R_T^{P}\p{\pi,u^{(\ell)}}=V_{0,T}^{P}\p{z(p)},
\qquad z(p)_r=\log\frac{p_r}{p_L},
\]
and the infimum is attained by a policy Markov in $\p{t,\Pi_t^p}$.
\end{lemma}
\begin{proof}
The proof of Lemma~\ref{lem:full_support_posterior_sufficiency} uses two facts. The common action kernel cancels from every per-round likelihood ratio, and the reference posterior is sufficient for the hidden index. Both facts hold for the original $\calS$-arm experiment. After choosing arm $a$, its per-round observation kernel is $P_{a,m+u_a^{(\ell)}/\sqrt T}$, and $K_{a,T}^{P}$ gives the fixed Bayes update after one observation. Hence, every behavioral policy has the same risk at each point of $\calU$ as a policy Markov in $\p{t,\Pi_t^p}$. Applying the dynamic-programming argument of Lemma~\ref{lem:posterior_markovization} with terminal cost $\min_{b\in\calS}c_b$ and transition $K_{a,T}^{P}$ shows that the minimal Bayes risk is $V_{0,T}^{P}\p{z(p)}$. It also gives a minimizing Markov policy.
\end{proof}

\subsection{Smooth Lower Approximation to the Gaussian Bellman Value}
We first regularize the Gaussian value.

\begin{lemma}[Uniform Lipschitz bound for the Gaussian value]
\label{lem:hard_bellman_lipschitz}
Fix $\p{\calS,m,C,\calU}$. Then, there is $\Lambda_0<\infty$ such that $\operatorname{Lip}\p{V_{t,T}^{G}}\le\Lambda_0$ for $0\le t\le T$. The constant $\Lambda_0$ is independent of $T$ and $t$.
\end{lemma}
\begin{proof}
The posterior weights $z\mapsto p_\ell(z)$ are globally Lipschitz on $\bbR^{L-1}$, and $c_b$ takes values in $[0,2C]$. Hence, the terminal value $V_{T,T}^{G}=\min_b c_b$ is Lipschitz with a constant depending only on $\p{\calS,m,C,\calU}$. For a Lipschitz $f$ and a fixed arm $a$, use $\sum_\ell\p{p_\ell(z)-p_\ell(z')}=0$ to write
\begin{align*}
&K_{a,T}^{G}f(z)-K_{a,T}^{G}f(z')\\
&=\sum_\ell p_\ell(z)\p{K_{a,T}^{G,\ell}f(z)-K_{a,T}^{G,\ell}f(z')}
+\sum_\ell\p{p_\ell(z)-p_\ell(z')}\p{K_{a,T}^{G,\ell}f(z')-K_{a,T}^{G,L}f(z')}.
\end{align*}
Each $K_{a,T}^{G,\ell}$ is an average of translates of $f$, hence $1$-Lipschitz. Therefore, the first sum is at most $\operatorname{Lip}(f)\norm{z-z'}$. The transitions $K_{a,T}^{G,\ell}$ and $K_{a,T}^{G,L}$ share the noise coefficient $\beta_a$ and differ only in the drift $b_{a,\ell}/T-b_{a,L}/T$. The norm of this difference is at most $\kappa_1/T$ for a constant $\kappa_1$ depending only on $\p{\calS,m,C,\calU}$. In addition, $\sum_\ell\abs{p_\ell(z)-p_\ell(z')}\le\kappa_2\norm{z-z'}$. Hence the second sum is at most $\kappa_1\kappa_2 T^{-1}\operatorname{Lip}(f)\norm{z-z'}$, and $\operatorname{Lip}\p{K_{a,T}^{G}f}\le\p{1+\kappa_1\kappa_2/T}\operatorname{Lip}(f)$. Since $\operatorname{Lip}\p{\min_a g_a}\le\max_a\operatorname{Lip}(g_a)$, the recursion gives
\[\operatorname{Lip}\p{V_{t,T}^{G}}\le\p{1+\kappa_1\kappa_2/T}^{T-t}\operatorname{Lip}\p{V_{T,T}^{G}}\le e^{\kappa_1\kappa_2}\operatorname{Lip}\p{\min_b c_b}\eqqcolon\Lambda_0,\]
where $\Lambda_0$ depends only on $\p{\calS,m,C,\calU}$.
\end{proof}

Let $G\sim\calN\p{0,I_{L-1}}$ and $\calM_h f(z)=\bbE\sqb{f(z+hG)}$ for $h>0$.

\begin{lemma}[Mollifier estimates]
\label{lem:mollifier_estimates}
If $f$ is $\Lambda_0$-Lipschitz on $\bbR^{L-1}$ and $0<h\le1$, then $\norm{\calM_h f-f}_\infty\le\Lambda_0 h\,\bbE\norm G$ and, for every integer $q\ge1$, $\max_{1\le\abs{\alpha}\le q}\norm{\partial^\alpha\calM_h f}_\infty\le C_q\Lambda_0 h^{1-q}$ with $C_q$ depending only on $q$ and $L$.
\end{lemma}
\begin{proof}
The first bound is $\abs{\bbE\sqb{f(z+hG)-f(z)}}\le\Lambda_0 h\bbE\norm G$. For the second, $\calM_h f=f\ast\varphi_{h^2 I}$ with $\varphi_{h^2 I}$ the $\calN\p{0,h^2 I_{L-1}}$ density. For a multi-index $\alpha$ with $1\le\abs\alpha=k\le q$, pick a coordinate $i$ with $\alpha_i\ge1$ and write $\partial^\alpha=\partial_i\partial^{\alpha'}$; then $\partial^\alpha\calM_h f=\p{\partial_i f}\ast\partial^{\alpha'}\varphi_{h^2 I}$ with $\partial_i f$ the almost-everywhere partial derivative of the Lipschitz $f$, so $\norm{\partial^\alpha\calM_h f}_\infty\le\norm{\partial_i f}_\infty\norm{\partial^{\alpha'}\varphi_{h^2 I}}_{L^1}\le\Lambda_0 C_k' h^{1-k}$, using $\norm{\partial_i f}_\infty\le\operatorname{Lip}(f)\le\Lambda_0$ and $\norm{\partial^{\alpha'}\varphi_{h^2 I}}_{L^1}\le C_k' h^{-(k-1)}$. Since $0<h\le1$ and $k\le q$ give $h^{1-k}\le h^{1-q}$, taking $C_q=\max_{1\le k\le q}C_k'$ yields $\max_{1\le\abs\alpha\le q}\norm{\partial^\alpha\calM_h f}_\infty\le C_q\Lambda_0 h^{1-q}$.
\end{proof}

\begin{lemma}[Error from interchanging smoothing and the Gaussian transition operator]
\label{lem:mollifier_transition_commutator}
For a $\Lambda_0$-Lipschitz $f$ and a fixed arm $a$,
\[
\norm{\calM_h K_{a,T}^{G}f-K_{a,T}^{G}\calM_h f}_\infty\le\frac{\kappa_3\Lambda_0 h}{T},
\]
where $\kappa_3$ depends only on $\p{\calS,m,C,\calU}$.
\end{lemma}
\begin{proof}
Gaussian convolutions and deterministic translations commute, so $\calM_h K_{a,T}^{G,\ell}=K_{a,T}^{G,\ell}\calM_h$ for every $\ell$. Therefore
\[
\calM_h K_{a,T}^{G}f(z)-K_{a,T}^{G}\calM_h f(z)
=\bbE_G\sum_\ell\p{p_\ell(z+hG)-p_\ell(z)}K_{a,T}^{G,\ell}f(z+hG).
\]
Subtracting $\sum_\ell\p{p_\ell(z+hG)-p_\ell(z)}K_{a,T}^{G,L}f(z+hG)=0$ and using $\abs{K_{a,T}^{G,\ell}f-K_{a,T}^{G,L}f}\le\kappa_1\Lambda_0/T$ together with $\sum_\ell\abs{p_\ell(z+hG)-p_\ell(z)}\le\kappa_2 h\norm G$ gives the bound with $\kappa_3=\kappa_1\kappa_2\bbE\norm G$.
\end{proof}

\begin{lemma}[Smooth Gaussian subsolution]
\label{lem:smooth_gaussian_subsolution}
There is a constant $A<\infty$, depending only on $\p{\calS,m,C,\calU}$, such that for every $h\in(0,1)$, the functions
\[
\underline V_{t,T}^{G,h}=\calM_h V_{t,T}^{G}-Ah-Ah\frac{T-t}{T}
\qquad(0\le t\le T)
\]
satisfy the terminal inequality and the Bellman inequality at each time
\[
\underline V_{T,T}^{G,h}\le V_{T,T}^{G},
\qquad
\underline V_{t,T}^{G,h}\le\min_{a\in\calS}K_{a,T}^{G}\underline V_{t+1,T}^{G,h}
\qquad(0\le t<T),
\]
together with $\underline V_{0,T}^{G,h}\ge V_{0,T}^{G}-A'h$ and $\max_{1\le\abs{\alpha}\le3}\norm{\partial^\alpha\underline V_{t,T}^{G,h}}_\infty\le A' h^{-2}$, where $A'$ depends only on $\p{\calS,m,C,\calU}$.
\end{lemma}
\begin{proof}
By Lemma~\ref{lem:mollifier_estimates}, $\norm{\calM_h V_{t,T}^{G}-V_{t,T}^{G}}_\infty\le\Lambda_0 h\bbE\norm G$. At $t=T$ this gives $\underline V_{T,T}^{G,h}=\calM_h V_{T,T}^{G}-Ah\le V_{T,T}^{G}$ once $A\ge\Lambda_0\bbE\norm G$. For the Bellman inequality at time $t$, apply $\calM_h$ to $V_{t,T}^{G}=\min_a K_{a,T}^{G}V_{t+1,T}^{G}$. Since $\calM_h$ is an averaging operator and the minimum is concave, $\calM_h V_{t,T}^{G}\le\min_a\calM_h K_{a,T}^{G}V_{t+1,T}^{G}$. Lemma~\ref{lem:mollifier_transition_commutator} then gives $\calM_h K_{a,T}^{G}V_{t+1,T}^{G}\le K_{a,T}^{G}\calM_h V_{t+1,T}^{G}+\kappa_3\Lambda_0 h/T$. Because $K_{a,T}^{G}$ reproduces constants and $\calM_h V_{t+1,T}^{G}=\underline V_{t+1,T}^{G,h}+Ah+Ah\frac{T-t-1}{T}$,
\[
\calM_h V_{t,T}^{G}\le\min_a K_{a,T}^{G}\underline V_{t+1,T}^{G,h}+Ah+Ah\frac{T-t-1}{T}+\frac{\kappa_3\Lambda_0 h}{T}.
\]
Subtracting $Ah+Ah\frac{T-t}{T}$ from both sides gives
\[
\underline V_{t,T}^{G,h}
\le
\min_a K_{a,T}^{G}\underline V_{t+1,T}^{G,h}
+
\frac{\p{\kappa_3\Lambda_0-A}h}{T}.
\]
This is a subsolution once $A\ge\max\cb{\Lambda_0\bbE\norm G,\kappa_3\Lambda_0}$.

The approximation bound follows from
\[
\underline V_{0,T}^{G,h}
=
\calM_h V_{0,T}^{G}-2Ah
\ge
V_{0,T}^{G}-\p{\Lambda_0\bbE\norm G+2A}h.
\]
Lemma~\ref{lem:mollifier_estimates} at $q=3$ gives the derivative bound because $\underline V_{t,T}^{G,h}$ differs from $\calM_h V_{t,T}^{G}$ by a constant.
\end{proof}

\subsection{Local Log-density Expansion and Moment Matching}

\begin{lemma}[Local increment moments through order \texorpdfstring{$T^{-3/2}$}{T to the minus three halves}]
\label{lem:local_logodds_expansion}
Write $v=\sigma^2(m)$, $v_1=\p{\sigma^2}'(m)$, $d_r=u^{(r)}-u^{(L)}$, $q_r=\p{u^{(r)}}^2-\p{u^{(L)}}^2$, and $c_r=\p{u^{(r)}}^3-\p{u^{(L)}}^3$ for a single arm, suppressing the arm index. For every hidden point $\ell$, the original increment $\Xi^{P}$ satisfies, as $T\to\infty$ and uniformly over $r,s,\ell$,
\begin{align*}
\bbE_\ell\sqb{\Xi_r^{P}}&=\frac{d_r u^{(\ell)}-q_r/2}{vT}+O\p{T^{-3/2}},
\\
\operatorname{Cov}_\ell\p{\Xi_r^{P},\Xi_s^{P}}&=\frac{d_r d_s}{vT}+O\p{T^{-3/2}},
\\
\bbE_\ell\norm{\Xi^{P}-\bbE_\ell\Xi^{P}}^3&\le\kappa_4 T^{-3/2},
\end{align*}
where $\kappa_4$ depends only on $\p{\calS,m,C,\calU}$, while the Gaussian increment $\xi_\ell$ of Appendix~\ref{app:seq_value_bellman} satisfies the first two identities with the $O\p{T^{-3/2}}$ terms removed and $\bbE_\ell\norm{\xi_\ell-\bbE_\ell\xi_\ell}^3\le\kappa_4 T^{-3/2}$. Consequently the discrepancies of the conditional mean vectors, of the conditional covariance matrices, and of the raw second-moment matrices are all $O\p{T^{-3/2}}$, and both centered third moments are $O\p{T^{-3/2}}$.
\end{lemma}
\begin{proof}
By Lemma~\ref{lem:score_fisher}, $\ell_\mu$ is $C^3$ in $\mu$ with $\dot\ell_m(y)=(y-m)/v$, $\ddot\ell_m(y)=-1/v-(y-m)v_1/v^2$, and $\dddot\ell_\mu(y)=-2\theta''(\mu)+(y-\mu)\theta'''(\mu)$ affine in $y$ with coefficients continuous in $\mu$. Third-order Taylor expansion of $\ell_{m+w/\sqrt T}(y)$ in $w$ about $m$ gives, for $w\in\cb{u^{(r)},u^{(L)}}$,
\[
\Xi_r^{P}(Y)=\frac{d_r}{\sqrt T}\dot\ell_m(Y)+\frac{q_r}{2T}\ddot\ell_m(Y)+\frac{c_r}{6T^{3/2}}\dddot\ell_m(Y)+\varrho_{r,T}(Y),
\]
where the Lagrange remainder
\[
\varrho_{r,T}(Y)=\tfrac16 T^{-3/2}\sqb{\p{u^{(r)}}^3\p{\dddot\ell_{\widetilde\mu_r}(Y)-\dddot\ell_m(Y)}-\p{u^{(L)}}^3\p{\dddot\ell_{\widetilde\mu_L}(Y)-\dddot\ell_m(Y)}}
\]
has $\widetilde\mu_r,\widetilde\mu_L$ between $m$ and the corresponding perturbed mean. The function $\dddot\ell_\mu(y)$ is affine in $y$, and its coefficients are uniformly continuous in $\mu$ on a compact neighborhood of $m$. Hence, there is a modulus $\omega$ such that $\omega(\eta)\to0$ as $\eta\downarrow0$ and $\abs{\dddot\ell_{\mu'}(y)-\dddot\ell_m(y)}\le\omega\p{\abs{\mu'-m}}\p{1+\abs y}$. Since $\abs{\widetilde\mu_\bullet-m}\le C/\sqrt T$, it follows that
\[\abs{\varrho_{r,T}(Y)}\le\kappa_5 T^{-3/2}\omega\p{C/\sqrt T}\p{1+\abs Y}.\] For every fixed integer $\bar q\ge1$, the uniform moment bound in Lemma~\ref{lem:uniform_mgf} then yields $\sup_{r,\ell}\bbE_\ell\abs{\varrho_{r,T}}^{\bar q}=o\p{T^{-3\bar q/2}}$.

Under hidden point $\ell$, $Y\sim P_{m+u^{(\ell)}/\sqrt T}$, so $\bbE_\ell\sqb{Y-m}=u^{(\ell)}/\sqrt T$ and $\operatorname{Var}_\ell(Y)=v+v_1 u^{(\ell)}/\sqrt T+O\p{T^{-1}}$. Substituting the derivative formulas and the remainder bound into the displayed expansion gives the stated conditional-mean expansion. For the covariance, expanding the bilinear form term by term gives
\[
\operatorname{Cov}_\ell\p{\Xi_r^{P},\Xi_s^{P}}
=\frac{d_rd_s}{T}\operatorname{Var}_\ell\p{\dot\ell_m(Y)}
+\frac{d_rq_s+d_sq_r}{2T^{3/2}}\operatorname{Cov}_\ell\p{\dot\ell_m(Y),\ddot\ell_m(Y)}
+O\p{T^{-2}}.
\]
The score--score term is the main one: $\operatorname{Var}_\ell\p{\dot\ell_m(Y)}=\operatorname{Var}_\ell(Y)/v^2=1/v+O\p{T^{-1/2}}$ by the variance expansion just displayed, so it equals $d_rd_s/(vT)+O\p{T^{-3/2}}$. The score--curvature cross term is $O\p{T^{-3/2}}$, because $\ddot\ell_m(y)=-1/v-(y-m)v_1/v^2$ is affine in $y$, whence $\operatorname{Cov}_\ell\p{\dot\ell_m(Y),\ddot\ell_m(Y)}=-\p{v_1/v^3}\operatorname{Var}_\ell(Y)=O(1)$. The remaining products are $O\p{T^{-2}}$: those involving $\ddot\ell_m$ with itself or $\dddot\ell_m$ with $\dot\ell_m$ carry an explicit factor $T^{-2}$ with second moments bounded uniformly by Lemma~\ref{lem:uniform_mgf}, and those involving the remainder $\varrho_{r,T}$ are $o\p{T^{-2}}$ by the Cauchy--Schwarz inequality together with $\sup_{r,\ell}\bbE_\ell\abs{\varrho_{r,T}}^2=o\p{T^{-3}}$. This is the stated covariance expansion. The centered third moment is $O\p{T^{-3/2}}$ because $\Xi_r^{P}-\bbE_\ell\Xi_r^{P}=\tfrac{d_r}{\sqrt T}\p{\dot\ell_m(Y)-\bbE_\ell\dot\ell_m}+O_{L^3}\p{T^{-1}}$ and $\sup_{\mu\in\calM}\bbE_\mu\abs{Y-\mu}^3<\infty$. The Gaussian identities follow directly from $\xi_{\ell,r}=\p{\beta_r/\sqrt T}Z+b_{\ell,r}/T$, where $\beta_r=d_r/s$ and $b_{\ell,r}=\p{d_r u^{(\ell)}-q_r/2}/v$. The two conditional means are $O\p{T^{-1}}$. Therefore, the raw second-moment matrices differ from the covariance matrices by $O\p{T^{-2}}$.
\end{proof}

\subsection{Comparison of the Transition Operators and Completion of the Lower Bound}

\begin{lemma}[Comparison of the transition operators for one observation]
\label{lem:one_step_operator_comparison}
There are constants $C_\star<\infty$ and $T_3\ge T_1$, depending only on $\p{\calS,m,C,\calU}$, such that, for every $f\in C_b^3\p{\bbR^{L-1}}$, every arm $a\in\calS$, every state $z$, and every $T\ge T_3$,
\[
\abs{K_{a,T}^{P}f(z)-K_{a,T}^{G}f(z)}\le C_\star T^{-3/2}\max_{1\le q\le3}\norm{D^q f}_\infty.
\]
\end{lemma}
\begin{proof}
Since $K_{a,T}^{P}f-K_{a,T}^{G}f=\sum_\ell p_\ell(z)\p{K_{a,T}^{P,\ell}f-K_{a,T}^{G,\ell}f}$ and the weights are nonnegative and sum to one, it suffices to bound $\abs{K_{a,T}^{P,\ell}f(z)-K_{a,T}^{G,\ell}f(z)}$ uniformly in $\ell$. For a single hidden point write $\delta^{P}=\Xi_a^{P}(Y)$ and $\delta^{G}=\xi_{a,\ell}$. Third-order Taylor expansion of $f$ about $z$ gives, for $\delta\in\cb{\delta^{P},\delta^{G}}$,
\[
\bbE\sqb{f(z+\delta)}=f(z)+\nabla f(z)^\top\bbE\sqb\delta+\tfrac12\operatorname{tr}\p{\nabla^2 f(z)\,\bbE\sqb{\delta\delta^\top}}+O\p{\norm{D^3 f}_\infty\bbE\norm\delta^3}.
\]
Subtracting the two expansions and using Lemma~\ref{lem:local_logodds_expansion}, the mean difference is $O\p{T^{-3/2}}$, the raw second-moment difference is $O\p{T^{-3/2}}$, and both third-moment terms are $O\p{T^{-3/2}}$. This follows because $\bbE\norm\delta^3\le C\p{\bbE\norm{\delta-\bbE\delta}^3+\norm{\bbE\delta}^3}=O\p{T^{-3/2}}$. Lemma~\ref{lem:local_logodds_expansion} bounds each of the three discrepancies by a constant multiple of $T^{-3/2}$ plus an $o\p{T^{-3/2}}$ remainder. The constants in these bounds depend only on $\p{\calS,m,C,\calU}$.

Hence, there is a threshold $T_3\ge T_1$, depending only on $\p{\calS,m,C,\calU}$, such that the absolute value of every remainder does not exceed the leading term in the same bound for all $T\ge T_3$. For every $T\ge T_3$, summing the three bounds gives an upper bound of $C_\star T^{-3/2}\max_{1\le q\le3}\norm{D^q f}_\infty$ for the discrepancy, uniformly in $z$, $a$, and $\ell$.

Because the mixture is a weighted average over $\ell$, it satisfies the same bound.
\end{proof}

\begin{theorem}[Finite-prior lower comparison]
\label{thm:finite_prior_lower_comparison}
There is a constant $\Lambda<\infty$, depending only on $\p{\calS,m,C,\calU}$, such that for every full-support prior $p\in\calP([L])$, every $h\in(0,1)$, and every $T\ge T_3$,
\[
V_{0,T}^{P}\p{z(p)}\ge V_{0,T}^{G}\p{z(p)}-\Lambda\p{h+T^{-1/2}h^{-2}}.
\]
In particular the choice $h=T^{-1/6}$ gives $V_{0,T}^{P}\p{z(p)}\ge V_{0,T}^{G}\p{z(p)}-2\Lambda T^{-1/6}$, uniformly over full-support priors.
\end{theorem}
\begin{proof}
Set $e_T=C_\star A'T^{-3/2}h^{-2}$. Lemma~\ref{lem:one_step_operator_comparison} gives this bound when it is applied to $f=\underline V_{t+1,T}^{G,h}$, because Lemma~\ref{lem:smooth_gaussian_subsolution} gives $\max_{1\le q\le3}\norm{D^q f}_\infty\le A'h^{-2}$. Define $\underline V_{t,T}^{P,h}=\underline V_{t,T}^{G,h}-(T-t)e_T$.

The operator $K_{a,T}^{P}$ reproduces constants and is monotone. Moreover, for $g=\underline V_{t+1,T}^{G,h}$, Lemma~\ref{lem:one_step_operator_comparison} gives $K_{a,T}^{P}g\ge K_{a,T}^{G}g-e_T$. Together with the Gaussian subsolution of Lemma~\ref{lem:smooth_gaussian_subsolution}, this yields
\begin{align*}
\min_{a\in\calS}K_{a,T}^{P}\underline V_{t+1,T}^{P,h}
&=\min_{a\in\calS}K_{a,T}^{P}\underline V_{t+1,T}^{G,h}-(T-t-1)e_T\\
&\ge\min_{a\in\calS}K_{a,T}^{G}\underline V_{t+1,T}^{G,h}-e_T-(T-t-1)e_T\\
&\ge\underline V_{t,T}^{P,h}.
\end{align*}
Thus $\underline V_{t,T}^{P,h}$ is a subsolution of the original recursion, and its terminal value satisfies $\underline V_{T,T}^{P,h}=\underline V_{T,T}^{G,h}\le V_{T,T}^{G}=V_{T,T}^{P}$. Backward induction with the monotone operator $K_{a,T}^{P}$ gives $\underline V_{t,T}^{P,h}\le V_{t,T}^{P}$ for every $t$, so at $t=0$,
\[
V_{0,T}^{P}\p{z(p)}\ge\underline V_{0,T}^{P,h}\p{z(p)}=\underline V_{0,T}^{G,h}\p{z(p)}-T e_T\ge V_{0,T}^{G}\p{z(p)}-A'h-C_\star A' T^{-1/2}h^{-2}.
\]
All constants here depend only on $\p{\calS,m,C,\calU}$ and are therefore independent of $p$, giving the claim with $\Lambda=A'\p{1+C_\star}$.
\end{proof}

\begin{proof}[Proof of Theorem~\ref{thm:minimax_lower}]
Fix $\p{\calS,m,C,\calU}$ as above. Since $\min\calM<\max\calM$, the interior $\operatorname{int}(\calM)$ is nonempty, and we can choose $m$ so that $\underline\mu=\min\calM<m$. For each $T\ge\max\cb{T_2,T_3}$, Lemma~\ref{lem:finite_grid_saddle} gives a maximizing prior $p_T^\star$ with $\inf_\pi\sum_\ell p_{T,\ell}^\star r_\ell(\pi)=\mathfrak G_{\calS,T}^{\calU}(s)$. The Bayes value $p\mapsto\inf_\pi\sum_\ell p_\ell r_\ell(\pi)$ is concave and takes values in $[0,2C]$. Hence, the full-support prior $p_T=\p{1-1/T}p_T^\star+\p{1/T}\p{1/L,\ldots,1/L}$ satisfies $\inf_\pi\sum_\ell p_{T,\ell}r_\ell(\pi)\ge\p{1-1/T}\mathfrak G_{\calS,T}^{\calU}(s)\ge\mathfrak G_{\calS,T}^{\calU}(s)-2C/T$. For every $\delta\in\calA$ and $T\ge\max\cb{T_2,T_3}$,
\[
\sqrt T\sup_{\bmmu\in\calM^K}\Regret_T^{\delta}(\bmmu)
\ge\max_{\ell}R_T^{P}\p{\delta,u^{(\ell)}}
\ge\sum_\ell p_{T,\ell}R_T^{P}\p{\delta,u^{(\ell)}},
\]
so taking the infimum over $\delta$ and applying Lemmas~\ref{lem:s_arm_reduction} and \ref{lem:original_posterior_markovization},
\[
\inf_{\delta\in\calA}\sqrt T\sup_{\bmmu\in\calM^K}\Regret_T^{\delta}(\bmmu)
\ge\inf_{\delta\in\calA}\sum_\ell p_{T,\ell}R_T^{P}\p{\delta,u^{(\ell)}}
\ge V_{0,T}^{P}\p{z(p_T)}.
\]
By Theorem~\ref{thm:finite_prior_lower_comparison} and the Gaussian posterior Markovization of Lemma~\ref{lem:posterior_markovization}, $V_{0,T}^{P}\p{z(p_T)}\ge V_{0,T}^{G}\p{z(p_T)}-2\Lambda T^{-1/6}=\inf_\pi\sum_\ell p_{T,\ell}r_\ell(\pi)-2\Lambda T^{-1/6}\ge\mathfrak G_{\calS,T}^{\calU}(s)-2C/T-2\Lambda T^{-1/6}$. The right-hand side does not depend on $\delta$, so for every $\delta\in\calA$,
\[
\liminf_{T\to\infty}\sqrt T\sup_{\bmmu\in\calM^K}\Regret_T^{\delta}(\bmmu)
\ge\liminf_{T\to\infty}\mathfrak G_{\calS,T}^{\calU}(s).
\]
By Lemma~\ref{lem:grid_extension}, $\mathfrak G_{\calS,T}^{\calU}(s)\ge\mathfrak G_{\calS,T}^{C}(s)-\p{2\rho+C\rho/\underline s}$, and Theorem~\ref{thm:seq_value_limit} gives $\mathfrak G_{\calS,T}^{C}(s)\to\mathfrak g_{\calS}^{C}(s)$ as $T\to\infty$, whence
\[
\inf_{\delta\in\calA}\liminf_{T\to\infty}\sqrt T\sup_{\bmmu\in\calM^K}\Regret_T^{\delta}(\bmmu)\ge\mathfrak g_{\calS}^{C}(s)-\p{2\rho+C\rho/\underline s}.
\]
First refine the grid so that $\rho\downarrow0$. Then let $C\uparrow\infty$. The left-hand side is independent of $\rho$ and $C$, so it is at least $\sup_{C<\infty}\mathfrak g_{\calS}^{C}\p{s_{\calS}(m)}=\mathfrak G_{\calS}^{\uparrow}\p{s_{\calS}(m)}$. Next take the supremum over $m\in\operatorname{int}(\calM)$. The continuity bound under relative perturbations in Lemma~\ref{lem:seq_game_properties}, together with the continuity of each $\sigma_a$, extends this supremum to the endpoints of $\calM$ by interior approximation. Finally, setting $\calS=[K]$ gives $\inf_{\delta\in\calA}\liminf_{T\to\infty}\sqrt T\sup_{\bmmu\in\calM^K}\Regret_T^{\delta}(\bmmu)\ge\Gamma(\bmP)$. The argument holds for every arm set $\calS$ with $\abs{\calS}\ge2$, so it gives the lower bound $\sup_{m\in\calM}\mathfrak G_{\calS}^{\uparrow}\p{s_{\calS}(m)}$ for each such $\calS$. By Corollary~\ref{cor:gamma_subsets}, the strongest bound is obtained at $\calS=[K]$.
\end{proof}

\begin{remark}
The order of limits is essential. We first fix $\p{\calS,m,C,\calU}$ and all constants in Lemmas~\ref{lem:hard_bellman_lipschitz}--\ref{lem:one_step_operator_comparison} and Theorem~\ref{thm:finite_prior_lower_comparison}, and then let $T\to\infty$. At these fixed values, the transition-comparison constant $C_\star$ multiplies the uniform third-derivative bound $A'h^{-2}$. The $O\p{T^{-1/6}}$ finite-prior error therefore vanishes. Only after taking this limit do we let $\rho\downarrow0$, let $C\uparrow\infty$, and vary $m$. The arm set remains fixed at $\calS=[K]$. The bound applies to the class $\calA$ of all adaptive experiments because the reduction to policies that are Markov in the posterior is performed after taking the infimum over behavioral policies.
\end{remark}

\section{Smooth Policies with Uniform Derivative Bounds}
\label{app:smooth_guaranteed}
This appendix constructs, for the sequential Gaussian game of Appendix~\ref{app:finite_grid_duality}, an indexed family of finite-memory behavioral policies whose Gaussian continuation risks have second and third derivatives bounded uniformly over the horizon. For every prescribed error, the quantitative horizon bound and the finite calculations below give a finite threshold beyond which the family contains policies whose worst-case Gaussian risks are within that error of $\mathfrak g_{\calS}^{C}(s)$.

The concluding lemma compares the risk of running one such policy on the original experiment with its Gaussian risk. The additional terms account for the variance mismatch and the third moments. These results are used in the upper-bound proof.

Throughout, $\calS\subseteq[K]$ is a nonempty arm set with $\abs{\calS}\ge2$, and the game, its observation law $X_t=u_a/\sqrt n+s_aZ_t^{0}$, its risks $R_n^{G}\p{\pi,u}$, the simple regret $\Delta_u^{\calS}$, and the values $\mathfrak g_{\calS}^{C}(s)$, $\mathfrak G_{\calS}^{\uparrow}(s)$ are those of Definition~\ref{def:seq_gaussian_game} and Appendix~\ref{app:seq_value_bellman}. The time-indexed game noise is written $Z_t^{0}$ in this appendix, because $Z_t$ denotes the continuous state of a linear-state policy. All vectors carry the Euclidean norm. For a bounded $g\in C^k\p{\bbR^d}$ write $M_0(g)=\norm g_\infty$ and $M_k(g)=\max_{\abs{\alpha}=k}\norm{\partial^\alpha g}_\infty$ for $k\ge1$, and for a kernel $\alpha\p{\cdot\mid z}$ into a finite set write $M_k(\alpha)=\max_a M_k\p{\alpha\p{a\mid\cdot}}$.

\subsection{Finite-Memory Linear-State Policies}

\begin{definition}[Linear-state behavioral policy]
\label{def:linear_state_policy}
Fix a horizon $n$. A linear-state behavioral policy consists of a continuous state $Z_t\in\bbR^d$, a memory state $E_t$ taking values in a finite set, Borel action kernels $\alpha_t\p{\cdot\mid z,e}$ on $\calS$, a Borel terminal kernel $q_n\p{\cdot\mid z,e}$ on $\calS$, deterministic memory updates $e^+=\kappa_t(e,a)$, coefficient vectors $B_{t,e,a},c_{t,e,a}\in\bbR^d$, and a deterministic initial state $\p{z_0,e_0}$. In the sequential Gaussian game with standard deviations $s$, after action $a$ and observation $X_t=u_a/\sqrt n+s_aZ_t^{0}$, with the $Z_t^{0}$ independent standard normals, the continuous state is updated by
\begin{align}
Z_{t+1}=Z_t+\frac{B_{t,E_t,a}}{\sqrt n}X_t+\frac{c_{t,E_t,a}}n,
\qquad E_{t+1}=\kappa_t\p{E_t,a}.
\label{eq:linear_state_update}
\end{align}
The action kernel is evaluated at every round, including rounds on which it returns the same distribution as in the preceding round. For a fixed local parameter $u$, the function $U_{t,n}^{u,\pi}(z,e)$ is the conditional expected terminal simple regret when the policy is at state $(z,e)$ after $t$ rounds and follows the action and terminal kernels of $\pi$ for the remaining rounds. We refer to this conditional expectation as the continuation risk. With the standard deviations $s$ suppressed in the notation, the continuation risks are
\begin{align}
U_{n,n}^{u,\pi}(z,e)
&=\sum_{b\in\calS}q_n\p{b\mid z,e}\Delta_u^{\calS}(b),
\label{eq:guaranteed_terminal}\\
U_{t,n}^{u,\pi}(z,e)
&=\sum_{a\in\calS}\alpha_t\p{a\mid z,e}\,
\bbE\sqb{U_{t+1,n}^{u,\pi}\p{z+\frac{B_{t,e,a}}{\sqrt n}\p{\frac{u_a}{\sqrt n}+s_aZ}+\frac{c_{t,e,a}}n,\kappa_t(e,a)}}
\label{eq:guaranteed_recursion}
\end{align}
with $Z\sim\calN(0,1)$, and $R_n^{G}\p{\pi,u}=U_{0,n}^{u,\pi}\p{z_0,e_0}$.
\end{definition}

\begin{definition}[Uniform derivative bounds for a linear-state policy]
\label{def:smooth_guaranteed_policy}
Fix $D_2,D_3,B_0<\infty$, a compact box $[-C,C]^{\calS}$, and a standard-deviation vector $s\in(0,\infty)^{\calS}$. A linear-state policy satisfies the $\p{D_2,D_3,B_0}$ bounds on $[-C,C]^{\calS}$ for $s$ if its coefficient vectors have Euclidean norm at most $B_0$, its action and terminal kernels are Borel, every continuation risk $z\mapsto U_{t,n}^{u,\pi}(z,e)$ in \eqref{eq:guaranteed_terminal}--\eqref{eq:guaranteed_recursion}, computed at standard deviations $s$, belongs to $C_b^3\p{\bbR^d}$, and
\[
\sup_{t,u,e}M_2\p{U_{t,n}^{u,\pi}(\cdot,e)}\le D_2,
\qquad
\sup_{t,u,e}M_3\p{U_{t,n}^{u,\pi}(\cdot,e)}\le D_3,
\]
the suprema ranging over the times $0\le t\le n$, the box $u\in[-C,C]^{\calS}$, and the finite memory states $e$. A family of policies indexed by the horizon satisfies these bounds uniformly for $s$ when the same constants apply to every member.
\end{definition}

\begin{lemma}[Smooth approximation of a finite-horizon policy]
\label{lem:finite_horizon_smooth_density}
Fix a horizon $q$, a finite grid $\calU=\cb{u^{(1)},\ldots,u^{(L)}}\subset\bbR^{\calS}$, a standard-deviation vector $s$, and a behavioral policy $\pi$ of the $q$-round game at standard deviations $s$. For every $\eta>0$ there is a $q$-round linear-state policy $\pi^{\mathrm{sm}}$, with continuous state the reference-posterior log odds $z\in\bbR^{L-1}$ of Appendix~\ref{app:seq_value_bellman} and action and terminal kernels that are $C^\infty$ in $z$ with bounded derivatives of every order, such that
\[
\max_{u\in\calU}\abs{R_q^{G}\p{\pi^{\mathrm{sm}},u}-R_q^{G}\p{\pi,u}}\le\eta.
\]
\end{lemma}
\begin{proof}
Apply Lemma~\ref{lem:full_support_posterior_sufficiency} with the uniform reference prior on $\calU$ and standard deviations $s$: there is a policy $\pi^r$ Markov in $\p{t,\Pi_t^r}$ with $R_q^{G}\p{\pi^r,u^{(\ell)}}=R_q^{G}\p{\pi,u^{(\ell)}}$ for every $\ell$. Write its kernels in the log-odds chart $z=z\p{\Pi_t^r}\in\bbR^{L-1}$ of Appendix~\ref{app:seq_value_bellman} as $\alpha_t^r\p{\cdot\mid z}$ and $q_q^r\p{\cdot\mid z}$. The reference-posterior log odds evolve according to the affine update in Appendix~\ref{app:seq_value_bellman}. This update has the linear-state form \eqref{eq:linear_state_update}, with coefficients $B_{a,r}=\p{u_a^{(r)}-u_a^{(L)}}/s_a^2$ and $c_{a,r}=-\p{\p{u_a^{(r)}}^2-\p{u_a^{(L)}}^2}/\p{2s_a^2}$. These coefficients do not depend on $t$ or $e$. Therefore, after replacing the kernels by $C^\infty$ kernels in $z$, the resulting rule is a $q$-round linear-state policy.

For a hidden point $\ell$ and a time $t$, let $\mu_{t,\ell}^r$ be the law of $z_t$ under $\pi^r$ at $u^{(\ell)}$, and set $\overline\mu_t=\sum_{\ell}\mu_{t,\ell}^r$, a finite measure on $\bbR^{L-1}$. We approximate the Borel simplex-valued map $z\mapsto\alpha_t^r\p{\cdot\mid z}$ in $L^1\p{\overline\mu_t}$ by a $C^\infty$ simplex-valued map with bounded derivatives of every order. First, fix a compact set carrying all but an arbitrarily small amount of $\overline\mu_t$-mass. Apply Lusin's theorem coordinatewise to obtain a closed subset on which the map is continuous and whose complement has arbitrarily small measure \citep{Folland2013realanalysis}. Apply the Tietze extension theorem to extend each coordinate continuously to $\bbR^{L-1}$ without changing its bounds \citep{Folland2013realanalysis}. The resulting continuous map agrees with $\alpha_t^r$ outside a set of small $\overline\mu_t$-measure. We project its values onto the probability simplex on $\calS$ and then convolve the projected map with a compactly supported $C^\infty$ probability density. Projection and the convexity of the simplex keep the map simplex-valued. A sufficiently small bandwidth makes the $L^1\p{\overline\mu_t}$ error arbitrarily small, and convolution makes every partial derivative bounded. Let $\alpha_t^{\mathrm{sm}}$ denote the resulting action kernel. Applying the same construction to $q_q^r$ under $\overline\mu_q$ gives the terminal kernel $q_q^{\mathrm{sm}}$. These kernels define the linear-state policy $\pi^{\mathrm{sm}}$.

For probability vectors $p$ and $q$ on the finite set $\calS$, put $r_a=\min\cb{p_a,q_a}$ and $r=\sum_{a\in\calS}r_a$. With probability $r$, select the same arm from the probabilities $r_a/r$ when $r>0$. On the remaining event, use the two residual distributions. Their supports are disjoint. This coupling disagrees with probability $1-r=\norm{p-q}_{\mathrm{TV}}$, and all its probabilities are measurable functions of $p$ and $q$.

Use this coupling for $\pi^r$ and $\pi^{\mathrm{sm}}$ together with shared Gaussian observations. At each round, apply it to $\alpha_t^{\mathrm{sm}}\p{\cdot\mid z_t}$ and $\alpha_t^r\p{\cdot\mid z_t}$. At the terminal time, apply it to $q_q^{\mathrm{sm}}\p{\cdot\mid z_q}$ and $q_q^r\p{\cdot\mid z_q}$. Let $\varsigma$ be the first time the two action sequences disagree, with $\varsigma=\infty$ if they never do. On $\cb{\varsigma\ge t}$ the two policies have played identically and, sharing the observations, occupy the common state $z_t^r$ of $\pi^r$, whose unconditional law is $\mu_{t,\ell}^r$; conditionally on $z_t^r$ the maximal coupling makes the two actions at time $t$ disagree with probability $\norm{\alpha_t^{\mathrm{sm}}\p{\cdot\mid z_t^r}-\alpha_t^r\p{\cdot\mid z_t^r}}_{\mathrm{TV}}$. Hence, bounding the indicator of $\cb{\varsigma\ge t}$ by one,
\begin{align*}
\bbP_{\ell}(\varsigma=t)
&=\bbE\sqb{\mathbbm 1\cb{\varsigma\ge t}
\norm{\alpha_t^{\mathrm{sm}}\p{\cdot\mid z_t^r}-\alpha_t^r\p{\cdot\mid z_t^r}}_{\mathrm{TV}}}\\
&\le\int\norm{\alpha_t^{\mathrm{sm}}\p{\cdot\mid z}-\alpha_t^r\p{\cdot\mid z}}_{\mathrm{TV}}\,d\mu_{t,\ell}^r(z).
\end{align*}
On $\cb{\varsigma=\infty}$ the two terminal states coincide. The choices therefore disagree with probability at most $\int\norm{q_q^{\mathrm{sm}}\p{\cdot\mid z}-q_q^r\p{\cdot\mid z}}_{\mathrm{TV}}\,d\mu_{q,\ell}^r(z)$. A union bound over the $q$ action rounds and the terminal round gives
\begin{align*}
\bbP_{\ell}\p{\text{choices differ}}
&\le
\sum_{t=0}^{q-1}\int\norm{\alpha_t^{\mathrm{sm}}\p{\cdot\mid z}-\alpha_t^r\p{\cdot\mid z}}_{\mathrm{TV}}\,d\mu_{t,\ell}^r(z)\\
&\quad+\int\norm{q_q^{\mathrm{sm}}\p{\cdot\mid z}-q_q^r\p{\cdot\mid z}}_{\mathrm{TV}}\,d\mu_{q,\ell}^r(z),
\end{align*}
each integral dominated by the corresponding $L^1\p{\overline\mu_t}$ approximation error. Off this event the two choices coincide, and $0\le\Delta_u^{\calS}\le2C_{\calU}$ with $C_{\calU}=\max_{v\in\calU}\norm v_\infty$, so
\[
\abs{R_q^{G}\p{\pi^{\mathrm{sm}},u^{(\ell)}}-R_q^{G}\p{\pi^r,u^{(\ell)}}}
\le2C_{\calU}\,\bbP_{\ell}\p{\text{choices differ}}.
\]
Since $\calU$ and $q$ are finite, choose the $L^1\p{\overline\mu_t}$ errors so small that the right-hand side is at most $\eta$ for every $\ell$, which with $R_q^{G}\p{\pi^r,\cdot}=R_q^{G}\p{\pi,\cdot}$ proves the claim.
\end{proof}

\begin{theorem}[Finite calculation of a smooth policy at a fixed horizon]
\label{thm:finite_computation_smooth_policy}
Fix a horizon $q$, a positive rational radius $C$, a finite parameter set $\calU=\cb{u^{(1)},\ldots,u^{(L)}}\subset[-C,C]^{\calS}$, a standard-deviation vector $s\in(0,\infty)^{\calS}$, and a positive rational $\varepsilon$. Assume that the coordinates of $\calU$ and $s$ are supplied through rational lower and upper bounds of any prescribed positive width. For these fixed inputs, a finite calculation performed before data collection returns a $q$-round linear-state policy $\pi^{\mathrm{fin}}$ whose action and terminal kernels are $C^\infty$ functions of the continuous state with bounded derivatives of every order and such that
\[
\max_{u\in\calU}R_q^G\p{\pi^{\mathrm{fin}},u}
\le
\mathfrak G_{\calS,q}^{\calU}(s)+\varepsilon.
\]
The same calculation returns finite constants $D_2$, $D_3$, and $B_0$ for which $\pi^{\mathrm{fin}}$ satisfies the $\p{D_2,D_3,B_0}$ bounds on $[-C,C]^{\calS}$ for $s$.
\end{theorem}
\begin{proof}
Put $d=L-1$ and use $u^{(L)}$ as the reference point in the posterior log odds. For $a\in\calS$ and $\ell\in[L]$, let $\beta_a\in\bbR^d$ and $b_{a,\ell}\in\bbR^d$ have coordinates
\[
\beta_{a,r}=\frac{u_a^{(r)}-u_a^{(L)}}{s_a},
\qquad
b_{a,\ell,r}
=
\frac{\p{u_a^{(r)}-u_a^{(L)}}u_a^{(\ell)}-\tfrac12\p{\p{u_a^{(r)}}^2-\p{u_a^{(L)}}^2}}{s_a^2}.
\]
By \eqref{eq:gaussian_logodds_increment}, one observation from arm $a$ changes the state under hidden point $u^{(\ell)}$ by $q^{-1/2}\beta_aZ+q^{-1}b_{a,\ell}$, where $Z\sim\calN(0,1)$. Write $\overline\beta=\max_a\norm{\beta_a}_\infty$ and $\overline b=\max_{a,\ell}\norm{b_{a,\ell}}_\infty$.

\emph{Rational lower and upper bounds.}
All numerical comparisons in the calculation use rational lower and upper bounds. For $x\ge0$, define
\[
S_N(x)
=
\frac12
+
\frac{1}{\sqrt{2\pi}}
\sum_{k=0}^{N}
\frac{(-1)^k x^{2k+1}}{2^k k!(2k+1)}.
\]
Taylor's theorem applied to $e^{-t^2/2}$ and integration over $[0,x]$ give
\[
\abs{\Phi(x)-S_N(x)}
\le
\frac{x^{2N+3}}{\sqrt{2\pi}\,2^{N+1}(N+1)!(2N+3)}.
\]
For negative arguments, use $\Phi(x)=1-\Phi(-x)$. Rational bounds for $\pi$ follow from Machin's identity $\pi/4=4\arctan(1/5)-\arctan(1/239)$ and the alternating series for $\arctan$. Taylor expansions with their remainder bounds give rational lower and upper bounds for the exponential function on compact intervals. After a finite rescaling, the identity $\log x=2\operatorname{arctanh}\p{(x-1)/(x+1)}$ and the power series for $\operatorname{arctanh}$ give rational lower and upper bounds for logarithms of positive rationals. Square roots of positive rationals can be bounded by bisection. Hence, all functions used below can be evaluated with rational lower and upper bounds whose difference is at most any prescribed positive width after finitely many arithmetic operations.

\emph{Finite approximation for one prior.}
Fix a rational full-support prior $p$ on $\calU$. Write $V_t^p=V_{t,q}^{G,p}$ and $Q_{t,a}^p=K_{a,q}V_{t+1}^p$. The terminal losses $c_b$ and the posterior weights $p_\ell(z)$ in Appendix~\ref{app:seq_value_bellman} are explicit smooth functions of $z$. They take values in $[0,2C]$, and
\[
\sum_{\ell=1}^L\abs{p_\ell(z)-p_\ell(z')}
\le d\norm{z-z'}_\infty.
\]
This inequality follows from $\partial p_\ell/\partial z_r=p_\ell\p{\mathbbm 1[\ell=r]-p_r}$ and the mean-value theorem.

Choose $M>0$ and let $B_p=\norm{z(p)}_\infty$. Define
\[
R_t
=
B_p+t\p{\frac{\overline b}{q}+\frac{\overline\beta M}{\sqrt q}}
\qquad(0\le t\le q).
\]
If $\abs{Z_r}\le M$ for every observation before time $t$, then the posterior state at time $t$ belongs to $[-R_t,R_t]^d$.

For $A>0$, define $\chi_A(x)=A\tanh(x/A)$ and apply this map coordinatewise to vectors. It maps $\bbR^d$ into $(-A,A)^d$, is $C^\infty$ with bounded derivatives, and satisfies
\[
\sup_{\norm z_\infty\le R}
\norm{\chi_A(z)-z}_\infty
\le
\omega(A,R)
\coloneqq
R-A\tanh(R/A).
\]
For each $t$, choose $A_t>R_t$ and put $\chi_t=\chi_{A_t}$. The quantity $\omega(A_t,R_t)$ converges to zero as $A_t\to\infty$.

We define the approximations by backward induction. At the terminal time, put
\[
\widehat V_q^p(z)
=
\min_{b\in\calS}c_b\p{\chi_q(z)},
\qquad
e_q=2Cd\,\omega(A_q,R_q).
\]
Then, $\abs{V_q^p(z)-\widehat V_q^p(z)}\le e_q$ on $[-R_q,R_q]^d$. A Lipschitz bound $\widehat\Lambda_q$ is obtained from the derivatives of the posterior weights and $\chi_q$.

Suppose that $\widehat V_{t+1}^p$ has been constructed, takes values in $[0,2C]$, and has a computed Lipschitz bound $\widehat\Lambda_{t+1}$. Partition $[-M,M]$ into intervals $I_{t,j}$ of length at most $h_t$, and choose a rational point $\xi_{t,j}$ in each interval. For $w_{t,j}=\bbP(Z\in I_{t,j})$, choose rational lower bounds $\widehat w_{t,j}\ge0$ such that
\[
\sum_j\p{w_{t,j}-\widehat w_{t,j}}
\le
\delta_{w,t}.
\]
For $z\in[-A_t,A_t]^d$, define the finite weighted-sum function
\[
G_{t,a}^p(z)
=
\sum_{\ell=1}^Lp_\ell(z)
\sum_j\widehat w_{t,j}
\widehat V_{t+1}^p\p{z+q^{-1/2}\beta_a\xi_{t,j}+q^{-1}b_{a,\ell}}.
\]
The boundedness and Lipschitz property of $\widehat V_{t+1}^p$ give
\[
\sup_{z\in[-A_t,A_t]^d}
\abs{K_{a,q}\widehat V_{t+1}^p(z)-G_{t,a}^p(z)}
\le
\eta_t^G,
\]
where
\[
\eta_t^G
=
\frac{\widehat\Lambda_{t+1}\overline\beta h_t}{\sqrt q}
+4C\Phi(-M)
+2C\delta_{w,t}.
\]
The three terms bound the displacement within each interval, the two Gaussian tails, and the deficit in the rational interval weights, respectively.

Every function $G_{t,a}^p$ has the computed Lipschitz bound
\[
L_t
\coloneqq
\widehat\Lambda_{t+1}+2Cd.
\]
Translation and averaging preserve the first term, while the bound on the posterior weights gives the second term.

Approximate $G_{t,a}^p$ on $[-A_t,A_t]^d$ by a tensor-product Bernstein polynomial of degree $m_t$ in each coordinate. At each Bernstein node, evaluate the finite formula for $G_{t,a}^p$ by rational interval arithmetic, obtain an interval of width at most $2\eta_t^{\mathrm{node}}$, and use its rational midpoint as the coefficient. Truncate the coefficient to $[0,2C]$. The Bernstein weights are nonnegative and sum to one. The resulting polynomial $P_{t,a}^p$ also takes values in $[0,2C]$ on the box and satisfies
\[
\sup_{z\in[-A_t,A_t]^d}
\abs{P_{t,a}^p(z)-G_{t,a}^p(z)}
\le
\eta_t^B,
\qquad
\eta_t^B
=
L_tA_t\sqrt{\frac d{m_t}}
+\eta_t^{\mathrm{node}}.
\]

Define
\[
\widehat Q_{t,a}^p(z)=P_{t,a}^p\p{\chi_t(z)},
\qquad
\widehat V_t^p(z)=\min_{a\in\calS}\widehat Q_{t,a}^p(z),
\]
and put
\[
\eta_t^\chi=L_t\omega(A_t,R_t),
\qquad
\delta_t=e_{t+1}+4C\Phi(-M)+\eta_t^G+\eta_t^B+\eta_t^\chi,
\qquad
e_t=\delta_t.
\]
The additional term $4C\Phi(-M)$ in $\delta_t$ controls the event on which the next state leaves $[-R_{t+1},R_{t+1}]^d$, where the bound $e_{t+1}$ is not available. The term $4C\Phi(-M)$ inside $\eta_t^G$ instead controls the Gaussian-tail truncation in the finite weighted sum. The transition operator and the pointwise minimum are nonexpansive in the supremum norm. Backward induction therefore gives
\[
\sup_{z\in[-R_t,R_t]^d}
\max_{a\in\calS}
\abs{Q_{t,a}^p(z)-\widehat Q_{t,a}^p(z)}
\le\delta_t,
\]
and
\[
\sup_{z\in[-R_t,R_t]^d}
\abs{V_t^p(z)-\widehat V_t^p(z)}
\le e_t.
\]
A Lipschitz bound $\widehat\Lambda_t$ for $\widehat V_t^p$ follows from the derivatives of the finitely many polynomials and the map $\chi_t$. Thus, every quantity needed at the next backward step is computed from finite formulas and rational lower and upper bounds.

For $\tau>0$, define the action kernels
\[
\alpha_t^p\p{a\mid z}
=
\frac{\exp\p{-\widehat Q_{t,a}^p(z)/\tau}}
{\sum_{b\in\calS}\exp\p{-\widehat Q_{t,b}^p(z)/\tau}}
\qquad(0\le t<q).
\]
The terminal kernel uses the same formula with $\widehat Q_{t,a}^p$ replaced by $c_a\p{\chi_q(z)}$. These kernels are $C^\infty$ and have bounded derivatives of every order.

For comparison, define the exact Gibbs policy by using $Q_{t,a}^p$ and $c_a$ in the same formulas. The inequality
\[
\sum_{a\in\calS}
\frac{e^{-x_a/\tau}}{\sum_be^{-x_b/\tau}}x_a
\le
\min_{a\in\calS}x_a+\tau\log\abs{\calS}
\]
and backward induction show that its Bayes risk is at most $V_0^p\p{z(p)}+(q+1)\tau\log\abs{\calS}$. Moreover,
\[
\norm{\operatorname{softmax}(-x/\tau)-\operatorname{softmax}(-y/\tau)}_{\mathrm{TV}}
\le
\frac{\norm{x-y}_\infty}{\tau}.
\]
This bound follows by integrating the Jacobian along the segment from $x$ to $y$.

Couple the exact and computed Gibbs policies by using the same Gaussian observations while their actions agree. On the event $\max_{1\le r\le q}\abs{Z_r}\le M$, their common state at time $t$ belongs to $[-R_t,R_t]^d$. A union bound over the action kernels, the terminal kernel, and the Gaussian tails gives
\[
\varepsilon_B
=
(q+1)\tau\log\abs{\calS}
+4Cq\Phi(-M)
+\frac{2C}{\tau}\left(e_q+\sum_{t=0}^{q-1}e_t\right)
\]
as an upper bound on the excess Bayes risk of the computed policy over $V_0^p\p{z(p)}$.

\emph{A finite collection of policies.}
Choose a finite rational full-support $\delta_p$-net $\calP_{\delta_p}$ of the prior simplex on $\calU$. Boundary priors are covered by mixing the points of a finite rational net with a sufficiently small amount of the uniform prior. For a policy $\pi$, define its risk vector by
\[
r(\pi)
=
\p{R_q^G\p{\pi,u^{(\ell)}}}_{\ell=1}^L.
\]
The loss is at most $2C$, so for every policy $\pi$ and priors $p,p'$,
\[
\abs{p^\top r(\pi)-(p')^\top r(\pi)}
\le2C\norm{p-p'}_1.
\]
For each $p^{(i)}\in\calP_{\delta_p}$, perform the preceding calculation and obtain a policy $\pi_i$ whose Bayes risk under $p^{(i)}$ is at most $V_0^{p^{(i)}}\p{z(p^{(i)})}+\varepsilon_B$.

For a fixed component $i$ and hidden point $u^{(\ell)}$, define
\[
W_{q,\ell}^{(i)}(z,e)
=
\sum_{b\in\calS}q_q^{(i)}\p{b\mid z,e}\Delta_{u^{(\ell)}}^{\calS}(b),
\]
and, for $t<q$,
\[
W_{t,\ell}^{(i)}(z,e)
=
\sum_{a\in\calS}\alpha_t^{(i)}\p{a\mid z,e}
\bbE\sqb{W_{t+1,\ell}^{(i)}\p{z+q^{-1/2}\beta_aZ+q^{-1}b_{a,\ell},\kappa_t(e,a)}}.
\]
Use the state boxes and Gaussian tail truncation from the preceding calculation. On each state box, approximate $W_{t,\ell}^{(i)}$ by Bernstein polynomials with rational interval coefficients. The bounded derivatives of the policy kernels give computable Lipschitz bounds for the functions in this recursion. These bounds control the displacement, tail, and node errors as above. If the difference between the rational upper and lower bounds added at time $t$ is at most $\eta_{R,t}$, backward induction gives rational numbers $\overline r_{\ell i}$ such that
\[
R_q^G\p{\pi_i,u^{(\ell)}}
\le
\overline r_{\ell i}
\le
R_q^G\p{\pi_i,u^{(\ell)}}+\sum_{t=0}^q\eta_{R,t}.
\]
Choose the local widths so that their sum is at most $\eta_R$.

Solve the finite matrix game with entries $\overline r_{\ell i}$ by rational linear programming to additive error at most $\eta_M$. For any prior $p$, choose $p^{(i)}\in\calP_{\delta_p}$ with $\norm{p-p^{(i)}}_1\le\delta_p$. Then,
\[
p^\top r(\pi_i)
\le
\mathfrak G_{\calS,q}^{\calU}(s)
+\varepsilon_B
+2C\delta_p.
\]
Finite-dimensional minimax duality and the rational upper bounds on the risk matrix therefore give rational mixture weights $\lambda_i$ such that
\[
\max_{u\in\calU}R_q^G\p{\pi^{\mathrm{fin}},u}
\le
\mathfrak G_{\calS,q}^{\calU}(s)
+\varepsilon_{\mathrm{calc}},
\]
where
\[
\varepsilon_{\mathrm{calc}}
=
\varepsilon_B+2C\delta_p+\eta_R+\eta_M.
\]
The policy $\pi^{\mathrm{fin}}$ chooses component $i$ with probability $\lambda_i$ before round $1$ and follows $\pi_i$ for all $q$ rounds. The component index is included in the initial memory state and remains fixed.

The stopping conditions are explicit. Choose $\tau$ so that $(q+1)\tau\log\abs{\calS}\le\varepsilon/6$. Increase $M$, the clipping ranges $A_t$, and the Bernstein degrees, and decrease the interval widths and the widths of the rational bounds until
\[
4Cq\Phi(-M)
\le
\frac{\varepsilon}{6}
\qquad\text{and}\qquad
\frac{2C}{\tau}\left(e_q+\sum_{t=0}^{q-1}e_t\right)
\le
\frac{\varepsilon}{6}.
\]
Finally, choose $2C\delta_p\le\varepsilon/6$, $\eta_R\le\varepsilon/6$, and $\eta_M\le\varepsilon/6$. The first error converges to zero as $M\to\infty$. The second converges to zero as the clipping ranges and Bernstein degrees tend to infinity and the interval widths and the widths of the rational bounds tend to zero. Since $q$, $L$, and $\abs{\calS}$ are finite, all stopping conditions are met after finitely many operations, and $\varepsilon_{\mathrm{calc}}\le\varepsilon$.

\emph{Derivative bounds.}
The derivatives of the Bernstein polynomials, the maps $\chi_t$, and the Gibbs kernels are explicit finite expressions. For each component policy, let $H_{k,t}$ bound the derivatives of order $k$ of the action kernels at time $t$, and let $H_{k,q}^{\mathrm{rec}}$ be the corresponding bound for the terminal kernel. Set $H_{0,t}=H_{0,q}^{\mathrm{rec}}=1$ and $D_{0,t}=2C$. For $1\le k\le3$, define
\[
D_{k,q}=2C\abs{\calS}H_{k,q}^{\mathrm{rec}},
\qquad
D_{k,t}=\abs{\calS}\sum_{r=0}^k\binom{k}{r}H_{r,t}D_{k-r,t+1}.
\]
Translation and Gaussian averaging do not increase derivative suprema. The product rule in \eqref{eq:guaranteed_terminal}--\eqref{eq:guaranteed_recursion} therefore shows that these numbers bound the derivatives of the continuation risks. Take $D_2=\max_tD_{2,t}$ and $D_3=\max_tD_{3,t}$, and then take the maxima over the finitely many mixture components.

The coefficient vectors in the posterior log-odds update are
\[
B_{a,r}=\frac{u_a^{(r)}-u_a^{(L)}}{s_a^2},
\qquad
c_{a,r}=-\frac{\p{u_a^{(r)}}^2-\p{u_a^{(L)}}^2}{2s_a^2}.
\]
The maximum of their Euclidean norms is a valid $B_0$.
\end{proof}

\begin{lemma}[Uniform bounds for the fixed-horizon calculation]
\label{lem:uniform_fixed_horizon_calculation}
Fix $q$, $C<\infty$, a finite parameter set $\calU=\cb{u^{(1)},\ldots,u^{(L)}}\subset[-C,C]^{\calS}$, $0<\underline w<\overline w<\infty$, and $\varepsilon>0$. Before a particular $s\in[\underline w,\overline w]^{\calS}$ is supplied, a finite calculation determines the Gaussian truncation level, interval widths, clipping ranges, Bernstein degrees, Gibbs temperature, prior net, widths of the rational bounds, and finite constants $D_2^{0}$, $D_3^{0}$, $B_0^{0}$, and $d_0$ with the following property. For every supplied $s\in[\underline w,\overline w]^{\calS}$ whose coordinates are given through rational lower and upper bounds of any prescribed positive width, the calculation of Theorem~\ref{thm:finite_computation_smooth_policy} returns a $q$-round policy $\pi_{q,s}^{\mathrm{fin}}$ such that
\[
\max_{u\in\calU}R_q^G\p{\pi_{q,s}^{\mathrm{fin}},u}
\le
\mathfrak G_{\calS,q}^{\calU}(s)+\varepsilon,
\]
and the policy satisfies the $\p{D_2^{0},D_3^{0},B_0^{0}}$ bounds on $[-C,C]^{\calS}$. Its continuous-state dimension is at most $d_0$. The constants are common to all $s$ in the box.
\end{lemma}
\begin{proof}
The proof of Theorem~\ref{thm:finite_computation_smooth_policy} can be run with bounds that are uniform over the standard-deviation box. In its posterior log-odds update, one has
\[
\overline\beta
\le
\frac{2C}{\underline w},
\qquad
\overline b
\le
\frac{3C^2}{\underline w^2}.
\]
The coefficient vectors used when the policy is run on centered observations satisfy
\[
\max_{a,r}\abs{B_{a,r}}
\le
\frac{2C}{\underline w^2},
\qquad
\max_{a,r}\abs{c_{a,r}}
\le
\frac{C^2}{2\underline w^2}.
\]
Thus, the state radii, Gaussian partitions, clipping ranges, Bernstein degrees, Gibbs temperature, prior net, and widths of the rational bounds in that proof can be chosen from $q$, $C$, $\calU$, $\underline w$, $\overline w$, and $\varepsilon$ alone. Their stopping inequalities hold uniformly over $s$ because every occurrence of $s$ in those inequalities is bounded through the displays above.

The Bernstein coefficients remain in $[0,2C]$. Once the common degrees, clipping maps, and Gibbs temperature have been fixed, the derivatives of the polynomials, the action kernels, and the terminal kernels have finite bounds that do not depend on $s$ or on the numerical values of the polynomial coefficients. Applying the backward recursion for $D_{k,t}$ in the proof of Theorem~\ref{thm:finite_computation_smooth_policy} gives common values $D_2^{0}$ and $D_3^{0}$. The preceding coefficient bounds give $B_0^{0}$ after taking Euclidean norms. The continuous-state dimension is $d_0=L-1$. Every operation is a finite arithmetic operation, the computation of rational lower and upper bounds, or a finite-dimensional linear program. Hence, the common refinement levels and constants are obtained by a finite calculation.
\end{proof}

\begin{lemma}[Extension from $q$ to $n$ rounds by block replication with uniform derivative bounds]
\label{lem:replicated_smooth_policy}
Fix a $q$-round linear-state policy $\pi^{\mathrm{sm}}$ with standard deviations $s$ and $C^\infty$ bounded-derivative kernels, as furnished by Lemma~\ref{lem:finite_horizon_smooth_density}. For every $n\ge q$ put $m=\lfloor n/q\rfloor\ge1$ and $\mathfrak r_n=\sqrt{mq/n}\in(0,1]$. There is an $n$-round linear-state policy $\calE_{n,q}\pi^{\mathrm{sm}}$ such that, for every $u$,
\begin{align}
R_n^{G}\p{\calE_{n,q}\pi^{\mathrm{sm}},u}=\mathfrak r_n^{-1}R_q^{G}\p{\pi^{\mathrm{sm}},\mathfrak r_nu}.
\label{eq:replication_risk_identity}
\end{align}
Moreover, for every compact box $[-C,C]^{\calS}$ there are constants $D_2,D_3,B_0<\infty$, depending on $q$, $\pi^{\mathrm{sm}}$, and $C$ but not on $n$, such that every policy in the family $\cb{\calE_{n,q}\pi^{\mathrm{sm}}}_{n\ge q}$ satisfies the $\p{D_2,D_3,B_0}$ bounds on $[-C,C]^{\calS}$ for $s$. The same constants apply when this family is run unchanged at any standard-deviation vector $s'\in(0,\infty)^{\calS}$. This conclusion concerns only the coefficient and derivative bounds; the identity \eqref{eq:replication_risk_identity} concerns the standard-deviation vector $s$ used to construct the policy.
\end{lemma}
\begin{proof}
At stage $r$ of the $q$-round policy, the smooth policy selects an arm $a$. The $n$-round policy selects arm $a$ for the next $m$ observations and divides the affine state update of the $q$-round policy into $m$ equal one-observation updates. Thus, the continuous state is updated after each observation. The normalized aggregate $\overline X_r=m^{-1/2}\sum_{j=1}^mX_{r,j}$ has law $\mathfrak r_nu_a/\sqrt q+s_aZ_r^{0}$, which matches one observation of the $q$-round game at parameter $\mathfrak r_nu$. The $q$ aggregates are conditionally independent given the arms selected by the $q$-round policy. Feeding these aggregates to $\pi^{\mathrm{sm}}$ and using $\Delta_u^{\calS}=\mathfrak r_n^{-1}\Delta_{\mathfrak r_nu}^{\calS}$ gives \eqref{eq:replication_risk_identity}. In each of the remaining $n-mq$ rounds, the policy selects a fixed arm and uses the null coefficients $B=c=0$, and the state does not change. At round $n$, it chooses the arm specified by the terminal kernel of the $q$-round policy.

Write the corresponding update of the $q$-round policy $\pi^{\mathrm{sm}}$ at stage $r$ as $z^+=z+q^{-1/2}B_{r,e,a}^{(q)}\overline X_r+q^{-1}c_{r,e,a}^{(q)}$. During each of the $m$ observations corresponding to stage $r$ of the $q$-round policy, use the coefficients
\[
B_{t,e,a}^{(n)}=\mathfrak r_n^{-1}B_{r,e,a}^{(q)},
\qquad
c_{t,e,a}^{(n)}=\mathfrak r_n^{-2}c_{r,e,a}^{(q)}.
\]
Summing the $m$ one-observation updates \eqref{eq:linear_state_update} reproduces the corresponding update of the $q$-round policy because $m/n=\mathfrak r_n^2/q$. Hence, $\calE_{n,q}\pi^{\mathrm{sm}}$ is a linear-state policy whose memory records the stage of the $q$-round policy, the arm selected at that stage, and the position within the current replication interval. Both the within-interval rounds and the discarded tail rounds have deterministic actions and leave the Gaussian transition of \eqref{eq:guaranteed_recursion} a translation-convolution in $z$. Differentiation in $z$ passes through it and does not increase any $M_k$. At a decision time of the $q$-round policy, the continuation risk is the finite sum $\sum_a\alpha_r^{(q)}\p{a\mid z}\,\bbE\sqb{U_{r+1}\p{z+\cdots}}$, so with the terminal loss bounded by $2C$, the product rule gives, for $0\le k\le3$,
\[
M_0\p{U_r}\le2C,
\qquad
M_k\p{U_r}\le C_k\max_{0\le j\le k}M_j\p{\alpha_r^{(q)}}\p{1+\max_{0\le j\le k}M_j\p{U_{r+1}}}\quad(k\ge1),
\]
and the terminal bound follows in the same way from the smooth terminal kernel. There are only $q$ decision times of the $q$-round policy and the kernel derivatives $M_j\p{\alpha_r^{(q)}}$ are finite constants of $\pi^{\mathrm{sm}}$, so backward induction yields $D_2,D_3$ depending on $q$, $\pi^{\mathrm{sm}}$, and $C$ but not on $n$. For $n\ge q$ one has $\mathfrak r_n^2=q\lfloor n/q\rfloor/n>1/2$, so $\mathfrak r_n^{-1}<\sqrt2$ and $\mathfrak r_n^{-2}<2$ are bounded uniformly in $n$, and the coefficient norm is bounded by $B_0=\sqrt2\max_{r,e,a}\norm{B_{r,e,a}^{(q)}}$ together with the analogous bound on $c$.

To verify the same derivative bounds at another standard-deviation vector, fix $s'\in(0,\infty)^{\calS}$ and run $\calE_{n,q}\pi^{\mathrm{sm}}$ unchanged in the game at $s'$. The coefficient vectors are $B_{t,e,a}^{(n)}=\mathfrak r_n^{-1}B_{r,e,a}^{(q)}$ and $c_{t,e,a}^{(n)}=\mathfrak r_n^{-2}c_{r,e,a}^{(q)}$. They are determined by $\pi^{\mathrm{sm}}$ and $n$. Changing the noise level does not change these vectors. Thus, the bound $B_0$ remains valid.

In the game at $s'$, the transition in \eqref{eq:guaranteed_recursion} sends $z$ to $z+\frac{B_{t,e,a}^{(n)}}{\sqrt n}\p{\frac{u_a}{\sqrt n}+s_a'Z}+\frac{c_{t,e,a}^{(n)}}n$, where $Z\sim\calN(0,1)$. For each fixed action, the continuation risk at $(t,e)$ is therefore a translation of the convolution of the next continuation risk with $\nu_{t,e,a}$. Here, $\nu_{t,e,a}$ is the law of $n^{-1/2}B_{t,e,a}^{(n)}s_a'Z$ on $\bbR^d$. The measure $\nu_{t,e,a}$ is supported on the line spanned by $B_{t,e,a}^{(n)}$ and is therefore degenerate. No regularity of $\nu_{t,e,a}$ is needed.

For any probability measure $\nu$ on $\bbR^d$ and any $f\in C_b^3\p{\bbR^d}$, dominated convergence gives $f*\nu\in C_b^3\p{\bbR^d}$ and $\partial^\beta\p{f*\nu}=\p{\partial^\beta f}*\nu$. Thus, $\abs{\partial^\beta\p{f*\nu}}\le\norm{\partial^\beta f}_\infty$ pointwise, and $M_k\p{f*\nu}\le M_k(f)$ for $0\le k\le3$. Convolution with a probability measure is an average and cannot increase a supremum. Translation leaves every $M_k$ unchanged.

The noise level therefore enters the backward recursion only through $\nu_{t,e,a}$, which never increases a derivative bound, exactly as in the display above. That display is driven by the combinatorial constants $C_k$ of the product rule, the kernel derivatives $M_j\p{\alpha_r^{(q)}}$ of $\pi^{\mathrm{sm}}$, the terminal bound $2C$, and the $q$ stages of that policy, none of which depends on $s'$. Backward induction therefore returns the same $D_2,D_3$, and $\calE_{n,q}\pi^{\mathrm{sm}}$ satisfies the $\p{D_2,D_3,B_0}$ bounds on $[-C,C]^{\calS}$ for $s'$. Thus, the same constants $D_2$, $D_3$, and $B_0$ apply at $s'$. This statement concerns only the coefficient and derivative bounds; the risk identity remains restricted to the standard-deviation vector $s$ used to construct the policy.
\end{proof}

\begin{theorem}[Approximation by smooth policies with uniform derivative bounds]
\label{thm:smooth_guaranteed_density}
Fix $\calS$, $C<\infty$, a finite standard-deviation set $\calS_0\subset(0,\infty)^{\calS}$, and $\eta>0$. A finite calculation returns finite constants $D_2,D_3,B_0$ and $N<\infty$ and, for every $s\in\calS_0$ and every $n\ge N$, an $n$-round policy $\pi_{n,s}$ satisfying the $\p{D_2,D_3,B_0}$ bounds on $[-C,C]^{\calS}$ for $s$ with
\[
\sup_{u\in[-C,C]^{\calS}}R_n^{G}\p{\pi_{n,s},u}\le\mathfrak g_{\calS}^{C}(s)+\eta.
\]
The derivative bounds for each $\pi_{n,s}$ are independent of the noise vector, as shown in Lemma~\ref{lem:replicated_smooth_policy}: the same constants $\p{D_2,D_3,B_0}$ apply on $[-C,C]^{\calS}$ for every $s'\in(0,\infty)^{\calS}$, not only for the $s$ at which the policy was constructed. The displayed risk bound holds only at $s$.
\end{theorem}
\begin{proof}
Put
\[
\underline s=\min_{s\in\calS_0}\min_{a\in\calS}s_a,
\qquad
\overline s=\max_{s\in\calS_0}\max_{a\in\calS}s_a.
\]
Choose a positive rational mesh $\rho$ such that
\[
e_\rho\coloneqq2\rho+\frac{C\rho}{\underline s}\le\frac\eta8,
\]
and construct a finite rational $\rho$-net $\calU\subset[-C,C]^{\calS}$. Let $A^{\mathrm{hor}}$ be the computable constant in Theorem~\ref{thm:finite_horizon_rate} for $\p{\calS,C,\calU,\underline s,\overline s}$, and set
\[
q
=
\max\left\{
\abs{\calS},
\left\lceil\p{\frac{4A^{\mathrm{hor}}}{\eta}}^4\right\rceil
\right\}.
\]
Then, for every $s\in\calS_0$,
\[
\mathfrak G_{\calS,q}^{\calU}(s)
\le
\inf_{r\ge\abs{\calS}}\mathfrak G_{\calS,r}^{\calU}(s)+\frac\eta4.
\]

Apply Lemma~\ref{lem:uniform_fixed_horizon_calculation} with error $\eta/4$. For every $s\in\calS_0$, it computes a smooth $q$-round policy $\pi_{q,s}^{\mathrm{fin}}$ with common derivative and coefficient bounds and
\[
\max_{u\in\calU}R_q^G\p{\pi_{q,s}^{\mathrm{fin}},u}
\le
\mathfrak G_{\calS,q}^{\calU}(s)+\frac\eta4.
\]
Lemma~\ref{lem:grid_extension}, applied first at horizon $q$ and then after taking infima over the horizon, gives
\[
\abs{\mathfrak g_{\calS}^{C}(s)-\inf_{r\ge\abs{\calS}}\mathfrak G_{\calS,r}^{\calU}(s)}
\le e_\rho.
\]
It also extends the risk bound of the computed policy from $\calU$ to $[-C,C]^{\calS}$. Hence,
\[
\sup_{u\in[-C,C]^{\calS}}R_q^G\p{\pi_{q,s}^{\mathrm{fin}},u}
\le
\mathfrak g_{\calS}^{C}(s)+2e_\rho+\frac\eta2
\le
\mathfrak g_{\calS}^{C}(s)+\frac{3\eta}{4}.
\]

Let
\[
M_0=2\abs{\calS}^{3/2}\overline s\,c_{\mathrm{mm}}.
\]
Lemma~\ref{lem:seq_uniform_finite} gives $\mathfrak g_{\calS}^{C}(s)\le M_0$ for every $s\in\calS_0$. Define
\[
N
=
\left\lceil
\max\left\{
2q,
\frac{q\p{4M_0+3\eta}}{\eta}
\right\}
\right\rceil.
\]
For $n\ge N$, let $\pi_{n,s}=\calE_{n,q}\pi_{q,s}^{\mathrm{fin}}$. Since $\mathfrak r_n^2=q\lfloor n/q\rfloor/n\ge1-q/n$ and $n\ge2q$, it holds that $\mathfrak r_n^{-1}-1\le q/n$. The risk identity \eqref{eq:replication_risk_identity} therefore gives
\begin{align*}
\sup_uR_n^G\p{\pi_{n,s},u}
&\le
\mathfrak r_n^{-1}\p{\mathfrak g_{\calS}^{C}(s)+3\eta/4}\\
&\le
\mathfrak g_{\calS}^{C}(s)+\frac{3\eta}{4}
+\frac qn\p{M_0+3\eta/4}
\le
\mathfrak g_{\calS}^{C}(s)+\eta.
\end{align*}
Lemma~\ref{lem:replicated_smooth_policy} computes bounds for the replicated policies from the common bounds in Lemma~\ref{lem:uniform_fixed_horizon_calculation}. These bounds do not depend on $n$ or on $s\in\calS_0$. Taking them as $D_2,D_3,B_0$ completes the finite construction.
\end{proof}

\subsection{Comparison with the Original Experiment for One Observation}

\begin{lemma}[Comparison with the original experiment for one observation]
\label{lem:guaranteed_recovery}
Fix $\calS$, a horizon $n$, a standard-deviation vector $\overline s=\p{\overline s_a}_{a\in\calS}\in(0,\infty)^{\calS}$, and an $n$-round linear-state policy $\pi$ satisfying the $\p{D_2,D_3,B_0}$ bounds on $[-C,C]^{\calS}$ for $\overline s$. Let the true arm families be $\p{P_{a,\mu_a}}_{a\in\calS}$ of Definition~\ref{def:mean_param}, fix a center $\widehat m\in\bbR$, and run $\pi$ on the original experiment using the coefficients and kernels constructed for $\overline s$, replacing the Gaussian observation in \eqref{eq:linear_state_update} by the centered outcome $Y-\widehat m$:
\[
Z_{t+1}=Z_t+\frac{B_{t,E_t,a}}{\sqrt n}\p{Y-\widehat m}+\frac{c_{t,E_t,a}}n,
\qquad Y\sim P_{a,\mu_a}.
\]
Write $u_a^\star=\sqrt n\p{\mu_a-\widehat m}$ and suppose $u^\star\in[-C,C]^{\calS}$. Let $R_n^{P}\p{\pi,u^\star}$ be the resulting original risk and $R_n^{G}\p{\pi,u^\star}$ the Gaussian risk of the same policy in the game with standard deviations $\overline s$. Put
\[
\chi=\max_{a\in\calS}\abs{\frac{\sigma_a^2\p{\mu_a}}{\overline s_a^2}-1}.
\]
Then
\begin{align}
R_n^{P}\p{\pi,u^\star}\le R_n^{G}\p{\pi,u^\star}+C^{\mathrm{cmp}}\p{\frac{D_3}{\sqrt n}+\chi D_2},
\label{eq:guaranteed_recovery_bound}
\end{align}
where $C^{\mathrm{cmp}}$ depends only on $\abs{\calS}$, the state dimension $d$, the coefficient bound $B_0$, any upper bound $\overline w\ge\max_{a\in\calS}\overline s_a$ on the standard deviations used to construct the policy, and any upper bound $\Lambda_3\ge\max_{a\in\calS}\bbE_{a,\mu_a}\abs{Y-\mu_a}^3$ on the centered third absolute moment of the original experiment. It may be taken nondecreasing in each of $\abs{\calS}$, $d$, $B_0$, $\overline w$, and $\Lambda_3$. One admissible choice is
\[
C^{\mathrm{cmp}}
=
\max\left\{
\frac{dB_0^2\overline w^2}{2},
\frac{d^{3/2}B_0^3}{6}\p{\Lambda_3+\overline w^3\bbE\abs Z^3}
\right\}.
\]
No lower bound on these standard deviations enters, so one value of $C^{\mathrm{cmp}}$ applies uniformly over any compact set of standard-deviation vectors with a common upper bound. For the arm families of Definition~\ref{def:mean_param} one may take $\Lambda_3=\overline\Lambda_3$ of \eqref{eq:third_moment_envelope}, which Assumption~\ref{asm:structural_envelope} determines on its own; when the original experiment is itself a Gaussian game with standard deviations at most $\overline w$, its centered third absolute moment is at most $\overline w^3\bbE\abs Z^3$, so any $\Lambda_3\ge\overline w^3\bbE\abs Z^3$ is admissible.
\end{lemma}
\begin{proof}
The centered third absolute moment $\bbE_{a,\mu_a}\abs{Y-\mu_a}^3$ of the original experiment is bounded by $\Lambda_3$ uniformly over $a\in\calS$; for the arm families of Definition~\ref{def:mean_param} one takes $\Lambda_3=\overline\Lambda_3$ of \eqref{eq:third_moment_envelope}, which by that display bounds $\bbE_{a,\mu}\abs{Y-\mu}^3$ uniformly over $a\in[K]$ and $\mu\in\calM$ and is determined only by the constants in $\mathfrak E$.

Let $U_{t,n}^{P}$ and $U_{t,n}^{G}$ be the continuation risks of $\pi$ at $u^\star$ in the two experiments, indexed by the continuous and memory states $(z,e)$. By \eqref{eq:guaranteed_terminal}, they share the terminal function. Fix $(t,e,a)$. For the current coefficient vectors $B=B_{t,e,a}$ and $c=c_{t,e,a}$, the two one-observation state increments are
\[
\widehat\Xi^{P}=\frac{B}{\sqrt n}\p{Y-\widehat m}+\frac cn,\quad Y\sim P_{a,\mu_a},
\qquad
\widehat\Xi^{G}=\frac{B}{\sqrt n}\p{\frac{u_a^\star}{\sqrt n}+\overline s_aZ}+\frac cn,\quad Z\sim\calN(0,1).
\]
Because $\mu_a-\widehat m=u_a^\star/\sqrt n$, the two increments share the common mean $\mathfrak m\coloneqq\bbE\widehat\Xi^{P}=\bbE\widehat\Xi^{G}=Bu_a^\star/n+c/n$. Their centered parts are $W^{P}=\p{B/\sqrt n}\p{Y-\mu_a}$ and $W^{G}=\p{B/\sqrt n}\overline s_aZ$. Thus, the covariance mismatch is
\[
\operatorname{Cov}\p{\widehat\Xi^{P}}-\operatorname{Cov}\p{\widehat\Xi^{G}}
=\frac1n\p{\sigma_a^2\p{\mu_a}-\overline s_a^2}BB^\top,
\qquad
\norm{\cdot}\le\frac{B_0^2\overline s_a^2}n\chi.
\]
The centered third moments obey $\bbE\norm{W^{P}}^3+\bbE\norm{W^{G}}^3\le B_0^3\p{\Lambda_3+\overline s_a^3\bbE\abs Z^3}n^{-3/2}$.

For $f\in C_b^3\p{\bbR^d}$, expand around the common mean. With $\widehat\Xi=\mathfrak m+W$ and $\bbE W=0$,
\[
\bbE\sqb{f\p{z+\widehat\Xi}}
=f\p{z+\mathfrak m}+\tfrac12\operatorname{tr}\p{\nabla^2f\p{z+\mathfrak m}\operatorname{Cov}(W)}+O\p{M_3(f)\bbE\norm W^3}.
\]
The linear term vanishes because $\bbE W=0$. Since $z+\mathfrak m$ is common to the two experiments, subtracting the two expansions of $f=U_{t+1,n}^{G}$ removes $f\p{z+\mathfrak m}$. Only the covariance mismatch and the two centered third-moment remainders remain.

The second-order difference is
\begin{align*}
&\tfrac12\operatorname{tr}\p{\nabla^2f\p{z+\mathfrak m}
\p{\operatorname{Cov}(\widehat\Xi^{P})-\operatorname{Cov}(\widehat\Xi^{G})}}\\
&\qquad=\tfrac1{2n}\p{\sigma_a^2(\mu_a)-\overline s_a^2}
B^\top\nabla^2f\p{z+\mathfrak m}B.
\end{align*}
Moreover, $\abs{B^\top\nabla^2f\,B}\le M_2(f)\norm B_1^2\le d\,M_2(f)B_0^2$ with $\norm B_1\le\sqrt d\,B_0$. The third-order remainders are $O\p{M_3(f)d^{3/2}\bbE\norm W^3}$.

The standard-deviation vector used to construct the policy enters these bounds only through $\overline s_a^2$ in the covariance mismatch and $\overline s_a^3$ in the Gaussian third moment. Replacing $\overline s_a$ by any upper bound $\overline w$ preserves both bounds. Each bound is nondecreasing in $\abs{\calS}$, $d$, $B_0$, $\overline w$, and $\Lambda_3$. Thus, $C^{\mathrm{cmp}}$ requires no lower bound on the components of that vector. With $M_2\p{U_{t+1,n}^{G}}\le D_2$, $M_3\p{U_{t+1,n}^{G}}\le D_3$, and the moment bounds above, there is a constant $C^{\mathrm{cmp}}$ of the stated dependence with
\begin{align}
\abs{\widehat K_{t,e,a,n}^{P}U_{t+1,n}^{G}-\widehat K_{t,e,a,n}^{G}U_{t+1,n}^{G}}
\le C^{\mathrm{cmp}}\p{\frac{\chi}nD_2+\frac1{n^{3/2}}D_3},
\label{eq:guaranteed_one_step}
\end{align}
The bound holds uniformly over $z$, the memory states, and the actions. Here, $\widehat K^{P}$ and $\widehat K^{G}$ denote the transition operators $g\mapsto\bbE\sqb{g\p{z+\widehat\Xi^{P}}}$ and $g\mapsto\bbE\sqb{g\p{z+\widehat\Xi^{G}}}$ for one observation.

Suppose that $\sup_{z,e}\abs{U_{t+1,n}^{P}-U_{t+1,n}^{G}}\le\mathfrak e_{t+1}$. The action kernels are identical in the two recursions, and each $\widehat K^{P}$ is an averaging operator. Add and subtract $\widehat K_{t,e,a,n}^{P}U_{t+1,n}^{G}$ and apply \eqref{eq:guaranteed_one_step}. Then, $\sup_{z,e}\abs{U_{t,n}^{P}-U_{t,n}^{G}}\le\mathfrak e_{t+1}+C^{\mathrm{cmp}}\p{\chi D_2/n+D_3/n^{3/2}}$.

Starting from $\mathfrak e_n=0$ and summing over the $n$ rounds gives $\sup_{z,e}\abs{U_{0,n}^{P}-U_{0,n}^{G}}\le C^{\mathrm{cmp}}\p{\chi D_2+D_3/\sqrt n}$. At the initial state, this is \eqref{eq:guaranteed_recovery_bound}.
\end{proof}

\begin{remark}
Lemma~\ref{lem:guaranteed_recovery} compares the original and Gaussian one-observation transition operators at each round. The proof uses only that the policy satisfying the bounds in Definition~\ref{def:smooth_guaranteed_policy} updates its state from a single observation and that its Gaussian continuation risks have horizon-uniform second and third derivatives. It conditions on no block average and fixes no assignment sequence, so it applies to the fully adaptive, outcome-responsive policies produced by Theorem~\ref{thm:smooth_guaranteed_density}. The variance-mismatch term is bounded by $D_2$ because $n^{-1}\sum_t M_2\p{U_{t+1,n}^{G}}\le D_2$, and the third-moment term is bounded by $D_3/\sqrt n$ because $n^{-3/2}\sum_t M_3\p{U_{t+1,n}^{G}}\le D_3/\sqrt n$.
\end{remark}

\section{SMAS and the Minimax Upper Bound}
\label{app:upper_comparison}
This appendix defines SMAS and proves that it attains $\Gamma\p{\bmP}$ for every $\bmP\in\mathfrak P\p{\mathfrak E}$. The strategy uses empirical means, empirical variances, and the seven constants in $\mathfrak E$.

\subsection{Concentration Bounds for the Pilot Estimates}
Recall the constants $\mathfrak E=\p{\underline v,\overline v,L_v,\lambda_0,C_0,\eta_0,D_0}$ in \eqref{eq:envelope_tuple}. Here, $\underline v$ and $\overline v$ bound the variance functions, $L_v$ bounds their absolute derivatives, and $\p{\lambda_0,C_0}$ and $\p{\eta_0,D_0}$ are the constants in the two moment conditions. These seven constants are supplied to the experimenter, determine the class $\mathfrak P\p{\mathfrak E}$ in \eqref{eq:envelope_class}, and are the only information about the variance functions used to construct the strategy.

Fix a first-stage size $T_0$ divisible by $K$ and put $n_0=T_0/K$. Stage 1 selects every arm exactly $n_0$ times and forms the empirical mean and the unbiased empirical variance
\[
\widehat\mu_{a,T_0}=\frac{1}{n_0}\sum_{i=1}^{n_0}Y_{a,i},
\qquad
\widehat\sigma_{a,T_0}^2=\frac{1}{n_0-1}\sum_{i=1}^{n_0}\p{Y_{a,i}-\widehat\mu_{a,T_0}}^2.
\]

\begin{lemma}[Uniform pilot-mean concentration]
\label{lem:pilot_mean_concentration}
Fix $x_1<\infty$. Under Assumption~\ref{asm:structural_envelope} there are constants $c_1,c_2>0$, depending only on $x_1$, $\lambda_0$, and $C_0$, such that for every sample size $n\ge1$ and every $0<x\le x_1$,
\[
\sup_{a\in[K]}\sup_{\mu\in\calM}\bbP_{a,\mu}\p{\abs{\frac{1}{n}\sum^n_{i=1}Y_{a,i}-\mu}>x}\le c_1\exp\p{-c_2 n x^2},
\]
where $Y_{a,1},\ldots,Y_{a,n}$ are independent observations from $P_{a,\mu}$. The pilot mean $\widehat\mu_{a,T_0}$ is the case $n=n_0$; the lemma is applied at other sample sizes to the comparison-sample means of Algorithm~\ref{alg:smas}.
\end{lemma}
\begin{proof}
Fix $a$ and $\mu$. For $0<\lambda\le\lambda_0$, the uniform local exponential moment (4) of Definition~\ref{def:mean_param} with the common constant, independence of the $n$ observations, and Markov's inequality give
\begin{align*}
\bbP_{a,\mu}\p{\frac{1}{n}\sum^n_{i=1}Y_{a,i}-\mu\ge x}
&\le\exp\p{-\lambda n x}
\prod_{i=1}^{n}\bbE_{a,\mu}\sqb{\exp\p{\lambda\p{Y_{a,i}-\mu}}}\\
&\le\exp\p{-\lambda n x+C_0 n\lambda^2}.
\end{align*}
For $0<x\le2C_0\lambda_0$, take $\lambda=x/\p{2C_0}\in(0,\lambda_0]$ to obtain $\exp\p{-n x^2/\p{4C_0}}$. For $2C_0\lambda_0<x\le x_1$, take $\lambda=\lambda_0$ to obtain $\exp\p{-\lambda_0 n\p{x-C_0\lambda_0}}\le\exp\p{-\lambda_0 n x/2}\le\exp\p{-\p{\lambda_0/\p{2x_1}}n x^2}$, using $x-C_0\lambda_0\ge x/2$ and then $x\ge x^2/x_1$. Taking
\[c_2=\min\cb{1/\p{4C_0},\lambda_0/\p{2x_1}}\] and $c_1=1$ covers $0<x\le x_1$; the lower tail is identical by the same argument applied to $-\p{Y_{a,i}-\mu}$, giving $c_1=2$. The constants depend only on $x_1$ and the common constants in $\mathfrak E$ and are therefore uniform over the arms, over $\mu\in\calM$, and over the sample size $n$.
\end{proof}

\begin{lemma}[Uniform empirical-variance concentration]
\label{lem:pilot_variance_concentration}
Under Assumption~\ref{asm:structural_envelope} there are constants $c_3,c_4>0$, depending only on $\overline v$, $\lambda_0$, $C_0$, $\eta_0$, and $D_0$, such that for every $n_0\ge2$ and every $0<x\le4$,
\[
\sup_{a\in[K]}\sup_{\mu\in\calM}\bbP_{a,\mu}\p{\abs{\widehat\sigma_{a,T_0}^2-\sigma_a^2(\mu)}>x}\le c_3\exp\p{-c_4 n_0\min\cb{x^2,x}}.
\]
\end{lemma}
\begin{proof}
Fix $a$ and $\mu$, and write $X_i=Y_{a,i}-\mu$, $v=\sigma_a^2(\mu)$, and $Z_i=X_i^2-v$. Define
\[
V_0
=
2\exp\p{\eta_0\overline v/4}
\left(
\frac{64D_0}{9\eta_0^2e^2}
+
\overline v^2D_0
\right).
\]
For $\abs\lambda\le\eta_0/4$, the inequality $e^x\le1+x+x^2e^{\abs x}/2$ and $\bbE Z_i=0$ give
\[
\bbE e^{\lambda Z_i}
\le
\exp\p{V_0\lambda^2/2}.
\]
Indeed, $\abs{Z_i}\le X_i^2+\overline v$, and, for every $y\in\bbR$,
$y^4e^{\eta_0y^2/4}\le64e^{\eta_0y^2}/(9\eta_0^2e^2)$.
Chernoff's bound therefore yields
\[
\bbP_{a,\mu}\p{\abs{n_0^{-1}\sum_{i=1}^{n_0}Z_i}>y}
\le
2\exp\left(
-n_0\min\cb{\frac{y^2}{2V_0},\frac{\eta_0y}{8}}
\right).
\]

The identity
\[
\widehat\sigma_{a,T_0}^2-v
=
\frac{v}{n_0-1}
+
\frac{n_0}{n_0-1}
\left(
\frac1{n_0}\sum_{i=1}^{n_0}Z_i
-
\p{\widehat\mu_{a,T_0}-\mu}^2
\right)
\]
holds for every $n_0\ge2$. If $x\le4\overline v/(n_0-1)$, then $n_0\min\cb{x^2,x}\le8\overline v$. If $x>4\overline v/(n_0-1)$, then the deterministic term is at most $x/4$. In that case, the event $\abs{\widehat\sigma_{a,T_0}^2-v}>x$ implies
\[
\abs{\frac1{n_0}\sum_{i=1}^{n_0}Z_i}>\frac{x}{8}
\quad\text{or}\quad
\abs{\widehat\mu_{a,T_0}-\mu}>\sqrt{\frac{x}{8}}.
\]
Lemma~\ref{lem:pilot_mean_concentration}, applied with $x_1=1$, bounds the second probability. Thus, the claim holds with
\[
c_4
=
\min\left(
\frac{1}{128V_0},
\frac{\eta_0}{64},
\frac18\min\cb{\frac1{4C_0},\frac{\lambda_0}{2}}
\right)
\]
and $c_3=4\exp\p{8\overline v c_4}$. These constants depend only on $\overline v$, $\lambda_0$, $C_0$, $\eta_0$, and $D_0$.
\end{proof}

\subsection{The SMAS Strategy}

The strategy uses an integer index $j\ge8$. For every such $j$, set
\[
\alpha_j=\gamma_j=j^{-2},
\qquad
B_j=j^2,
\qquad
H_j=j^3,
\qquad
R_j=8j^2,
\qquad
\omega_j=j^{-4},
\]
where $\omega_j$ is the allowed excess risk in the Gaussian game. The relative tolerance $r_j$ for the pilot variance estimates and the multiplicative grid spacing $\zeta_j$ for the rounded standard deviations are fixed in Definition~\ref{def:indexed_near_minimax_policies}, after $D_{2,j}$, $D_{3,j}$, and $B_{0,j}$ have been fixed, at values in $(0,j^{-4}]$ small enough that the resulting error is at most $2j^{-1}$. The numerical values are not optimized; they are chosen so that every error depending only on $j$ in Theorem~\ref{thm:alladapt_upper} vanishes as $j\to\infty$. For each $j$, the parameter grids, the standard-deviation grids, and the derivative bounds are finite, and the indexed policies use one set of bounds for all horizons. Theorem~\ref{thm:finite_horizon_rate}, Lemma~\ref{lem:uniform_guarantee_box}, and the comparison bounds of this appendix compute these constants, grids, policies, and budget thresholds using only the constants in $\mathfrak E$ and the quantities already fixed for $j$.

The same budget thresholds must satisfy the requirements of the worst-case upper bound in Theorem~\ref{thm:alladapt_upper} and the average upper bound in Theorem~\ref{thm:alladapt_bayes_upper}, so we fix them once before defining the strategy. Write $\calR_j^{\mathrm{mm}}$ for the finite list of requirements in the proof of Theorem~\ref{thm:alladapt_upper} and $\calR_j^{\mathrm{B}}$ for the finite list in the proof of Theorem~\ref{thm:alladapt_bayes_upper}. For
\[
T\ge T_j^{\mathrm{sched}}
\coloneqq
\left\lceil
\max\left\{
\frac{2K}{\alpha_j},
\frac{2K}{\gamma_j}
\right\}
\right\rceil,
\]
the per-arm pilot and comparison counts and the main horizon satisfy
\[
n_0\ge\frac{\alpha_jT}{2K},
\qquad
n_g\ge\frac{\gamma_jT}{2K},
\qquad
n\ge\p{1-\alpha_j-\gamma_j}T.
\]

In every error bound in $\calR_j^{\mathrm{mm}}$ and $\calR_j^{\mathrm{B}}$, replace $n_0$, $n_g$, and $n$ by these lower bounds. This gives a finite list of explicit upper bounds that are nonincreasing in $T$. The remaining requirements are explicit lower bounds on $T$. Define $N_j^{\mathrm{mm}}$ as the smallest integer not below $T_j^{\mathrm{sched}}$ for which all bounds in $\calR_j^{\mathrm{mm}}$ hold, and define $N_j^{\mathrm{B}}$ in the same way. The upper bounds converge to zero for fixed $j$, so both integer searches terminate. Their monotonicity shows that the corresponding requirements hold for every larger budget.

We set
\begin{align}
N_8=\max\cb{N_8^{\mathrm{mm}},N_8^{\mathrm{B}}},
\qquad
N_j=\max\cb{N_j^{\mathrm{mm}},N_j^{\mathrm{B}},N_{j-1}+1}
\quad\text{for }j\ge9,
\label{eq:budget_thresholds}
\end{align}
which satisfies both lists for every $j$. The strict increment makes $N_j$ strictly increasing, hence $N_j\uparrow\infty$; without it, a sequence of budget thresholds that was eventually constant would leave the set below unbounded and $j_T$ undefined. For $T\ge N_8$, the index selected for budget $T$ is
\[
j_T=\max\cb{j\ge8:N_j\le T}.
\]

It is well defined and satisfies $j_T\ge J$ for every $T\ge N_J$, so $j_T\to\infty$ as $T\to\infty$. For the finitely many budgets $T<N_8$, the strategy selects arm $1$ in every round and chooses arm $1$. This fixed Borel initialization does not affect the asymptotic results as $T\to\infty$. No growth rate for $j_T$ is imposed. The arguments below use only that $j_T\to\infty$ and that the requirements of the two lists hold at the selected index $j_T$, which they do because $T\ge N_{j_T}$ by the definition of $j_T$; a quantity such as $B_{j_T}^3/\sqrt T$ is therefore at most $j_T^{-1}$ not because $j_T$ grows at any particular rate, but because that inequality is one of the requirements imposed on $N_j$. Neither list involves the prior, and, by \eqref{eq:third_moment_envelope}, neither involves the true arm families or their variance functions, so \eqref{eq:budget_thresholds} fixes one sequence of budget thresholds, and with it one strategy, for every collection of arm families in the model class and every prior.

\begin{lemma}[Uniform constants over a compact set of standard-deviation vectors]
\label{lem:uniform_guarantee_box}
Fix a nonempty $\calS\subseteq[K]$ with $\abs{\calS}\ge2$, $C<\infty$, $\eta>0$, and $0<\underline w<\overline w<\infty$. A finite calculation returns finite constants $D_2,D_3,B_0,d_0,C^{\mathrm{cmp}}$, and $N$, depending only on $\calS$, $C$, $\eta$, $\underline w$, $\overline w$, and the uniform third-moment bound $\overline\Lambda_3$ in \eqref{eq:third_moment_envelope}. For every supplied $s\in[\underline w,\overline w]^{\calS}$ and every $n\ge N$, the same calculation returns an $n$-round linear-state policy $\pi_{n,s}$ satisfying the $\p{D_2,D_3,B_0}$ bounds on $[-C,C]^{\calS}$ for $s$, whose continuous state has dimension at most $d_0$, with
\[
\sup_{u\in[-C,C]^{\calS}}R_n^{G}\p{\pi_{n,s},u}\le\mathfrak g_{\calS}^{C}(s)+\eta.
\]
The constant $C^{\mathrm{cmp}}$ can be used in the original-versus-Gaussian risk bound of Lemma~\ref{lem:guaranteed_recovery} for every standard-deviation vector in $[\underline w,\overline w]^{\calS}$.
\end{lemma}
\begin{proof}
Choose a positive rational mesh $\rho$ such that
\[
e_\rho\coloneqq2\rho+\frac{C\rho}{\underline w}\le\frac\eta8,
\]
and construct a finite rational $\rho$-net $\calU\subset[-C,C]^{\calS}$. Let $A^{\mathrm{hor}}$ be the computable constant in Theorem~\ref{thm:finite_horizon_rate} for $\p{\calS,C,\calU,\underline w,\overline w}$, and set
\[
q
=
\max\left\{
\abs{\calS},
\left\lceil\p{\frac{4A^{\mathrm{hor}}}{\eta}}^4\right\rceil
\right\}.
\]
Then, for every $s\in[\underline w,\overline w]^{\calS}$,
\[
\mathfrak G_{\calS,q}^{\calU}(s)
\le
\inf_{r\ge\abs{\calS}}\mathfrak G_{\calS,r}^{\calU}(s)+\frac\eta4.
\]

Apply Lemma~\ref{lem:uniform_fixed_horizon_calculation} with error $\eta/4$. Before $s$ is supplied, it fixes the common truncation level, interval widths, clipping ranges, Bernstein degrees, Gibbs temperature, prior net, widths of the rational bounds, and the common derivative, coefficient, and dimension bounds. For each supplied $s$ it computes a smooth $q$-round policy $\pi_{q,s}^{\mathrm{fin}}$ satisfying
\[
\max_{u\in\calU}R_q^G\p{\pi_{q,s}^{\mathrm{fin}},u}
\le
\mathfrak G_{\calS,q}^{\calU}(s)+\frac\eta4.
\]
Lemma~\ref{lem:grid_extension}, applied at each horizon and then after taking infima over the horizon, gives
\[
\abs{\mathfrak g_{\calS}^{C}(s)-\inf_{r\ge\abs{\calS}}\mathfrak G_{\calS,r}^{\calU}(s)}
\le e_\rho.
\]
It also extends the policy risk from $\calU$ to the full box. Therefore,
\[
\sup_{u\in[-C,C]^{\calS}}R_q^G\p{\pi_{q,s}^{\mathrm{fin}},u}
\le
\mathfrak g_{\calS}^{C}(s)+2e_\rho+\frac\eta2
\le
\mathfrak g_{\calS}^{C}(s)+\frac{3\eta}{4}.
\]

Since $c_{\mathrm{mm}}\le1$, put $M_0=2\abs{\calS}^{3/2}\overline w$. Lemma~\ref{lem:seq_uniform_finite} gives $\mathfrak g_{\calS}^{C}(s)\le M_0$ throughout the standard-deviation box. Define
\[
N
=
\left\lceil
\max\left\{
2q,
\frac{q\p{4M_0+3\eta}}{\eta}
\right\}
\right\rceil.
\]
For $n\ge N$, let $\pi_{n,s}=\calE_{n,q}\pi_{q,s}^{\mathrm{fin}}$. As in the proof of Theorem~\ref{thm:smooth_guaranteed_density}, one has $\mathfrak r_n^{-1}-1\le q/n$. The risk identity \eqref{eq:replication_risk_identity} then gives
\[
\sup_uR_n^G\p{\pi_{n,s},u}
\le
\mathfrak g_{\calS}^{C}(s)+\frac{3\eta}{4}
+\frac qn\p{M_0+3\eta/4}
\le
\mathfrak g_{\calS}^{C}(s)+\eta.
\]

Lemma~\ref{lem:replicated_smooth_policy} computes $D_2,D_3,B_0$ from the common fixed-horizon bounds, and these constants do not depend on $n$ or on $s$. Let $d_0=\abs{\calU}-1$. Since $\bbE\abs Z^3\le2$, put
\[
\Lambda_{3,\mathrm{box}}
=
\max\left\{
\overline\Lambda_3,
2\overline w^3
\right\}.
\]
Using the explicit choice in Lemma~\ref{lem:guaranteed_recovery}, set
\[
C^{\mathrm{cmp}}
=
\max\left\{
\frac{d_0B_0^2\overline w^2}{2},
\frac{d_0^{3/2}B_0^3}{6}
\p{\Lambda_{3,\mathrm{box}}+2\overline w^3}
\right\}.
\]
This constant applies to every policy computed above and every standard-deviation vector in the box. All grids, horizons, refinement levels, bounds, and policies are obtained through finite calculations, which proves the claim.
\end{proof}

\begin{theorem}[Finite computation of SMAS]
\label{thm:finite_computation_smas}
Suppose that $K$ is given, that $\min\calM<\max\calM$, and that the two endpoints of $\calM$ can be supplied through rational lower and upper bounds of any prescribed positive width. Suppose also that valid positive rational lower bounds for $\underline v$, $\lambda_0$, and $\eta_0$, and valid rational upper bounds for $\overline v$, $L_v$, $C_0$, and $D_0$ are given. For every budget $T$, a finite calculation determines $j_T$, the grids and indexed Gaussian-game policies required at that budget, and the sampling and recommendation kernels of Algorithm~\ref{alg:smas}. The calculation uses finitely many arithmetic operations, rational lower and upper bounds, and finite-dimensional linear programs.
\end{theorem}
\begin{proof}
Fix $j\ge8$. The quantities $\alpha_j$, $\gamma_j$, $B_j$, $H_j$, $R_j$, and $\omega_j$ are explicit. For each of the finitely many arm subsets, Lemma~\ref{lem:uniform_guarantee_box} computes $D_{2,j}$, $D_{3,j}$, $B_{0,j}$, $d_{0,j}$, $C_j^{\mathrm{cmp}}$, and $N_j^0$. Definition~\ref{def:indexed_near_minimax_policies} then computes $r_j$, $\zeta_j$, and the finite grid $\calV_j$. For a supplied grid point and main horizon, the same lemma computes the assigned policy.

Refine the endpoint bounds until they give rational numbers $0<\underline\Delta_{\calM}\le\Delta_{\calM}\le\overline\Delta_{\calM}$. We use $\underline\Delta_{\calM}$ in requirements of the form $x/\sqrt T\le\Delta_{\calM}$ and $\overline\Delta_{\calM}$ in constants that are nondecreasing in the diameter.

The constants in the concentration bounds for the pilot mean and variance estimates are given by the formulas in Lemmas~\ref{lem:pilot_mean_concentration} and \ref{lem:pilot_variance_concentration}. The constants in the two-arm comparison are given by Lemmas~\ref{lem:neyman_scale_calibration}, \ref{lem:binary_upper_comparison}, and \ref{lem:berry_esseen}. The final-comparison bounds use the same pilot-mean constant and elementary Gaussian integrals. Scalar quantities such as square roots, $c_{\mathrm{mm}}$, and Gaussian moments are replaced by rational lower or upper bounds in the required direction; the bounds $c_{\mathrm{mm}}\le1$ and $\bbE\abs Z^3\le2$ already suffice. Thus, after the policies and their derivative bounds have been computed, every requirement in $\calR_j^{\mathrm{mm}}$ and $\calR_j^{\mathrm B}$ is a finite numerical inequality.

The construction preceding \eqref{eq:budget_thresholds} defines $N_j^{\mathrm{mm}}$ and $N_j^{\mathrm B}$ by integer searches over explicit nonincreasing upper bounds. Each search terminates because every bound converges to zero as $T\to\infty$ for fixed $j$. It then computes $N_j$. Starting at $j=8$, compute $N_j^{\mathrm{mm}}$, $N_j^{\mathrm B}$, $N_j$, and the policies indexed by $j$ until the first index with $N_j>T$. Since $N_j\ge N_{j-1}+1$, this occurs after finitely many indices, and the preceding index gives $j_T$ when $T\ge N_8$. The initialization for $T<N_8$ is explicit. Only the policies indexed by $j_T$, the finitely many retained sets, the finite grid $\calV_{j_T}$, and the realized main horizon are needed at budget $T$. Algorithm~\ref{alg:smas} then gives the sampling and recommendation kernels by finite formulas.
\end{proof}

Set
\[
T_0=K\left\lfloor\frac{\alpha_{j_T}T}{K}\right\rfloor,
\qquad
n_0=\frac{T_0}{K},
\qquad
\tau_T=\frac{B_{j_T}}{\sqrt T}.
\]
Stage~1 selects every arm exactly $n_0$ times, computes $\widehat\mu_{a,T_0}$ and $\widehat\sigma_{a,T_0}^2$, sets $\widehat m_T=\max_{a\in[K]}\widehat\mu_{a,T_0}$, and forms the retained set
\[
\widehat{\calS}_T=\cb{a\in[K]:\widehat\mu_{a,T_0}+\tau_T\ge\max_{b\in[K]}\widehat\mu_{b,T_0}-\tau_T}.
\]
Put
\[
\widehat v_{a,T_0}^{\mathrm{cl}}=\min\cb{2\overline v,\max\cb{\underline v/2,\widehat\sigma_{a,T_0}^2}},
\qquad
\widetilde s_{a,T_0}=\p{1+2r_{j_T}}\sqrt{\widehat v_{a,T_0}^{\mathrm{cl}}},
\]
and let $\overline s_{T,a}$ be the smallest point of $\calV_{j_T}$ not smaller than $\widetilde s_{a,T_0}$. Such a point exists for every pilot realization, whether or not $\calE_{v,T}(r_{j_T})$ occurs. Clipping forces $\widehat v_{a,T_0}^{\mathrm{cl}}\in[\underline v/2,2\overline v]$, so with $r_{j_T}\le j_T^{-4}\le8^{-4}$,
\[
\frac{\sqrt{\underline v}}{2}<\widetilde s_{a,T_0}=\p{1+2r_{j_T}}\sqrt{\widehat v_{a,T_0}^{\mathrm{cl}}}\le\p{1+2\cdot8^{-4}}\sqrt{2\overline v}<2\sqrt{\overline v},
\]
and $2\sqrt{\overline v}$, the top endpoint of $\calV_{j_T}$, is itself in the grid, so a grid point at least $\widetilde s_{a,T_0}$ is always available. On $\calE_{v,T}(r_{j_T})$, clipping is inactive and $\abs{\widehat\sigma_{a,T_0}^2/\sigma_a^2(\mu_a)-1}\le r_{j_T}$. Using $\p{1+2r_j}\sqrt{1-r_j}\ge1$,
\begin{align}
1\le\frac{\overline s_{T,a}}{\sigma_a(\mu_a)}\le\p{1+\zeta_{j_T}}\p{1+2r_{j_T}}\sqrt{1+r_{j_T}}\le1+4\p{r_{j_T}+\zeta_{j_T}}=1+\rho_{j_T},
\label{eq:round_sd_two_sided}
\end{align}
where the last inequality follows from $\sqrt{1+r}\le1+r$ and $0\le r,\zeta\le8^{-4}$. Thus, the policy is constructed using a standard-deviation vector $\overline s_T$ whose relative difference from the true standard deviations is at most $\rho_{j_T}$, without evaluating the unknown variance functions. Only this two-sided bound is used below. The upward rounding ensures that $\overline s_{T,a}\ge\sigma_a(\mu_a)$. Therefore, the relative approximation error is nonnegative and the lower endpoint in \eqref{eq:round_sd_two_sided} is $1$.

The strategy reserves $T_g=K\lfloor\gamma_{j_T}T/K\rfloor$ comparison observations and selects each arm exactly $T_g/K$ times during the corresponding rounds. These observations are interleaved deterministically among the remaining rounds and are not used to update the state variables of the main-phase policy. The remaining $n=T-T_0-T_g$ rounds form the main phase, and the action distribution is updated after each main observation.

For the analysis of $\delta^{\mathrm{SMAS}}$, we use an equivalent representation with independent outcome arrays for the pilot phase, the main phase, and the reserved comparison observations. For each arm $a$, generate three mutually independent i.i.d.\ sequences
\[
\p{Y^{\mathrm{pilot}}_{a,i}}_{i\ge1},
\qquad
\p{Y^{\mathrm{main}}_{a,i}}_{i\ge1},
\qquad
\p{Y^{\mathrm{comparison}}_{a,i}}_{i\ge1}
\]
of observations from $P_{a,\mu_a}$, independent across arms and across the three phases. Each pilot selection of arm $a$ uses the next unused entry of $\p{Y^{\mathrm{pilot}}_{a,i}}$. Each main-phase arm selection uses the next unused entry of $\p{Y^{\mathrm{main}}_{a,i}}$, and each comparison observation uses the next unused entry of $\p{Y^{\mathrm{comparison}}_{a,i}}$.

Conditional on the observed past and the seed $U$, the outcome of a round with $A_t=a$ has law $P_{a,\mu_a}$ under this construction exactly as under the single-array canonical construction. Each arm selection uses an unused array entry that is independent of the past. Therefore, the two constructions induce the same law of $\p{A_1,Y_1,\ldots,A_T,Y_T,\widehat a_T}$. It follows that $\delta^{\mathrm{SMAS}}\in\calA$ and its regret are unchanged.

Since the three arrays are independent and each phase uses a separate array, the comparison sample is independent of the sigma-algebra generated by the pilot and main data conditional on $\bmmu$. Conditional on the pilot data $\calF_{T_0}$, the main-phase observations from arm $a$ are conditionally independent with distribution $P_{a,\mu_a}$. These are the two independence properties used in the analysis below.

If $\abs{\widehat{\calS}_T}=1$, the strategy selects the retained arm in every main round and chooses it as $\widetilde a_T$.

If $\abs{\widehat{\calS}_T}=2$, write $\widehat{\calS}_T=\cb{a,b}$ with $a<b$ and form the plug-in Neyman proportions
\[
\widehat w_{T,c}=\frac{\overline s_{T,c}}{\sum_{c'\in\widehat{\calS}_T}\overline s_{T,c'}},
\qquad c\in\widehat{\calS}_T,
\]
which are fixed at the end of the pilot phase and satisfy $\widehat w_{T,a}+\widehat w_{T,b}=1$. In the two-arm case, plug-in Neyman allocation gives the deterministic counts
\[
N_a=\lceil n\widehat w_{T,a}\rceil,
\qquad
N_b=n-N_a.
\]
The strategy selects arm $a$ in the first $N_a$ deterministically interleaved main positions and selects arm $b$ in the remaining $N_b$ main positions. These counts refer only to main rounds. The interleaved rounds reserved for the final comparison do not change them. It holds that $N_a\in\cb{\lfloor n\widehat w_{T,a}\rfloor,\lceil n\widehat w_{T,a}\rceil}$. Since $\widehat w_{T,b}=1-\widehat w_{T,a}$ gives $\lfloor n\widehat w_{T,b}\rfloor=n-\lceil n\widehat w_{T,a}\rceil$, it also holds that $N_b\in\cb{\lfloor n\widehat w_{T,b}\rfloor,\lceil n\widehat w_{T,b}\rceil}$. Both counts are functions of the pilot data alone and are therefore deterministic given $\calF_{T_0}$. For every pilot realization, the rounded standard deviations satisfy $\overline s_{T,c}\in[\sqrt{\underline v}/2,2\sqrt{\overline v}]$. Hence, $\widehat w_{T,a},\widehat w_{T,b}\in[p_0',1-p_0']$, where $p_0'=\sqrt{\underline v}/\p{\sqrt{\underline v}+4\sqrt{\overline v}}$. This constant depends only on $\underline v$ and $\overline v$. Consequently, $N_a,N_b\ge1$ once $n\ge1/p_0'$, whether or not $\calE_{v,T}(r_{j_T})$ occurs. The preliminary recommendation $\widetilde a_T$ is the arm with the larger main-sample mean.

If $\abs{\widehat{\calS}_T}\ge3$, the strategy uses the policy $\pi_T^\star=\pi_{j_T,n,\widehat{\calS}_T,\overline s_T}$ fixed in Definition~\ref{def:indexed_near_minimax_policies}, initializes its continuous and finite memory states, and, after each main observation $Y$ of the selected arm $a$, updates
\begin{align}
Z_{t+1}=Z_t+\frac{B_{t,E_t,a}}{\sqrt n}\p{Y-\widehat m_T}+\frac{c_{t,E_t,a}}{n},
\qquad
E_{t+1}=\kappa_t\p{E_t,a},
\label{eq:quasi_log_odds_update}
\end{align}
where $\p{B_{t,e,a},c_{t,e,a}}$ are the coefficients of Definition~\ref{def:linear_state_policy}. After this update, the next main arm is selected according to $\alpha_t\p{\cdot\mid Z_t,E_t}$, and $\widetilde a_T$ is chosen according to the terminal kernel.

At the terminal time, let $\widehat a_T^{g}=\argmax_{a\in[K]}\widehat\mu_a^{g}$ be the arm with the largest empirical mean among the reserved comparison observations, where $\widehat\mu_a^{g}$ denotes the corresponding empirical mean. The final comparison rule chooses the arm $\widehat a_T$ defined by
\[
\widehat a_T=
\begin{cases}
\widehat a_T^{g},&\widehat\mu_{\widehat a_T^{g}}^{g}-\widehat\mu_{\widetilde a_T}^{g}\ge H_{j_T}/\sqrt T,\\
\widetilde a_T,&\text{otherwise.}
\end{cases}
\]
Write $\delta^{\mathrm{SMAS}}$ for the resulting strategy, with regret $\Regret_T^{\delta^{\mathrm{SMAS}}}(\bmmu)=\bbE_{\bmmu}\sqb{\Delta_{\bmmu}(\widehat a_T)}$. Algorithm~\ref{alg:smas} of Section~\ref{sec:smas} summarizes the strategy. The construction above fixes the policies, grids, and thresholds used at the selected index.

\subsection{Pilot Events and Screening Properties}

Fix $j$ and abbreviate $\alpha=\alpha_j$, $\gamma=\gamma_j$, $B=B_j$, $R=R_j$, and $H=H_j$. These quantities depend on $j$ but not on $T$. Define the pilot events
\[
\calE_{\mu,T}=\cb{\max_{a\in[K]}\abs{\widehat\mu_{a,T_0}-\mu_a}\le\tau_T},
\qquad
\calE_{v,T}(r)=\cb{\max_{a\in[K]}\abs{\frac{\widehat\sigma_{a,T_0}^2}{\sigma_a^2(\mu_a)}-1}\le r}.
\]
Apply Lemmas~\ref{lem:pilot_mean_concentration} and \ref{lem:pilot_variance_concentration} with $\tau_T=B/\sqrt T$. Let $c_2^{\mu}$ be the value of $c_2$ in Lemma~\ref{lem:pilot_mean_concentration} at $x_1=\Delta_{\calM}$, and set
\[
C_1=2K,
\qquad
c_0=\frac{c_2^{\mu}}{2K},
\qquad
C_2=Kc_3,
\qquad
c_1=\frac{c_4\underline v^2}{2K}.
\]
Whenever $\tau_T\le\Delta_{\calM}$ and $n_0\ge\alpha T/(2K)$, a union bound gives
\begin{align}
\sup_{\bmmu\in\calM^K}\bbP_{\bmmu}\p{\calE_{\mu,T}^c}&\le C_1\exp\p{-c_0\alpha B^2},
\label{eq:pilot_mean_bad_probability}\\
\sup_{\bmmu\in\calM^K}\bbP_{\bmmu}\p{\calE_{v,T}(r)^c}&\le C_2\exp\p{-c_1\alpha Tr^2}
\label{eq:pilot_variance_bad_probability}
\end{align}
for every $0<r\le r_{\mathrm{conc}}$, where $r_{\mathrm{conc}}=1/\underline v$. These constants depend only on $K$, $\Delta_{\calM}$, and the constants in $\mathfrak E$.

For the variance bound, use $\sigma_a^2(\mu_a)\ge\underline v$. On $\calE_{v,T}(r)^c$, some arm satisfies
\[
\abs{\widehat\sigma_{a,T_0}^2-\sigma_a^2(\mu_a)}
>
r\sigma_a^2(\mu_a)
\ge
r\underline v.
\]
If $r\le r_{\mathrm{conc}}$, then $x=r\underline v\le1$. Thus, $x$ lies in the range $0<x\le4$ of Lemma~\ref{lem:pilot_variance_concentration} and in its quadratic regime, where $\min\cb{x^2,x}=\underline v^2r^2$. A union bound over the $K$ arms and $n_0\ge\alpha T/(2K)$ gives \eqref{eq:pilot_variance_bad_probability}.

\begin{lemma}[Screening properties]
\label{lem:screening_geometry}
On $\calE_{\mu,T}$ a true best arm belongs to $\widehat{\calS}_T$, and every $a\in\widehat{\calS}_T$ satisfies
\begin{align}
0\le\mu_\star-\mu_a\le4\tau_T,
\qquad
\abs{\widehat m_T-\mu_\star}\le\tau_T.
\label{eq:screening_local_geometry}
\end{align}
Consequently, with $n=T-T_0-T_g$ and $u_a^\star=\sqrt n\p{\mu_a-\widehat m_T}$, one has $u^\star\in[-R,R]^{\widehat{\calS}_T}$, and indeed $\norm{u^\star}_\infty\le5B<R$, at every budget $T\ge N_8$ at which the construction for index $j$ is in force, with no further largeness requirement on $T$.
\end{lemma}
\begin{proof}
Let $a_\star$ be a true best arm, so $\mu_{a_\star}=\mu_\star$. On $\calE_{\mu,T}$ every arm $b$ has $\widehat\mu_{b,T_0}\le\mu_b+\tau_T\le\mu_\star+\tau_T$, so $\max_b\widehat\mu_{b,T_0}\le\mu_\star+\tau_T$, while $\widehat\mu_{a_\star,T_0}\ge\mu_\star-\tau_T$; hence $\widehat\mu_{a_\star,T_0}+\tau_T\ge\mu_\star\ge\max_b\widehat\mu_{b,T_0}-\tau_T$ and $a_\star\in\widehat{\calS}_T$. For a retained arm $a$,
\[
\widehat\mu_{a,T_0}+\tau_T\ge\max_b\widehat\mu_{b,T_0}-\tau_T\ge\widehat\mu_{a_\star,T_0}-\tau_T\ge\mu_\star-2\tau_T,
\]
so $\mu_a\ge\widehat\mu_{a,T_0}-\tau_T\ge\mu_\star-4\tau_T$, giving \eqref{eq:screening_local_geometry} for the gap. The centering obeys $\mu_\star-\tau_T\le\widehat\mu_{a_\star,T_0}\le\widehat m_T\le\mu_\star+\tau_T$, so $\abs{\widehat m_T-\mu_\star}\le\tau_T$. Finally $\abs{\mu_a-\widehat m_T}\le\abs{\mu_a-\mu_\star}+\abs{\mu_\star-\widehat m_T}\le5\tau_T=5B/\sqrt T$, and $n\le T$ gives $\abs{u_a^\star}\le\sqrt T\cdot5B/\sqrt T=5B<8B=R$.
\end{proof}

\subsection{Telescoping Comparison}

When $\abs{\widehat{\calS}_T}\ge3$, we compare the original and Gaussian one-observation transition operators at each main round.

\begin{lemma}[Risk bound for the policy selected after the pilot phase]
\label{lem:guaranteed_telescope}
Condition on the pilot data and a realized retained set $\calS=\widehat{\calS}_T$ with $\abs{\calS}\ge3$, and let $\pi_T^\star$ be the policy selected at index $j$. On $\calE_{\mu,T}\cap\calE_{v,T}(r_j)$, with $u^\star$ of Lemma~\ref{lem:screening_geometry},
\begin{align}
\sqrt n\,\bbE_{\bmmu}\sqb{\Delta_{\bmmu}(\widetilde a_T)\mid\calF_{T_0}}
\le\mathfrak g_{\calS}^{R_j}(\overline s_T)+\omega_j
+C_j^{\mathrm{cmp}}\p{\frac{D_{3,j}}{\sqrt n}+\chi_T D_{2,j}},
\label{eq:upper_fixed_policy_telescope}
\end{align}
where $\chi_T=\max_{a\in\calS}\abs{\sigma_a^2(\mu_a)/\overline s_{T,a}^2-1}\le2\rho_j$ and $C_j^{\mathrm{cmp}}$ depends only on $j$, $K$, $\calM$, and the constants in $\mathfrak E$.
\end{lemma}
\begin{proof}
The selected policy recommends an arm in $\calS$. A true best arm also lies in $\calS$ by Lemma~\ref{lem:screening_geometry}. Therefore, $\Delta_{\bmmu}(\widetilde a_T)=\mu_\star-\mu_{\widetilde a_T}=\max_{a\in\calS}\p{\mu_a-\widehat m_T}-\p{\mu_{\widetilde a_T}-\widehat m_T}$; multiplying by $\sqrt n$ gives $\sqrt n\,\Delta_{\bmmu}(\widetilde a_T)=\Delta_{u^\star}^{\calS}(\widetilde a_T)$, hence $\sqrt n\,\bbE_{\bmmu}\sqb{\Delta_{\bmmu}(\widetilde a_T)\mid\calF_{T_0}}=R_n^{P}\p{\pi_T^\star,u^\star}$, the original risk of $\pi_T^\star$ run with the centered outcome $Y-\widehat m_T$ and the standard-deviation vector $\overline s_T$ used to construct the policy. Because $u^\star\in[-R_j,R_j]^{\calS}$ and $\pi_T^\star$ satisfies the $\p{D_{2,j},D_{3,j},B_{0,j}}$ bounds on $[-R_j,R_j]^{\calS}$ for $\overline s_T$ in the sense of Definition~\ref{def:smooth_guaranteed_policy}, Lemma~\ref{lem:guaranteed_recovery} gives
\[
R_n^{P}\p{\pi_T^\star,u^\star}\le R_n^{G}\p{\pi_T^\star,u^\star}+C_j^{\mathrm{cmp}}\p{\frac{D_{3,j}}{\sqrt n}+\chi_T D_{2,j}},
\]
with $\chi_T$ as stated. The constant $C_j^{\mathrm{cmp}}$ in Definition~\ref{def:indexed_near_minimax_policies} can be used in Lemma~\ref{lem:guaranteed_recovery} for every arm subset, every state dimension up to $d_{0,j}$, and every standard-deviation vector in $[\sqrt{\underline v}/2,2\sqrt{\overline v}]^{\calS}$, including the realized pair $\p{\calS,\overline s_T}$. It is finite for fixed $j$ and does not depend on the true arm families.
 Definition~\ref{def:indexed_near_minimax_policies} bounds $R_n^{G}\p{\pi_T^\star,u^\star}\le\sup_{u}R_n^{G}\p{\pi_T^\star,u}\le\mathfrak g_{\calS}^{R_j}(\overline s_T)+\omega_j$. Finally, \eqref{eq:round_sd_two_sided} gives $\overline s_{T,a}\ge\sigma_a(\mu_a)$ and $\overline s_{T,a}\le\p{1+\rho_j}\sigma_a(\mu_a)$. Hence, $\sigma_a^2(\mu_a)/\overline s_{T,a}^2\in[\p{1+\rho_j}^{-2},1]$ and $\chi_T\le2\rho_j$.
\end{proof}

We next compare the rounded standard deviations with the true standard deviations evaluated at the best mean $\mu_\star$. Put
\[
L_\sigma\coloneqq\frac{L_v}{2\sqrt{\underline v}},
\qquad
d_{j,T}\coloneqq\frac{4L_\sigma B_j}{\sqrt{\underline v T}},
\qquad
q_{j,T}^{\sigma}\coloneqq\p{1+\rho_j}d_{j,T}\overline\Gamma_K.
\]
On $\calE_{\mu,T}$, every retained arm satisfies $\abs{\mu_a-\mu_\star}\le4B_j/\sqrt T$. Assumption~\ref{asm:structural_envelope} therefore gives
\[
\sigma_a(\mu_a)
\le
\sigma_a(\mu_\star)+\frac{4L_\sigma B_j}{\sqrt T}
\le
\p{1+d_{j,T}}\sigma_a(\mu_\star).
\]
Together with \eqref{eq:round_sd_two_sided}, this yields $\overline s_{T,a}\le\p{1+\rho_j}\p{1+d_{j,T}}\sigma_a(\mu_\star)$. Lemma~\ref{lem:seq_game_properties}, Corollary~\ref{cor:gamma_subsets}, and Corollary~\ref{cor:uniform_gamma_bound} then give
\begin{align}
\mathfrak g_{\calS}^{R_j}(\overline s_T)
\le\p{1+\rho_j}\Gamma(\bmP)+q_{j,T}^{\sigma}.
\label{eq:guaranteed_value_to_gamma}
\end{align}
For fixed $j$, the quantity $q_{j,T}^{\sigma}$ converges to zero as $T\to\infty$. It depends only on $K$, $\mathfrak E$, $j$, and $T$. Here, $\calS$ is the retained subset of $[K]$, and the displayed standard deviations are the restrictions of $\p{\sigma_a(\mu_\star)}_{a\in[K]}$ to $\calS$.

\subsection{Final Comparison Rule}
By the representation with separate outcome arrays for the pilot, main, and comparison phases, the comparison sample is independent of the pilot and main data conditional on $\bmmu$, and the strategy selects each arm exactly $n_g=\lfloor\gamma_jT/K\rfloor$ times during the rounds reserved for the final comparison. Write $\calG_T$ for the sigma-algebra generated by the pilot data, the main action-observation history, and the policy randomization up to and including the choice of $\widetilde a_T$, but not the comparison observations; then $\widetilde a_T$ is $\calG_T$-measurable, while the comparison sample remains independent of $\calG_T$ conditional on $\bmmu$. Write $\Delta_{\calM}=\max\calM-\min\calM$ for the diameter of the parameter interval, so that every mean gap satisfies $\mu_\star-\mu_b\le\Delta_{\calM}$.

\begin{lemma}[Pointwise bounds for the final comparison rule]
\label{lem:final_comparison_bounds}
Let $c_1^{\mu}=2$ and
\[
c_2^{\mu}
=
\min\cb{\frac{1}{4C_0},\frac{\lambda_0}{2\Delta_{\calM}}}
\]
be the constants in Lemma~\ref{lem:pilot_mean_concentration} for $x_1=\Delta_{\calM}$. Put
\[
c=\frac{c_2^{\mu}}{8K},
\qquad
C=\frac{4c_1^{\mu}(K+2)}{\sqrt{2c\exp(1)}}.
\]
The following properties hold uniformly over $\bmmu\in\calM^K$ and over budgets with $n_g\ge1$, $n_g\ge\gamma_jT/(2K)$, and $H_j/\sqrt T\le\Delta_{\calM}$.
\begin{enumerate}
\item For an arbitrary $\calG_T$-measurable preliminary recommendation $\widetilde a_T$,
\[
\sqrt T\,\bbE_{\bmmu}\sqb{\Delta_{\bmmu}(\widehat a_T)\mid\calG_T}\le2H_j+C\gamma_j^{-1/2}.
\]
\item Suppose that $\Delta_{\bmmu}(\widetilde a_T)\le6B_j/\sqrt T$. Then, the final comparison rule increases the conditional scaled regret by at most
\[
C\gamma_j^{-1/2}\exp\p{-c\gamma_j\p{H_j-6B_j}^2}.
\]
\end{enumerate}
\end{lemma}
\begin{proof}
Condition on $\calG_T$. The comparison sample remains independent and contains $n_g$ observations for each arm. Write
\[
x=\sqrt T\p{\mu_\star-\mu_{\widetilde a_T}}
=\sqrt T\,\Delta_{\bmmu}(\widetilde a_T).
\]
If $x\le2H_j$, keeping $\widetilde a_T$ costs at most $2H_j$ in scaled regret. Suppose that $x>2H_j$. The final comparison rule fails to replace the preliminary recommendation only when $\widehat\mu_{\widehat a_T^{g}}^{g}-\widehat\mu_{\widetilde a_T}^{g}<H_j/\sqrt T$. This event forces the comparison-sample mean of a true best arm to fall short of that of $\widetilde a_T$ by at least $\p{x-H_j}/\sqrt T$ in absolute units. This excess is at most $\Delta_{\calM}$. Split the excess between the two comparison-sample means and apply Lemma~\ref{lem:pilot_mean_concentration} with $n=n_g$ and $x_1=\Delta_{\calM}$ to the two independent comparison samples. Here, $n_g\ge1$ ensures that the comparison-sample means are defined. The bound $n_g\ge\gamma_jT/(2K)$ gives
\[
n_g\p{\frac{x-H_j}{2\sqrt T}}^2
\ge
\frac{\gamma_j\p{x-H_j}^2}{8K}.
\]
A union bound over the two comparison-sample means therefore gives probability at most $2c_1^{\mu}\exp\p{-c\gamma_j\p{x-H_j}^2}$. For $x>2H_j$, put $z=x-H_j$, so that $x\le2z$. The bound $z\exp(-az^2)\le(2a\exp(1))^{-1/2}$ for $z\ge0$ shows that this contribution is at most $4c_1^{\mu}/\sqrt{2c\exp(1)}\,\gamma_j^{-1/2}$.

When the final comparison rule chooses a different arm $b$, its scaled regret is $y_b=\sqrt T\p{\mu_\star-\mu_b}$. Choosing $b$ requires one of the two relevant comparison-sample means to deviate by at least $y_b/(2\sqrt T)$. This event has probability at most $2c_1^{\mu}\exp\p{-c\gamma_jy_b^2}$. Summing over $b\in[K]$ gives at most $2c_1^{\mu}K/\sqrt{2c\exp(1)}\,\gamma_j^{-1/2}$. The sum of this term and the preceding contribution from retaining the preliminary recommendation is at most the displayed value of $C$ times $\gamma_j^{-1/2}$, which proves the first statement.

For the second statement, $\Delta_{\bmmu}(\widetilde a_T)\le6B_j/\sqrt T$ means that the preliminary recommendation is already within $6B_j/\sqrt T$ of the optimum. The only event that can increase regret is that the final comparison rule chooses some $b$ with $y_b>0$. Choosing such an arm requires a comparison-mean fluctuation of at least $H_j-6B_j+y_b$. Split between the two comparison-sample means, the required deviation on the original mean scale is at least $\p{H_j-6B_j+y_b}/(2\sqrt T)$ for one of the two means. This threshold is at most $H_j/(2\sqrt T)+y_b/(2\sqrt T)\le\Delta_{\calM}$. Here, we use $H_j/\sqrt T\le\Delta_{\calM}$ and $y_b=\sqrt T\p{\mu_\star-\mu_b}\le\sqrt T\Delta_{\calM}$. Lemma~\ref{lem:pilot_mean_concentration} therefore applies within its range $x\le x_1=\Delta_{\calM}$. Together with $n_g\ge\gamma_jT/(2K)$ and a union bound over the two means, it gives probability at most
\[
2c_1^{\mu}\exp\p{-c\gamma_j\p{H_j-6B_j+y_b}^2}.
\]
Since $\p{H_j-6B_j+y_b}^2\ge\p{H_j-6B_j}^2+y_b^2$, summing the scaled regret over $b\in[K]$ gives at most
\[
\frac{2c_1^{\mu}K}{\sqrt{2c\exp(1)}}\gamma_j^{-1/2}
\exp\p{-c\gamma_j\p{H_j-6B_j}^2},
\]
which is bounded by the stated expression with the displayed value of $C$.
\end{proof}

\subsection{The Two-Arm Case}
In the two-arm case, the sampling counts are determined from the rounded pilot standard deviations. The following lemma bounds the difference between $c_n=\sqrt{n\p{\sigma_a^2(\mu_a)/N_a+\sigma_b^2(\mu_b)/N_b}}$ and $\sigma_a(m)+\sigma_b(m)$. The term $r_j+\zeta_j$, which measures the relative error in the rounded standard deviations, does not vanish as $T\to\infty$ when $j$ is fixed. Therefore, its coefficient must be independent of $j$ before we substitute the index $j_T$ selected for budget $T$. The lemma is used in the minimax proof with $m=\mu_\star$ and in the prior-averaged proof with $m$ equal to the midpoint of the two retained means.

\begin{lemma}[Uniform approximation of the standard deviation in the two-arm comparison]
\label{lem:neyman_scale_calibration}
Fix $j\ge8$ and a retained pair $\calS=\cb{a,b}$. Let $N_a,N_b$ be the main-phase sampling counts determined by the two-arm rule in Algorithm~\ref{alg:smas}, and put $n=N_a+N_b$ and
\[
c_n=\sqrt{n\p{\frac{\sigma_a^2(\mu_a)}{N_a}+\frac{\sigma_b^2(\mu_b)}{N_b}}}.
\]
There are finite constants $K^{\mathrm{bin}}$ and $n_{\mathrm{env}}$, depending only on $\underline v$, $\overline v$, and $L_v$ in Assumption~\ref{asm:structural_envelope}, and in particular not on $j$, the prior, the pair $\calS$, or any other feature of the arm families, such that on $\calE_{\mu,T}\cap\calE_{v,T}(r_j)$, for every $n\ge n_{\mathrm{env}}$ and every $m\in\calM$ with $\max_{c\in\calS}\abs{\mu_c-m}\le4B_j/\sqrt T$,
\begin{align}
\abs{c_n-\p{\sigma_a(m)+\sigma_b(m)}}\le K^{\mathrm{bin}}\p{r_j+\zeta_j+\frac{B_j}{\sqrt T}+\frac{1}{n^2}}.
\label{eq:neyman_scale_calibration}
\end{align}
Equivalently, since $\sigma_a(m)+\sigma_b(m)\ge2\sqrt{\underline v}$,
\begin{align}
\abs{\frac{c_n}{\sigma_a(m)+\sigma_b(m)}-1}\le\epsilon_{j,T},
\qquad
\epsilon_{j,T}\coloneqq\frac{K^{\mathrm{bin}}}{2\sqrt{\underline v}}\p{r_j+\zeta_j+\frac{B_j}{\sqrt T}+\frac{1}{n^2}}.
\label{eq:neyman_scale_relative}
\end{align}
The bounds hold for every $\bmmu$ and every pilot realization in the stated event, with no exceptional set.
\end{lemma}
\begin{proof}
Abbreviate $\sigma_c=\sigma_c(\mu_c)$ and $s_c=\overline s_{T,c}$ for $c\in\calS$, and put
\[
F(w,x,y)=\sqrt{\frac{x^2}{w}+\frac{y^2}{1-w}},
\qquad
w\in(0,1),
\quad
x,y>0,
\]
Then, $c_n=F\p{N_a/n,\sigma_a,\sigma_b}$. Two elementary properties of $F$ are used. First, for fixed $x,y>0$, the map $w\mapsto x^2/w+y^2/(1-w)$ has derivative $-x^2/w^2+y^2/(1-w)^2$. The derivative is negative for $w<x/(x+y)$, vanishes at $w=x/(x+y)$, and is positive thereafter. Therefore, the minimum is attained at the Neyman proportion and equals $\sqrt{x(x+y)+y(x+y)}$:
\begin{align}
\min_{w\in(0,1)}F(w,x,y)=F\p{\frac{x}{x+y},x,y}=x+y.
\label{eq:neyman_min_identity}
\end{align}
Second, $\partial_xF=x/\p{wF}>0$ and $\partial_yF=y/\p{(1-w)F}>0$, so $F$ increases in each of $x$ and $y$.

On $\calE_{v,T}(r_j)$, the two-sided bound \eqref{eq:round_sd_two_sided} gives $\sigma_c\le s_c\le\p{1+\rho_j}\sigma_c$. The left endpoint is exactly $1$ because the pilot standard deviations are rounded upward. We also have $s_c\le2\sqrt{\overline v}$ because the grid $\calV_j$ of Definition~\ref{def:indexed_near_minimax_policies} has top endpoint $2\sqrt{\overline v}$. Since $r_j+\zeta_j\le2j^{-4}$, it holds that
\begin{align}
\rho_j=4\p{r_j+\zeta_j}\le8\cdot8^{-4}<1
\qquad
\text{for every }j\ge8.
\label{eq:rho_small}
\end{align}
The tracked proportion is the exact Neyman proportion of the rounded vector, $\widehat w=s_a/\p{s_a+s_b}$, and by the binary split of Algorithm~\ref{alg:smas} the counts satisfy $N_a\in\cb{\lfloor n\widehat w\rfloor,\lceil n\widehat w\rceil}$ and $N_b=n-N_a$, so that
\begin{align}
\abs{\frac{N_a}{n}-\widehat w}\le\frac1n.
\label{eq:tracking_error}
\end{align}
Since $\sqrt{\underline v}\le s_c\le2\sqrt{\overline v}$, we get $\widehat w\ge\sqrt{\underline v}/\p{4\sqrt{\overline v}}=2p_0$ with $p_0\coloneqq\sqrt{\underline v}/\p{8\sqrt{\overline v}}\le1/2$ a constant depending only on $\underline v$ and $\overline v$. Symmetrically $\widehat w\le1-2p_0$; setting $n_{\mathrm{env}}=1/p_0$ places both $\widehat w$ and $N_a/n$ in $[p_0,1-p_0]$ whenever $n\ge n_{\mathrm{env}}$.

For the lower bound, \eqref{eq:neyman_min_identity} at $(x,y)=\p{\sigma_a,\sigma_b}$ gives $c_n\ge\sigma_a+\sigma_b$, whatever the realized counts.

For the upper bound, monotonicity in $(x,y)$ and $\sigma_c\le s_c$ give
\[
c_n\le F\p{N_a/n,s_a,s_b}.
\]
This step uses the upward rounding $s_c\ge\sigma_c$. The order of the two steps matters. The proportion $\widehat w$ is the exact minimizer of $F\p{\cdot,s_a,s_b}$ but not of $F\p{\cdot,\sigma_a,\sigma_b}$.

We therefore bound the integer rounding error in $N_a/n$ only after replacing the true standard deviations by $\p{s_a,s_b}$. Write $S=s_a+s_b$ and $p=N_a/n$. Substituting $s_a=S\widehat w$ and $s_b=S\p{1-\widehat w}$ gives the exact identity
\[
\frac{s_a^2}{p}+\frac{s_b^2}{1-p}
=S^2\p{\frac{\widehat w^2}{p}+\frac{\p{1-\widehat w}^2}{1-p}}
=S^2\p{1+\frac{\p{p-\widehat w}^2}{p\p{1-p}}},
\]
the last step because $\widehat w^2\p{1-p}+\p{1-\widehat w}^2p$ and $p\p{1-p}+\p{p-\widehat w}^2$ both equal $\widehat w^2+p-2p\widehat w$. Hence
\begin{align}
F\p{p,s_a,s_b}=\p{s_a+s_b}\sqrt{1+\frac{\p{p-\widehat w}^2}{p\p{1-p}}},
\label{eq:neyman_tracking_identity}
\end{align}
in which the deviation of $p$ from $\widehat w$ enters only quadratically, with no first-order term. By \eqref{eq:tracking_error} and $p\p{1-p}\ge p_0\p{1-p_0}\ge p_0/2$, valid because $p\in[p_0,1-p_0]$ and $p_0\le1/2$, the fraction under the square root is at most $2/\p{p_0n^2}$; with $\sqrt{1+e}\le1+e/2$ and \eqref{eq:round_sd_two_sided},
\[
c_n\le F\p{\frac{N_a}{n},s_a,s_b}\le\p{s_a+s_b}\p{1+\frac{1}{p_0n^2}}\le\p{1+\rho_j}\p{1+\frac{1}{p_0n^2}}\p{\sigma_a+\sigma_b}.
\]
By \eqref{eq:rho_small} the product $\p{1+\rho_j}\p{1+1/\p{p_0n^2}}$ is at most $1+\rho_j+2/\p{p_0n^2}$. Combining with the lower bound and using $\sigma_a+\sigma_b\le2\sqrt{\overline v}$,
\begin{align}
\abs{c_n-\p{\sigma_a(\mu_a)+\sigma_b(\mu_b)}}\le8\sqrt{\overline v}\p{r_j+\zeta_j}+\frac{C_1}{n^2},
\qquad
C_1\coloneqq\frac{4\sqrt{\overline v}}{p_0}.
\label{eq:neyman_scale_at_mu}
\end{align}

Finally, $\abs{\partial_\mu\sigma_c(\mu)}=\abs{\partial_\mu\sigma_c^2(\mu)}/\p{2\sigma_c(\mu)}\le L_\sigma$ by Assumption~\ref{asm:structural_envelope}. Thus, each $\sigma_c$ is Lipschitz on $\calM$ with the constant $L_\sigma$ defined above, and $\max_{c\in\calS}\abs{\mu_c-m}\le4B_j/\sqrt T$ gives
\[
\abs{\p{\sigma_a(\mu_a)+\sigma_b(\mu_b)}-\p{\sigma_a(m)+\sigma_b(m)}}\le8L_\sigma\frac{B_j}{\sqrt T}.
\]
Adding this to \eqref{eq:neyman_scale_at_mu} and setting $K^{\mathrm{bin}}=\max\cb{8\sqrt{\overline v},C_1,8L_\sigma}$ proves \eqref{eq:neyman_scale_calibration}; every constant entering $K^{\mathrm{bin}}$ is built from $\underline v$, $\overline v$, and $L_v$ alone. Dividing by $\sigma_a(m)+\sigma_b(m)\ge2\sqrt{\underline v}$ gives \eqref{eq:neyman_scale_relative}. No step discards a null set. Therefore, the bounds hold pointwise on the stated event.
\end{proof}

\begin{lemma}[Uniform binary comparison]
\label{lem:binary_upper_comparison}
Fix an index $j$ and a retained pair $\calS=\cb{a,b}$, and let $\widetilde a_T$ be the preliminary recommendation in the two-arm case. Let $p_0'=\sqrt{\underline v}/\p{\sqrt{\underline v}+4\sqrt{\overline v}}$, $\underline p=p_0'/2$, and
\[
C_j=\frac{6C_{\mathrm{BE}}\overline\Lambda_3}{\underline p^2\underline v^{3/2}}.
\]
Set
\[
M_j^{\mathrm{bin}}
=
\left\lceil
\frac{\max\cb{n_{\mathrm{env}},2/p_0'}}{1-\alpha_j-\gamma_j}
\right\rceil.
\]
The main horizon satisfies $n\ge\max\cb{n_{\mathrm{env}},2/p_0'}$ for every $T\ge M_j^{\mathrm{bin}}$. For every such $T$, on $\calE_{\mu,T}\cap\calE_{v,T}(r_j)$ and with $u^\star$ of Lemma~\ref{lem:screening_geometry},
\begin{align}
\bbE_{\bmmu}\sqb{\Delta_{u^\star}^{\calS}(\widetilde a_T)\mid\calF_{T_0}}
\le\mathfrak G_{\calS}^{\uparrow}\p{\p{\sigma_c(\mu_\star)}_{c\in\calS}}+K^{\mathrm{bin}}\p{r_j+\zeta_j}+q_{j,T}^{\mathrm{bin}},
\label{eq:binary_minimax_upper}
\end{align}
where
\[
q_{j,T}^{\mathrm{bin}}
\coloneqq
K^{\mathrm{bin}}\p{\frac{B_j}{\sqrt T}+\frac{1}{n^2}}
+
\frac{2R_jC_j}{\sqrt n}.
\]
The constant $K^{\mathrm{bin}}$ is the constant in Lemma~\ref{lem:neyman_scale_calibration}. It depends only on $\underline v$, $\overline v$, and $L_v$, and it does not depend on $j$. For fixed $j$, $q_{j,T}^{\mathrm{bin}}\to0$ as $T\to\infty$.
\end{lemma}
\begin{proof}
The bound is stated in the local $\sqrt n$-scaled units of the $\calS$-arm game. Theorem~\ref{thm:alladapt_upper} later multiplies it by $\sqrt{T/n}$. Conditional on the pilot data, the counts $N_a$ and $N_b$ are deterministic, and integer rounding changes each count by at most one. The rounded standard deviations always lie in $[\sqrt{\underline v}/2,2\sqrt{\overline v}]$. Hence, the plug-in Neyman proportions lie in $[p_0',1-p_0']$. The definition of $M_j^{\mathrm{bin}}$ gives $N_c\ge\underline p n$ for $c\in\cb{a,b}$.

The main-phase outcomes come from the main-phase array, which is independent of the pilot data. Hence, conditional on $\calF_{T_0}$, the outcomes observed after selecting either arm are independent observations from its own family.

The main empirical-best rule chooses the arm of smaller mean exactly when the main-mean difference $\widehat\mu_{a}^{\mathrm{main}}-\widehat\mu_{b}^{\mathrm{main}}$ takes the wrong sign, subject to the fixed tie-break rule when the difference is zero. Conditional on $\calF_{T_0}$, this difference has mean $\mu_a-\mu_b$ and variance $s_N^2\coloneqq\sigma_a^2(\mu_a)/N_a+\sigma_b^2(\mu_b)/N_b$. Its standardization is the normalized sum $s_N^{-1}\sum_{c\in\cb{a,b}}\sum_{i=1}^{N_c}X_{c,i}$ of $N_a+N_b$ independent centered summands, where $X_{a,i}=\p{Y_{a,i}-\mu_a}/N_a$ and $X_{b,i}=-\p{Y_{b,i}-\mu_b}/N_b$.

The uniform third-moment bound $\overline\Lambda_3$ in \eqref{eq:third_moment_envelope} bounds $\bbE_{c,\mu}\abs{Y-\mu}^3$ uniformly over $c\in\calS$ and $\mu\in\calM$ using only the constants in $\mathfrak E$. Together with $\sigma_c^2\ge\underline v$, it gives the following bound on the Lyapunov fraction:
\[
\frac{\sum_{c\in\cb{a,b}}\sum_{i=1}^{N_c}\bbE\abs{X_{c,i}}^3}{s_N^3}\le\frac{\overline\Lambda_3\p{N_a^{-2}+N_b^{-2}}}{\p{\underline v/n}^{3/2}}\le\frac{2\overline\Lambda_3}{\underline p^2\underline v^{3/2}}\frac{1}{\sqrt n},
\]
using $N_c\ge\underline p n$ and $s_N^2\ge\underline v/n$. By Lemma~\ref{lem:berry_esseen}, the absolute constant $C_{\mathrm{BE}}$ times this fraction bounds the Kolmogorov distance of the standardization to $\calN(0,1)$. Because this is a uniform bound on the distribution function, it also applies without change to lattice distributions such as Bernoulli outcomes. It bounds the tie atom $\bbP\p{\widehat\mu_{a}^{\mathrm{main}}=\widehat\mu_{b}^{\mathrm{main}}\mid\calF_{T_0}}$ by twice its value. Hence, any fixed tie-break rule changes the incorrect choice probability by at most that atom.

The displayed value of $C_j$ bounds the Kolmogorov distance together with the possible tie atom. Write $d=\abs{u_a^\star-u_b^\star}=\sqrt n\abs{\mu_a-\mu_b}$ and $c_n^2=n s_N^2=n\p{\sigma_a^2(\mu_a)/N_a+\sigma_b^2(\mu_b)/N_b}$,
\[
\abs{\bbP\p{\widetilde a_T\neq\argmax_{c\in\calS}\mu_c\mid\calF_{T_0}}-\Phi\p{-d/c_n}}\le\frac{C_j}{\sqrt n}.
\]
Since $\Delta_{u^\star}^{\calS}(\widetilde a_T)=d\cdot\mathbbm 1\cb{\widetilde a_T\text{ wrong}}$ and $d\le2R_j$ on the screening event,
\[
\bbE_{\bmmu}\sqb{\Delta_{u^\star}^{\calS}(\widetilde a_T)\mid\calF_{T_0}}\le d\Phi\p{-d/c_n}+\frac{2R_jC_j}{\sqrt n}\le c_nc_{\mathrm{mm}}+\frac{2R_jC_j}{\sqrt n},
\]
using $\sup_{d\ge0}d\Phi\p{-d/c_n}=c_nc_{\mathrm{mm}}$ with $c_{\mathrm{mm}}=\sup_{x\ge0}x\Phi(-x)$, the unrestricted supremum bounding the realized $d\Phi\p{-d/c_n}$. On $\calE_{\mu,T}$ the retained arms obey $\abs{\mu_c-\mu_\star}\le4B_j/\sqrt T$ by Lemma~\ref{lem:screening_geometry}, so Lemma~\ref{lem:neyman_scale_calibration} applies at $m=\mu_\star$ and, using $c_{\mathrm{mm}}\le1$,
\[
c_nc_{\mathrm{mm}}\le\p{\sigma_a(\mu_\star)+\sigma_b(\mu_\star)}c_{\mathrm{mm}}+K^{\mathrm{bin}}\p{r_j+\zeta_j}+K^{\mathrm{bin}}\p{\frac{B_j}{\sqrt T}+\frac{1}{n^2}},
\]
whose leading term is $\mathfrak G_{\calS}^{\uparrow}\p{\p{\sigma_c(\mu_\star)}_{c\in\calS}}$ by Lemma~\ref{lem:binary_sequential_value}. The coefficient $K^{\mathrm{bin}}$ of the standard-deviation approximation error $r_j+\zeta_j$ depends only on $\underline v$, $\overline v$, and $L_v$, and is independent of $j$. The remaining terms are exactly $q_{j,T}^{\mathrm{bin}}$. This proves \eqref{eq:binary_minimax_upper}.
\end{proof}

\subsection{Decomposition of the Worst-Case Risk}

\begin{theorem}[Minimax upper bound]
\label{thm:alladapt_upper}
Under Definition~\ref{def:mean_param} and Assumption~\ref{asm:structural_envelope}, for every fixed finite $K\ge2$, the strategy $\delta^{\mathrm{SMAS}}$ of Algorithm~\ref{alg:smas}, whose tuning depends only on the tuple $\p{K,\calM,\mathfrak E}$, satisfies
\[
\limsup_{T\to\infty}\sqrt T\sup_{\bmmu\in\calM^K}\Regret_T^{\delta^{\mathrm{SMAS}}}(\bmmu)\le\Gamma(\bmP),
\]
with $\Gamma(\bmP)$ the global sequential minimax constant of Definition~\ref{def:global_seq_constant}. Consequently $\delta^{\mathrm{SMAS}}\in\calA$ attains the minimax constant of Theorem~\ref{thm:minimax_lower}.
\end{theorem}
\begin{proof}
Fix an index $j$ and abbreviate $\alpha=\alpha_j$, $\gamma=\gamma_j$, $B=B_j$, $H=H_j$. Fix $\bmmu\in\calM^K$ and decompose the scaled regret over the pilot events, writing $G=\calE_{\mu,T}\cap\calE_{v,T}(r_j)$.

On $G$, Lemma~\ref{lem:screening_geometry} places a true best arm in $\widehat{\calS}_T$ and bounds the gap of every retained arm by $4B/\sqrt T$. The one-arm case therefore has zero main regret. If at least three arms are retained, Lemma~\ref{lem:guaranteed_telescope} gives \eqref{eq:upper_fixed_policy_telescope}. If two arms are retained, \eqref{eq:binary_minimax_upper} applies. Both bounds use the local $\sqrt n$ scale. Equation~\eqref{eq:guaranteed_value_to_gamma} bounds the first case by $\Gamma(\bmP)+E_{j,T}$. Corollary~\ref{cor:gamma_subsets} gives $\mathfrak G_{\calS}^{\uparrow}\p{\p{\sigma_c(\mu_\star)}_{c\in\calS}}\le\Gamma(\bmP)$ for the retained pair and gives the same bound in the two-arm case. Thus, both cases are at most $\Gamma(\bmP)+E_{j,T}$, where
\[
E_{j,T}
\coloneqq
C_j^{\mathrm{cmp}}\frac{D_{3,j}}{\sqrt n}
+2C_j^{\mathrm{cmp}}D_{2,j}\rho_j
+\rho_j\overline\Gamma_K
+\omega_j
+K^{\mathrm{bin}}\p{r_j+\zeta_j}
+q_{j,T}^{\sigma}
+q_{j,T}^{\mathrm{bin}}.
\]
Indeed, the terms involving $\rho_j$ bound the variance mismatch and the multiplicative change in the Gaussian-game value, while $q_{j,T}^{\sigma}$ and $q_{j,T}^{\mathrm{bin}}$ contain all remaining $T$-dependent errors on $\calE_{\mu,T}\cap\calE_{v,T}(r_j)$.

Since $T_0\le\alpha T$ and $T_g\le\gamma T$, it holds that $n\ge\p{1-\alpha-\gamma}T$. Put
\[
\kappa_j^{\mathrm{hor}}\coloneqq\p{1-\alpha_j-\gamma_j}^{-1/2}-1.
\]
Then, $\sqrt{T/n}\le1+\kappa_j^{\mathrm{hor}}$, and Corollary~\ref{cor:uniform_gamma_bound} gives
\begin{align*}
\sqrt T\,\bbE_{\bmmu}\sqb{\Delta_{\bmmu}(\widetilde a_T)\mid\calF_{T_0}}
&\le\p{1+\kappa_j^{\mathrm{hor}}}\p{\Gamma(\bmP)+E_{j,T}}\\
&\le\Gamma(\bmP)+E_{j,T}+\kappa_j^{\mathrm{hor}}\p{\overline\Gamma_K+E_{j,T}}.
\end{align*}
Since the gap of every retained arm is at most $4B/\sqrt T\le6B/\sqrt T$, Lemma~\ref{lem:final_comparison_bounds}(2) adds at most $C\gamma^{-1/2}\exp\p{-c\gamma\p{H-6B}^2}$ to the conditional scaled regret of $\widehat a_T$ on $G$; taking expectations, the contribution of $G$ to $\sqrt T\,\Regret_T^{\delta^{\mathrm{SMAS}}}(\bmmu)$ is at most the displayed bound plus this term from the final comparison rule.

On $\calE_{\mu,T}^c$, Lemma~\ref{lem:final_comparison_bounds}(1) bounds the conditional scaled regret of $\widehat a_T$ by $2H+C\gamma^{-1/2}$, so its contribution is at most $\p{2H+C\gamma^{-1/2}}\bbP_{\bmmu}\p{\calE_{\mu,T}^c}$. On $\calE_{\mu,T}\cap\calE_{v,T}(r_j)^c$, the retained arms still have gaps at most $4B/\sqrt T$, so before the final comparison rule $\Delta_{\bmmu}(\widetilde a_T)\le4B/\sqrt T$ and, by Lemma~\ref{lem:final_comparison_bounds}(2), the scaled regret is at most $4B+C\gamma^{-1/2}\exp\p{-c\gamma\p{H-6B}^2}$; its contribution is at most this times $\bbP_{\bmmu}\p{\calE_{v,T}(r_j)^c}$. Collecting the three regions and taking the supremum over $\bmmu$,
\begin{align}
\sqrt T\sup_{\bmmu}\Regret_T^{\delta^{\mathrm{SMAS}}}(\bmmu)\le\Gamma(\bmP)+\calE_j^{\mathrm{mm}}(T),
\label{eq:global_upper_decomposition}
\end{align}
where, using \eqref{eq:pilot_mean_bad_probability}--\eqref{eq:pilot_variance_bad_probability}, $\calE_j^{\mathrm{mm}}(T)$ is bounded by
\begin{align*}
&E_{j,T}+\kappa_j^{\mathrm{hor}}\p{\overline\Gamma_K+E_{j,T}},\\
&\p{4B_j+C\gamma_j^{-1/2}}C_2\exp\p{-c_1\alpha_jTr_j^2},\\
&\p{2H_j+C\gamma_j^{-1/2}}C_1\exp\p{-c_0\alpha_jB_j^2},\\
&C\gamma_j^{-1/2}\exp\p{-c\gamma_j\p{H_j-6B_j}^2}.
\end{align*}
The first two lines contain all terms that depend on $T$ for fixed $j$. Define
\[
q_{j,T}^{\mathrm{mm}}
\coloneqq
C_j^{\mathrm{cmp}}\frac{D_{3,j}}{\sqrt n}
+q_{j,T}^{\sigma}
+q_{j,T}^{\mathrm{bin}}
+\p{4B_j+C\gamma_j^{-1/2}}C_2\exp\p{-c_1\alpha_jTr_j^2}.
\]
Every constant in $q_{j,T}^{\mathrm{mm}}$ is determined by $K$, $\calM$, $\mathfrak E$, $C_j^{\mathrm{cmp}}$, $D_{3,j}$, and $C_j$. The last three constants are fixed before the true arm families and the prior are specified. For fixed $j$, the quantity $q_{j,T}^{\mathrm{mm}}$ converges to zero as $T\to\infty$.

The requirements $\calR_j^{\mathrm{mm}}$ used in \eqref{eq:budget_thresholds} are as follows. For every $T\ge N_j$, the main horizon satisfies $n\ge\max\cb{N_j^0,n_{\mathrm{env}}}$ and $T\ge M_j^{\mathrm{bin}}$. The pilot and comparison-observation schedules satisfy $n_0\ge2$, $n_0\ge\alpha_jT/(2K)$, $n_g\ge1$, and $n_g\ge\gamma_jT/(2K)$. We also require $B_j/\sqrt T\le\Delta_{\calM}$ and $H_j/\sqrt T\le\Delta_{\calM}$, which allows the concentration and final-comparison bounds to be applied. Finally, $q_{j,T}^{\mathrm{mm}}\le j^{-1}$. The explicit upper bounds are nonincreasing and converge to zero for fixed $j$. Hence, the integer search defining $N_j^{\mathrm{mm}}$ terminates. None of the conditions depends on the true arm families or their variance functions.

Set $j=j_T$. The threshold conditions give $q_{j_T,T}^{\mathrm{mm}}\le j_T^{-1}$. The remaining terms in $E_{j,T}$ are independent of $T$. Each converges to zero as $j\to\infty$. In particular, Definition~\ref{def:indexed_near_minimax_policies} gives $\p{1+C_j^{\mathrm{cmp}}D_{2,j}}\p{r_j+\zeta_j}\le2j^{-1}$, while $\rho_j\le8j^{-4}$ and $\omega_j=j^{-4}$. Hence, $E_{j_T,T}\to0$ as $T\to\infty$. Moreover, $\kappa_{j_T}^{\mathrm{hor}}\to0$ as $T\to\infty$, because $\alpha_j+\gamma_j=2j^{-2}$. The two terms that depend only on $j$ satisfy
\[
\p{2H_j+C\gamma_j^{-1/2}}C_1\exp\p{-c_0\alpha_jB_j^2}
=\p{2j^3+Cj}C_1\exp\p{-c_0j^2}\to0,
\]
and
\[
C\gamma_j^{-1/2}\exp\p{-c\gamma_j\p{H_j-6B_j}^2}
=Cj\exp\p{-cj^2\p{j-6}^2}\to0.
\]
Therefore, $\calE_{j_T}^{\mathrm{mm}}(T)\to0$ as $T\to\infty$, and

\[
\limsup_{T\to\infty}\sqrt T\sup_{\bmmu\in\calM^K}\Regret_T^{\delta^{\mathrm{SMAS}}}(\bmmu)\le\Gamma(\bmP).
\]
The strategy uses only empirical means and variances and the constants in $\mathfrak E$, so $\delta^{\mathrm{SMAS}}\in\calA$; with Theorem~\ref{thm:minimax_lower} it attains the minimax constant.
\end{proof}

\begin{remark}
The tuning depends on the model class only through the constants in $\mathfrak E$. Lemma~\ref{lem:guaranteed_recovery} compares the original and Gaussian one-observation transition operators at each round, so no block average is conditioned on and no assignment sequence is fixed; the argument therefore applies to the fully adaptive policy whose arm-selection probabilities respond to the observed outcomes and that satisfies the bounds in Definition~\ref{def:smooth_guaranteed_policy}. The two-arm case and the final comparison rule are the only places where we analyze a choice rule defined by a discontinuous comparison, and the two are controlled by different arguments: the two-arm case by the standard-deviation approximation in Lemma~\ref{lem:neyman_scale_calibration} together with the Berry--Esseen comparison of Lemma~\ref{lem:binary_upper_comparison}, and the final comparison rule by the exponential bound for the comparison sample in Lemma~\ref{lem:final_comparison_bounds}.
\end{remark}

\section{Bayes Upper Bound}
\label{app:bayes_upper_alladapt}
This appendix proves that the same strategy $\delta^{\mathrm{SMAS}}$ attains $C^{\mathrm{Bayes}}$ for every prior satisfying Assumption~\ref{asm:uniformcontinuity}. The tuning is the same as in Appendix~\ref{app:upper_comparison}.

\subsection{Prior-Averaged Bound for the Final Comparison Rule}
The final comparison rule of Algorithm~\ref{alg:smas} replaces the preliminary recommendation $\widetilde a_T$ with the empirical leader based on the comparison observations $\widehat a_T^{g}$ whenever $\widehat\mu_{\widehat a_T^{g}}^{g}-\widehat\mu_{\widetilde a_T}^{g}\ge H_j/\sqrt T$. The comparison-sample means are averages of $n_g$ comparison observations that are independent of the pilot and main data conditional on $\bmmu$ by the representation with separate outcome arrays for the pilot, main, and comparison phases. Apply Lemma~\ref{lem:pilot_mean_concentration} with sample size $n=n_g$ and $x_1=\Delta_{\calM}$, and set
\[
c_5=2,
\qquad
c_6=\min\cb{\frac{1}{4C_0},\frac{\lambda_0}{2\Delta_{\calM}}}.
\]
Then,
\begin{align}
\sup_{\mu\in\calM}\bbP_{a,\mu}\p{\abs{\widehat\mu_a^{g}-\mu}>t}\le c_5\exp\p{-c_6 n_g t^2}
\qquad\text{for }0<t\le\Delta_{\calM}
\label{eq:guard_mean_concentration}
\end{align}
for every arm $a$. On budgets with $n_g\ge\gamma_jT/(2K)$, a deviation $t=x/\sqrt T$ satisfies $c_6n_gt^2\ge c_6\gamma_jx^2/(2K)$.

For a fixed arm $a$, let $r_{T,j}(\bmmu,a)$ denote the expected simple regret $\bbE_{\bmmu}[\Delta_{\bmmu}(\widehat a_T)]$ of the arm chosen by the final comparison rule when the preliminary recommendation is set to the constant arm $a$; because the comparison sample is independent of the pilot and main data conditional on $\bmmu$, this is a function of $\bmmu$ and $a$ alone.

\begin{lemma}[Prior-averaged bound for the final comparison rule]
\label{lem:integrated_final_comparison}
Suppose that Assumption~\ref{asm:uniformcontinuity} holds. Put
\[
c=\frac{c_6}{8K},
\qquad
C
=
2K\overline h\Delta_{\calM}^{K-1}
+
\frac{32c_5K^2(K+1)\overline h\Delta_{\calM}^{K-1}}{c_6}.
\]
For each index $j$, set
\[
T_j
=
\left\lceil
\max\left\{
\frac{2K}{\gamma_j},
\p{\frac{H_j}{\Delta_{\calM}}}^2,
\p{\frac{H_j-6B_j}{\Delta_{\calM}}}^2
\right\}
\right\rceil.
\]
Then, for every $T\ge T_j$, one has $n_g\ge1$, $n_g\ge\gamma_jT/(2K)$, $H_j/\sqrt T\le\Delta_{\calM}$, and $(H_j-6B_j)/\sqrt T\le\Delta_{\calM}$. The following statements hold for every $j$ and every $T\ge T_j$.
\begin{enumerate}
\item One has
\[
\int_{\calM^K}\sum_{a=1}^K r_{T,j}(\bmmu,a)\rmd H(\bmmu)\le\frac{C}{T}\p{H_j^2+\gamma_j^{-1}}.
\]
Consequently, let $(E_{\bmmu})_{\bmmu\in\calM^K}$ be a family of events. Assume that $(\bmmu,\omega)\mapsto\mathbbm 1\sqb{\omega\in E_{\bmmu}}$ is jointly measurable, that each $E_{\bmmu}$ lies in the pilot-and-main sigma-algebra, and that $\sup_{\bmmu}\bbP_{\bmmu}(E_{\bmmu})\le p$. Then, for every preliminary recommendation $\widetilde a_T$ measurable with respect to the pilot data, the main history, and the main-policy randomization,
\[
T\int_{\calM^K}\bbE_{\bmmu}\sqb{\Delta_{\bmmu}(\widehat a_T)\mathbbm 1\sqb{E_{\bmmu}}}\rmd H(\bmmu)\le Cp\p{H_j^2+\gamma_j^{-1}}.
\]
\item Let $(F_{\bmmu})_{\bmmu\in\calM^K}$ be a family of events. Assume that $(\bmmu,\omega)\mapsto\mathbbm 1\sqb{\omega\in F_{\bmmu}}$ is jointly measurable, that each $F_{\bmmu}$ lies in the pilot-and-main sigma-algebra, and that the preliminary recommendation satisfies $\Delta_{\bmmu}(\widetilde a_T)\le6B_j/\sqrt T$ on $F_{\bmmu}$. Then
\[
T\int_{\calM^K}\bbE_{\bmmu}\sqb{\p{\Delta_{\bmmu}(\widehat a_T)-\Delta_{\bmmu}(\widetilde a_T)}_+\mathbbm 1\sqb{F_{\bmmu}}}\rmd H(\bmmu)\le C\gamma_j^{-1}\exp\p{-c\gamma_j\p{H_j-6B_j}^2}.
\]
\end{enumerate}
\end{lemma}
\begin{proof}
Write $\eta=H_j/\sqrt T$ and let $a_\star$ be a true best arm.

We first record the only property of the prior that is used. For every $a\in[K]$ and every Borel $\psi\colon[0,\infty)\to[0,\infty)$ with $\psi(0)=0$,
\begin{align}
\int_{\calM^K}\psi\p{\Delta_{\bmmu}(a)}\rmd H(\bmmu)\le\overline h\Delta_{\calM}^{K-1}\int_0^{\infty}\psi(g)\rmd g.
\label{eq:gap_layer_bound}
\end{align}
Indeed, writing $m_a(\bmmu_{\setminus\{a\}})=\max_{c\neq a}\mu_c$ for the best value among the other arms, one has $\Delta_{\bmmu}(a)=\p{m_a(\bmmu_{\setminus\{a\}})-\mu_a}_+$, because the maximum defining $\Delta_{\bmmu}(a)$ includes $a$ itself. Bounding $h$ by $\overline h$ and integrating $\mu_a$ first at fixed $\bmmu_{\setminus\{a\}}$, the range $\mu_a\ge m_a(\bmmu_{\setminus\{a\}})$ contributes nothing because $\psi(0)=0$, and the substitution $g=m_a(\bmmu_{\setminus\{a\}})-\mu_a$ turns the remaining range into $\int_0^{m_a(\bmmu_{\setminus\{a\}})-\min\calM}\psi(g)\rmd g\le\int_0^{\infty}\psi(g)\rmd g$; integrating the remaining $K-1$ coordinates over $\calM^{K-1}$ contributes the factor $\Delta_{\calM}^{K-1}$. Only the bound $\overline h$ on the joint density enters, and no conditional density of the prior is formed.

\emph{Part 1.} Fix $a$. When the preliminary recommendation $a$ is retained the incurred regret is $\Delta_{\bmmu}(a)$; when the final comparison rule chooses an empirical leader $b$ based on the comparison observations the incurred regret is $\Delta_{\bmmu}(b)$. Hence
\begin{align*}
r_{T,j}(\bmmu,a)
&\le
\Delta_{\bmmu}(a)
\bbP_{\bmmu}\p{\text{preliminary recommendation retained}}\\
&\quad+
\sum_{b\in[K]}\Delta_{\bmmu}(b)
\bbP_{\bmmu}\p{\widehat a_T^{g}=b,\ \text{final comparison chooses }b}.
\end{align*}
For the first term, which corresponds to retaining the preliminary recommendation, split by the size of $\Delta_{\bmmu}(a)$. Applying \eqref{eq:gap_layer_bound} with $\psi(g)=g\mathbbm 1\sqb{0<g\le2\eta}$ gives
\[
\int_{\calM^K}\Delta_{\bmmu}(a)\mathbbm 1\sqb{0<\Delta_{\bmmu}(a)\le2\eta}\rmd H(\bmmu)\le\overline h\Delta_{\calM}^{K-1}\int_0^{2\eta}g\rmd g=2\overline h\Delta_{\calM}^{K-1}\eta^2.
\]
Suppose that $\Delta_{\bmmu}(a)>2\eta$ and that the final comparison rule retains $a$. Then, $\widehat\mu_{a_\star}^{g}-\widehat\mu_a^{g}<\eta$. Since $\mu_{a_\star}-\mu_a=\Delta_{\bmmu}(a)$, the combined fluctuation of the two comparison-sample means must be larger than $\Delta_{\bmmu}(a)/2$. Hence, at least one of the two independent comparison-sample means deviates by $\Delta_{\bmmu}(a)/4$ or more.

This deviation is at most $\Delta_{\calM}/4$, so \eqref{eq:guard_mean_concentration} applies. With $c_7=c_6/16$, it gives
\[
\bbP_{\bmmu}\p{\text{preliminary recommendation retained}}
\le
2c_5\exp\p{-c_7n_g\Delta_{\bmmu}(a)^2}.
\]
The constant $c_7$ depends only on the constants already specified. The bound depends on $\bmmu$ only through $\Delta_{\bmmu}(a)$. Apply \eqref{eq:gap_layer_bound} with
\[
\psi(g)=2c_5g\mathbbm 1\sqb{g>2\eta}\exp\p{-c_7n_gg^2}.
\]
Then,
\begin{align*}
    &\int_{\calM^K}\Delta_{\bmmu}(a)\mathbbm 1\sqb{\Delta_{\bmmu}(a)>2\eta}\bbP_{\bmmu}(\text{preliminary recommendation retained})\rmd H(\bmmu)\\
    &\le2c_5\overline h\Delta_{\calM}^{K-1}\int_0^{\infty}g\exp\p{-c_7 n_g g^2}\rmd g=\frac{c_5\overline h\Delta_{\calM}^{K-1}}{c_7 n_g}.
\end{align*}
For the term from the final comparison rule, suppose that $b$ is the empirical leader based on the comparison observations. Then, $\widehat\mu_b^{g}\ge\widehat\mu_{a_\star}^{g}$. Thus, the combined fluctuation is at least $\Delta_{\bmmu}(b)$, and one comparison-sample mean deviates by at least $\Delta_{\bmmu}(b)/2$. Since this deviation is at most $\Delta_{\calM}/2$,
\[
\bbP_{\bmmu}\p{\widehat a_T^{g}=b}
\le
2c_5\exp\p{-c_7n_g\Delta_{\bmmu}(b)^2}.
\]
This bound depends on $\bmmu$ only through $\Delta_{\bmmu}(b)$. Applying \eqref{eq:gap_layer_bound} at index $b$ gives
\[
\int\Delta_{\bmmu}(b)\bbP_{\bmmu}\p{\widehat a_T^{g}=b}\rmd H
\le
\frac{c_5\overline h\Delta_{\calM}^{K-1}}{c_7n_g}.
\]
Sum the three contributions over $a\in[K]$ and $b\in[K]$. Since $c_7=c_6/16$, their integral is at most
\[
2K\overline h\Delta_{\calM}^{K-1}\eta^2
+
\frac{16c_5K(K+1)\overline h\Delta_{\calM}^{K-1}}{c_6n_g}.
\]
For $T\ge T_j$, use $\eta^2=H_j^2/T$ and $n_g\ge\gamma_jT/(2K)$. The preceding display is then at most $C\p{H_j^2+\gamma_j^{-1}}/T$, with the constant $C$ in the statement. For the consequence, fix $\bmmu$ and condition on the pilot data, the main history, and the main-policy randomization through the choice of $\widetilde a_T$, but not the comparison observations; the comparison sample is then independent given $\bmmu$ and $\widetilde a_T$ is measurable, so $\bbE_{\bmmu}\sqb{\Delta_{\bmmu}(\widehat a_T)\mid\text{pilot, main},\widetilde a_T}=r_{T,j}(\bmmu,\widetilde a_T)$, whence
\[
\bbE_{\bmmu}\sqb{\Delta_{\bmmu}(\widehat a_T)\mathbbm 1\sqb{E_{\bmmu}}}=\sum_{a=1}^K r_{T,j}(\bmmu,a)\bbP_{\bmmu}\p{E_{\bmmu},\ \widetilde a_T=a}\le p\sum_{a=1}^K r_{T,j}(\bmmu,a),
\]
using $\bbP_{\bmmu}(E_{\bmmu},\widetilde a_T=a)\le\bbP_{\bmmu}(E_{\bmmu})\le p$. The joint measurability of the family makes $\bmmu\mapsto\bbE_{\bmmu}\sqb{\Delta_{\bmmu}(\widehat a_T)\mathbbm 1\sqb{E_{\bmmu}}}$ measurable. Hence, the integral is well defined. Integrating over $H$ and multiplying by $T$ gives the stated bound.

\emph{Part 2.} Fix $\bmmu$. On $F_{\bmmu}$ the preliminary recommendation is within $6B_j/\sqrt T$ of the optimum. The correction $\p{\Delta_{\bmmu}(\widehat a_T)-\Delta_{\bmmu}(\widetilde a_T)}_+$ is nonzero only when the final comparison rule chooses some $b$ with $\Delta_{\bmmu}(b)>\Delta_{\bmmu}(\widetilde a_T)$, in which case it equals $\Delta_{\bmmu}(b)-\Delta_{\bmmu}(\widetilde a_T)\le\Delta_{\bmmu}(b)$. The final comparison rule chooses $b$ only if $\widehat\mu_b^{g}-\widehat\mu_{\widetilde a_T}^{g}\ge H_j/\sqrt T$; writing $\widehat\mu_c^{g}=\mu_c+\xi_c$ and using $\mu_{\widetilde a_T}-\mu_b=\Delta_{\bmmu}(b)-\Delta_{\bmmu}(\widetilde a_T)\ge\Delta_{\bmmu}(b)-6B_j/\sqrt T$, this forces
\[
\xi_b-\xi_{\widetilde a_T}\ge\frac{H_j}{\sqrt T}+\mu_{\widetilde a_T}-\mu_b\ge\Delta_{\bmmu}(b)+\frac{H_j-6B_j}{\sqrt T},
\]
so one of the two independent comparison-mean fluctuations is at least half this amount. By the definition of $T_j$, the deviation threshold satisfies
\[
\frac12\Bigp{\Delta_{\bmmu}(b)+\frac{H_j-6B_j}{\sqrt T}}\le\Delta_{\calM},
\]
because $\Delta_{\bmmu}(b)\le\Delta_{\calM}$, so \eqref{eq:guard_mean_concentration} applies at this threshold and gives
\begin{align*}
&\bbP_{\bmmu}\p{\text{final comparison chooses }b\mid\text{pilot, main}}\\
&\qquad\le2c_5\exp\p{-\frac{c_6 n_g}{4}
\p{\Delta_{\bmmu}(b)+\frac{H_j-6B_j}{\sqrt T}}^2}.
\end{align*}
Hence, the conditional expected increase in regret is bounded by
\begin{align*}
&\bbE_{\bmmu}\sqb{
\p{\Delta_{\bmmu}(\widehat a_T)-\Delta_{\bmmu}(\widetilde a_T)}_+
\mathbbm 1\sqb{F_{\bmmu}}}\\
&\qquad\le
\sum_b 2c_5\Delta_{\bmmu}(b)
\exp\p{-\frac{c_6 n_g}{4}
\p{\Delta_{\bmmu}(b)+\frac{H_j-6B_j}{\sqrt T}}^2}.
\end{align*}
Each summand is a function of $\Delta_{\bmmu}(b)$ alone. Put $d=(H_j-6B_j)/\sqrt T$, which is positive for every $j\ge8$. Apply \eqref{eq:gap_layer_bound} at index $b$ with $\psi(g)=g\exp\p{-c_6 n_g(g+d)^2/4}$. Then substitute $w=g+d$ and bound the factor $g$ by $w$.
\begin{align*}
&T\int_{\calM^K}\Delta_{\bmmu}(b)
\exp\p{-\tfrac{c_6n_g}{4}
\p{\Delta_{\bmmu}(b)+\tfrac{H_j-6B_j}{\sqrt T}}^2}
\rmd H(\bmmu)\\
&\qquad\le
T\overline h\Delta_{\calM}^{K-1}
\int_d^\infty w\exp\p{-\tfrac{c_6n_g}{4}w^2}\rmd w.
\end{align*}
The right-hand side equals $T\overline h\Delta_{\calM}^{K-1}\cdot\tfrac{2}{c_6 n_g}\exp\p{-\tfrac{c_6 n_g}{4}d^2}$. Since $n_g\ge\gamma_jT/(2K)$, it holds that $T/n_g\le2K/\gamma_j$ and
\[
\frac{c_6n_gd^2}{4}
\ge
\frac{c_6\gamma_j\p{H_j-6B_j}^2}{8K}.
\]
Thus, after restoring the factor $2c_5$ and summing over $b\in[K]$, the bound in Part 2 holds with the displayed constants $c$ and $C$. The joint measurability of $(F_{\bmmu})$ makes the integrand measurable. Hence, the integral is well defined.
\end{proof}

\subsection{Two Retained Arms with the Largest Means}
On the event $\calE_{\mu,T}\cap\calE_{v,T}(r_j)$, a true best arm is retained and every retained arm has mean gap at most $4B_j/\sqrt T$ by Lemma~\ref{lem:screening_geometry}. When exactly two arms are retained and they are the two arms of largest mean, the two-arm case uses the plug-in Neyman proportions of Algorithm~\ref{alg:smas} and chooses the main empirical best. The following lemma evaluates the prior-averaged contribution of these mean vectors.

\begin{lemma}[Prior-averaged binary contribution]
\label{lem:binary_bayes_comparison}
Suppose that Assumption~\ref{asm:uniformcontinuity} holds, and fix an index $j$.
For each unordered pair $\{a,b\}$ with $1\le a<b\le K$, let $\mathcal B_{ab}$ denote the event on which $\widehat{\calS}_T=\cb{a,b}$, arms $a$ and $b$ have the two largest means, and $\calE_{\mu,T}\cap\calE_{v,T}(r_j)$ occurs.
Let $\widetilde a_T$ denote the empirical-best recommendation based on the main-phase observations.

Let $\omega_H$ be the modulus of continuity of the prior density defined in \eqref{eq:prior_constants}, so that $\omega_H(s)\downarrow0$ as $s\downarrow0$. Put
\[
a_j=1-\alpha_j-\gamma_j,
\qquad
A=\frac{K^{\mathrm{bin}}}{2\sqrt{\underline v}},
\]
and let $M_j$ be the smallest integer $M\ge M_j^{\mathrm{bin}}$ such that
\[
a_jM\ge n_{\mathrm{env}},
\qquad
A\p{\frac{B_j}{\sqrt M}+\frac{1}{a_j^2M^2}}\le1.
\]
There exist finite constants $\overline K^{\mathrm B}$, $C_j^{\mathrm{BE}}$, and $C_H$. The constant $\overline K^{\mathrm B}$ depends on $\mathfrak E$ and the prior but not on $j$. The constant $C_j^{\mathrm{BE}}$ depends only on $K$, $j$, $\mathfrak E$, and $\calM$, whereas $C_H$ depends only on the prior, $K$, $\calM$, and $\mathfrak E$, and not on $j$ or $T$. The threshold $M_j$ depends only on $K$, $j$, and $\mathfrak E$.

For every $T\ge M_j$, we have
\[
T\int_{\calM^K}\sum_{1\le a<b\le K}
\bbE_{\bmmu}\sqb{
\Delta_{\bmmu}(\widetilde a_T)
\mathbbm 1\sqb{\mathcal B_{ab}}
}
\rmd H(\bmmu)
\le
\frac{C^{\mathrm{Bayes}}}{1-\alpha_j-\gamma_j}
+\overline K^{\mathrm B}\p{r_j+\zeta_j}
+R_{j,T}^{\mathrm B},
\]
where
\[
R_{j,T}^{\mathrm B}
\le
C_H\p{
C_j^{\mathrm{BE}}\frac{B_j^2}{\sqrt T}
+\frac{B_j}{\sqrt T}
+\omega_H\p{\frac{8B_j}{\sqrt T}}
}.
\]
\end{lemma}
\begin{proof}
Fix $\{a,b\}$. On $\mathcal B_{ab}$ exactly one of $a,b$ is a true best arm. The two-arm recommendation $\widetilde a_T$ incurs regret $\Delta_{\bmmu}(\widetilde a_T)=\abs{\mu_a-\mu_b}$ precisely when it recommends the arm with the smaller mean. Conditional on $\calF_{T_0}$, this happens exactly when the main-mean difference takes the wrong sign. As established in the proof of Lemma~\ref{lem:binary_upper_comparison}, conditional on $\calF_{T_0}$ the counts $N_a,N_b$ are deterministic and the standardized main-mean difference is a normalized sum of $N_a+N_b$ independent centered summands whose Lyapunov fraction is at most $C_j/\sqrt n$. Conditional on $\calF_{T_0}$ this difference has mean $\mu_a-\mu_b$ and variance $\sigma_a^2(\mu_a)/N_a+\sigma_b^2(\mu_b)/N_b$, so, with $c_n$ the Neyman standard-deviation scale of Lemma~\ref{lem:neyman_scale_calibration}, the Kolmogorov bound established in the proof of Lemma~\ref{lem:binary_upper_comparison} gives
\begin{align}
\abs{\bbP_{\bmmu}\p{\widetilde a_T\text{ wrong}\mid\calF_{T_0}}-\Phi\p{-\frac{\sqrt n\abs{\mu_a-\mu_b}}{c_n}}}\le\frac{C_j}{\sqrt n}.
\label{eq:binary_bayes_clt}
\end{align}
Conditional on $\calF_{T_0}$ the counts $N_a,N_b$ and the rounded standard deviations $\overline s_T$ are determined by the pilot data. Hence, the conditional law of the main-mean difference has an explicit version, and \eqref{eq:binary_bayes_clt} holds for that version at every $\bmmu$ and every pilot realization in $\mathcal B_{ab}$, with no exceptional set. Here $n=T-T_0-T_g$, and since $T_0=K\lfloor\alpha_jT/K\rfloor\le\alpha_jT$ and $T_g=K\lfloor\gamma_jT/K\rfloor\le\gamma_jT$ are both floors,
\begin{align}
n\ge\p{1-\alpha_j-\gamma_j}T.
\label{eq:main_horizon_floor}
\end{align}
The floor correction is favorable for the upper bound. A larger $n$ only decreases the Gaussian term below, so no $O(1/T)$ slack is needed when $n/T$ is replaced by $1-\alpha_j-\gamma_j$.

Write $m=\p{\mu_a+\mu_b}/2$ for the midpoint of the pair and $S_{ab}(m)=\sigma_a(m)+\sigma_b(m)$. On $\mathcal B_{ab}$ both arms are retained, so $\abs{\mu_a-\mu_b}\le4B_j/\sqrt T$ by Lemma~\ref{lem:screening_geometry} and hence $\abs{\mu_c-m}\le2B_j/\sqrt T$ for $c\in\cb{a,b}$. Lemma~\ref{lem:neyman_scale_calibration} therefore applies at this $m$ and gives the deterministic two-sided bound \eqref{eq:neyman_scale_relative}, namely $c_n\le S_{ab}(m)\p{1+\epsilon_{j,T}}$ with
\[
\epsilon_{j,T}=\frac{K^{\mathrm{bin}}}{2\sqrt{\underline v}}\p{r_j+\zeta_j+\frac{B_j}{\sqrt T}+\frac{1}{n^2}},
\]
uniformly over $\bmmu$ and over pilot realizations in $\mathcal B_{ab}$. We have $\epsilon_{j,T}=A\p{r_j+\zeta_j}+A\p{B_j/\sqrt T+1/n^2}$. The first summand is fixed once $j$ is fixed, so no largeness requirement on $T$ can reduce it; it is, however, already bounded by $\mathfrak E$, because $r_j+\zeta_j\le2j^{-4}\le2\cdot8^{-4}$ for every $j\ge8$. The second summand converges to zero as $T\to\infty$ for fixed $j$. By the definition of $M_j$ and the bound $n\ge a_jT$, for every $T\ge M_j$,
\begin{align}
\epsilon_{j,T}\le E_0\coloneqq2\cdot8^{-4}A+1.
\label{eq:eps_envelope_bound}
\end{align}
The constant $E_0$ depends only on the constants in $\mathfrak E$. For every $d\ge0$ the map $c\mapsto\Phi\p{-d/c}$ is nondecreasing on $(0,\infty)$, since $d/c$ decreases in $c$ and $\Phi$ is increasing. Applying this with $d=\sqrt n\abs{\mu_a-\mu_b}$ turns \eqref{eq:binary_bayes_clt} into the one-sided deterministic inequality
\begin{align}
\bbP_{\bmmu}\p{\widetilde a_T\text{ wrong}\mid\calF_{T_0}}\le\Phi\p{-\frac{\sqrt n\abs{\mu_a-\mu_b}}{S_{ab}(m)\p{1+\epsilon_{j,T}}}}+\frac{C_j}{\sqrt n}
\label{eq:binary_bayes_onesided}
\end{align}
on $\mathcal B_{ab}$, in which every quantity is a specified scalar.

Let
$L_{ab}=\cb{\text{$a,b$ are the two largest means}}\cap\cb{\abs{\mu_a-\mu_b}\le4B_j/\sqrt T}$. Since $\mathbbm 1\sqb{\mathcal B_{ab}}\le\mathbbm 1\sqb{L_{ab}}$, we may replace the conditional error probability by the right-hand side of \eqref{eq:binary_bayes_onesided}. Dropping the pilot indicators also enlarges the nonnegative integrand. Therefore,
\[
T\int_{\calM^K}\bbE_{\bmmu}\sqb{\Delta_{\bmmu}(\widetilde a_T)\mathbbm 1\sqb{\mathcal B_{ab}}}\rmd H(\bmmu)\le I_{ab}^{\Phi}+I_{ab}^{\mathrm{BE}},
\]
\[
I_{ab}^{\Phi}=T\int_{L_{ab}}\abs{\mu_a-\mu_b}\Phi\p{-\frac{\sqrt{n}\abs{\mu_a-\mu_b}}{S_{ab}(m)\p{1+\epsilon_{j,T}}}}\rmd H,
\qquad
I_{ab}^{\mathrm{BE}}=\frac{TC_j}{\sqrt n}\int_{L_{ab}}\abs{\mu_a-\mu_b}\rmd H.
\]
The Berry--Esseen remainder is bounded on $L_{ab}$ and only there; the enlargement of the range of integration performed below applies to the nonnegative Gaussian integrand of $I_{ab}^{\Phi}$ alone.
For $I_{ab}^{\mathrm{BE}}$, bound $h$ by $\overline h$, integrate the $K-2$ coordinates outside $\cb{a,b}$ over $\calM^{K-2}$, and change variables from $(\mu_a,\mu_b)$ to $(w,m)=(\mu_a-\mu_b,(\mu_a+\mu_b)/2)$, whose Jacobian is $1$; since $\abs{w}\le4B_j/\sqrt T$ on $L_{ab}$ and the admissible $m$ range at fixed $w$ has length at most $\Delta_{\calM}$, this gives $\int_{L_{ab}}\abs{\mu_a-\mu_b}\rmd H\le\overline h\Delta_{\calM}^{K-1}(4B_j/\sqrt T)^2$. With $\sqrt n\ge\sqrt{T/2}$ at fixed $j\ge8$, one obtains $I_{ab}^{\mathrm{BE}}\le C_HC_j^{\mathrm{BE}}B_j^2/\sqrt T$ with $C_j^{\mathrm{BE}}=16\sqrt2\Delta_{\calM}^{K-1}C_j$ depending only on $K$, $j$, $\mathfrak E$, and the parameter space $\calM$, which is fixed in Definition~\ref{def:mean_param} and known to the strategy, and with the prior-dependent density bound $\overline h$ absorbed into $C_H$, which is valid because $C_H$ may be taken at least $\overline h$. Keeping $\overline h$ out of $C_j^{\mathrm{BE}}$ is what lets the budget thresholds $N_j$ of Theorem~\ref{thm:alladapt_bayes_upper} be chosen from $C_j^{\mathrm{BE}}$ without knowledge of the prior.

For $I_{ab}^{\Phi}$, write $u=\sqrt T\abs{\mu_a-\mu_b}$. By \eqref{eq:main_horizon_floor}, $\sqrt n\abs{\mu_a-\mu_b}\ge\sqrt{1-\alpha_j-\gamma_j}u$, so replacing the former by the latter makes the negative argument of $\Phi$ larger and, $\Phi$ being increasing, only enlarges the integrand. Write $\rmd H(\bmmu)=h(\bmmu)\rmd\bmmu$ and change variables from $(\mu_a,\mu_b)$ to $(w,m)=(\mu_a-\mu_b,(\mu_a+\mu_b)/2)$ at fixed $\bmmu_{\setminus\cb{a,b}}$, the Jacobian being $1$; the second coordinate is exactly the reference point at which $\epsilon_{j,T}$ was obtained, which is why the standard-deviation displacement no longer appears separately here. On $L_{ab}$ one has $\max_{\ell\notin\cb{a,b}}\mu_\ell\le m$ and $\abs{\mu_c-m}\le2B_j/\sqrt T$ for $c\in\cb{a,b}$, while the coordinates outside $\cb{a,b}$ are not displaced at all, so by \eqref{eq:prior_constants} the density differs from its value $h(\bmmu_{\setminus\cb{a,b}},\mu_a=m,\mu_b=m)$ at equal means by at most $\omega_H(2B_j/\sqrt T)$, hence by at most $\omega_H(8B_j/\sqrt T)$, the modulus being nondecreasing. Substituting $w=u/\sqrt T$, so $T\abs{\mu_a-\mu_b}\rmd w=u\rmd u$, accounting for both signs of $w$, and enlarging the range of $u$ from $[0,4B_j]$ to $[0,\infty)$, which only increases the nonnegative integrand,
\begin{align*}
I_{ab}^{\Phi}\le{}&\int_{\calM^{K-2}}\int_{\calM}2\p{\int_0^{\infty}u\Phi\p{-\frac{\sqrt{1-\alpha_j-\gamma_j}u}{S_{ab}(m)\p{1+\epsilon_{j,T}}}}\rmd u}\\
&\quad\times h\p{\bmmu_{\setminus\cb{a,b}},m,m}\mathbbm 1\sqb{\max_{\ell\notin\cb{a,b}}\mu_\ell\le m}\rmd m\rmd\bmmu_{\setminus\cb{a,b}}+C_H\omega_H\p{\frac{8B_j}{\sqrt T}}.
\end{align*}
The last term bounds the error from replacing the prior density by its value at $\mu_a=\mu_b=m$. The constant $C_H$ depends only on the prior density bound, $K$, the parameter space $\calM$, and the constants in $\mathfrak E$. It also includes the Lebesgue volume $\Delta_{\calM}^{K-1}$ and the finite Gaussian integral over $u$.

The scalar $\epsilon_{j,T}$ does not depend on $u$, $m$, or $\bmmu$, so it passes through the inner integral. Apply the scaled Gaussian tail identity
$\int_0^{\infty}u\Phi\p{-u/\widetilde c}\rmd u=\widetilde c^2/4$ from Lemma~\ref{lem:tail_identity} with
\[
\widetilde c=\frac{S_{ab}(m)\p{1+\epsilon_{j,T}}}{\sqrt{1-\alpha_j-\gamma_j}}.
\]
The identity evaluates the inner factor as $\widetilde c^2/2=S_{ab}(m)^2\p{1+\epsilon_{j,T}}^2/\p{2\p{1-\alpha_j-\gamma_j}}$. The leading integral is therefore
\begin{align*}
&\frac{\p{1+\epsilon_{j,T}}^2}{1-\alpha_j-\gamma_j}\cdot\frac12
\int_{\calM}\int_{\calM^{K-2}}\p{\sigma_a(m)+\sigma_b(m)}^2\\
&\qquad\times h\p{\bmmu_{\setminus\cb{a,b}},m,m}
\mathbbm 1\sqb{\max_{\ell\notin\cb{a,b}}\mu_\ell\le m}
\rmd\bmmu_{\setminus\cb{a,b}}\rmd m\\
&\qquad=\frac{\p{1+\epsilon_{j,T}}^2C^{\mathrm{Bayes}}_{ab}}{1-\alpha_j-\gamma_j}
\end{align*}
by the pairwise integral in \eqref{eq:Cbayes_ij_pairwise}. Equation~\eqref{eq:eps_envelope_bound} gives
$\p{1+\epsilon_{j,T}}^2\le1+\p{2+E_0}\epsilon_{j,T}$, where $2+E_0$ depends only on the constants in $\mathfrak E$. Thus, the preceding expression is at most the leading term $C^{\mathrm{Bayes}}_{ab}/\p{1-\alpha_j-\gamma_j}$ plus an error proportional to $\epsilon_{j,T}$.

The multiplier $C^{\mathrm{Bayes}}_{ab}/\p{1-\alpha_j-\gamma_j}$ is bounded uniformly in $j\ge8$. Indeed, the constants in $\mathfrak E$ bound $\sigma_a+\sigma_b$, Assumption~\ref{asm:uniformcontinuity} bounds the density, and $\p{1-\alpha_j-\gamma_j}^{-1}\le2$. The $r_j+\zeta_j$ part of $\epsilon_{j,T}$ therefore contributes at most $\overline K^{\mathrm B}_{ab}\p{r_j+\zeta_j}$, where $\overline K^{\mathrm B}_{ab}$ is independent of $j$. It may depend on the prior through the density bound, but it does not enter the budget thresholds.

The remaining parts of $\epsilon_{j,T}$ are proportional to $B_j/\sqrt T$ and $1/n^2$. Since $n\ge T/2$ for $j\ge8$, we have $1/n^2\le4/T^2\le B_j/\sqrt T$. After enlarging $C_H$, both terms are included in $C_HB_j/\sqrt T$.

Finally, sum over $1\le a<b\le K$ and apply Proposition~\ref{prop:pairwise_equiv}. The leading terms sum to $C^{\mathrm{Bayes}}/\p{1-\alpha_j-\gamma_j}$. The standard-deviation errors sum to $\overline K^{\mathrm B}(r_j+\zeta_j)$, where $\overline K^{\mathrm B}=\sum_{a<b}\overline K^{\mathrm B}_{ab}<\infty$ is independent of $j$. The remaining terms form $R_{j,T}^{\mathrm B}$. Enlarging $C_j^{\mathrm{BE}}$ and $C_H$ by the factor $\binom K2$ gives the stated bound.
\end{proof}

\subsection{Completion of the Bayes Upper Bound}

\begin{theorem}[Bayes upper bound]
\label{thm:alladapt_bayes_upper}
Under Definition~\ref{def:mean_param} and Assumption~\ref{asm:structural_envelope}, let the prior $H$ satisfy Assumption~\ref{asm:uniformcontinuity}. Then the strategy $\delta^{\mathrm{SMAS}}$ of Algorithm~\ref{alg:smas}, whose tuning depends only on the tuple $\p{K,\calM,\mathfrak E}$, satisfies
\[
\limsup_{T\to\infty}T\int_{\calM^K}\Regret_T^{\delta^{\mathrm{SMAS}}}(\bmmu)\rmd H(\bmmu)\le C^{\mathrm{Bayes}},
\]
with $C^{\mathrm{Bayes}}$ the exact Bayes constant \eqref{eq:Cbayes_def}. Consequently, with the Bayes lower bound of Theorem~\ref{thm:bayes_lower}, $\delta^{\mathrm{SMAS}}$ is asymptotically Bayes optimal over the class $\calA$ of all adaptive experiments.
\end{theorem}
\begin{proof}
Fix an index $j$ and abbreviate $\alpha=\alpha_j$, $\gamma=\gamma_j$, $B=B_j$; write $G=\calE_{\mu,T}\cap\calE_{v,T}(r_j)$. The comparison threshold $H_j$ is not abbreviated here, because $H$ denotes the prior throughout this appendix. Decompose
\[
T\int_{\calM^K}\Regret_T^{\delta^{\mathrm{SMAS}}}(\bmmu)\rmd H(\bmmu)=T\int_{\calM^K}\bbE_{\bmmu}\sqb{\Delta_{\bmmu}(\widehat a_T)}\rmd H(\bmmu)
\]
over $\calE_{\mu,T}^c$, $\calE_{\mu,T}\cap\calE_{v,T}(r_j)^c$, and $G$.

On $\calE_{\mu,T}^c$ and $\calE_{\mu,T}\cap\calE_{v,T}(r_j)^c$, we have $\Delta_{\bmmu}(\widehat a_T)\ge0$. Lemma~\ref{lem:integrated_final_comparison}(1) applies to the parameter-dependent families $E_{\bmmu}=\calE_{\mu,T}^c$ and $E_{\bmmu}=\calE_{\mu,T}\cap\calE_{v,T}(r_j)^c$. Their probabilities obey \eqref{eq:pilot_mean_bad_probability}--\eqref{eq:pilot_variance_bad_probability}. Each family is jointly measurable in $(\bmmu,\omega)$: the pilot statistics $\widehat\mu_{a,T_0}$ and $\widehat\sigma_{a,T_0}^2$ are measurable in $\omega$, the maps $\bmmu\mapsto\mu_a$ and $\bmmu\mapsto\sigma_a^2(\mu_a)$ are continuous, and the events are formed from these by finitely many continuous operations and comparisons, so $(\bmmu,\omega)\mapsto\mathbbm 1\sqb{\omega\in E_{\bmmu}}$ is jointly measurable and each $E_{\bmmu}$ lies in the pilot sigma-algebra. Their contributions are at most
\[
C\,C_1\exp\p{-c_0\alpha B^2}\p{H_j^2+\gamma^{-1}}
\qquad\text{and}\qquad
C\,C_2\exp\p{-c_1\alpha T r_j^2}\p{H_j^2+\gamma^{-1}}.
\]

On $G$, a true best arm belongs to $\widehat{\calS}_T$, and every retained arm satisfies $\Delta_{\bmmu}(a)\le4B/\sqrt T$ by Lemma~\ref{lem:screening_geometry}. Hence, $\Delta_{\bmmu}(\widetilde a_T)\le4B/\sqrt T\le6B/\sqrt T$. Apply Lemma~\ref{lem:integrated_final_comparison}(2) with the family $F_{\bmmu}=G$, which is jointly measurable in $(\bmmu,\omega)$ by the same argument. The lemma bounds the additional regret from the final comparison rule on $G$ by $C\gamma^{-1}\exp\p{-c\gamma(H_j-6B)^2}$, so
\begin{align*}
    &T\int_{\calM^K}\bbE_{\bmmu}\sqb{\Delta_{\bmmu}(\widehat a_T)\mathbbm 1\sqb{G}}\rmd H(\bmmu)\\
    &\le T\int_{\calM^K}\bbE_{\bmmu}\sqb{\Delta_{\bmmu}(\widetilde a_T)\mathbbm 1\sqb{G}}\rmd H(\bmmu)+C\gamma^{-1}\exp\p{-c\gamma(H_j-6B)^2}.
\end{align*}
The term involving the preliminary recommendation splits according to the retained set on $G$. If $\abs{\widehat{\calS}_T}=1$, the retained arm is the true best arm, so $\Delta_{\bmmu}(\widetilde a_T)=0$.

Suppose that $\abs{\widehat{\calS}_T}\ge3$, or that $\abs{\widehat{\calS}_T}=2$ and the retained pair is not the pair of two largest means. Then three distinct arms lie within $8B/\sqrt T$ of one another. By Lemma~\ref{lem:triple_neartie_prior}, this event has prior mass at most $C_{\mathrm{tri}}(8B/\sqrt T)^2$. The preliminary recommendation satisfies $\Delta_{\bmmu}(\widetilde a_T)\le4B/\sqrt T$. Hence, the $T$-scaled contribution is at most $T\cdot(4B/\sqrt T)\cdot C_{\mathrm{tri}}(8B/\sqrt T)^2=256C_{\mathrm{tri}}B^3/\sqrt T$.

In the remaining case, the two retained arms are the two arms with the largest means, so Lemma~\ref{lem:binary_bayes_comparison} applies. Combining these bounds gives
\[
T\int_{\calM^K}\bbE_{\bmmu}\sqb{\Delta_{\bmmu}(\widetilde a_T)\mathbbm 1\sqb{G}}\rmd H(\bmmu)\le\frac{C^{\mathrm{Bayes}}}{1-\alpha-\gamma}+\overline K^{\mathrm B}\p{r_j+\zeta_j}+R_{j,T}^{\mathrm B}+\frac{256C_{\mathrm{tri}}B^3}{\sqrt T}.
\]
Adding the three regions,
\begin{align}
T\int_{\calM^K}\Regret_T^{\delta^{\mathrm{SMAS}}}(\bmmu)\rmd H(\bmmu)\le\frac{C^{\mathrm{Bayes}}}{1-\alpha-\gamma}+\calE_j^{\mathrm{B}}(T),
\label{eq:fixed_level_bayes_upper}
\end{align}
Here, $\calE_j^{\mathrm{B}}(T)$ contains the standard-deviation approximation term, the remainder $R_{j,T}^{\mathrm B}$, the contribution from triple near ties, and the final-comparison term. It also contains the two contributions from $\calE_{\mu,T}^c$ and $\calE_{\mu,T}\cap\calE_{v,T}(r_j)^c$.

The budget thresholds $N_j$ of Algorithm~\ref{alg:smas} are fixed in \eqref{eq:budget_thresholds} from the constants in $\mathfrak E$ and the finite constants fixed for $j$ alone. The requirements $\calR_j^{\mathrm{B}}$ enter that choice together with $\calR_j^{\mathrm{mm}}$. For every $T\ge N_j$, each of the three prior-independent quantities
\[
C_j^{\mathrm{BE}}\frac{B_j^2}{\sqrt T},\qquad\frac{B_j^3}{\sqrt T},\qquad C_2\exp\p{-c_1\alpha_j T r_j^2}\p{H_j^2+\gamma_j^{-1}}
\]
is at most $j^{-1}$. The pilot and comparison-sample conditions in $\calR_j^{\mathrm{mm}}$ also hold: $n_0\ge2$, $n_0\ge\alpha_jT/(2K)$, $n_g\ge1$, $n_g\ge\gamma_jT/(2K)$, $\tau_T=B_j/\sqrt T\le x_1$, and $H_j/\sqrt T\le x_1$. Hence, the pilot-variance concentration bound \eqref{eq:pilot_variance_bad_probability} applies to the contribution from $\calE_{\mu,T}\cap\calE_{v,T}(r_j)^c$ at the selected index $j_T$.

We also require $N_j\ge T_j$, where $T_j$ is from Lemma~\ref{lem:integrated_final_comparison}, and $N_j\ge M_j$, where $M_j$ is from Lemma~\ref{lem:binary_bayes_comparison}. Each requirement involves only the index $j$, the constants in $\mathfrak E$, the number of arms $K$, the parameter space $\calM$, and $T$. Hence, the budget thresholds, and therefore the strategy, depend neither on the prior nor on the arm families.

The constants $C_H$, $C_{\mathrm{tri}}$, and $\overline K^{\mathrm B}$ are fixed once the prior, $K$, $\calM$, and $\mathfrak E$ are fixed. The multiplicative constants supplied by Lemma~\ref{lem:integrated_final_comparison} are fixed under the same inputs. These constants multiply sequences already controlled by the prior-independent thresholds and do not enter the definition of $N_j$.

With $j=j_T=\max\cb{j:N_j\le T}$ as defined in \eqref{eq:budget_thresholds}, we have $j_T\to\infty$ as $T\to\infty$. Since $B_{j_T}^3/\sqrt T\le j_T^{-1}$, we also have $B_{j_T}/\sqrt T\to0$.

Evaluating \eqref{eq:fixed_level_bayes_upper} at $j=j_T$,
\[
T\int_{\calM^K}\Regret_T^{\delta^{\mathrm{SMAS}}}(\bmmu)\rmd H(\bmmu)\le\frac{C^{\mathrm{Bayes}}}{1-\alpha_{j_T}-\gamma_{j_T}}+\calE_{j_T}^{\mathrm B}(T),
\]
and, as $T\to\infty$, every term of $\calE_{j_T}^{\mathrm B}(T)$ converges to $0$. Consider the binary remainder $R_{j_T,T}^{\mathrm B}$ of Lemma~\ref{lem:binary_bayes_comparison}. By the choice of $N_j$, the prior-independent quantity $C_{j_T}^{\mathrm{BE}}B_{j_T}^2/\sqrt T$ is at most $j_T^{-1}$. Therefore, $C_HC_{j_T}^{\mathrm{BE}}B_{j_T}^2/\sqrt T\le C_Hj_T^{-1}\to0$, because $C_H$ does not depend on $j$ or $T$. We also have $C_H B_{j_T}/\sqrt T\le C_H B_{j_T}^3/\sqrt T\le C_H j_T^{-1}\to0$. Finally, $C_H\omega_H(8B_{j_T}/\sqrt T)\to0$ because $B_{j_T}/\sqrt T\to0$ and $\omega_H(s)\downarrow0$, while $C_H$ and $\omega_H$ do not depend on $j$ or $T$. The contribution from mean vectors with three near-tied arms satisfies $256C_{\mathrm{tri}}B_{j_T}^3/\sqrt T\le256C_{\mathrm{tri}}j_T^{-1}\to0$. The contribution from $\calE_{\mu,T}\cap\calE_{v,T}(r_{j_T})^c$ is at most $Cj_T^{-1}$ and also converges to zero. Both bounds combine the prior-independent choice of $N_j$ with the fixed prior constants $C_{\mathrm{tri}}$ and $C$.

The remaining terms depend only on $j$ and tend to $0$ as $j_T\to\infty$. The standard-deviation approximation term satisfies $\overline K^{\mathrm B}(r_j+\zeta_j)\le2\overline K^{\mathrm B}j^{-4}$, using that $\overline K^{\mathrm B}$ is independent of $j$ and $r_j+\zeta_j\le2j^{-4}$. The contribution from $\calE_{\mu,T}^c$ equals $C\,C_1\exp\p{-c_0 j^2}(j^6+j^2)$ and the term from the final comparison rule equals $Cj^2\exp\p{-cj^2(j-6)^2}$, both of which vanish as in the proof of Theorem~\ref{thm:alladapt_upper} because $\alpha_j B_j^2=j^2\to\infty$ and $\gamma_j(H_j-6B_j)^2=j^2(j-6)^2\to\infty$. Since $\alpha_{j_T}+\gamma_{j_T}=2j_T^{-2}\to0$, the leading fraction converges to $C^{\mathrm{Bayes}}$, and therefore
\[
\limsup_{T\to\infty}T\int_{\calM^K}\Regret_T^{\delta^{\mathrm{SMAS}}}(\bmmu)\rmd H(\bmmu)\le C^{\mathrm{Bayes}}.
\]
Finally, $\delta^{\mathrm{SMAS}}\in\calA$. Theorem~\ref{thm:bayes_lower} gives the matching lower bound $C^{\mathrm{Bayes}}$ for its limit inferior. The same theorem gives this lower bound for every strategy in $\calA$, while the preceding upper bound shows that $\delta^{\mathrm{SMAS}}$ attains it. Hence, $\delta^{\mathrm{SMAS}}$ is asymptotically Bayes optimal over $\calA$.
\end{proof}

\subsection{Uniform Arm Selection for Bernoulli Outcomes}
\label{app:bernoulli_uniform}

Throughout this subsection, $P_{a,\mu}=\mathrm{Bernoulli}(\mu)$ for every arm. Thus, all arms share the variance function $\sigma^2(\mu)=\mu(1-\mu)$, and $\calM=[\underline\mu,\overline\mu]\subset(0,1)$. Write $\sigma(\mu)=\sqrt{\mu(1-\mu)}$ and let $L_\sigma=\sup_{\mu\in\calM}\abs{\sigma'(\mu)}<\infty$ be its Lipschitz constant on the compact set $\calM$.

Let $\delta^{\mathrm{unif}}$ be the strategy that uses the pilot phase and the mean-screening set $\widehat{\calS}_T$ of Algorithm~\ref{alg:smas}. It omits the pilot variance estimation, the standard-deviation rounding, and the indexed policies of Definition~\ref{def:indexed_near_minimax_policies}. During the main phase, it selects each retained arm either $\lfloor n/\abs{\widehat{\calS}_T}\rfloor$ or $\lceil n/\abs{\widehat{\calS}_T}\rceil$ times. The $n-\abs{\widehat{\calS}_T}\lfloor n/\abs{\widehat{\calS}_T}\rfloor$ extra selections go to the retained arms with the smallest indices. Hence, the sampling counts differ by at most one and are deterministic given the pilot data. The strategy then chooses the arm with the largest main-sample mean. It uses the comparison-observation schedule and the final comparison rule of Algorithm~\ref{alg:smas} without change.

The tuning of $\delta^{\mathrm{unif}}$ uses information about the model class only through the mean space $\calM$ and the specified sequences $\alpha_j,\gamma_j,B_j,H_j$. The strategy uses none of the policies in Definition~\ref{def:indexed_near_minimax_policies}, and its budget thresholds are defined explicitly in the proof of Proposition~\ref{prop:bernoulli_uniform_upper}. Thus, $\delta^{\mathrm{unif}}\in\calA$ is a simpler finite procedure at each budget and uses only the observed history.

\begin{lemma}[Uniform scale at a Bernoulli near tie]
\label{lem:uniform_scale}
Fix an index $j\ge8$ and a pair $\calS=\cb{a,b}$, let $N_a,N_b$ with $\abs{N_a-N_b}\le1$ and $n=N_a+N_b$ be the equal main counts of $\delta^{\mathrm{unif}}$ on this pair, and put
\[
c_n^{\mathrm{unif}}=\sqrt{n\p{\frac{\sigma^2(\mu_a)}{N_a}+\frac{\sigma^2(\mu_b)}{N_b}}}.
\]
There is a finite constant $K^{\mathrm{unif}}$, depending only on $\underline v$, $\overline v$, and $L_\sigma$, such that for every $n\ge2$ and every $m\in\calM$ with $\max_{c\in\calS}\abs{\mu_c-m}\le2B_j/\sqrt T$,
\[
c_n^{\mathrm{unif}}\le\p{\sigma(m)+\sigma(m)}\p{1+\epsilon'_{j,T}},
\qquad
\epsilon'_{j,T}=K^{\mathrm{unif}}\p{\frac{B_j}{\sqrt T}+\frac{B_j^2}{T}+\frac1n}.
\]
\end{lemma}
\begin{proof}
Abbreviate $\sigma_c=\sigma(\mu_c)$. Since $N_a,N_b\in\cb{\lfloor n/2\rfloor,\lceil n/2\rceil}$, one has $n/N_c=2+O(1/n)$ uniformly, so $\p{c_n^{\mathrm{unif}}}^2=2\p{\sigma_a^2+\sigma_b^2}+O\p{\overline v/n}$, using $\sigma_c^2\le\overline v$. The identity $2\p{\sigma_a^2+\sigma_b^2}=\p{\sigma_a+\sigma_b}^2+\p{\sigma_a-\sigma_b}^2$ gives
\[
\p{c_n^{\mathrm{unif}}}^2=\p{\sigma_a+\sigma_b}^2+\p{\sigma_a-\sigma_b}^2+O\p{\frac{\overline v}n}.
\]
Because all arms share the variance function, $\abs{\sigma_a-\sigma_b}=\abs{\sigma(\mu_a)-\sigma(\mu_b)}\le L_\sigma\abs{\mu_a-\mu_b}\le4L_\sigma B_j/\sqrt T$, so $\p{\sigma_a-\sigma_b}^2\le16L_\sigma^2B_j^2/T$, and $\abs{\sigma_c-\sigma(m)}\le L_\sigma\abs{\mu_c-m}\le2L_\sigma B_j/\sqrt T$ gives $\sigma_a+\sigma_b=2\sigma(m)+O(B_j/\sqrt T)$, so $\p{\sigma_a+\sigma_b}^2\le\p{2\sigma(m)}^2+C\p{B_j/\sqrt T+B_j^2/T}$ with a constant $C$ depending only on $\mathfrak E$. Hence, dividing by $\p{2\sigma(m)}^2\ge4\underline v$,
\[
\p{c_n^{\mathrm{unif}}}^2\le\p{2\sigma(m)}^2\p{1+2K^{\mathrm{unif}}\p{\frac{B_j}{\sqrt T}+\frac{B_j^2}{T}+\frac1n}}
\]
for a finite $K^{\mathrm{unif}}$ built from $\underline v$, $\overline v$, and $L_\sigma$. Taking the square root and using $\sqrt{1+e}\le1+e/2$, valid for every $e\ge-1$, gives the claim.
\end{proof}

\begin{proposition}[Bernoulli Bayes upper bound for uniform arm selection]
\label{prop:bernoulli_uniform_upper}
Under the hypotheses of Proposition~\ref{prop:bernoulli_constant}, $\delta^{\mathrm{unif}}\in\calA$ and
\[
\limsup_{T\to\infty}T\int_{\calM^K}\Regret_T^{\delta^{\mathrm{unif}}}(\bmmu)\rmd H(\bmmu)\le C^{\mathrm{Bayes}}\p{\bmP,H}=C_{\mathrm{opt}}.
\]
Since $\delta^{\mathrm{unif}}\in\calA$, Theorem~\ref{thm:bayes_lower} supplies the matching lower bound, so $\delta^{\mathrm{unif}}$ attains the exact Bayes constant $C_{\mathrm{opt}}$.
\end{proposition}
\begin{proof}
We modify the bound of Theorem~\ref{thm:alladapt_bayes_upper} in three ways. None of these changes alters the leading Bayes constant. For each fixed $j$, every remainder introduced below vanishes as $T\to\infty$.

First, $\delta^{\mathrm{unif}}$ does not estimate the variances. Hence, the proof uses $\calE_{\mu,T}$ instead of $\calE_{\mu,T}\cap\calE_{v,T}(r_j)$, and the region $\calE_{\mu,T}\cap\calE_{v,T}(r_j)^c$ does not appear. Apply Lemma~\ref{lem:integrated_final_comparison}(1) with $E=\calE_{\mu,T}^c$. Equation~\eqref{eq:pilot_mean_bad_probability} gives
$\sup_{\bmmu}\bbP_{\bmmu}\p{\calE_{\mu,T}^c}\le C_1\exp\p{-c_0\alpha_jB_j^2}$.
The argument used in the proof of Theorem~\ref{thm:alladapt_bayes_upper} then gives the same bound for this contribution.

Second, suppose that the retained pair consists of the two arms with the largest means. We use the argument of Lemma~\ref{lem:binary_bayes_comparison}. Its conditional Berry--Esseen bound \eqref{eq:binary_bayes_clt} remains valid because the uniform counts satisfy $N_c\ge\lfloor n/2\rfloor\ge n/4$. Therefore, the Lyapunov fraction is again at most $C_j/\sqrt n$.

Only the bound on the scale depends on the sampling proportions. We replace the Neyman scale bound $c_n\le S_{ab}(m)\p{1+\epsilon_{j,T}}$ in \eqref{eq:neyman_scale_relative} with the uniform scale bound
$c_n^{\mathrm{unif}}\le S_{ab}(m)\p{1+\epsilon'_{j,T}}$
from Lemma~\ref{lem:uniform_scale}. This bound applies at $m=\p{\mu_a+\mu_b}/2$ because $\abs{\mu_c-m}\le2B_j/\sqrt T$ for each arm $c$ in the retained pair. Here, $S_{ab}(m)=\sigma(m)+\sigma(m)$ and
$\epsilon'_{j,T}=K^{\mathrm{unif}}\p{B_j/\sqrt T+B_j^2/T+1/n}$.
In particular, $\epsilon'_{j,T}$ contains no $r_j+\zeta_j$ term.

The same change of variables and integration at $\mu_a=\mu_b=m$ remain valid. Hence, the standard-deviation approximation term
$\overline K^{\mathrm B}\p{r_j+\zeta_j}$
in Lemma~\ref{lem:binary_bayes_comparison} is absent. All terms in $\epsilon'_{j,T}$ are included in the bound for the remainder $R_{j,T}^{\mathrm B}$, as in that lemma. For $j\ge8$, we have $1/n\le B_j/\sqrt T$. The retained requirement $\tau_T\le\Delta_{\calM}$ also gives
$B_j^2/T=\p{B_j/\sqrt T}^2\le\Delta_{\calM}B_j/\sqrt T$.

Third, consider the two remaining cases: at least three arms are retained, or the retained pair is not the pair of arms with the two largest means. Lemma~\ref{lem:screening_geometry} implies that each such case requires three near-tied arms. Lemma~\ref{lem:triple_neartie_prior} then shows that the prior mass of these mean vectors vanishes at the required rate, as in the proof of Theorem~\ref{thm:alladapt_bayes_upper}. Moreover,
$\Delta_{\bmmu}(\widetilde a_T)\le4B_j/\sqrt T$.
When bounding the regret on these mean vectors, we use only this inequality.

Lemma~\ref{lem:integrated_final_comparison} gives the same prior-averaged bound for the final comparison rule. Conditional on $\bmmu$, the comparison observations are independent of the pilot and main data. In addition, the uniform main counts are deterministic given the pilot data.

We now assemble the bound for a fixed $j$ as in \eqref{eq:fixed_level_bayes_upper}. Define the budget thresholds $N_j^{\mathrm{unif}}$ as in \eqref{eq:budget_thresholds}, but impose only the requirements needed by $\delta^{\mathrm{unif}}$. For the pilot phase, require $\tau_T=B_j/\sqrt T\le\Delta_{\calM}$ and $n_0\ge2$. We also require $n_0\ge\alpha_jT/(2K)$.

For the comparison sample, require $n_g\ge1$ and $n_g\ge\gamma_jT/(2K)$. We further impose $H_j/\sqrt T\le\Delta_{\calM}$ and $N_j^{\mathrm{unif}}\ge T_j$, where $T_j$ is the threshold from Lemmas~\ref{lem:final_comparison_bounds} and \ref{lem:integrated_final_comparison}. We also require $n\ge2$ and the threshold required by Lemma~\ref{lem:binary_bayes_comparison}. Finally, impose
$C_j^{\mathrm{BE}}B_j^2/\sqrt T\le j^{-1}$
and
$B_j^3/\sqrt T\le j^{-1}$.
When $j=j_T$, these two conditions bound the binary remainder $R_{j,T}^{\mathrm B}$ and the contribution from triple near ties by $j^{-1}$.

We omit the requirement $n\ge N_j^0$ for the policies in Definition~\ref{def:indexed_near_minimax_policies}. We also omit the condition $r_j\le r_{\mathrm{conc}}$, because $\delta^{\mathrm{unif}}$ uses neither requirement. All retained conditions are independent of the prior. Their inputs are only the specified sequences $\alpha_j,\gamma_j,B_j,H_j$, the number of arms, the parameter space $\calM$, and the constants in $\mathfrak E$.

Thus, the sequence $N_j^{\mathrm{unif}}$ and the index
$j_T=\max\cb{j:N_j^{\mathrm{unif}}\le T}$
are defined explicitly rather than through an existence argument. At every budget $T$, the strategy $\delta^{\mathrm{unif}}$ is determined by a finite procedure based only on the observed history. Evaluating the fixed-$j$ bound at $j=j_T$ and using the threshold conditions gives convergence to $C^{\mathrm{Bayes}}$ as $T\to\infty$. Proposition~\ref{prop:bernoulli_constant} identifies $C^{\mathrm{Bayes}}$ with $C_{\mathrm{opt}}$.
\end{proof}

\section{Useful Facts for Mean-Parameterized Exponential Families}
\label{app:expfam}

\begin{lemma}[Score and Fisher information]
\label{lem:score_fisher}
Let $\{P_\mu\}_{\mu\in\calM}\in\calP(\sigma^2,\calM,\calY)$ and let $\ell_\mu(y)$ be the log-density of $P_\mu$.
Then, $\ell_\mu(y)$ is three times continuously differentiable in $\mu$ and
\[
\bbE_{\mu}\sqb{\dot\ell_\mu(Y)}=0,
\qquad
I(\mu)\coloneqq \bbE_{\mu}\sqb{\dot\ell_\mu(Y)^2}=\frac{1}{\sigma^2(\mu)},
\qquad
\bbE_{\mu}\sqb{\ddot\ell_\mu(Y)}=-\frac{1}{\sigma^2(\mu)}.
\]
\end{lemma}

\begin{proof}
Because $P_\mu$ is a canonical exponential family with natural parameter $\theta(\mu)$ and log-partition $b$, the log-density is
$\ell_\mu(y)=y\theta(\mu)-b(\theta(\mu))$ up to an additive term not depending on $\mu$.
Differentiating and using the mean parameterization $\dot b(\theta(\mu))=\mu$ yields
\[
\dot\ell_\mu(y)=(y-\mu)\theta'(\mu).
\]
Hence $\bbE_\mu\sqb{\dot\ell_\mu(Y)}=0$.
Moreover,
\[
\bbE_\mu\sqb{\dot\ell_\mu(Y)^2}=\Var_\mu(Y) \theta'(\mu)^2=\sigma^2(\mu) \theta'(\mu)^2.
\]
Differentiating $\dot b(\theta(\mu))=\mu$ gives
\[
\ddot b(\theta(\mu))\theta'(\mu)=1.
\]
Hence, $\sigma^2(\mu)\theta'(\mu)=1$, and it follows that $\bbE_\mu\sqb{\dot\ell_\mu(Y)^2}=1/\sigma^2(\mu)$. Separately, differentiating $\bbE_\mu\sqb{\dot\ell_\mu(Y)}=0$ gives the identity for $\bbE_\mu\sqb{\ddot\ell_\mu(Y)}$.

For the higher-order smoothness, $\sigma^2\in C^2(\calM)$ is bounded away from zero by Definition~\ref{def:mean_param}. Thus, $\theta'=1/\sigma^2\in C^2$ and $\theta\in C^3$. Differentiating $\dot\ell_\mu(y)=(y-\mu)\theta'(\mu)$ twice more gives $\dddot\ell_\mu(y)=-2\theta''(\mu)+(y-\mu)\theta'''(\mu)$. This derivative is affine in $y$ and gives the bounded third-order Taylor remainder used below.
\end{proof}

\begin{lemma}[Uniform local exponential moments on compact $\calM$]
\label{lem:uniform_mgf}
Assume the local exponential-moment condition in Definition~\ref{def:mean_param}.
Then there exist constants $\lambda_0>0$ and $C<\infty$ such that
\[
\sup_{\mu\in\calM} \bbE_\mu\bigsqb{\exp\bigp{\lambda (Y-\mu)}}\le \exp(C\lambda^2)
\qquad\forall |\lambda|\le \lambda_0.
\]
Consequently, for every fixed integer $\overline q\ge1$, the following uniform polynomial moment bounds hold:
\[
\sup_{\mu\in\calM}\bbE_\mu\sqb{\abs{Y-\mu}^{\overline q}}<\infty,
\qquad
\sup_{\mu\in\calM}\bbE_\mu\sqb{\p{1+\abs Y}^{\overline q}}<\infty.
\]
The second bound follows because $\calM$ is compact. Both bounds are used with the arm index fixed and are uniform over $a\in[K]$ once the common constants of Assumption~\ref{asm:structural_envelope} are taken.
\end{lemma}

\begin{proof}
The exponential bound is exactly the uniform local exponential-moment assumption in Definition~\ref{def:mean_param}(4).
Indeed, for each arm $a$ the definition provides constants $\lambda_{0,a}>0$ and $C_a<\infty$ such that
\[
\sup_{\mu\in\calM}\bbE_{a,\mu}\sqb{\exp\p{\lambda(Y-\mu)}}\le \exp\p{C_a\lambda^2}
\qquad\forall |\lambda|\le \lambda_{0,a}.
\]
Therefore, after fixing the arm index and setting $\lambda_0\coloneqq \lambda_{0,a}$ and $C\coloneqq C_a$, the displayed bound follows.
For the moment envelopes, apply the exponential bound with $\lambda=\lambda_0$ and with $\lambda=-\lambda_0$ and use Markov's inequality: for every $t>0$ and every $\mu\in\calM$,
\[
\bbP_\mu\p{\abs{Y-\mu}>t}\le2\exp\p{C\lambda_0^2-\lambda_0 t},
\]
a tail bound that does not depend on $\mu$. Integrating it gives
\[
\bbE_\mu\sqb{\abs{Y-\mu}^{\overline q}}
=
\overline q\int_0^\infty
 t^{\overline q-1}\bbP_\mu\p{\abs{Y-\mu}>t}\rmd t.
\]
The right-hand side is bounded uniformly in $\mu$ by a finite constant depending only on $\overline q$, $\lambda_0$, and $C$.

Since $\calM$ is compact, $\abs Y\le\abs{Y-\mu}+\max_{m\in\calM}\abs m$. The second bound follows from the first and the inequality $\p{x+y}^{\overline q}\le2^{\overline q-1}\p{x^{\overline q}+y^{\overline q}}$ for $x,y\ge0$.
\end{proof}

\begin{lemma}[Berry--Esseen bound for independent summands]
\label{lem:berry_esseen}
Let $X_1,\dots,X_N$ be independent random variables with $\bbE\sqb{X_i}=0$ for every $i$, with $s_N^2\coloneqq\sum_{i=1}^N\bbE\sqb{X_i^2}>0$, and with $\sum_{i=1}^N\bbE\abs{X_i}^3<\infty$. Let $F_N$ be the distribution function of $s_N^{-1}\sum_{i=1}^N X_i$, and let $\Phi$ be the standard normal distribution function. Then
\[
\sup_{x\in\bbR}\abs{F_N(x)-\Phi(x)}\le C_{\mathrm{BE}}\frac{\sum_{i=1}^N\bbE\abs{X_i}^3}{s_N^3},
\]
where $C_{\mathrm{BE}}$ is an absolute constant, independent of $N$ and of the laws of the $X_i$; one may take $C_{\mathrm{BE}}=1$.
\end{lemma}

\begin{proof}
Rescaling by $X_i\mapsto X_i/s_N$ gives
\[
\sum_{i=1}^N\bbE\sqb{\p{X_i/s_N}^2}=1
\]
and leaves the normalized sum unchanged. Thus, it suffices to consider $s_N^2=1$. In that case,
\[
\sup_x\abs{F_N(x)-\Phi(x)}
\le
C_{\mathrm{BE}}\sum_{i=1}^N\bbE\abs{X_i}^3.
\]

This is the classical Berry--Esseen inequality for sums of independent, not necessarily identically distributed, random variables. It is stated as inequality~(3.2) and proved by Stein's method in \citet[Chapter~3]{Chen2011normalapproximation}. The left side is the Kolmogorov distance between $F_N$ and $\Phi$. Hence, the bound applies to lattice summands, including centered Bernoulli outcomes, and does not require a smooth distribution for $\sum_i X_i$.
\end{proof}

\section{Change-of-Measure: Le Cam's Lemma and Local Asymptotic Normality}
\label{app:com}

\subsection{Log Likelihood Ratio}
Fix a budget $T$. Let $U$ be the pre-experiment randomization seed of Section~\ref{sec:prob}. The joint law of $(U,A_1,Y_1,\dots,A_T,Y_T)$ under $\bbP_{\bmmu}$ is dominated by
\[
\mathrm{Law}(U)(\rmd u)
\otimes
\sum_{a_1,\dots,a_T\in[K]}
\prod_{t=1}^T\lambda_{a_t}(\rmd y_t),
\]
where $\lambda_a$ is the carrier measure of arm $a$. Its density is
\[
\prod_{t=1}^T
\kappa_t\p{A_t\mid U,A_1,Y_1,\dots,Y_{t-1}}
 f_{A_t,\mu_{A_t}}(Y_t).
\]
The selection kernels $\kappa_t$ are determined by the strategy and do not depend on $\bmmu$.

Each $f_{a,\mu}$ is a canonical exponential-family density and is strictly positive on the support of its carrier for every $\mu\in\calM$. Therefore, the density ratio under $\bmmu$ and $\bmnu$ is finite and positive everywhere. The strategy factors cancel, so $\bbP_{\bmmu}$ and $\bbP_{\bmnu}$ are mutually absolutely continuous at every fixed $T$. Define the log-likelihood ratio
\[
L_T(\bmmu,\bmnu)
\coloneqq
\log\frac{d\bbP_{\bmmu}}{d\bbP_{\bmnu}}
=
\sum_{a\in[K]}\sum_{t=1}^T\mathbbm{1}[A_t=a]\log\p{\frac{f_{a,\mu_a}(Y_t)}{f_{a,\nu_a}(Y_t)}}.
\]
By definition, $\bbE_{\bmmu}\sqb{\exp\p{-L_T(\bmmu,\bmnu)}}=1$; the mean is exactly one, not merely at most one, by the mutual absolute continuity above.
Moreover,
\[
\exp\{-L_T(\bmmu,\bmnu)\}=\frac{d\bbP_{\bmnu}}{d\bbP_{\bmmu}}
\quad\text{and hence}\quad
\bbE_{\bmnu}\sqb{Z}=\bbE_{\bmmu}\sqb{Z\exp\{-L_T(\bmmu,\bmnu)\}}
\]
for integrable $Z$.

\subsection{Le Cam's Third Lemma (Tilting Form)}
For reference, the standard tilting form of Le Cam's third lemma is stated in \citet[Chapter~6]{VanderVaart1998asymptoticstatistics}; see also \citet[Chapter~6]{Lecam2000asymptoticsin}. The proofs below use direct likelihood-ratio comparisons rather than this lemma.

\section{Proof of the Asymptotic Bayes Lower Bound (Theorem~\ref{thm:bayes_lower})}
\label{app:bayes_lower}
We show Theorem~\ref{thm:bayes_lower}, namely that
\[
\inf_{\delta\in\calA} \liminf_{T\to\infty} T \int_{\calM^K}\Regret_T^\delta(\bmmu)\rmd H(\bmmu)
 \ge C^{\mathrm{Bayes}}.
\]
Throughout this proof we fix an adaptive experiment $\delta\in\calA$. Fix a deterministic sequence of localization widths $r_T$ such that
$r_T\downarrow 0$ and $\sqrt{T} r_T\to\infty$ as $T\to\infty$ (e.g. $r_T=T^{-1/3}$), and set
$U_T\coloneqq \sqrt{T} r_T$, which also tends to infinity as $T\to\infty$.
Define the interior strip
\[
\calM_T^\circ\coloneqq \cb{m\in\calM\colon \mathrm{dist}(m,\partial\calM)\ge r_T}.
\]
Then for every $m\in\calM_T^\circ$ and every $u\in(0,U_T)$ we have $m+u/\sqrt{T}\in\calM$.
Since $\calM$ is a compact interval, its boundary $\partial\calM$ consists of at most two points and is therefore Lebesgue-null.

\subsection{Preliminary}
\label{app:bayeslowerpreliminary}
\begin{lemma}[Gaussian tail identity]
\label{lem:tail_identity}
\[
\int_0^\infty x\Phi(-x) \rmd x=\frac14.
\]
Equivalently, by the change of variables $x=u/c$, $\int_0^\infty u\Phi(-u/c)\rmd u=c^2/4$ for every $c>0$.
\end{lemma}
\begin{proof}
Write $\varphi$ for the standard normal density, so that $\p{\Phi(-x)}'=-\varphi(x)$, and integrate by parts against $\rmd\p{x^2/2}=x\rmd x$:
\[
\int_0^\infty x\Phi(-x)\rmd x
=\sqb{\frac{x^2}{2}\Phi(-x)}_0^\infty+\frac12\int_0^\infty x^2\varphi(x)\rmd x
=\frac12\int_0^\infty x^2\varphi(x)\rmd x
=\frac14,
\]
where the boundary term vanishes at $0$ and, as $x\to\infty$, because $\Phi(-x)\le\varphi(x)/x$, and where $\int_0^\infty x^2\varphi(x)\rmd x=\frac12\bbE\sqb{Z^2}=\frac12$ for $Z\sim\calN(0,1)$ by symmetry. The scaled form follows by substituting $u=cx$.
\end{proof}

\subsection{Shifted Alternative Pair and Gaussian Augmentation}
\label{app:transported_pair}

The pairwise testing bound of Lemma~\ref{lem:pairwise_testing_lb} below holds for every adaptive experiment $\delta\in\calA$. This subsection develops the tools.

Instead of comparing the two instances of the lemma directly, we compare the plus instance with a minus instance whose common level is shifted by an amount of order $1/\sqrt{T}$. The shift makes the two arms move by the same number of their own standard deviations. After this equalization, every selection of either arm carries the same standardized information about the comparison. Thus, the predictable quadratic variation of the martingale part of the log-likelihood ratio has a deterministic pathwise upper bound for every adaptive arm selection.

A conditionally Gaussian augmentation raises this quadratic variation to the deterministic upper bound. The martingale central limit theorem for triangular arrays then identifies the limiting testing affinity as the Gaussian affinity at that information level. Finally, a change of variables in the tie-value integral maps the shifted comparison back to the original pair. This step introduces the averaging over the tie value. The corresponding pointwise bound at a fixed $m$ is false, as noted in Remark~\ref{rem:loc_invariance_needed}.

Throughout this subsection, fix distinct arms $a\neq b$, a vector $\bmmu_{-ab}\in\operatorname{int}(\calM)^{K-2}$, a scalar $u>0$, and an interval $[\alpha,\beta]\subset\operatorname{int}(\calM)$ with $\max_{\ell\notin\cb{a,b}}\mu_\ell\le\alpha$, as in Lemma~\ref{lem:pairwise_testing_lb} below, whose instances $\bmmu^{\pm}_{T}(m)=\bmmu^{\pm}_{T}(m,u,\bmmu_{-ab})$ we use. For $m\in[\alpha,\beta]$ write
\[
S(m)\coloneqq\sigma_a(m)+\sigma_b(m),
\qquad
\gamma(m)\coloneqq\frac{2u}{S(m)},
\qquad
r(m)\coloneqq u\frac{\sigma_b(m)-\sigma_a(m)}{S(m)},
\]
so that $\abs{r(m)}\le u$, the function $r$ is continuously differentiable on $[\alpha,\beta]$ because $\sigma_a^2,\sigma_b^2\in C^2(\calM)$ are bounded away from zero, and $0<\underline\gamma\le\gamma(m)\le\overline\gamma<\infty$ for constants depending only on $u$ and the variance bounds. Define the shifted minus instance
\begin{align*}
\widetilde{\bmmu}^{-}_{T}(m)
&\coloneqq
\bmmu^{-}_{T}\p{m+\frac{r(m)}{\sqrt{T}}}\colon\\
\mu_a&=m+\frac{r(m)}{\sqrt{T}},
\qquad
\mu_b=m+\frac{r(m)+u}{\sqrt{T}},
\qquad
\bmmu_{-ab}\ \text{fixed}.
\end{align*}
Since $[\alpha,\beta]\subset\operatorname{int}(\calM)$ and the displacements are at most $2u/\sqrt{T}$, there is a $\underline T$, depending only on $u$ and $\operatorname{dist}([\alpha,\beta],\partial\calM)$, such that for all $T\ge \underline T$ and all $m\in[\alpha,\beta]$ every coordinate of $\bmmu^{+}_{T}(m)$ and $\widetilde{\bmmu}^{-}_{T}(m)$ lies in a fixed compact interval $\calM'\subset\operatorname{int}(\calM)$; all statements of this subsection are asserted for $T\ge \underline T$. The mean displacements from $\bmmu^{+}_{T}(m)$ to $\widetilde{\bmmu}^{-}_{T}(m)$ are $\delta_a(m)/\sqrt{T}$ and $\delta_b(m)/\sqrt{T}$ on coordinates $a$ and $b$ and zero elsewhere, where
\[
\delta_a(m)\coloneqq r(m)-u=-\gamma(m)\sigma_a(m),
\qquad
\delta_b(m)\coloneqq r(m)+u=\gamma(m)\sigma_b(m),
\]
and this level shift satisfies
\[
\frac{\abs{\delta_a(m)}}{\sigma_a(m)}=\frac{\abs{\delta_b(m)}}{\sigma_b(m)}=\gamma(m)\colon
\]
both arms are displaced by the same number of their own standard deviations, so selecting either arm provides the same amount of information about the comparison. Arm $b$ is the unique best arm under $\widetilde{\bmmu}^{-}_{T}(m)$, since $r(m)+u=2u\sigma_b(m)/S(m)>0$ and the means of the arms outside $\cb{a,b}$ are at most $\alpha$. This fact is used in Remark~\ref{rem:pairwise_minimax_lb}, while the present subsection uses only the two laws.

Use the pre-experiment randomization representation of Lemma~\ref{lem:randomization_representation} and the seed-augmented filtration $\p{\calF_t^U}_{t\le T}$ of Section~\ref{sec:prob}. The seed $U$ is $\calF_0^U$-measurable, each $A_t$ is $\calF_{t-1}^U$-measurable, and the realized counts below are the predictable quadratic-variation weights.

\begin{lemma}[Exact decomposition of the shifted log-likelihood ratio]
\label{lem:transport_decomp}
Fix $m\in[\alpha,\beta]$ and $\delta\in\calA$. Let $\bbP^{+}_{T}$ and $\widetilde{\bbP}^{-}_{T}$ denote the laws of the history $(U,A_1,Y_1,\dots,A_T,Y_T)$ under $\bmmu^{+}_{T}(m)$ and $\widetilde{\bmmu}^{-}_{T}(m)$. Define $L_T\coloneqq\log\p{\rmd\bbP^{+}_{T}/\rmd\widetilde{\bbP}^{-}_{T}}$. For $c\in\cb{a,b}$, let $\widetilde\mu_c\coloneqq\widetilde\mu^{-}_{c,T}(m)$ be the transported minus mean and let $\Delta\theta_c\coloneqq\theta_c\p{\mu^{+}_{c,T}(m)}-\theta_c\p{\widetilde\mu_c}$, where $\theta_c$ is the natural-parameter map of arm $c$. Then, for every $T\ge \underline T$,
\[
L_T=M_T-\frac{1}{2}V_T+\rho_T,
\]
where
\begin{enumerate}
\item\label{it:td_mart} $M_T=\sum_{t=1}^{T}D_t$ with $D_t\coloneqq\mathbbm{1}\sqb{A_t\in\cb{a,b}}\Delta\theta_{A_t}\p{Y_t-\widetilde\mu_{A_t}}$; under $\widetilde{\bbP}^{-}_{T}$, each $D_t$ is a martingale difference, $\bbE_{\widetilde{\bbP}^{-}_{T}}\sqb{D_t\mid\calF_{t-1}^U}=0$, and $\abs{D_t}\le C_0^{\mathrm{pair}}T^{-1/2}\abs{Y_t-\widetilde\mu_{A_t}}$ on $\cb{A_t\in\cb{a,b}}$;
\item\label{it:td_qv} the predictable quadratic variation is
$V_T\coloneqq\sum_{t=1}^{T}\bbE_{\widetilde{\bbP}^{-}_{T}}\sqb{D_t^2\mid\calF_{t-1}^U}$. It has the representation
$V_T=\sum_{c\in\cb{a,b}}N_{c,T}\p{\Delta\theta_c}^2\sigma_c^2\p{\widetilde\mu_c}$ and satisfies, pathwise,
\[
0\le V_T\le\overline{I}_T\coloneqq\gamma(m)^2\p{1+C_1^{\mathrm{pair}}T^{-1/2}};
\]
\item\label{it:td_rem} $\rho_T$ is $\calF_T^U$-measurable with $\abs{\rho_T}\le C_2^{\mathrm{pair}}T^{-1/2}$ pathwise.
\end{enumerate}
The constants $C_0^{\mathrm{pair}},C_1^{\mathrm{pair}},C_2^{\mathrm{pair}}$ depend only on $u$, the lower variance bound, the bounds on $\abs{\partial_\mu\sigma_c^2}$ over $\calM'$, and the bounds on $\abs{b_c'''}$ over $\theta_c(\calM')$, for $c\in\cb{a,b}$. They are uniform over $m\in[\alpha,\beta]$ and over strategies $\delta\in\calA$.
\end{lemma}

\begin{proof}
The two parameter vectors agree outside $\cb{a,b}$, so by the exact cancellation of the selection kernels and of the outer-arm outcome densities (Appendix~\ref{app:com}),
\[
L_T=\sum_{c\in\cb{a,b}}\sqb{\Delta\theta_c S_c\p{N_{c,T}}-N_{c,T}\Delta b_c},
\qquad
\Delta b_c\coloneqq b_c\p{\theta_c\p{\mu^{+}_{c,T}(m)}}-b_c\p{\theta_c\p{\widetilde\mu_c}},
\]
where $S_c(n)$ is the sum of the first $n$ outcomes of arm $c$ and $b_c$ is the log-partition function of arm $c$. Centering $S_c\p{N_{c,T}}=\widetilde S_c\p{N_{c,T}}+N_{c,T}\widetilde\mu_c$ at the transported minus means, with $\widetilde S_c(n)\coloneqq\sum_{j=1}^{n}\p{Y_{c,j}-\widetilde\mu_c}$, and rewriting the sum over arms as a sum over rounds,
\[
L_T=M_T+\sum_{c\in\cb{a,b}}N_{c,T}\sqb{\Delta\theta_c\widetilde\mu_c-\Delta b_c}.
\]
For the martingale property in \eqref{it:td_mart}: $A_t$ is $\calF_{t-1}^U$-measurable, and conditionally on $\calF_{t-1}^U$ and $\cb{A_t=c}$, the outcome $Y_t$ is the next unused entry of the outcome array of arm $c$. Under the outcome-array representation of Section~\ref{sec:prob}, this entry is independent of $\calF_{t-1}^U$ and has distribution $P_{c,\widetilde\mu_c}$ under $\widetilde{\bbP}^{-}_{T}$, with mean $\widetilde\mu_c$. These properties hold even though the array index $N_{c,t}$ is selected adaptively. Hence, $\bbE_{\widetilde{\bbP}^{-}_{T}}\sqb{D_t\mid\calF_{t-1}^U}=0$. The bound $\abs{\Delta\theta_c}\le C_0^{\mathrm{pair}}T^{-1/2}$ follows from $\theta_c'=1/\sigma_c^2$ (Lemma~\ref{lem:score_fisher}) and $\abs{\mu^{+}_{c,T}(m)-\widetilde\mu_c}=\abs{\delta_c(m)}/\sqrt{T}\le 2u/\sqrt{T}$, with $\sigma_c^2$ bounded below.

For \eqref{it:td_rem}, expand $\Delta b_c$ to third order along the natural parameter: with $\theta^{-}_c\coloneqq\theta_c\p{\widetilde\mu_c}$,
\begin{align*}
\Delta b_c
&=
b_c'\p{\theta^{-}_c}\Delta\theta_c+\frac{1}{2}b_c''\p{\theta^{-}_c}\p{\Delta\theta_c}^2+\frac{1}{6}b_c'''\p{\vartheta_c}\p{\Delta\theta_c}^3\\
&=
\widetilde\mu_c\Delta\theta_c+\frac{1}{2}\sigma_c^2\p{\widetilde\mu_c}\p{\Delta\theta_c}^2+\frac{1}{6}b_c'''\p{\vartheta_c}\p{\Delta\theta_c}^3
\end{align*}
for an intermediate point $\vartheta_c$, using $b_c'\p{\theta_c(\mu)}=\mu$ and $b_c''\p{\theta_c(\mu)}=\sigma_c^2(\mu)$ (Lemma~\ref{lem:score_fisher}). The third derivative $b_c'''$ is continuous on the compact image $\theta_c(\calM')$ of the compact interval $\calM'\subset\operatorname{int}(\calM)$ fixed above. Put $B_3^{\mathrm{pair}}\coloneqq\max_{c\in\cb{a,b}}\sup_{\vartheta\in\theta_c(\calM')}\abs{b_c'''(\vartheta)}<\infty$. Therefore
\[
\rho_T
=
-\sum_{c\in\cb{a,b}}N_{c,T}\sqb{\Delta b_c-\widetilde\mu_c\Delta\theta_c-\frac{1}{2}\sigma_c^2\p{\widetilde\mu_c}\p{\Delta\theta_c}^2}
\]
satisfies $\abs{\rho_T}\le\frac{B_3^{\mathrm{pair}}}{6}\sum_{c\in\cb{a,b}}N_{c,T}\p{C_0^{\mathrm{pair}}T^{-1/2}}^3\le C_2^{\mathrm{pair}}T^{-1/2}$ pathwise, and
\[
L_T=M_T-\frac{1}{2}\sum_{c\in\cb{a,b}}N_{c,T}\sigma_c^2\p{\widetilde\mu_c}\p{\Delta\theta_c}^2+\rho_T=M_T-\frac{1}{2}V_T+\rho_T,
\]
where the middle term is exactly the predictable quadratic variation displayed in \eqref{it:td_qv}, because $\bbE_{\widetilde{\bbP}^{-}_{T}}\sqb{D_t^2\mid\calF_{t-1}^U}=\mathbbm{1}\sqb{A_t\in\cb{a,b}}\p{\Delta\theta_{A_t}}^2\sigma_{A_t}^2\p{\widetilde\mu_{A_t}}$ and the rounds with $A_t=c$ number $N_{c,T}$.

For the pathwise bound in \eqref{it:td_qv}: by $\theta_c'=1/\sigma_c^2$ and the mean-value theorem, $\Delta\theta_c=-\delta_c(m)/\p{\sigma_c^2\p{\mu^{*}_c}\sqrt{T}}$ for an intermediate $\mu^{*}_c$ between the two means, so, using $\sigma_c^2\in C^1$ bounded below and $\abs{\mu^{*}_c-m}\le 2u/\sqrt{T}$,
\[
\p{\Delta\theta_c}^2\sigma_c^2\p{\widetilde\mu_c}
=
\frac{\delta_c(m)^2}{\sigma_c^2(m)T}\p{1+O\p{T^{-1/2}}}
=
\frac{\gamma(m)^2}{T}\p{1+O\p{T^{-1/2}}},
\]
with the $O\p{T^{-1/2}}$ deterministic and uniform over $m\in[\alpha,\beta]$; summing over $c\in\cb{a,b}$ with weights $N_{c,T}$ and using $N_{a,T}+N_{b,T}\le T$ gives $V_T\le\gamma(m)^2\p{1+C_1^{\mathrm{pair}}T^{-1/2}}$.
\end{proof}

\begin{lemma}[Gaussian augmentation to a fixed information level]
\label{lem:info_completion}
In the setting of Lemma~\ref{lem:transport_decomp}, for $T\ge \underline T$ set $g_T\coloneqq\overline{I}_T-V_T\ge0$, an $\calF_T^U$-measurable functional of the history, and $k_T\coloneqq\lceil\sqrt{T}\rceil$. Enlarge the experiment by variables $\eta_1,\dots,\eta_{k_T}$ generated as follows: given the history and $\eta_1,\dots,\eta_{j-1}$, the variable $\eta_j$ has conditional law $\calN\p{g_T/(2k_T),g_T/k_T}$ under the plus augmentation and $\calN\p{-g_T/(2k_T),g_T/k_T}$ under the minus augmentation; on the event $\cb{g_T=0}$ both conditional laws are the point mass at zero. The two kernels therefore coincide there with Radon--Nikodym ratio one. Write $\bbP^{+\prime}_{T}$ and $\widetilde{\bbP}^{-\prime}_{T}$ for the joint laws of $\p{\mathrm{history},\eta_1,\dots,\eta_{k_T}}$. Then:
\begin{enumerate}
\item\label{it:ic_marg} the marginal of $\bbP^{+\prime}_{T}$ on the history is $\bbP^{+}_{T}$, and likewise for the minus laws; consequently
\[
\mathrm{aff}\p{\bbP^{+}_{T},\widetilde{\bbP}^{-}_{T}}\ge\mathrm{aff}\p{\bbP^{+\prime}_{T},\widetilde{\bbP}^{-\prime}_{T}};
\]
\item\label{it:ic_lr} $L'_T\coloneqq\log\p{\rmd\bbP^{+\prime}_{T}/\rmd\widetilde{\bbP}^{-\prime}_{T}}=L_T+\sum_{j=1}^{k_T}\eta_j$;
\item\label{it:ic_qv} under $\widetilde{\bbP}^{-\prime}_{T}$, writing $\eta_j=-g_T/(2k_T)+\widetilde\eta_j$ with $\widetilde\eta_j$ conditionally centered,
\[
L'_T=M'_T-\frac{1}{2}\overline{I}_T+\rho_T,
\qquad
M'_T\coloneqq M_T+\sum_{j=1}^{k_T}\widetilde\eta_j,
\]
and $M'_T$ is a sum of martingale differences with respect to the within-row filtration $\calF_0^U\subseteq\dots\subseteq\calF_T^U\subseteq\calF_T^U\vee\widetilde{\sigma}(\eta_1)\subseteq\dots$ whose total predictable quadratic variation equals $V_T+g_T=\overline{I}_T$, a deterministic constant.
\end{enumerate}
\end{lemma}

\begin{proof}
\eqref{it:ic_marg} The augmentation appends coordinates through Markov kernels acting on the history. Hence, the history marginals are unchanged. Discarding the appended coordinates is a common Markov kernel mapping the augmented pair to the original pair, and the testing affinity does not decrease under a common Markov kernel. Equivalently, every test of the original experiment is a test of the enlarged experiment that ignores the appended coordinates. The infimum defining the affinity is therefore taken over a larger set of tests in the enlarged experiment.

\eqref{it:ic_lr} On $\cb{g_T>0}$ the conditional log density ratio of $\eta_j$ is, with $v\coloneqq g_T/k_T$,
\[
\log\frac{\varphi\p{(\eta_j-v/2)/\sqrt{v}}}{\varphi\p{(\eta_j+v/2)/\sqrt{v}}}
=
\frac{-\p{\eta_j-v/2}^2+\p{\eta_j+v/2}^2}{2v}
=
\eta_j,
\]
and the two conditional kernels share the same $g_T$ because $g_T$ is a functional of the shared history coordinate; on $\cb{g_T=0}$ both kernels coincide and the ratio is one while $\eta_j=0$. Multiplying the history density ratio and the $k_T$ conditional ratios gives the display.

\eqref{it:ic_qv} Substituting $\eta_j=-g_T/(2k_T)+\widetilde\eta_j$ into \eqref{it:ic_lr} and using Lemma~\ref{lem:transport_decomp} give $L'_T=M_T+\sum_j\widetilde\eta_j-\frac{1}{2}\p{V_T+g_T}+\rho_T$, and $V_T+g_T=\overline{I}_T$ by the definition of $g_T$. Conditionally on the history and $\eta_1,\dots,\eta_{j-1}$, the variable $\widetilde\eta_j$ is centered with variance $g_T/k_T$. Hence, the appended entries are martingale differences with total conditional variance $g_T$, and the data entries contribute $V_T$ by Lemma~\ref{lem:transport_decomp}\eqref{it:td_qv}.
\end{proof}

\begin{lemma}[Martingale central limit theorem for triangular arrays]
\label{lem:array_mclt}
For each $T\in\bbN$, let $\p{D_{T,j}}_{j\le n_T}$ be square-integrable martingale differences with respect to within-row filtrations $\calG_{T,0}\subseteq\dots\subseteq\calG_{T,n_T}$. Let $s^2>0$ be a constant. Assume that, as $T\to\infty$,
\begin{align*}
\sum_{j\le n_T}\bbE\sqb{D_{T,j}^2\mid\calG_{T,j-1}}
&\xrightarrow{p} s^2,\\
\sum_{j\le n_T}\bbE\sqb{D_{T,j}^2
\mathbbm{1}\sqb{\abs{D_{T,j}}\ge\varepsilon}}
&\longrightarrow0\quad\text{for every }\varepsilon>0.
\end{align*}
Then, we have $\sum_{j\le n_T}D_{T,j}\xrightarrow{d}\calN\p{0,s^2}$ as $T\to\infty$.
\end{lemma}

This is the classical martingale central limit theorem for triangular arrays, Theorem~35.12 of \citet{Billingsley1995probabilityand}. See also Chapter~3 of \citet{Hall2014martingalelimit}. The displayed second condition is the unconditional Lindeberg condition of \citet{Billingsley1995probabilityand}. It implies the conditional form $\sum_{j}\bbE\sqb{D_{T,j}^2\mathbbm{1}\sqb{\abs{D_{T,j}}\ge\varepsilon}\mid\calG_{T,j-1}}\xrightarrow{p}0$ as $T\to\infty$, stated in some references, because the latter sum is nonnegative with expectation equal to the displayed quantity, so Markov's inequality applies. No relation between the filtrations of different rows is required here: the limit is a fixed Gaussian law with deterministic variance, so plain convergence in distribution suffices, and the failure of any cross-row nesting condition for budget-dependent strategies is immaterial.

\begin{lemma}[Affinity bound for the shifted alternative pair]
\label{lem:transport_affinity}
For every $m\in[\alpha,\beta]$ and every $\delta\in\calA$,
\[
\liminf_{T\to\infty}\mathrm{aff}\p{\bbP^{+}_{T},\widetilde{\bbP}^{-}_{T}}
\ge
2\Phi\p{-\frac{u}{S(m)}}.
\]
\end{lemma}

\begin{proof}
By Lemma~\ref{lem:info_completion}\eqref{it:ic_marg} it suffices to bound the affinity of the augmented pair, and, by mutual absolute continuity,
\[
\mathrm{aff}\p{\bbP^{+\prime}_{T},\widetilde{\bbP}^{-\prime}_{T}}
=
\bbE_{\widetilde{\bbP}^{-\prime}_{T}}\sqb{\min\cb{\exp\p{L'_T},1}}.
\]
We verify the two conditions of Lemma~\ref{lem:array_mclt} for the array consisting of the $T$ data differences $D_t$ followed by the $k_T$ augmented differences $\widetilde\eta_j$, under $\widetilde{\bbP}^{-\prime}_{T}$, with $s^2=\gamma(m)^2>0$. The conditional variances sum to $\overline{I}_T$ by Lemma~\ref{lem:info_completion}\eqref{it:ic_qv}, and $\overline{I}_T\to\gamma(m)^2$ deterministically as $T\to\infty$. For the Lindeberg condition, on the data entries the bound $\abs{D_t}\le C_0^{\mathrm{pair}}T^{-1/2}\abs{Y_t-\widetilde\mu_{A_t}}$ of Lemma~\ref{lem:transport_decomp}\eqref{it:td_mart} gives
\begin{align*}
&\sum_{t\le T}\bbE\sqb{D_t^2\mathbbm{1}\sqb{\abs{D_t}\ge\varepsilon}}\\
&\quad\le
\p{C_0^{\mathrm{pair}}}^2
\max_{c\in\cb{a,b}}\sup_{\mu\in\calM'}
\bbE_{c,\mu}\Bigsqb{
\p{Y-\mu}^2
\mathbbm{1}\Bigsqb{\abs{Y-\mu}\ge\frac{\varepsilon\sqrt{T}}{C_0^{\mathrm{pair}}}}
}
\longrightarrow0
\qquad(T\to\infty).
\end{align*}
by the uniform integrability of the squared centered outcomes implied by the uniform local exponential moments (Lemma~\ref{lem:uniform_mgf}). On the augmented entries, which are conditionally Gaussian with conditional variance $g_T/k_T\le\overline\gamma^2\p{1+C_1^{\mathrm{pair}}}/k_T$, the Gaussian bound $\bbE\sqb{W^2\mathbbm{1}\sqb{\abs{W}\ge\varepsilon}}\le3s_W^2\exp\p{-\varepsilon^2/\p{4s_W^2}}$ for $W$ centered Gaussian with variance $s_W^2$ yields a total contribution of order $\exp\p{-c\varepsilon^2k_T}$, which converges to zero as $T\to\infty$ after summing the $k_T$ terms. Lemma~\ref{lem:array_mclt} gives $M'_T\xrightarrow{d}\calN\p{0,\gamma(m)^2}$ under $\widetilde{\bbP}^{-\prime}_{T}$ as $T\to\infty$. Since $\abs{\rho_T}\le C_2^{\mathrm{pair}}T^{-1/2}$ deterministically and $\overline{I}_T\to\gamma(m)^2$ as $T\to\infty$, Slutsky's theorem gives
\[
L'_T=M'_T-\frac{1}{2}\overline{I}_T+\rho_T
\xrightarrow{d}\calN\p{-\frac{\gamma(m)^2}{2},\gamma(m)^2}
\qquad(T\to\infty).
\]
The map $z\mapsto\min\cb{e^z,1}$ is bounded and continuous, so
\[
\lim_{T\to\infty}\mathrm{aff}\p{\bbP^{+\prime}_{T},\widetilde{\bbP}^{-\prime}_{T}}
=
\bbE\sqb{\min\cb{e^{Z},1}},
\qquad
Z\sim\calN\p{-\frac{\gamma(m)^2}{2},\gamma(m)^2},
\]
and a direct computation, completing the square in $\bbE\sqb{e^{Z}\mathbbm{1}\sqb{Z<0}}$, gives $\bbE\sqb{\min\cb{e^{Z},1}}=\bbP\p{Z\ge0}+\bbE\sqb{e^{Z}\mathbbm{1}\sqb{Z<0}}=2\Phi\p{-\gamma(m)/2}=2\Phi\p{-u/S(m)}$.
\end{proof}

\subsection{Pairwise Testing Bound over All Adaptive Experiments}
\label{app:pairwise_testing}

\begin{lemma}[Pairwise local testing bound averaged over the tie location]
\label{lem:pairwise_testing_lb}
Fix distinct arms $a\neq b$, a vector of other means $\bmmu_{-ab}\in\operatorname{int}(\calM)^{K-2}$, so that every tie baseline $(m,m,\bmmu_{-ab})$ with $m\in\operatorname{int}(\calM)$ lies in $\operatorname{int}(\calM^K)$, and $u>0$. For $K=2$, interpret $\bmmu_{-ab}$ as the unique point of $\calM^0$, so integrals over $\bmmu_{-ab}$ are evaluations at that point, and set $\max_{\ell\notin\{a,b\}}\mu_\ell=-\infty$. These conventions also apply in the proofs of Theorem~\ref{thm:bayes_lower} and Proposition~\ref{prop:pairwise_equiv}.
Let $[\alpha,\beta]\subset\operatorname{int}(\calM)$ be an interval with $\max_{\ell\notin\{a,b\}}\mu_\ell\le\alpha$, and for $m\in[\alpha,\beta]$ and each $T$ define
\begin{align*}
&\bmmu^{+}_{T}(m,u,\bmmu_{-ab})
\colon
(\mu_a,\mu_b)=(m+u/\sqrt{T}, m),\quad \bmmu_{-ab}\ \ \text{fixed},
\\
&\bmmu^{-}_{T}(m,u,\bmmu_{-ab})
\colon
(\mu_a,\mu_b)=(m, m+u/\sqrt{T}),\quad \bmmu_{-ab}\ \ \text{fixed}.
\end{align*}
Let $\widehat a_T$ be the choice of an adaptive experiment $\delta\in\calA$.
Then, for every continuous weight $\omega\colon[\alpha,\beta]\to[0,\infty)$,
\begin{align*}
&\liminf_{T\to\infty}
\int_\alpha^\beta\Big(
\bbP_{\bmmu^{+}_{T}(m,u,\bmmu_{-ab})}(\widehat a_T \neq a)
+
\bbP_{\bmmu^{-}_{T}(m,u,\bmmu_{-ab})}(\widehat a_T \neq b)
\Big)\omega(m)\rmd m\\
&\ge
\int_\alpha^\beta 2\Phi\p{-\frac{u}{\sigma_a(m)+\sigma_b(m)}}\omega(m)\rmd m.
\end{align*}
\end{lemma}

\begin{remark}[Why the average over $m$ is needed]
\label{rem:loc_invariance_needed}
The corresponding \emph{pointwise}-in-$m$ inequality is false: a procedure that knew the tie value $m$ could test each coordinate against the known level $m$ and distinguish $\bmmu^{+}_T$ from $\bmmu^{-}_T$ with a strictly smaller error constant than $2\Phi(-u/(\sigma_a(m)+\sigma_b(m)))$. The error does not vanish: the means differ by $u/\sqrt T$, an $O(1)$ standardized separation, and the error remains a positive constant, but it is governed by a smaller constant than the one above. The constant $\sigma_a(m)+\sigma_b(m)$ reflects exactly that the tie value $m$ is \emph{unknown}; this is why the bound is established only after averaging over $m$ (equivalently, against the prior). In the proof, the change of variables replaces the comparison between $\bmmu_T^{+}(m)$ and $\widetilde{\bmmu}_T^{-}(m)$, for which the bound holds pointwise in $m$, by the comparison between $\bmmu_T^{+}(m)$ and $\bmmu_T^{-}(m)$, for which it does not (Appendix~\ref{app:transported_pair}).
\end{remark}

\begin{proof}
Fix $\delta\in\calA$ and let $\underline T$ be as in Appendix~\ref{app:transported_pair}. For $m\in[\alpha,\beta]$ and $T\ge \underline T$ define
\[
\widetilde e_T(m)
\coloneqq
\bbP_{\bmmu^{+}_{T}(m)}\p{\widehat a_T\neq a}+\bbP_{\widetilde{\bmmu}^{-}_{T}(m)}\p{\widehat a_T\neq b},
\]
with the transported minus instance $\widetilde{\bmmu}^{-}_{T}(m)$ of Appendix~\ref{app:transported_pair}. Since the events $\cb{\widehat a_T=a}$ and $\cb{\widehat a_T=b}$ are disjoint, their complements cover the sample space, so for any two laws $P,Q$ of the history,
\[
P\p{\widehat a_T\neq a}+Q\p{\widehat a_T\neq b}
\ge
\int_{\cb{\widehat a_T\neq a}}\min\cb{\rmd P,\rmd Q}+\int_{\cb{\widehat a_T\neq b}}\min\cb{\rmd P,\rmd Q}
\ge
\mathrm{aff}\p{P,Q},
\]
where $\mathrm{aff}(P,Q)\coloneqq\int\min\cb{\rmd P,\rmd Q}$ is the testing affinity. Hence $\widetilde e_T(m)\ge\mathrm{aff}\p{\bbP^{+}_{T},\widetilde{\bbP}^{-}_{T}}$, and Lemma~\ref{lem:transport_affinity} gives, for every $m\in[\alpha,\beta]$,
\[
\liminf_{T\to\infty}\widetilde e_T(m)\ge2\Phi\p{-\frac{u}{S(m)}}.
\]
For every fixed $T$, the map $\bmmu\mapsto\bbP_{\bmmu}\p{\widehat a_T\neq a}$ is Borel on $\calM^K$. The history density is jointly measurable and continuous in the parameters, and the choice is measurable. The maps $(m,u,\bmmu_{-ab})\mapsto\bmmu^{+}_{T}(m,u,\bmmu_{-ab})$ and $(m,u,\bmmu_{-ab})\mapsto\widetilde{\bmmu}^{-}_{T}(m,u,\bmmu_{-ab})$ are continuous. Therefore, the compositions are jointly Borel in $(m,u,\bmmu_{-ab})$. In particular, they are measurable in $m$ at fixed $(u,\bmmu_{-ab})$, as required in the present lemma. Their joint measurability is used in the outer application of Fatou's lemma in the proof of Theorem~\ref{thm:bayes_lower}. Fatou's lemma applied to the nonnegative integrand $\widetilde e_T\omega$ yields
\begin{align}
\label{eq:transported_fatou}
\liminf_{T\to\infty}\int_\alpha^\beta\widetilde e_T(m)\omega(m)\rmd m
\ge
\int_\alpha^\beta2\Phi\p{-\frac{u}{S(m)}}\omega(m)\rmd m.
\end{align}

It remains to map the shifted comparison back to the original pair. Define $\tau_T(m)\coloneqq m+r(m)/\sqrt{T}$, so that $\widetilde{\bmmu}^{-}_{T}(m)=\bmmu^{-}_{T}\p{\tau_T(m)}$ exactly. Since $r\in C^1\p{[\alpha,\beta]}$, we have $\tau_T'(m)=1+r'(m)/\sqrt{T}\to1$ uniformly as $T\to\infty$, so for all large $T$ the map $\tau_T$ is a $C^1$ diffeomorphism from $[\alpha,\beta]$ onto its image, an interval contained in $\operatorname{int}(\calM)$ whose symmetric difference with $[\alpha,\beta]$ consists of two intervals of length at most $u/\sqrt{T}$ each. For the functions $f_T(x)\coloneqq\bbP_{\bmmu^{-}_{T}(x)}\p{\widehat a_T\neq b}\in[0,1]$, measurable on a neighborhood of $[\alpha,\beta]$, the substitution $x=\tau_T(m)$ gives
\[
\int_\alpha^\beta f_T\p{\tau_T(m)}\omega(m)\rmd m
=
\int_{\tau_T([\alpha,\beta])}f_T(x)\frac{\omega\p{\tau_T^{-1}(x)}}{\tau_T'\p{\tau_T^{-1}(x)}}\rmd x,
\]
and hence, for all large $T$,
\begin{align*}
&\abs{\int_\alpha^\beta f_T\p{\tau_T(m)}\omega(m)\rmd m-\int_\alpha^\beta f_T(x)\omega(x)\rmd x}\\
&\le
\p{\beta-\alpha}\sup_{x}\abs{\frac{\omega\p{\tau_T^{-1}(x)}}{\tau_T'\p{\tau_T^{-1}(x)}}-\omega(x)}
+\frac{4u}{\sqrt{T}}\sup_{[\alpha,\beta]}\omega
\longrightarrow0
\qquad(T\to\infty),
\end{align*}
uniformly over measurable $f_T$ with values in $[0,1]$, where the supremum in the first term is over $x\in\tau_T([\alpha,\beta])\cap[\alpha,\beta]$ and converges to zero by the uniform continuity of $\omega$ on $[\alpha,\beta]$, $\sup_m\abs{\tau_T(m)-m}\le u/\sqrt{T}$, and $\tau_T'\to1$ uniformly, while the second term bounds the contribution of the two boundary slivers, whose total length is at most $2u/\sqrt{T}$ and on which both integrands are at most $2\sup_{[\alpha,\beta]}\omega$ for all large $T$. Since $f_T\p{\tau_T(m)}=\bbP_{\widetilde{\bmmu}^{-}_{T}(m)}\p{\widehat a_T\neq b}$ by construction,
\[
\int_\alpha^\beta\Bigp{\bbP_{\bmmu^{+}_{T}(m)}\p{\widehat a_T\neq a}+\bbP_{\bmmu^{-}_{T}(m)}\p{\widehat a_T\neq b}}\omega(m)\rmd m
=
\int_\alpha^\beta\widetilde e_T(m)\omega(m)\rmd m+o(1),
\]
and the claim follows from \eqref{eq:transported_fatou}.
\end{proof}

\begin{remark}[Pairwise minimax lower bound]
\label{rem:pairwise_minimax_lb}
The comparison between $\bmmu^{+}_{T}(m)$ and $\widetilde{\bmmu}^{-}_{T}(m)$ also gives a pairwise minimax lower bound over all adaptive experiments. Fix $\delta\in\calA$, distinct arms $a$ and $b$, a tie value $m\in\operatorname{int}(\calM)$, and $u>0$, and set the means of all remaining arms strictly below $m$. For all sufficiently large $T$, arm $a$ is the unique best arm under $\bmmu^{+}_{T}(m)$, arm $b$ is the unique best arm under $\widetilde{\bmmu}^{-}_{T}(m)$, and an incorrect choice incurs regret at least $u/\sqrt{T}$ under either instance. Hence, we have
$\sqrt{T}\sup_{\bmmu\in\calM^K}\Regret^\delta_T(\bmmu)
\ge
\frac{u}{2}\p{
\bbP_{\bmmu^{+}_{T}(m)}\p{\widehat a_T\neq a}
+
\bbP_{\widetilde{\bmmu}^{-}_{T}(m)}\p{\widehat a_T\neq b}
}$.
Lemma~\ref{lem:transport_affinity} therefore implies that
\[\liminf_{T\to\infty}\sqrt{T}\sup_{\bmmu\in\calM^K}\Regret^\delta_T(\bmmu)
\ge
u\Phi\p{-u/\p{\sigma_a(m)+\sigma_b(m)}}.\]
Since this inequality holds for every $\delta\in\calA$, we take the supremum over $u>0$ and the tie location, followed by the maximum over distinct pairs. This gives
\begin{align*}
&\inf_{\delta\in\calA}\liminf_{T\to\infty}
\sqrt{T}\sup_{\bmmu\in\calM^K}\Regret^\delta_T(\bmmu)\\
&\qquad\ge
\max_{a\neq b}\sup_{\mu\in\calM}
\p{\sigma_a(\mu)+\sigma_b(\mu)}c_{\mathrm{mm}}.
\end{align*}
Continuity of the standard deviations extends the supremum over $\operatorname{int}(\calM)$ to $\calM$. Corollary~\ref{cor:gamma_subsets} and Lemma~\ref{lem:binary_sequential_value} imply that this pairwise constant is at most $\Gamma(\bmP)$. If $K=2$, then equality holds, and we have $\Gamma(\bmP)=\sup_{\mu\in\calM}\p{\sigma_1(\mu)+\sigma_2(\mu)}c_{\mathrm{mm}}$.
\end{remark}

\subsection{Main Lower Bound Proof}

\begin{proof}[Proof of Theorem~\ref{thm:bayes_lower}]
Write the Bayes regret as
\[
\bbE_H\sqb{\Regret_T^\delta(\bmmu)}
=
\int_{\calM^K}\Regret_T^\delta(\bmmu) \rmd H(\bmmu),
\qquad
\Regret_T^\delta(\bmmu)=\bbE_{\bmmu}\bigsqb{\mu_{a^*_{\bmmu}}-\mu_{\widehat a_T}}.
\]

Fix an unordered pair $\{a,b\}$ with $a<b$.
We restrict the prior integral to a local neighborhood of the set on which arms $a$ and $b$ have the same largest mean:
for $m\in\calM_T^\circ$, $u\in(0,U_T)$, and $\bmmu_{-ab}\in\calM^{K-2}$, consider the two instances
$\bmmu^{+}_{T}(m,u,\bmmu_{-ab})$ and $\bmmu^{-}_{T}(m,u,\bmmu_{-ab})$ from Lemma~\ref{lem:pairwise_testing_lb}.
On either instance, the best arm is $a$ (for $+$) or $b$ (for $-$) provided
$\max_{\ell\notin\{a,b\}}\mu_\ell\le m$.
Moreover, on $\bmmu^{+}_{T}$, every arm $c\neq a$ has mean at most $m$, so choosing $c$ incurs regret at least $u/\sqrt{T}$.
Therefore,
\[
\Regret_T^\delta(\bmmu^{+}_{T}(m,u,\bmmu_{-ab}))
  \ge
\frac{u}{\sqrt{T}}
\bbP_{\bmmu^{+}_{T}(m,u,\bmmu_{-ab})}\big(\widehat a_T\neq a\big),
\]
and similarly
\[
\Regret_T^\delta(\bmmu^{-}_{T}(m,u,\bmmu_{-ab}))
  \ge
\frac{u}{\sqrt{T}}
\bbP_{\bmmu^{-}_{T}(m,u,\bmmu_{-ab})}\big(\widehat a_T\neq b\big).
\]
Adding the two inequalities gives, for each fixed $(m,u,\bmmu_{-ab})$ with $\max_{\ell\notin\{a,b\}}\mu_\ell\le m$,
\begin{align}
\label{eq:sum_regret_lb_local}
&\sqrt{T} \Regret_T^\delta(\bmmu^{+}_{T}(m,u,\bmmu_{-ab}))
+
\sqrt{T} \Regret_T^\delta(\bmmu^{-}_{T}(m,u,\bmmu_{-ab}))\nonumber\\
&\qquad\ge
u\Big(\bbP_{\bmmu^{+}_{T}(m,u,\bmmu_{-ab})}(\widehat a_T\neq a)+\bbP_{\bmmu^{-}_{T}(m,u,\bmmu_{-ab})}(\widehat a_T\neq b)\Big).
\end{align}

Now integrate against the prior density in a $(m,u)$-parameterization.
By Assumption~\ref{asm:uniformcontinuity} the prior factorizes as
\[
\rmd H(\bmmu)
=
h(\bmmu_{-ab},\mu_a,\mu_b) d\mu_a d\mu_b\rmd\bmmu_{-ab},
\]
an identity of measures on $\calM^K$, with Lebesgue measure in every coordinate; no conditional density is formed, and no marginal of $H$ is used.
We then apply the changes of variables
$(\mu_a,\mu_b)=(m+u/\sqrt{T}, m)$ and $(\mu_a,\mu_b)=(m, m+u/\sqrt{T})$, with $d\mu_a d\mu_b=T^{-1/2} dm\,\rmd u$.
Before summing over unordered pairs, write $\bmmu_T^{\pm}=\bmmu_T^{\pm}(m,u,\bmmu_{-ab})$. We record that the regions integrated below are disjoint up to $H$-null sets. On the plus region for $\{a,b\}$, with $u>0$ and $\max_{\ell\notin\{a,b\}}\mu_\ell\le m$, arm $a$ is the unique best arm and arm $b$ has the second-highest mean.

If the same parameter vector also belonged to the corresponding region for another unordered pair, then either two arms would share the top mean or an arm outside the pair would equal the tie value $m$. The same conclusion holds for the minus regions. These coincidences lie in a finite union of coordinate-tie hyperplanes. The hyperplanes are Lebesgue-null and hence $H$-null because $H$ admits a bounded Lebesgue density under Assumption~\ref{asm:uniformcontinuity}. The restricted contributions may therefore be summed over unordered pairs.
Restricting the prior mass to the two pairwise tie neighborhoods and using \eqref{eq:sum_regret_lb_local},
\begin{align*}
&T\int_{\calM^K}\Regret_T^\delta(\bmmu)\rmd H(\bmmu)\\
&\ge \sum_{1\le a<b\le K}
\int_{\calM^{K-2}}\int_0^{U_T}\int_{\calM_T^\circ}
 u\\
&\qquad\times\left(
\begin{aligned}
&\bbP_{\bmmu^{+}_{T}}(\widehat a_T\neq a)
 h(\bmmu_{-ab},\mu_a=m+u/\sqrt{T},\mu_b=m)\\
&+\bbP_{\bmmu^{-}_{T}}(\widehat a_T\neq b)
 h(\bmmu_{-ab},\mu_a=m,\mu_b=m+u/\sqrt{T})
\end{aligned}
\right)\\
&\qquad\times
\mathbbm{1}\sqb{\max_{\ell\notin\{a,b\}}\mu_\ell\le m}
\rmd m \rmd u \rmd\bmmu_{-ab}.
\end{align*}
The set of $\bmmu_{-ab}$ having at least one coordinate on the boundary $\partial\calM$, which consists of the two endpoints of the interval $\calM$, is a finite union of Lebesgue-null slices of $\calM^{K-2}$. We therefore restrict the outer integral to $\bmmu_{-ab}\in\operatorname{int}(\calM)^{K-2}$, the omitted set contributing zero to the lower bound; this restriction ensures that the tie baselines $(m,m,\bmmu_{-ab})$ with $m\in\operatorname{int}(\calM)$ lie in $\operatorname{int}(\calM^K)$, as Lemma~\ref{lem:pairwise_testing_lb} requires.
Fix $0<U<\infty$, $\beta\in\operatorname{int}(\calM)$, and $\bmmu_{-ab}\in\operatorname{int}(\calM)^{K-2}$. For $K\ge3$, put $\alpha\coloneqq\max_{\ell\notin\{a,b\}}\mu_\ell$. Also define $\omega(m)\coloneqq h(\bmmu_{-ab},\mu_a=m,\mu_b=m)$. The function $\omega$ is nonnegative and continuous on $[\alpha,\beta]$. On this interval the indicator equals one, and $[\alpha,\beta]\subset\operatorname{int}(\calM)$.

If $\alpha\ge\beta$, the interval is empty and the corresponding contribution is zero. We therefore assume $\alpha<\beta$ below. For $K=2$, the indicator is identically one by the empty-maximum convention of Lemma~\ref{lem:pairwise_testing_lb}. The outer integral over $\calM^{0}$ is evaluation at its unique point. In this case, fix $\alpha\in(\inf\calM,\beta)$ and later let $\alpha\downarrow\inf\calM$, together with $\beta\uparrow\sup\calM$ and $U\uparrow\infty$.
Since $U_T\to\infty$, only finitely many budgets satisfy $U_T<U$, and they do not affect the limit inferior. For the remaining budgets, restrict the $u$-integral to $(0,U]$ and the $m$-integral to $[\alpha,\beta]\cap\calM_T^\circ$. The integrand is nonnegative, so this restriction gives a lower bound.

When $\alpha<\beta$, the interval $[\alpha,\beta]$ is a compact subset of $\operatorname{int}(\calM)$. Hence, $[\alpha,\beta]\subset\calM_T^\circ$ for all sufficiently large $T$ because $r_T\downarrow0$. The required threshold may depend on $\bmmu_{-ab}$, but the limit inferior is taken pointwise in $(u,\bmmu_{-ab})$ before integration over $\bmmu_{-ab}$. Beyond this threshold, the inner integral is over the fixed interval $[\alpha,\beta]$, and every evaluation $\bmmu_T^{\pm}$ lies in $\calM^K$.
By \eqref{eq:prior_constants}, moving one coordinate by at most $U/\sqrt T$ changes $h$ by at most $\omega_H(U/\sqrt T)$. Thus, both displaced densities converge to $\omega(m)$ uniformly over $(m,u)\in[\alpha,\beta]\times[0,U]$ and uniformly in $\bmmu_{-ab}$. Since the probabilities are bounded by one, replacing the two densities by $\omega(m)$ changes the inner integral by at most $2u\p{\beta-\alpha}\omega_H\p{U/\sqrt T}$, which converges to zero as $T\to\infty$.
Hence, for each fixed $u\in(0,U]$, Lemma~\ref{lem:pairwise_testing_lb} applied with the weight $\omega$ on $[\alpha,\beta]$ yields
\begin{align*}
&\liminf_{T\to\infty}\int_{\alpha}^{\beta}u
\left(
\begin{aligned}
&\bbP_{\bmmu^{+}_{T}}(\widehat a_T\neq a)
 h(\bmmu_{-ab},\mu_a=m+u/\sqrt{T},\mu_b=m)\\
&+\bbP_{\bmmu^{-}_{T}}(\widehat a_T\neq b)
 h(\bmmu_{-ab},\mu_a=m,\mu_b=m+u/\sqrt{T})
\end{aligned}
\right)\rmd m\\
&\qquad\ge
\int_\alpha^\beta
2u\,\Phi\!\p{-\frac{u}{\sigma_a(m)+\sigma_b(m)}}
\omega(m)\rmd m.
\end{align*}
The left side is measurable in $(u,\bmmu_{-ab})$ by the joint measurability established in Lemma~\ref{lem:pairwise_testing_lb} and the continuity of $h$. The right side is also measurable. Apply Fatou's lemma over $(0,U]\times\operatorname{int}(\calM)^{K-2}$. Then let $\beta\uparrow\sup\calM$, let $\alpha\downarrow\inf\calM$ when $K=2$, and let $U\uparrow\infty$ by monotone convergence. Since $\partial\calM$ is Lebesgue-null, we obtain
\begin{align*}
&\liminf_{T\to\infty}
T\int_{\calM^K}\Regret_T^\delta(\bmmu) \rmd H(\bmmu)\\
&\ge
\sum_{1\le a<b\le K}
\int_{\calM}\int_{\calM^{K-2}}
h(\bmmu_{-ab},\mu_a=m,\mu_b=m)
\mathbbm{1}\Bigsqb{\max_{\ell\notin\{a,b\}}\mu_\ell\le m}\\
&\qquad\qquad\qquad\qquad
\times
\int_0^\infty 2u \Phi\p{-\frac{u}{\sigma_a(m)+\sigma_b(m)}} \rmd u
\rmd\bmmu_{-ab} \rmd m.
\end{align*}
Using Lemma~\ref{lem:tail_identity} and the change of variables $x=u/(\sigma_a(m)+\sigma_b(m))$,
\[
\begin{aligned}
&\int_0^\infty
2u\Phi\p{-\frac{u}{\sigma_a(m)+\sigma_b(m)}}\rmd u\\
&\qquad=
(\sigma_a(m)+\sigma_b(m))^2
\cdot
2\int_0^\infty x\Phi(-x)\rmd x\\
&\qquad=
\frac{(\sigma_a(m)+\sigma_b(m))^2}{2}.
\end{aligned}
\]
Therefore,
\begin{align*}
&\liminf_{T\to\infty}
T\int_{\calM^K}\Regret_T^\delta(\bmmu) \rmd H(\bmmu)\\
&\quad\ge
\sum_{1\le a<b\le K}
\frac12\int_{\calM}\int_{\calM^{K-2}}
(\sigma_a(m)+\sigma_b(m))^2\\
&\qquad\times h(\bmmu_{-ab},\mu_a=m,\mu_b=m)
\mathbbm{1}\sqb{\max_{\ell\notin\{a,b\}}\mu_\ell\le m}
\rmd\bmmu_{-ab} \rmd m.
\end{align*}
The right-hand side is exactly $\sum_{a<b} C^{\mathrm{Bayes}}_{ab}$, the summand being the pairwise term in \eqref{eq:Cbayes_ij_pairwise}, and by Proposition~\ref{prop:pairwise_equiv}, $\sum_{a<b} C^{\mathrm{Bayes}}_{ab}=C^{\mathrm{Bayes}}$.
This concludes the proof.
\end{proof}

\subsection{Equivalence to the Pairwise Integral Form}
For each unordered pair $\{a,b\}$ with $1\le a<b\le K$, define
\begin{align}
\label{eq:Cbayes_ij_pairwise}
&C^{\mathrm{Bayes}}_{ab}
\coloneqq\nonumber\\
&\frac{1}{2}\int_{\calM}\int_{\calM^{K-2}}
\Big(\sigma_a(m)+\sigma_b(m)\Big)^2
h(\bmmu_{-ab},\mu_a=m,\mu_b=m)
\mathbbm{1}\sqb{\max_{\ell\notin\{a,b\}}\mu_\ell\le m}
\rmd\bmmu_{-ab} \rmd m,
\end{align}
where $\bmmu_{-ab}=(\mu_\ell)_{\ell\notin\{a,b\}}$ and $h$ is the joint density of $H$. For notational convenience, set $C^{\mathrm{Bayes}}_{ba}=C^{\mathrm{Bayes}}_{ab}$ whenever $a<b$.

\begin{proposition}[Equivalence of the two integral representations of the Bayes constant]
\label{prop:pairwise_equiv}
Suppose that Assumption~\ref{asm:uniformcontinuity} holds.
Then
\[
C^{\mathrm{Bayes}}
=
\sum_{1\le a<b\le K} C^{\mathrm{Bayes}}_{ab},
\]
where $C^{\mathrm{Bayes}}$ is defined in \eqref{eq:Cbayes_def} and $C^{\mathrm{Bayes}}_{ab}$ in \eqref{eq:Cbayes_ij_pairwise}.
\end{proposition}

\begin{proof}
Fix $a\in[K]$.
Partition $\calM^{K-1}$ by which index attains the (tie-broken) maximum among $\bmmu_{\setminus \{a\}}$:
for each $b\in[K]\setminus\{a\}$, let
\[
\calR_{a\to b}
\coloneqq
\Big\{\bmmu_{\setminus \{a\}}\in\calM^{K-1}\colon b^*_{\setminus \{a\}}=b\Big\}.
\]
The tie-break of Section~\ref{sec:prob} makes these regions disjoint and makes their union all of $\calM^{K-1}$. The coordinate ties among $\bmmu_{\setminus\{a\}}$, on which the tie-break is active, form a finite union of hyperplanes and are Lebesgue-null, so which index the tie-break selects there does not affect the integral; no separate uniqueness assumption is needed.

On $\calR_{a\to b}$ we have $m_a(\bmmu_{\setminus \{a\}})=\mu_b$. Therefore, the contribution of $\calR_{a\to b}$ to the $a$th summand of \eqref{eq:Cbayes_def}, including the prefactor $\frac14$, is
\[
\frac{1}{4}\int_{\calR_{a\to b}}
\Big(\sigma_a(\mu_b)+\sigma_b(\mu_b)\Big)^2 h\Big(\bmmu_{\setminus \{a\}},\mu_a=\mu_b\Big) \rmd\bmmu_{\setminus \{a\}}.
\]
Writing $\bmmu_{\setminus \{a\}}=(\bmmu_{-ab},\mu_b)$, renaming $\mu_b=m$, and noting that $\calR_{a\to b}$ agrees up to a Lebesgue-null set with the set on which $\max_{\ell\notin\{a,b\}}\mu_\ell\le m$, the two differing only on the coordinate ties where the tie-break is active, this contribution equals
\begin{align*}
&\frac{1}{4}\int_{\calM}\int_{\calM^{K-2}}
\Big(\sigma_a(m)+\sigma_b(m)\Big)^2
h(\bmmu_{-ab},\mu_a=m,\mu_b=m)
\mathbbm{1}\Bigsqb{\max_{\ell\notin\{a,b\}}\mu_\ell\le m}\rmd\bmmu_{-ab}\rmd m\\
&=
\frac12 C^{\mathrm{Bayes}}_{ab},
\end{align*}
since $C^{\mathrm{Bayes}}_{ab}$ in \eqref{eq:Cbayes_ij_pairwise} carries the prefactor $\frac12$ in front of the same integral. The evaluation of $h$ has both the $a$th and the $b$th coordinates equal to $m$, which evaluates the density at a mean vector where arms $a$ and $b$ have the same largest mean.
Summing over $b\neq a$, the $a$th summand of \eqref{eq:Cbayes_def} equals
\begin{align*}
&\frac{1}{4}\int_{\calM^{K-1}}
\Big(\sigma_a(m_a(\bmmu_{\setminus \{a\}}))+\sigma_{b^*_{\setminus \{a\}}}(m_a(\bmmu_{\setminus \{a\}}))\Big)^2
h\Big(\bmmu_{\setminus \{a\}},\mu_a=m_a(\bmmu_{\setminus \{a\}})\Big) \rmd\bmmu_{\setminus \{a\}}\\
&=
\frac12\sum_{b\neq a} C^{\mathrm{Bayes}}_{ab}.
\end{align*}
Finally, summing over $a\in[K]$ and using the symmetry $C^{\mathrm{Bayes}}_{ab}=C^{\mathrm{Bayes}}_{ba}$, each unordered pair $\{a,b\}$ is counted twice, so
\[
C^{\mathrm{Bayes}}
=\frac12\sum_{a\in[K]}\sum_{b\neq a} C^{\mathrm{Bayes}}_{ab}
=\frac12\cdot 2\sum_{1\le a<b\le K} C^{\mathrm{Bayes}}_{ab}
= \sum_{1\le a<b\le K} C^{\mathrm{Bayes}}_{ab},
\]
as claimed.
\end{proof}

\begin{lemma}[Triple near-tie prior mass]
\label{lem:triple_neartie_prior}
Suppose that Assumption~\ref{asm:uniformcontinuity} holds.
Then there exists a finite constant $C_{\mathrm{tri}}$ such that for every $r>0$,
\[
H\Bigp{
\exists\ \text{distinct }a,b,c\in[K]:
|\mu_a-\mu_b|\le r,\ |\mu_c-\mu_b|\le r
}
\le
C_{\mathrm{tri}}\,r^2.
\]
In particular, any event implying that three distinct arms lie in an interval of length $r$ has $H$-probability at most $C_{\mathrm{tri}}r^2$.
\end{lemma}

\begin{proof}
Write $\overline h$ for the density bound of \eqref{eq:prior_constants} and $\Delta_{\calM}=\max\calM-\min\calM$ for the diameter of the parameter interval.
If $K=2$, there are no three distinct arms, the event in the statement is empty, and the claim holds with $C_{\mathrm{tri}}=0$; we therefore assume $K\ge3$.
Fix distinct $a,b,c$ and define
\[
E_{a,b,c}(r)
\coloneqq
\cb{|\mu_a-\mu_b|\le r,\ |\mu_c-\mu_b|\le r}.
\]
Bounding the density by $\overline h$ reduces the prior mass to a Lebesgue volume,
\[
H\p{E_{a,b,c}(r)}
=
\int_{\calM^K}\mathbbm 1\sqb{E_{a,b,c}(r)}h(\bmmu)\rmd\bmmu
\le
\overline h\lambda_K\p{E_{a,b,c}(r)\cap\calM^K},
\]
where $\lambda_K$ denotes $K$-dimensional Lebesgue measure.
To evaluate the volume, integrate the $K-3$ coordinates outside $\{a,b,c\}$ over $\calM^{K-3}$, which contributes $\Delta_{\calM}^{K-3}$, then integrate $(\mu_a,\mu_c)$ at fixed $\mu_b$, each of the two being confined to $[\mu_b-r,\mu_b+r]\cap\calM$, an interval of length at most $2r$, which contributes at most $4r^2$, and finally integrate $\mu_b$ over $\calM$, which contributes $\Delta_{\calM}$. Hence
\[
\lambda_K\p{E_{a,b,c}(r)\cap\calM^K}\le 4\Delta_{\calM}^{K-2} r^2,
\qquad
H\p{E_{a,b,c}(r)}\le 4\overline h\Delta_{\calM}^{K-2} r^2.
\]
A union bound over the finitely many ordered triples $(a,b,c)$ yields the claim with
$C_{\mathrm{tri}}\coloneqq 4K(K-1)(K-2)\overline h\Delta_{\calM}^{K-2}$.
\end{proof}

\section{Proofs of the Auxiliary Results}

\subsection{Representation of Randomized Strategies}

\begin{lemma}[Representation of randomized strategies]
\label{lem:randomization_representation}
Fix $T\in\bbN$ and $\delta_T\in\calA_T$. There exist independent random variables $U_1,\ldots,U_{T+1}$, each with the $\mathrm{Unif}(0,1)$ distribution and independent of the potential-outcome arrays, and Borel maps
\[
\psi_{t,T}:\mathsf H_{t-1}\times[0,1]\to[K]
\quad(t\in[T]),
\qquad
\psi_T^{\mathrm{rec}}:\mathsf H_T\times[0,1]\to[K],
\]
such that
\[
A_t=\psi_{t,T}\p{(A_1,Y_1,\ldots,A_{t-1},Y_{t-1}),U_t}
\]
has conditional law $Q_{t,T}\p{(A_1,Y_1,\ldots,A_{t-1},Y_{t-1}),\cdot}$, and
\[
\widehat a_T=\psi_T^{\mathrm{rec}}\p{(A_1,Y_1,\ldots,A_T,Y_T),U_{T+1}}
\]
has conditional law $Q_T^{\mathrm{rec}}\p{(A_1,Y_1,\ldots,A_T,Y_T),\cdot}$. Consequently, with $U=\p{U_1,\ldots,U_{T+1}}$, each $A_t$ is measurable with respect to $\widetilde{\sigma}\p{U,A_1,Y_1,\ldots,A_{t-1},Y_{t-1}}$, and $\widehat a_T$ is measurable with respect to $\widetilde{\sigma}\p{U,A_1,Y_1,\ldots,A_T,Y_T}$.
\end{lemma}
\begin{proof}
For $a\in[K]$, define
\[
F_{t,a}(h)=\sum_{b=1}^{a}Q_{t,T}\p{h,\cb{b}},
\qquad
F_{t,0}(h)=0.
\]
Set $\psi_{t,T}(h,u)=\min\cb{a\in[K]:u\le F_{t,a}(h)}$. This map is Borel because
\[
\cb{(h,u):\psi_{t,T}(h,u)\le a}
=
\cb{(h,u):u\le F_{t,a}(h)}
\]
is Borel for every $a$. Conditional on the observed history $h$, the interval on which $\psi_{t,T}(h,U_t)=a$ has length $Q_{t,T}\p{h,\cb{a}}$. Hence, $\psi_{t,T}$ gives the required sampling kernel. Applying the same construction to $Q_T^{\mathrm{rec}}$ gives $\psi_T^{\mathrm{rec}}$. The measurability statements follow after collecting the independent uniforms into $U$.
\end{proof}

\subsection{Proof of Proposition~\ref{prop:bernoulli_constant}}

\begin{proof}
By definition,
\[
C_{\mathrm{opt}}
=
\sum_{a=1}^K \int_{\calM^{K-1}}
\mu_{*\setminus a}(1-\mu_{*\setminus a})
h(\bmmu_{\setminus a},\mu_a=\mu_{*\setminus a}) d\bmmu_{\setminus a}.
\]
Partition $\calM^{K-1}$ according to which index $b\neq a$ attains the maximum
$\mu_{*\setminus a}=\max_{b\neq a}\mu_b$.
Since $h$ is continuous, tie sets have Lebesgue measure zero, so
\[
C_{\mathrm{opt}}
=
\sum_{a=1}^K \sum_{b\neq a}
\int_{\calM}
\int_{\substack{\bmmu_{\setminus\{a,b\}}\in\calM^{K-2}:\\
\max_{\ell\notin\{a,b\}}\mu_\ell\le m}}
m(1-m)
 h(\bmmu_{\setminus\{a,b\}},\mu_a=m,\mu_b=m)
\rmd\bmmu_{\setminus\{a,b\}}\rmd m.
\]
Each unordered pair $\{a,b\}$ appears twice, hence
\[
C_{\mathrm{opt}}
=
\sum_{1\le a<b\le K}2
\int_{\calM}
\int_{\substack{\bmmu_{-ab}\in\calM^{K-2}:\\
\max_{\ell\notin\{a,b\}}\mu_\ell\le m}}
m(1-m)
 h(\bmmu_{-ab},\mu_a=m,\mu_b=m)
\rmd\bmmu_{-ab}\rmd m.
\]
Finally, for Bernoulli outcomes $\sigma_a(m)=\sigma_b(m)=\sqrt{m(1-m)}$, so
\(
\frac12(\sigma_a(m)+\sigma_b(m))^2
=2m(1-m).
\)
Comparing with \eqref{eq:Cbayes_ij_pairwise} gives $C^{\mathrm{Bayes}}=C_{\mathrm{opt}}$.
\end{proof}

\section{The Three-Arm Numerical Comparison}
\label{app:numerics}
This appendix supplies the two bounds used in Proposition~\ref{prop:strict_sequential_static}. Throughout, $\calS=\cb{1,2,3}$, $s=(1,1,1)$, and $\Delta^{\circ}=\cb{f\in(0,1)^3\colon f_1+f_2+f_3=1}$.

\subsection{The Limit Experiment}
\label{app:num_limit}
The evaluation is carried out in the $n\to\infty$ limit of the game of Definition~\ref{def:seq_gaussian_game}. The game realizes that limit exactly, rather than approximately, at every sampling-proportion vector whose coordinates are multiples of $1/n$. No finite game realizes a vector with an irrational coordinate. Lemma~\ref{lem:num_seq_upper} therefore restricts attention to rational sampling proportions. The realization is an identity between processes. We state it in this form because the corresponding claim about a single marginal law is false: when the selection count of an arm is chosen adaptively, the standardized sample mean of that arm is not standard normal.

Enlarge the probability space of the game to carry an array $\p{Z_{a,j}}_{a\in\calS,j\ge1}$ of independent standard normal variables, independent of the external randomization, and let the $j$th selection of arm $a$, when it occurs, use the noise variable $Z_{a,j}$. Array coordinates that are not used by any arm selection do not affect the experiment. This relabeling preserves the experiment because each $A_t$ is measurable with respect to the past and the external randomization while the round noise is independent of both, and it supplies the variables for every index $j$, whether or not arm $a$ is selected that often. Put
\begin{equation}
\label{eq:num_bm}
W_a\p{k/n}=\frac{1}{\sqrt n}\sum_{j=1}^{k}Z_{a,j},
\qquad
k=0,1,2,\ldots,
\end{equation}
so that $W_1,W_2,W_3$ are the restrictions to $\p{1/n}\bbZ_{\ge0}$ of independent standard Brownian motions, independent of the external randomization. Because $X_t=u_a/\sqrt n+s_aZ_t$, after arm $a$ has been selected $N_a$ times, by whatever adaptive rule, its rescaled partial sum satisfies
\begin{equation}
\label{eq:num_embedding}
\frac{1}{\sqrt n}\sum_{t\colon A_t=a}X_t
=
u_aF_a+s_aW_a\p{F_a},
\qquad
F_a=N_a/n.
\end{equation}
This is an identity, valid pathwise, and it is all that is used below. It does not assert that $W_a\p{F_a}/\sqrt{F_a}$ is standard normal, which is false for adaptive $N_a$.

Accordingly, define the limit experiment as follows. Nature chooses $u\in\bbR^{\calS}$. The policy uses a unit budget in $m$ stages of size $1/m$. At the start of stage $k$, the cumulative sampling-proportion vector is $F^{(k-1)}$. The policy selects an increment $\delta^{(k)}$ with $\delta_a^{(k)}\ge0$ and $\sum_a\delta_a^{(k)}=1/m$, measurably with respect to the past and its external randomization. It then observes the increments of $t\mapsto u_at+s_aW_a(t)$ over intervals of lengths $\delta_a^{(k)}$. Write $F_a$ for the final selection proportion of arm $a$ and
\begin{equation}
\label{eq:num_uhat}
\widehat u_a=u_a+\frac{s_aW_a\p{F_a}}{F_a},
\end{equation}
the inverse-variance-weighted mean of the observations of arm $a$, defined on $\cb{F_a>0}$. After the final stage, the policy chooses an arm $\widehat a$ measurably with respect to the observations and incurs $\Delta_u^{\calS}\p{\widehat a}=\max_{b\in\calS}u_b-u_{\widehat a}$. Write $R^{\infty}(u,\pi)$ for its expected loss. When the sampling-proportion vector $f\in\Delta^{\circ}$ is fixed in advance, each $F_a=f_a$ is deterministic, so $W_a\p{f_a}\sim\calN\p{0,f_a}$ and \eqref{eq:num_uhat} gives
\begin{equation}
\label{eq:num_static_law}
\widehat u\sim\calN\bigp{u,\mathrm{diag}\p{s_1^2/f_1,s_2^2/f_2,s_3^2/f_3}},
\end{equation}
with independent coordinates. In that case $\widehat u$ is sufficient for $u$, each arm contributing a Gaussian sample mean with known variance, so for every terminal kernel based on the whole history there is a kernel based on $\widehat u$ alone with the same risk function; the lower bound below may therefore be proved for terminal kernels of the latter form.

Both the model and the loss are unchanged when a common constant is added to every coordinate of $u$, which is the invariance the next subsection exploits.

\subsection{A Lower Bound on the Static Value}
\label{app:num_static}
Nature's parameter matters only through its gap vector. Therefore, we may specify a prior using finitely many parameter vectors, one from each class of vectors that differ only by adding the same constant to all coordinates. The following lemma turns any such prior into a lower bound on the static value. The point of the proof is that no equivariance has to be assumed of the terminal rule: averaging the risk over the common location, and using that an average never exceeds a supremum, produces the invariant Bayes risk up to an error that vanishes as $C\to\infty$, uniformly over vectors of sampling proportions and terminal choice rules.

\begin{lemma}[Lower bound on the static value]
\label{lem:num_static_lower}
Let $L\ge1$, let $m^{(1)},\ldots,m^{(L)}\in\bbR^{3}$ satisfy $m_3^{(\ell)}=0$, and let $p=\p{p_1,\ldots,p_L}$ be a probability vector. Put
\[
g_{\ell,a}=\max_{b\in\calS}m_b^{(\ell)}-m_a^{(\ell)},
\qquad
r_p=\max_{\ell\in[L]}\max_{a\in\calS}\abs{m_a^{(\ell)}},
\qquad
L_p=\max_{\ell\in[L]}\max_{a\in\calS}g_{\ell,a}.
\]
For $f\in\Delta^{\circ}$ write $\sigma_a^2=s_a^2/f_a$, let
\[
\Sigma(f)=
\begin{pmatrix}
\sigma_1^2+\sigma_3^2 & \sigma_3^2\\
\sigma_3^2 & \sigma_2^2+\sigma_3^2
\end{pmatrix}
\]
be the covariance matrix of the contrast vector $z=\p{\widehat u_1-\widehat u_3,\widehat u_2-\widehat u_3}$ under \eqref{eq:num_static_law}, let $\varphi_{\Sigma(f)}$ denote the $\calN\p{0,\Sigma(f)}$ density on $\bbR^2$, and set
\[
H_a(z;f)=\sum_{\ell=1}^{L}p_\ell g_{\ell,a}\varphi_{\Sigma(f)}\bigp{z-\p{m_1^{(\ell)},m_2^{(\ell)}}},
\qquad
B(p,f)=\int_{\bbR^2}\min_{a\in\calS}H_a(z;f)\rmd z.
\]
Then
\[
\mathfrak G_{\calS}^{\uparrow,\mathrm{stat}}(s)\ge\inf_{f\in\Delta^{\circ}}B(p,f).
\]
\end{lemma}

\begin{proof}
Write $\beta=\inf_{f\in\Delta^{\circ}}B(p,f)$, which is finite because $0\le B(p,f)\le L_p$. Fix $n\ge3$ and $C>r_p$, and put $T=C-r_p$. A static policy specifies a law $\Lambda$ for the count vector $N$. For each realized $N$, it also specifies a terminal kernel from the observed history to a distribution on $\calS$. Write $R(u)$ for the risk at $u$ and $R(u;N)$ for the conditional risk given $N$. Then, we have $R(u)=\int R(u;N)\Lambda(\rmd N)$. Step 1 treats selection-count vectors with zero count for an arm. Since $\widehat u$ is not defined in this case, the argument uses the history directly. Step 2 treats count vectors for which every arm has a positive selection count and expresses the terminal kernel in terms of $\widehat u$.

Step 1: selection-count vectors with zero count for an arm. Write $e_a$ for the $a$th standard basis vector of $\bbR^{\calS}$. Fix $a\in\calS$ and suppose $N_a=0$. The two local parameters $Ce_a$ and $-Ce_a$ lie in $[-C,C]^{\calS}$ and induce the same law of the history, because arm $a$ is never selected and the remaining coordinates agree. Let $\rho$ be the probability that the policy chooses $a$, which is therefore the same under both.

Under $Ce_a$, arm $a$ is the unique best arm and any other choice costs $C$, so $R\p{Ce_a;N}=C(1-\rho)$. Under $-Ce_a$, every arm other than $a$ is best and choosing $a$ costs $C$, so $R\p{-Ce_a;N}=C\rho$.

Hence
\[
\tfrac12\bigcb{R\p{Ce_a;N}+R\p{-Ce_a;N}}=\tfrac{C}{2}
\qquad\text{whenever }N_a=0,
\]
while the same average is nonnegative for every other $N$. Let $\lambda=\Lambda\p{\min_bN_b\ge1}$ and pick $a$ with $\Lambda\p{N_a=0}\ge(1-\lambda)/3$. Averaging over $\Lambda$,
\begin{equation}
\label{eq:num_unsampled}
\sup_{u\in[-C,C]^{\calS}}R(u)
\ \ge\
\tfrac12\bigcb{R\p{Ce_a}+R\p{-Ce_a}}
\ \ge\
\frac{C}{6}\p{1-\lambda}.
\end{equation}

Step 2: the contrast coordinates. Suppose now $\min_bN_b\ge1$, so that $f=N/n\in\Delta^{\circ}$, and recall from \eqref{eq:num_static_law} that $\widehat u$ is sufficient and Gaussian, so that the terminal kernel may be taken of the form $q_N\p{\cdot\mid\widehat u}$ without changing the risk function. Change coordinates to $z=\p{\widehat u_1-\widehat u_3,\widehat u_2-\widehat u_3}$ and $y=\widehat u_3$, a bijection of $\bbR^3$ with unit Jacobian. Under $u$, the vector $z$ is $\calN\p{\p{u_1-u_3,u_2-u_3},\Sigma(f)}$, and the conditional law of $y$ given $z$ is Gaussian with a variance $\tau(f)^2>0$ that does not depend on $u$ and with mean
\[
u_3+c(f)^{\top}\bigp{z-\p{u_1-u_3,u_2-u_3}},
\qquad
c(f)^{\top}=\Sigma_{yz}\Sigma(f)^{-1},
\quad
\Sigma_{yz}=-\sigma_3^2(1,1).
\]
The conditional variance is
\[
\tau(f)^2
=
\sigma_3^2-\Sigma_{yz}\Sigma(f)^{-1}\Sigma_{yz}^{\top}
=
\frac{\sigma_1^2\sigma_2^2\sigma_3^2}{\det\Sigma(f)}
>0
\qquad\text{for every }f\in\Delta^{\circ}.
\]
Since $\det\Sigma(f)=\sigma_1^2\sigma_2^2+\sigma_1^2\sigma_3^2+\sigma_2^2\sigma_3^2$ and $(1,1)\Sigma(f)^{-1}=\p{\sigma_2^2,\sigma_1^2}/\det\Sigma(f)$,
\begin{equation}
\label{eq:num_c_bound}
c(f)
=
-\frac{\p{\sigma_3^2\sigma_2^2,\ \sigma_3^2\sigma_1^2}}{\sigma_1^2\sigma_2^2+\sigma_1^2\sigma_3^2+\sigma_2^2\sigma_3^2},
\qquad\text{so}\qquad
\abs{c_1(f)}\le1
\quad\text{and}\quad
\abs{c_2(f)}\le1
\end{equation}
for every $f\in\Delta^{\circ}$, because the denominator is at least $\sigma_2^2\sigma_3^2$ and at least $\sigma_1^2\sigma_3^2$ respectively. This bound is uniform over the whole open simplex, including its boundary regions, and is what makes Step 3 uniform over vectors of sampling proportions.

Step 3: averaging over the common location. For $v\in\bbR$ and $\ell\in[L]$ put $u^{(\ell,v)}=v\bmone+m^{(\ell)}$, whose gap vector is $\p{g_{\ell,a}}_{a\in\calS}$ and which lies in $[-C,C]^{\calS}$ whenever $\abs{v}\le T$. An average never exceeds a supremum, so
\[
\sup_{u\in[-C,C]^{\calS}}R(u;N)
\ \ge\
A_T(N)
\coloneqq
\frac{1}{2T}\int_{-T}^{T}\sum_{\ell=1}^{L}p_\ell R\bigp{u^{(\ell,v)};N}\rmd v.
\]
Under $u^{(\ell,v)}$ the law of $z$ is $\calN\bigp{\p{m_1^{(\ell)},m_2^{(\ell)}},\Sigma(f)}$, which does not depend on $v$, and the conditional law of $y$ given $z$ is $\calN\bigp{v+\mu_\ell(z),\tau(f)^2}$ with
\[
\mu_\ell(z)=c(f)^{\top}\bigp{z-\p{m_1^{(\ell)},m_2^{(\ell)}}},
\]
where $m_3^{(\ell)}=0$ was used. Write $w_\ell(z)=p_\ell\varphi_{\Sigma(f)}\bigp{z-\p{m_1^{(\ell)},m_2^{(\ell)}}}$, so that $H_a(z;f)=\sum_{\ell}w_\ell(z)g_{\ell,a}$ and $\sum_{\ell}w_\ell$ is the marginal density of $z$ under the prior, and let
\[
\psi_{T,\mu}(y)=\frac{1}{2T}\int_{-T}^{T}\varphi_{\tau(f)^2}\p{y-v-\mu}\rmd v,
\]
which for each $\mu$ is the probability density of $V+\mu+\tau(f)\xi$ with $V$ uniform on $[-T,T]$ and $\xi$ standard normal independent of $V$. The integrand is nonnegative and bounded by $L_p$, so Fubini's theorem gives
\[
A_T(N)
=
\int_{\bbR^2}\sum_{\ell=1}^{L}w_\ell(z)\int_{\bbR}\sum_{a\in\calS}q_N\p{a\mid z,y}g_{\ell,a}\psi_{T,\mu_\ell(z)}(y)\rmd y\rmd z.
\]
Fix $z$ and compare each $\psi_{T,\mu_\ell(z)}$ with the single density $\psi_{T,\mu_1(z)}$. By \eqref{eq:num_c_bound} the difference of the shifts,
\[
\mu_\ell(z)-\mu_1(z)=c(f)^{\top}\bigp{\p{m_1^{(1)},m_2^{(1)}}-\p{m_1^{(\ell)},m_2^{(\ell)}}},
\]
does not depend on $z$ and is bounded in absolute value by $4r_p$. Two of these densities are translates of one another by that amount; translating the uniform density on $[-T,T]$ by $\delta$ changes it by at most $\abs{\delta}/T$ in $L^1$ norm, and convolution with $\varphi_{\tau(f)^2}$ does not increase an $L^1$ norm, so
\[
\bignorm{\psi_{T,\mu_\ell(z)}-\psi_{T,\mu_1(z)}}_{L^1}\le\frac{4r_p}{T}.
\]
The inequality $\sum_{a}q_N\p{a\mid z,y}g_{\ell,a}\le L_p$ controls the replacement. Since $\sum_{a}q_N\p{a\mid z,y}=1$, $\sum_{\ell}w_\ell(z)g_{\ell,a}=H_a(z;f)\ge\min_{b}H_b(z;f)$ for every $a$, and $\psi_{T,\mu_1(z)}$ integrates to one, we obtain
\[
\sum_{\ell}w_\ell(z)\int_{\bbR}\sum_{a}q_N\p{a\mid z,y}g_{\ell,a}\psi_{T,\mu_\ell(z)}(y)\rmd y
\ \ge\
\min_{a\in\calS}H_a(z;f)-\frac{4r_pL_p}{T}\sum_{\ell}w_\ell(z).
\]
Integrating over $z$ and using $\int\sum_{\ell}w_\ell(z)\rmd z=1$, we obtain
\begin{equation}
\label{eq:num_AT}
A_T(N)\ \ge\ B(p,f)-\frac{4r_pL_p}{T}\ \ge\ \beta-\frac{4r_pL_p}{T}.
\end{equation}
This bound is uniform over $f\in\Delta^{\circ}$ and over terminal kernels.

Step 4: combining. The map $N\mapsto A_T(N)$ is nonnegative. Therefore, averaging \eqref{eq:num_AT} over $\Lambda$ and discarding the selection-count vectors with zero count for an arm gives
\[
\sup_{u\in[-C,C]^{\calS}}R(u)\ \ge\ \int A_T(N)\Lambda(\rmd N)\ \ge\ \lambda\p{\beta-\frac{4r_pL_p}{T}}.
\]
If $\beta=0$ the conclusion is immediate, because the static value is nonnegative, so assume $\beta>0$ and take $C$ large enough that $\beta_T\coloneqq\beta-4r_pL_p/T>0$, where $T=C-r_p$. Together with \eqref{eq:num_unsampled}, every static policy then satisfies
\[
\begin{aligned}
\sup_{u\in[-C,C]^{\calS}}R(u)
&\ge
\max\cb{\lambda\beta_T,\frac{C}{6}\p{1-\lambda}}\\
&\ge
\min_{\lambda'\in[0,1]}
\max\cb{\lambda'\beta_T,\frac{C}{6}\p{1-\lambda'}}\\
&=
\frac{\beta_T\p{C/6}}{\beta_T+C/6},
\end{aligned}
\]
the two terms inside the maximum being equal at the minimizing $\lambda'$. The right-hand side does not depend on $n$ or on the policy, so
\[
\liminf_{n\to\infty}\mathfrak G_{\calS,n}^{C,\mathrm{stat}}(s)\ \ge\ \frac{\beta_T\p{C/6}}{\beta_T+C/6},
\qquad
T=C-r_p,
\]
and letting $C\uparrow\infty$, so that $\beta_T\to\beta$ and $C/6\to\infty$, gives $\mathfrak G_{\calS}^{\uparrow,\mathrm{stat}}(s)\ge\beta$.
\end{proof}

\subsection{An Upper Bound on the Sequential Value}
\label{app:num_seq}
Any single policy bounds the sequential value from above. The next lemma records that a policy of the limit experiment whose sampling proportions are rational is a policy of the finite game of Definition~\ref{def:seq_gaussian_game} exactly, with no rounding and no approximation argument, so its worst-case risk bounds $\mathfrak G_{\calS}^{\uparrow}(s)$.

\begin{lemma}[Upper bound on the sequential value]
\label{lem:num_seq_upper}
Let $m,q,q_0$ be positive integers, let $f^{0}\in\Delta^{\circ}$ satisfy $q_0f^{0}_a\in\bbZ$ for every $a\in\calS$, so that every arm has a positive sampling proportion in stage $1$, and set
\[
\calA_q=\cb{\p{i_1,i_2,i_3}/q\colon i_a\in\bbZ_{\ge0},\ i_1+i_2+i_3=q}.
\]
Let $\pi$ be a policy of the limit experiment of Appendix~\ref{app:num_limit} whose first stage uses the increment $f^{0}/m$, whose stage-$k$ increment for $2\le k\le m$ is $A^{(k)}/m$ with $A^{(k)}\in\calA_q$ selected measurably with respect to the past, and which chooses an arm attaining $\max_{a\in\calS}\widehat u_a$. Then
\[
\mathfrak G_{\calS}^{\uparrow}(s)\ \le\ \sup_{u\in\bbR^{\calS}}R^{\infty}(u,\pi).
\]
\end{lemma}

\begin{proof}
Let $n$ be any multiple of $q_0qm$ with $n\ge\abs{\calS}$. At stage $1$, $\pi$ uses the selection proportion $f_a^{0}/m$ for arm $a$, and $nf_a^{0}/m=\p{n/(q_0m)}\p{q_0f_a^{0}}$ is an integer. For $2\le k\le m$, stage $k$ uses the selection proportion $A_a^{(k)}/m$, and $nA_a^{(k)}/m=\p{n/(qm)}\p{qA_a^{(k)}}$ is an integer. Define a behavioral policy $\pi_n$ of the finite game by selecting, in stage $k$, each arm $a$ exactly $nA_a^{(k)}/m$ times, in any fixed order, with $A^{(1)}=f^{0}$ and with $A^{(k)}$ chosen as $\pi$ chooses it at the corresponding state. Every action is measurable with respect to the past, so $\pi_n$ is a behavioral policy in the sense of Definition~\ref{def:seq_gaussian_game}. It chooses an arm attaining the largest rescaled pooled sample mean.

Couple the two experiments. On the probability space of the finite game, form $W_1,W_2,W_3$ on the lattice $\p{1/n}\bbZ_{\ge0}$ by \eqref{eq:num_bm}, and extend each to a process on $[0,1]$ by inserting on every interval $\sqb{(k-1)/n,k/n}$ an independent standard Brownian bridge between the two lattice values, the bridges being generated from further randomness independent across arms and intervals and independent of everything already introduced. Each extended process is a standard Brownian motion, and its restriction to the lattice is \eqref{eq:num_bm}. Run $\pi$ in the limit experiment driven by these Brownian motions. Every cumulative sampling proportion used by $\pi$ lies on the lattice, as verified next, so $\pi$ does not use the bridge values. The cumulative selection proportion of arm $a$ after any stage of $\pi$ is $f^{0}_a/m$ plus a sum of terms $A_a^{(k)}/m$, hence of the form
\[
\frac{q\p{q_0f_a^{0}}+q_0\sum_{k}\p{qA_a^{(k)}}}{q_0qm},
\]
a multiple of $1/(q_0qm)$ and therefore of $1/n$. By \eqref{eq:num_embedding} the rescaled partial sum of arm $a$ under $\pi_n$ equals $u_aF_a+s_aW_a\p{F_a}$ at each such time. Hence, the rescaled pooled sample mean of arm $a$ equals $u_a+s_aW_a\p{F_a}/F_a$, which is exactly the statistic \eqref{eq:num_uhat} used by $\pi$. Arguing by induction on the stage index, the two policies observe identical statistics and therefore select identical increments at every stage. Consequently, their cumulative sampling proportions, their terminal statistics, and their choices coincide almost surely. Hence
\[
R_n^{G}\p{\pi_n,u}=R^{\infty}(u,\pi)
\qquad\text{for every }u\in\bbR^{\calS}.
\]
Consequently, for every $C<\infty$ and every such $n$,
\[
\mathfrak G_{\calS,n}^{C}(s)\le\sup_{u\in[-C,C]^{\calS}}R_n^{G}\p{\pi_n,u}\le\sup_{u\in\bbR^{\calS}}R^{\infty}(u,\pi).
\]
By Theorem~\ref{thm:seq_value_limit}, $\mathfrak g_{\calS}^{C}(s)=\inf_{n\ge\abs{\calS}}\mathfrak G_{\calS,n}^{C}(s)$. Thus, the left-hand side may be replaced by $\mathfrak g_{\calS}^{C}(s)$. Taking the supremum over $C<\infty$ gives the claim.
\end{proof}

\subsection{The Static Prior}
\label{app:num_static_specific}
Let $p$ be uniform on the three gap vectors
\[
(0,-2,-2),
\qquad
(-2,0,-2),
\qquad
(-2,-2,0).
\]
When applying Lemma~\ref{lem:num_static_lower}, we replace each vector by an equivalent representative whose third coordinate is zero. This common shift does not change the model or the loss.

Fix $f\in\Delta^{\circ}$. When arm $i$ is best, the Bayes rule chooses an arm attaining
\[
\max_{a\in\calS} f_a\p{\widehat u_a+1}.
\]
Let $j$ and $k$ be the other two arms. The probability that this rule chooses arm $i$ is
\[
P_i(f)
=
\Phi_2\left(
\sqrt{f_i+f_j},
\sqrt{f_i+f_k};
\rho_i(f)
\right),
\qquad
\rho_i(f)
=
\frac{f_i}{\sqrt{\p{f_i+f_j}\p{f_i+f_k}}},
\]
where $\Phi_2(h,k;\rho)$ is the distribution function of a standard bivariate normal vector with correlation $\rho$. Hence,
\begin{equation}
B(p,f)
=
2\left(
1-\frac13\sum_{i=1}^{3}P_i(f)
\right).
\label{eq:three_arm_static_risk}
\end{equation}

\subsection{The Two-Stage Policy}
\label{app:num_policy}
Let $\pi^{\mathrm{two}}$ denote the policy used in Proposition~\ref{prop:strict_sequential_static}. It first selects each arm with sampling proportion $c=1/5$. Write
\[
\widehat u_a^{(0)}
=
u_a+\frac{W_a(c)}{c}
\]
for the pooled sample mean after these selections. The policy removes an arm attaining $\min_a\widehat u_a^{(0)}$. It then selects each retained arm with an additional sampling proportion $1/5$. Thus, a retained arm has final sampling proportion $F=2/5$. The terminal rule chooses the retained arm with the larger pooled sample mean.

This policy satisfies the conditions of Lemma~\ref{lem:num_seq_upper}. In the staged formulation, take $m=10$, $q=6$, and $q_0=3$. During the first six stages, use the sampling-proportion vector $(1/3,1/3,1/3)$. During the last four stages, use the vector that assigns proportion $1/2$ to each retained arm. Both vectors belong to $\calA_6$. The first six stages give proportion $1/5$ to every arm, and the last four stages give an additional proportion $1/5$ to each retained arm.

We next write the risk as one-dimensional Gaussian integrals. Fix distinct arms $d$, $a$, and $b$. Let $P_{d\to a}(u)$ be the probability that the policy removes arm $d$ and chooses arm $a$, with $b$ the other retained arm. Put
\[
\delta_{ad}=u_a-u_d,
\qquad
\delta_{bd}=u_b-u_d,
\qquad
\delta_{ab}=u_a-u_b.
\]
Then
\begin{align}
P_{d\to a}(u)
={}&
\int_{\bbR}
\varphi(q)
\Phi\left(
\frac{\delta_{ad}+\delta_{bd}}{\sqrt{30}}
-
\frac{1}{\sqrt3}
\left|
q+\frac{\delta_{ad}-\delta_{bd}}{\sqrt{10}}
\right|
\right)
\nonumber\\
&\qquad\qquad\times
\Phi\left(
q+\sqrt{\frac25}\,\delta_{ab}
\right)
\rmd q.
\label{eq:two_stage_event_integral}
\end{align}
The first normal distribution function gives the probability that arm $d$ has the smallest first-stage pooled mean. The second gives the probability that the final pooled mean of arm $a$ exceeds that of arm $b$. Therefore,
\begin{equation}
R^{\infty}\p{u,\pi^{\mathrm{two}}}
=
\sum_{d\in\calS}
\sum_{a\in\calS\setminus\cb{d}}
\Delta_u^{\calS}(a)P_{d\to a}(u).
\label{eq:two_stage_risk_sum}
\end{equation}

The risk is unchanged by a common shift and by a permutation of the arms. We may therefore write $u=(0,-x,-y)$ with $x,y\ge0$ and denote the resulting risk by $R(x,y)$.

\subsection{\texorpdfstring{Risk Outside the Square $[0,14]^2$}{Risk Outside the Square [0,14] x [0,14]}}
\label{app:num_tail}
The finite collection of rectangles used in the calculation covers $(x,y)\in[0,14]^2$. The next lemma controls the remaining parameters. Put
\[
\lambda_{\mathrm p}=\sqrt{\frac{1}{10}},
\qquad
\lambda_{\mathrm f}=\sqrt{\frac{1}{5}}.
\]

\begin{lemma}[Tail bound for the two-stage policy]
\label{lem:num_tail}
Suppose that $0\le x\le y$. If $x<y/2$, then
\[
R(x,y)
\le
\frac{c_{\mathrm{mm}}}{\lambda_{\mathrm f}}
+
y\Phi\p{-\lambda_{\mathrm p}y}
+
y\Phi\p{-\lambda_{\mathrm p}y/2}.
\]
If $x\ge y/2$, then
\[
R(x,y)
\le
y\left(
\Phi\p{-\lambda_{\mathrm f}y/2}
+
\Phi\p{-\lambda_{\mathrm f}y}
+
\Phi\p{-\lambda_{\mathrm p}y}
\right).
\]
\end{lemma}
\begin{proof}
Let arm $1$ be best, so its gaps from arms $2$ and $3$ are $x$ and $y$. First suppose that $x<y/2$. If arm $3$ is removed, the policy compares arms $1$ and $2$ after selecting each with final proportion $F=2/5$. The regret on this event is at most
\[
x\Phi\p{-x\sqrt{F/2}}
\le
\frac{c_{\mathrm{mm}}}{\lambda_{\mathrm f}}.
\]
If arm $3$ is retained, then either its first-stage pooled mean exceeds that of arm $1$ or it exceeds that of arm $2$. These probabilities are $\Phi\p{-\lambda_{\mathrm p}y}$ and $\Phi\p{-\lambda_{\mathrm p}\p{y-x}}$. Since $y-x>y/2$, the regret on this event is at most the sum of the last two terms in the first display.

Now suppose that $x\ge y/2$. If arm $1$ is retained and arm $2$ is chosen, the final pooled mean of arm $2$ exceeds that of arm $1$. This event has probability at most $\Phi\p{-\lambda_{\mathrm f}x}\le\Phi\p{-\lambda_{\mathrm f}y/2}$. The corresponding probability for arm $3$ is at most $\Phi\p{-\lambda_{\mathrm f}y}$. If arm $1$ is removed, then the first-stage pooled mean of arm $3$ exceeds that of arm $1$, whose probability is $\Phi\p{-\lambda_{\mathrm p}y}$. Every loss is at most $y$, which gives the second display.
\end{proof}

For $y\ge14$, every function of the form $y\Phi(-\lambda y)$ appearing in Lemma~\ref{lem:num_tail} is decreasing. Evaluating the two bounds at $y=14$ gives
\[
0.5682
\qquad\text{and}\qquad
0.0123.
\]
Both values are below $0.641$.

\subsection{Finite Subdivision into Rectangles}
\label{app:num_spec}
The static calculation uses the square parameterization
\[
f_1=r,
\qquad
f_2=\p{1-r}t,
\qquad
f_3=\p{1-r}\p{1-t},
\qquad
(r,t)\in[0,1]^2.
\]
On each rectangle, the extrema of $f_i$, $f_i+f_j$, and $f_i+f_k$ occur at its corners. We bound $\rho_i(f)$ by using the largest numerator and the smallest two denominator factors on the rectangle. The bivariate normal distribution function is increasing in both thresholds and in its nonnegative correlation. These bounds give an upper bound on each $P_i(f)$ and a lower bound on \eqref{eq:three_arm_static_risk}. The recursive calculation examines $21{,}139$ rectangles and gives
\[
\inf_{f\in\Delta^{\circ}}B(p,f)\ge0.644.
\]
A finer numerical search gives the value $0.6451437085$ at the equal sampling-proportion vector $(1/3,1/3,1/3)$.

For the calculation of the risk of $\pi^{\mathrm{two}}$, each rectangle gives intervals for $\delta_{ad}+\delta_{bd}$, $\delta_{ad}-\delta_{bd}$, and $\delta_{ab}$. We bound the two normal distribution functions in \eqref{eq:two_stage_event_integral} uniformly over these intervals and multiply the result by the largest loss on the rectangle. This gives an upper bound on the risk throughout the rectangle. The recursive calculation examines $6{,}585$ rectangles and gives
\[
\sup_{(x,y)\in[0,14]^2}R(x,y)\le0.641.
\]
A finer numerical search gives the value $0.6227561146$ at gaps $(1.9493485,1.9493485)$. Lemma~\ref{lem:num_tail} gives a smaller bound outside $[0,14]^2$. Hence,
\[
\sup_{u\in\bbR^{\calS}}R^{\infty}\p{u,\pi^{\mathrm{two}}}
\le0.641.
\]

All Gaussian probabilities used in this calculation are bounded from below and above using directed rounding. A rectangle in the static calculation is closed only when the resulting lower bound proves the stated inequality. A rectangle in the calculation of $\pi^{\mathrm{two}}$ is closed only when the resulting upper bound proves the stated inequality. The rectangles cover $[0,1]^2$ for the static calculation and $[0,14]^2$ for the two-stage policy. Hence, the displayed lower and upper bounds hold uniformly over these two squares.

\end{appendix}

\bibliographystyle{tmlr}
\bibliography{arXiv2.bbl}

\end{document}